\documentclass[12pt]{article}

\usepackage{amsthm}
\theoremstyle{plain}

\theoremstyle{definition}

\theoremstyle{proposition}

\theoremstyle{lemma}

\theoremstyle{remark}

\usepackage{amsmath}
\usepackage{amssymb}
\usepackage{cite}
\usepackage[pdftex]{graphicx}

\makeatletter
 
  \@addtoreset{equation}{section}
 \makeatother

\newcommand {\tr}{\mbox{tr}}

\begin{document}
\setlength{\oddsidemargin}{0cm}
\setlength{\baselineskip}{7mm}

\begin{titlepage}

~~\\

\vspace*{0cm}
    \begin{large}
       \begin{center}
         {String Geometry and Non-perturbative Formulation of String Theory}
       \end{center}
    \end{large}
\vspace{1cm}

\begin{center}
           Matsuo S{\sc ato}\footnote
           {
e-mail address : msato@hirosaki-u.ac.jp}\\
      \vspace{1cm}
       
         {\it Graduate School of Science and Technology, Hirosaki University, Bunkyo-cho 3, Hirosaki, Aomori 036-8561, Japan}

\end{center}

\hspace{5cm}

\begin{abstract}
\noindent
We define string geometry: spaces of superstrings including the interactions, their topologies, charts, and metrics. Trajectories in asymptotic processes on a space of strings reproduce the right moduli space of the super Riemann surfaces in a target manifold. Based on the string geometry, we define Einstein-Hilbert action coupled with gauge fields, and formulate superstring theory non-perturbatively by summing over metrics and the gauge fields on the spaces of strings. This theory does not depend on backgrounds. The theory has a supersymmetry as a part of the diffeomorphisms symmetry on the superstring manifolds. We derive the all-order  perturbative scattering amplitudes that possess the super moduli in type IIA, type IIB and SO(32) type I superstring theories from the single theory, by considering fluctuations around fixed backgrounds representing type IIA, type IIB and SO(32) type I perturbative vacua, respectively. The theory predicts that we can see a string if we microscopically observe not only a particle but also a point in the space-time. That is, this theory unifies particles and the space-time.

\end{abstract}

\vfill
\end{titlepage}
\vfil\eject

\setcounter{footnote}{0}

\tableofcontents

\section{Introduction}
\setcounter{equation}{0}

In the T-duality and its generalization, the mirror symmetry, there is a coincidence between geometric invariants of two different manifolds. It is thought that the reason for this is that the spaces observed by the strings are the same although they are in the different target manifolds. Therefore, the space observed by the strings, which is invariant under the T-duality and mirror transformations, will be a geometric principle of string theory. A moduli space in a target manifold, which is a collection of on-shell embedding functions of the Riemann surfaces $X^{\mu}(\sigma, \tau)$, is invariant under the T-duality transformations. Actually, the T-duality rule $\partial_a X^{\mu}(\sigma, \tau)=i \epsilon_{ab} \partial^b X^{'\mu}(\sigma, \tau)$ gives an one-to-one correspondence between the on-shell embedding functions of the Riemann surfaces $X^{\mu}(\sigma, \tau)$ and $X^{' \mu}(\sigma, \tau)$. Moreover, the Riemann surfaces in the target manifold can be generated by trajectories in a space of strings. Therefore, a space of strings will be the geometric principle\footnote{Recently, in the homological mirror symmetry \cite{HMS}, it is shown in \cite{MSato} that the moduli space of the pseudo holomorphic curves in the A-model on a symplectic torus is homeomorphic to a moduli space of Feynman diagrams in the configuration space of the morphisms in the B-model on the corresponding elliptic curve. Therefore, a dynamical and non-perturbative generalization of the moduli space of the pseudo holomorphic curves will be the geometric principle of string theory. Here we discuss this generalization. First, a moduli space of pseudo holomorphic curves is defined even in closed string theory in \cite{Gromov}. Moreover, the moduli space can be defined by restricting a moduli space of curves, which is not necessarily holomorphic, to the holomorphic sector \cite{Polyfold1}. Because this is a restriction to the topological string theory, the moduli space of curves is the dynamical generalization. Furthermore, the curves can be generated by trajectories in a space of strings. Therefore, a space of strings will be the dynamical and non-perturbative generalization, that is the geometric principle.}.

Furthermore, string theory as a quantum gravity also suggests that a space of strings will be a geometric principle of string theory as follows.  It has not succeeded to obtain ordinary relativistic quantum gravity that is defined by a path integral over metrics on a space representing the spacetime itself because of ultraviolet divergences. The reason would be impossibility to regard points as fundamental constituents of the spacetime because the spacetime itself fluctuates at the Plank scale.  Thus, it is reasonable to define quantum gravity by a path integral over metrics on a space that consists of strings, by making a point have a structure of strings. In fact, perturbative strings are shown to suppress the ultraviolet divergences in quantum gravity.



In this paper, we geometrically define a space of superstrings including the effect of interactions. For this purpose, here we first review how such spaces of strings are defined in string field theories. In these theories, after a free loop space of strings are prepared, interaction terms of strings in actions are defined. In other words, the spaces of strings are defined by deforming the ring on the free loop space. Geometrically, the space of strings is defined by deformation quantization of the free loop space as a noncommutative geometry. Actually, in Witten's cubic open string field theory \cite{WittenCubic}, the interaction term is defined by using the $*$-product of noncommutative geometry. 
On the other hand, we adopt different approach, namely (infinite-dimensional) manifold theory\footnote{See \cite{RiemannianGeometry} as an example of text books for infinite-dimensional manifolds}. We do not start with a free loop space, but we define a space of strings including the effect of interactions from the beginning. The criterion to define a topology, which represents how near the strings are, is that trajectories in asymptotic processes on the space of strings reproduce the right moduli space of the super Riemann surfaces in a target manifold. We need Riemannian geometry naturally for fields on the space of strings because it is not flat\footnote{The spaces of strings in string field theories are different with those in string geometry because non-commutative geometry and Riemannian geometry describe different spaces.}.

By adopting the space of superstrings as geometric principle, we formulate superstring theory non-perturbatively. That is, we formulate the theory by summing over metrics on the space of strings. As a result, the theory is independent of backgrounds.

The organization of the paper is as follows. In section 2, we define string geometry and its Einstein-Hilbert action coupled with gauge fields. In section 3, we solve the equations of motion and obtain a string geometry solution that represents perturbative string vacuum. We derive the propagator of the fluctuations around the solution. Then, we move to the first quantization formalism, and we derive the all-order perturbative  scattering amplitudes that possess its moduli in  string theory. We extend these results to a supersymmetric theory in section 4, to a theory including open strings in section 5, and to a supersymmetric theory including open superstrings in section 6. The theory in section 6 is a non-perturbatively formulated superstring theory.  We derive the all-order  perturbative scattering amplitudes that possess the super moduli in type IIA, type IIB and SO(32) type I superstring theories  from the single theory, by considering fluctuations around fixed backgrounds representing type IIA, type IIB and SO(32) type I perturbative vacua, respectively. In section 7, we discuss a relation between the superstring geometry and supersymmetric matrix models of a new type. In section 8, based on superstring geometry, we formulate and study a theory that manifestly possesses the $SO(32)$ and $E_8 \times E_8$ heterotic perturbative vacua. We expect that this theory is equivalent to the theory in section 6, which manifestly possesses the type IIA, type IIB and $SO(32)$ type I perturbative vacua. We conclude in section 9 and discuss in section 10.

\vspace{1cm}

\section{String geometry}
\setcounter{equation}{0}

Let us define unique global time on a Riemann surface $\bar{\Sigma}$ with punctures $P^i$ ($i=1, \cdots, N$) in order to define string states by world-time constant lines. On $\bar{\Sigma}$, 
there exists an unique Abelian differential $dp$ that has simple poles with residues $f^i$ at $P^i$ where $\sum_i f^i=0$, if it is normalized to have purely imaginary periods with respect to all contours to fix ambiguity of adding holomorphic differentials. A global time is defined by $\bar{w}=\bar{\tau}+i\bar{\sigma}:=\int^{P} dp$ at any point $P$ on $\bar{\Sigma}$ \cite{KricheverNovikov1, KricheverNovikov2}. $\bar{\tau}$ takes the same value at the same point even if different contours are chosen in $\int^P dp$, because the real parts of the periods are zero by definition of the normalization. In particular, $\bar{\tau}=-\infty$ at $P^i$ with negative $f^i$ and $\bar{\tau}=\infty$ at $P^i$ with positive $f^i$.  A contour integral on $\bar{\tau}$ constant line around $P^i$: $i \Delta \bar{\sigma}=\oint dp=2\pi i f^i$ indicates that the $\bar{\sigma}$ region around $P^i$ is $2\pi f^i$. This means that $\bar{\Sigma}$ around $P^i$ represents a semi-infinite cylinder with radius $f^i$. The condition $\sum_i f^i=0$ means that the total $\bar{\sigma}$ region of incoming cylinders equals to that of outgoing ones if we choose the outgoing direction as positive. That is, the total $\bar{\sigma}$ region is conserved. In order to define the above global time uniquely, we fix the $\bar{\sigma}$ regions $2\pi f^i$ around $P^i$. We divide N $P^i$s to arbitrary two sets consist of $N_{-}$ and $N_{+}$ $P^i$s, respectively ($N_{-}+N_{+}=N$), then we divide  -1 to $N_{-}$ $f^i\equiv\frac{-1}{N_{-}}$ and 1 to $N_{+}$ $f^i\equiv\frac{1}{N_{+}}$ equally for all $i$.

Thus, under a conformal transformation, one obtains a Riemann surface $\bar{\Sigma}$ that has coordinates composed of the global time $\bar{\tau}$ and the position $\bar{\sigma}$. Because $\bar{\Sigma}$ can be a moduli of Riemann surfaces, any two-dimensional Riemannian manifold $\Sigma$ can be obtained by $\Sigma=\psi(\bar{\Sigma})$ where $\psi$ is a diffeomorphism times Weyl transformation.

\begin{figure}[htbp]
\begin{center}
\includegraphics[height=5cm, keepaspectratio, clip]{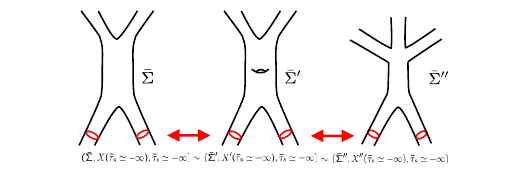}\end{center}\caption{An equivalence class of a string state $[\bar{\Sigma}, X(\bar{\tau}_s\simeq -\infty), \bar{\tau}_s\simeq -\infty]$. If the cylinders and the embedding functions are the same at $\bar{\tau} \simeq -\infty$, the states of strings at $\bar{\tau} \simeq -\infty$ specified by the red lines $(\bar{\Sigma}, X(\bar{\tau}_s\simeq -\infty), \bar{\tau}_s\simeq -\infty)$, $(\bar{\Sigma}', X'(\bar{\tau}_s\simeq -\infty), \bar{\tau}_s\simeq -\infty)$, and $(\bar{\Sigma}'', X''(\bar{\tau}_s\simeq -\infty), \bar{\tau}_s\simeq -\infty)$ should be identified.}\label{EquivalentClass}
\end{figure}

\begin{figure}[htbp]
\begin{center}
\includegraphics[height=5cm, keepaspectratio, clip]{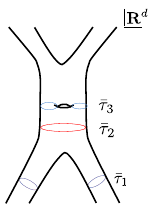}\end{center}\caption{Various string states. The red and blue lines represent one-string and two-string states, respectively.}
\label{states}
\end{figure}

Next, we will define the model space $E$.  We consider a state $(\bar{\Sigma}, X(\bar{\tau}_s), \bar{\tau}_s)$ determined by $\bar{\Sigma}$, a $\bar{\tau}=\bar{\tau}_s$ constant line and an arbitrary map $X(\bar{\tau}_s)$ from $\bar{\Sigma}|_{\bar{\tau}_s}$ to the Euclidean space $\bold{R}^d$. $\bar{\Sigma}$ is a union of $N_{\pm}$ cylinders with radii $f_i$ at $\bar{\tau}\simeq \pm \infty$. Thus, we define a string state as an equivalence class $[\bar{\Sigma}, X(\bar{\tau}_s\simeq \pm \infty), \bar{\tau}_s\simeq \pm \infty]$ by a relation $(\bar{\Sigma}, X(\bar{\tau}_s\simeq \pm \infty), \bar{\tau}_s\simeq \pm \infty) \sim (\bar{\Sigma}', X'(\bar{\tau}_s\simeq \pm \infty), \bar{\tau}_s\simeq \pm \infty)$ if $N_{\pm}=N'_{\pm}$, $f_i=f'_i$, and $X(\bar{\tau}_s\simeq \pm \infty)=X'(\bar{\tau}_s\simeq \pm \infty)$ as in Fig. \ref{EquivalentClass}. Because $\bar{\Sigma}|_{\bar{\tau}_s} \simeq S^1 \cup S^1 \cup \cdots \cup S^1$ and $X(\bar{\tau}_s): \Sigma|_{\bar{\tau}_s} \to M$, $[\bar{\Sigma}, X(\bar{\tau}_s), \bar{\tau}_s]$ represent many-body states of strings in $\bold{R}^d$ as in Fig. \ref{states}. The model space $E$ is defined by a collection of all the string states $\{[\bar{\Sigma}, X(\bar{\tau}_s), \bar{\tau}_s]\}$.

Here, we will define topologies of $E$. We define an $\epsilon$-open neighborhood of $[\bar{\Sigma}, X_s(\bar{\tau}_s), \bar{\tau}_s]$ by
\begin{eqnarray}
U([\bar{\Sigma}, X_s(\bar{\tau}_s), \bar{\tau}_s], \epsilon)
:=
\left\{[\bar{\Sigma},  X(\bar{\tau}), \bar{\tau}] \bigm| \sqrt{|\bar{\tau}-\bar{\tau}_s|^2+  \| X(\bar{\tau})-X_s(\bar{\tau}_s) \|^2} <\epsilon   \right\},
\label{neighbour}
\end{eqnarray}
where
\begin{equation}
\| X(\bar{\tau})-X_s(\bar{\tau}_s) \|^2:=\int_0^{2\pi}  d\bar{\sigma} 
|X(\bar{\tau}, \bar{\sigma})-X_s(\bar{\tau}_s, \bar{\sigma})|^2.
\label{norm}
\end{equation}
$U([\bar{\Sigma}, X(\bar{\tau}_s\simeq \pm \infty), \bar{\tau}_s\simeq \pm \infty], \epsilon)=
U([\bar{\Sigma}', X'(\bar{\tau}_s\simeq \pm \infty), \bar{\tau}_s\simeq \pm \infty], \epsilon)$ consistently if $N_{\pm}=N'_{\pm}$, $f_i=f'_i$, and $X(\bar{\tau}_s\simeq \pm \infty)=X'(\bar{\tau}_s\simeq \pm \infty)$, and $\epsilon$ is small enough, because the $\bar{\tau}_s\simeq \pm \infty$ constant line traverses only cylinders overlapped by $\bar{\Sigma}$ and $\bar{\Sigma}'$. $U$ is defined to be an open set of $E$ if there exists $\epsilon$ such that $U([\bar{\Sigma}, X(\bar{\tau}_s), \bar{\tau}_s], \epsilon) \subset U$ for an arbitrary point $[\bar{\Sigma}, X(\bar{\tau}_s), \bar{\tau}_s] \in U$.

Let $\mathcal{U}$ be a collection of all the open sets $U$. The topology of $E$ satisfies the axiom of topology $(i)$, $(ii)$, and $(iii)$.
\begin{eqnarray}
&&(i) \quad \emptyset, E \in \mathcal{U}
\nonumber \\
&&(ii) \quad U_1, U_2 \in \mathcal{U} \Rightarrow U_1 \cap U_2 \in \mathcal{U}
\nonumber \\
&&(iii) \quad U_{\lambda} \in \mathcal{U} \Rightarrow \cup_{\lambda \in \Lambda} U_{\lambda} \in \mathcal{U}. \nonumber
\end{eqnarray}
\begin{proof}
(i) Clear.

(ii) If $U_1 \cap U_2= \emptyset$, it is clear. Let us consider the case $U_1 \cap U_2 \neq \emptyset$. Because $U_1, U_2 \subset \mathcal{U}$, there exist $\epsilon$ and $\epsilon'$ such that $U([\bar{\Sigma}, X(\bar{\tau}_s), \bar{\tau}_s], \epsilon) \subset U_1$ and $U([\bar{\Sigma}, X(\bar{\tau}_s), \bar{\tau}_s], \epsilon') \subset U_2$ for all $[\bar{\Sigma}, X(\bar{\tau}_s), \bar{\tau}_s]\in U_1 \cap U_2$.  Let $\epsilon'':=min (\epsilon, \epsilon')$. Because $U([\bar{\Sigma}, X(\bar{\tau}_s), \bar{\tau}_s], \epsilon'') \subset U([\bar{\Sigma}, X(\bar{\tau}_s), \bar{\tau}_s], \epsilon) \subset U_1$ and $U([\bar{\Sigma}, X(\bar{\tau}_s), \bar{\tau}_s], \epsilon'') \subset U([\bar{\Sigma}, X(\bar{\tau}_s), \bar{\tau}_s], \epsilon') \subset U_2$, there exists $\epsilon''$ such that $U([\bar{\Sigma}, X(\bar{\tau}_s), \bar{\tau}_s], \epsilon'') \subset U_1 \cap U_2$ for all $[\bar{\Sigma}, X(\bar{\tau}_s), \bar{\tau}_s]\in U_1 \cap U_2$.

(iii) If $\cup_{\lambda \in \Lambda} U_{\lambda} = \emptyset$, it is clear. Let us consider the case $\cup_{\lambda \in \Lambda} U_{\lambda} \neq \emptyset$. For all $[\bar{\Sigma}, X(\bar{\tau}_s), \bar{\tau}_s] \in \cup_{\lambda \in \Lambda} U_{\lambda}$, there exists $\lambda_0$ such that $[\bar{\Sigma}, X(\bar{\tau}_s), \bar{\tau}_s] \in U_{\lambda_0}$. Because $U_{\lambda_0} \in \mathcal{U}$, there is $\epsilon$ such that $U([\bar{\Sigma}, X(\bar{\tau}_s), \bar{\tau}_s], \epsilon) \subset U_{\lambda_0}$. Then, $U([\bar{\Sigma}, X(\bar{\tau}_s), \bar{\tau}_s], \epsilon) \subset \cup_{\lambda \in \Lambda} U_{\lambda}$ for all $[\bar{\Sigma}, X(\bar{\tau}_s), \bar{\tau}_s] \in \cup_{\lambda \in \Lambda} U_{\lambda}$. 
\end{proof}
Although the model space is defined by using the coordinates $[\bar{\Sigma},  X(\bar{\tau}_s), \bar{\tau}_s]$, the model space does not depend on the coordinates, because the model space is a topological space.

In the following, we denote $[\bar{h}_{ mn}, X(\bar{\tau}), \bar{\tau}]$, where $\bar{h}_{ mn} (\bar{\sigma}, \bar{\tau})$ ($m, n =0,1$) is the worldsheet metric of $\bar{\Sigma}$, instead of $[\bar{\Sigma}, X(\bar{\tau}), \bar{\tau}]$, because giving a Riemann surface is equivalent to giving a  metric up to diffeomorphism and Weyl transformations.

\begin{figure}[htbp]
\begin{center}
\includegraphics[height=4cm, keepaspectratio, clip]{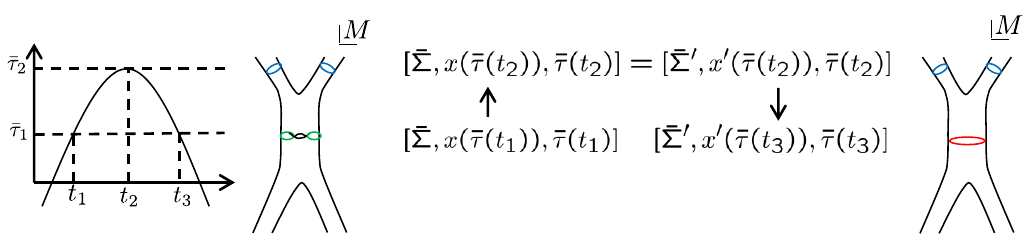}
\end{center}
\caption{A continuous trajectory. 
In case of general $\bar{\tau}(t)$ as in the left graph, string states on different  Riemann surfaces can be connected continuously in $\mathcal{M}_D$ as $[\bar{\Sigma}, x(\bar{\tau}(t_1)), \bar{\tau}(t_1)]$ and $[\bar{\Sigma}',x'(\bar{\tau}(t_3)), \bar{\tau}(t_3)]$ on the pictures.
}
\label{6Connected}
\end{figure}

Next, in order to define structures of string manifold, we consider how generally we can define general coordinate transformations between $[\bar{h}_{ mn}, X(\bar{\tau}), \bar{\tau}]$ and $[\bar{h}'_{ mn}, X'(\bar{\tau}'), \bar{\tau}']$ where $[\bar{h}_{ mn}, X(\bar{\tau}), \bar{\tau}] \in U \subset E$ and $[\bar{h}'_{ mn}, X'(\bar{\tau}'), \bar{\tau}'] \in U'\subset E$. $\bar{h}_{ mn}$ does not transform to $\bar{\tau}$ and $X(\bar{\tau})$ and vice versa, because $\bar{\tau}$ and $X(\bar{\tau})$ are continuous variables, whereas $\bar{h}_{ mn}$ is a discrete variable: $\bar{\tau}$ and $X(\bar{\tau})$ vary continuously, whereas $\bar{h}_{ mn}$ varies discretely in a trajectory on $E$ by definition of the neighborhoods. $\bar{\tau}$ and $\bar{\sigma}$ do not transform to each other because the string states are defined by $\bar{\tau}$ constant lines. Under these restrictions, the most general coordinate transformations are given by 
\begin{eqnarray}
[\bar{h}_{ mn}(\bar{\sigma}, \bar{\tau}), X^{\mu}(\bar{\sigma}, \bar{\tau}), \bar{\tau}] 
\mapsto 
[\bar{h}_{ mn}'(\bar{\sigma}'(\bar{\sigma}), \bar{\tau}'(\bar{\tau}, X(\bar{\tau}))), X'^{\mu}(\bar{\sigma}', \bar{\tau}')(\bar{\tau}, X(\bar{\tau})), \bar{\tau}'(\bar{\tau}, X(\bar{\tau}))],
\label{GeneralCoordTrans}
\end{eqnarray}
where $\bar{h}_{ mn} \mapsto \bar{h}_{ mn}'$ represent world-sheet diffeomorphism transformations\footnote{
We extend the model space from $E=\{[\bar{h}_{ mn}(\bar{\sigma}, \bar{\tau}), X^{\mu}(\bar{\sigma}, \bar{\tau}), \bar{\tau}]\}$ to $E=\{[\bar{h}_{ mn}'(\bar{\sigma}', \bar{\tau}'), X'^{\mu}(\bar{\sigma}', \bar{\tau}'), \bar{\tau}']\}$ by including the points generated by the diffeomorphisms $\bar{\sigma} \mapsto \bar{\sigma}'(\bar{\sigma})$ and $\bar{\tau} \mapsto \bar{\tau}'(\bar{\tau})$.}. 
$X'^{\mu}(\bar{\tau}, X(\bar{\tau}))$ and $\bar{\tau}'(\bar{\tau}, X(\bar{\tau}))$ are functionals of $\bar{\tau}$ and $X(\bar{\tau})$. $\mu=0, 1, \cdots d-1$.  
We consider all the manifolds which are constructed by patching open sets of the model space $E$ by the general coordinate transformations (\ref{GeneralCoordTrans}) and call them string manifolds $\mathcal{M}$\footnote{These coordinate transformations are diffeomorphisms,  and  functions over the string manifolds, for example metrics, are differentiable functions. These statements are justified mathematically by formulating the string manifolds as polyfolds. The  reference volume of polyfolds is given in \cite{Polyfold2}. For example, see  Example 1.8.  in  \cite{Polyfold3}. Indeed, for example, at the interaction point where two strings becomes one string, a cusp formed where the two strings touch, smoothly connects with the one string in the framework of polyfolds.}.

Here, we give an example of string manifolds: $\mathcal{M}_D:= \{ [\bar{\Sigma}, x(\bar{\tau}), \bar{\tau}] \}$, where $D$ represents a target manifold $M$. $x(\bar{\tau}): \bar{\Sigma}|_{\bar{\tau}} \to M$,  where the image of the embedding function $x(\bar{\tau})$ has a metric: 
$ds^2= dx^{\mu}(\bar{\tau}, \bar{\sigma})
dx^{\nu}(\bar{\tau}, \bar{\sigma})
G_{\mu \nu}(x(\bar{\tau}, \bar{\sigma}))$. 

We will show that $ \mathcal{M}_D$ has a structure of manifold, that is there exists a homeomorphism between the sufficiently small neighborhood around an arbitrary point $[\bar{\Sigma}, x_s(\bar{\tau}_s), \bar{\tau}_s] \in \mathcal{M}_D$ and an open set of $E$. There exists a general coordinate transformation $X(x)$ that satisfies 
$ds^2= dx^{\mu}
dx^{\nu}
G_{\mu \nu}(x)
=
dX^{\mu}
dX^{\nu}
\eta_{\mu \nu}$
on an arbitrary point $x$ in the $\epsilon_{\bar{\sigma}}$ open neighborhood around $x_s(\bar{\tau}_s, \bar{\sigma}) \in M$, if $\epsilon_{\bar{\sigma}}$ is sufficiently small. An arbitrary point $[\bar{\Sigma}, x(\bar{\tau}), \bar{\tau}]$ in the $\epsilon$ open neighborhood around $[\bar{\Sigma}, x_s(\bar{\tau}_s), \bar{\tau}_s] $ satisfies 
\begin{equation}
\int_0^{2\pi}  d\bar{\sigma} 
|x(\bar{\tau}, \bar{\sigma})-x_s(\bar{\tau}_s, \bar{\sigma})|^2
< \epsilon^2- |\bar{\tau}-\bar{\tau}_s|^2 \leqq \epsilon^{2}, \label{neigh}
\end{equation}
and thus 
\begin{equation}
|x(\bar{\tau}, \bar{\sigma})-x_s(\bar{\tau}_s, \bar{\sigma})|
< \epsilon_{\bar{\sigma}}'
\end{equation}
on an arbitrary $\bar{\sigma}$.
$\epsilon_{\bar{\sigma}}'< \epsilon_{\bar{\sigma}}$ is satisfied on an arbitrary $\bar{\sigma}$ if $\epsilon$ is taken to be sufficiently small. 
Then, there exists a transformation  $X^{\mu}(\bar{\tau}, \bar{\sigma}):=X^{\mu}(x(\bar{\tau}, \bar{\sigma}))$, which satisfies
\begin{equation}
ds^2= dx^{\mu}(\bar{\tau}, \bar{\sigma})
dx^{\nu}(\bar{\tau}, \bar{\sigma})
G_{\mu \nu}(x(\bar{\tau}, \bar{\sigma}))
=
dX^{\mu}(\bar{\tau}, \bar{\sigma})
dX^{\nu}(\bar{\tau}, \bar{\sigma})
\eta_{\mu \nu}. \label{LocalLorents}
\end{equation}
Because the tangent vector $X(\bar{\tau}, \bar{\sigma})$ exists for each $x(\bar{\tau}, \bar{\sigma})$, there exists a vector bundle $X(\bar{\tau})$ for $0 \leqq \bar{\sigma} < 2\pi$. $x(\bar{\tau})$ and $X(\bar{\tau})$ satisfy (\ref{LocalLorents}) on each $\bar{\sigma}$, that is $X(\bar{\tau}): \bar{\Sigma}|_{\bar{\tau}} \to \bold{R}^d$. Therefore, there exists a homeomorphism between the sufficiently small neighborhood around an arbitrary point $[\bar{\Sigma}, x_s(\bar{\tau}_s), \bar{\tau}_s] \in \mathcal{M}_D$ and an open set of $E$: $[\bar{\Sigma}, x(\bar{\tau}), \bar{\tau}] \mapsto [\bar{\Sigma}, X(\bar{\tau}), \bar{\tau}]$.
Actually, a map from the image $x(\bar{\tau}, \bar{\sigma})$ to the image $X(\bar{\tau}, \bar{\sigma})$ is explicitly given by an exponential map
\begin{equation}
x(\bar{\tau}, \bar{\sigma})
=
\exp_{x_s(\bar{\tau}_s, \bar{\sigma})}
X(\bar{\tau}, \bar{\sigma})
\simeq
x_s(\bar{\tau}_s, \bar{\sigma})
+X(\bar{\tau}, \bar{\sigma}).
\end{equation}
If we substitute this to an $\epsilon$ open neighborhood around an arbitrary point $[\bar{\Sigma}, x_s(\bar{\tau}_s), \bar{\tau}_s] \in \mathcal{M}_D$ (\ref{neigh}), we obtain an $\epsilon$ open neighborhood around $[\bar{\Sigma}, 0, \bar{\tau}_s] \in E$,
\begin{equation}
\int_0^{2\pi}  d\bar{\sigma} 
|X(\bar{\tau}, \bar{\sigma})|^2 +  |\bar{\tau}-\bar{\tau}_s|^2 < \epsilon^{2}.
\end{equation}

By definition of the $\epsilon$-open neighborhood, arbitrary two string states on a connected  Riemann surface in $\mathcal{M}_D$ are connected continuously. Thus, there is an one-to-one correspondence between a Riemann surface with punctures in $M$ and a curve  parametrized by $\bar{\tau}$ from $\bar{\tau}=-\infty$ to $\bar{\tau}=\infty$ on $\mathcal{M}_D$. That is, curves that represent asymptotic processes on $\mathcal{M}_D$ reproduce the right moduli space of the Riemann surfaces in the target manifold.

By a general curve parametrized by $t$ on $\mathcal{M}_D$, string states on different Riemann surfaces that have even different genera, can be connected continuously, for example see Fig. \ref{6Connected}, whereas different Riemann surfaces that have different genera cannot be connected continuously in the moduli space of the Riemann surfaces in the target space.

The tangent space is spanned by $\frac{\partial}{\partial\bar{\tau}}$ and $\frac{\partial}{\partial X^{\mu}(\bar{\sigma}, \bar{\tau})}$  as one can see from the $\epsilon$-open neighborhood (\ref{neighbour}). We should note that $\frac{\partial}{\partial \bar{h}_{ mn}}$  cannot be a part of basis that span the tangent space, because $\bar{h}_{ mn}$ is just a discrete variable in $E$. The index of $\frac{\partial}{\partial X^{\mu}(\bar{\sigma}, \bar{\tau}) }$ can be $(\mu \, \bar{\sigma})$.
We define a summation over $ \bar{\sigma}$ by $\int d\bar{\sigma}\bar{e} (\bar{\sigma}, \bar{\tau})$, where $\bar{e}:=\sqrt{\bar{h}_{ \bar{\sigma} \bar{\sigma}}}$.
This summation is invariant under $\bar{\sigma} \mapsto \bar{\sigma}'(\bar{\sigma})$ and transformed as a scalar under $\bar{\tau} \mapsto \bar{\tau}'(\bar{\tau}, X(\bar{\tau}))$.

Riemannian string manifold is obtained by defining a metric, which is a section of an inner product on the tangent space. The general form of a metric is given by
\begin{eqnarray}
&&ds^2(\bar{h}, X(\bar{\tau}), \bar{\tau}) \nonumber \\
=&&G(\bar{h}, X(\bar{\tau}), \bar{\tau})_{dd} (d\bar{\tau})^2 +2 d\bar{\tau} \int d\bar{\sigma}  \bar{e} (\bar{\sigma}, \bar{\tau})  \sum_{\mu}  G(\bar{h}, X(\bar{\tau}), \bar{\tau})_{d \; (\mu \bar{\sigma})} d X^{\mu}(\bar{\sigma}, \bar{\tau}) \nonumber \\
&&+\int d\bar{\sigma}   \bar{e} (\bar{\sigma}, \bar{\tau}) \int d\bar{\sigma}' \bar{e} (\bar{\sigma}', \bar{\tau})  \sum_{\mu, \mu'} G(\bar{h}, X(\bar{\tau}), \bar{\tau})_{ \; (\mu \bar{\sigma})  \; (\mu' \bar{\sigma}')} d X^{\mu}(\bar{\sigma}, \bar{\tau}) d X^{\mu'}(\bar{\sigma}', \bar{\tau}). \nonumber \\
\end{eqnarray}
We summarize the vectors as $dX^I$ ($I=d,(\mu\bar{\sigma})$), where  $dX^d:=d\bar{\tau}$ and $d X^{(\mu\bar{\sigma})}:=dX^{\mu}(\bar{\sigma}, \bar{\tau})$. Then, the components of the metric are summarized as $G_{IJ}(\bar{h}, X(\bar{\tau}), \bar{\tau})$. The inverse of the metric $G^{IJ}(\bar{h}, X(\bar{\tau}), \bar{\tau})$ is defined by $G_{IJ}G^{JK}=G^{KJ}G_{JI}=\delta_I^K$, where $\delta_d^d=1$ and $\delta_{\mu\bar{\sigma}}^{\mu'\bar{\sigma}'}=\frac{1}{\bar{e} (\bar{\sigma}, \bar{\tau}) } \delta_{\mu}^{\mu'}\delta(\bar{\sigma}-\bar{\sigma}')$. The components of the Riemannian curvature tensor are given by $R^I_{JKL}$ in the basis $\frac{\partial}{\partial X^I}$. The components of the Ricci tensor are $R_{IJ}:=R^K_{IKJ}=R^d_{IdJ}+\int d\bar{\sigma} \bar{e}  R^{(\mu \bar{\sigma})}_{I \; (\mu \bar{\sigma}) \; J}$. The scalar curvature is 
\begin{eqnarray}
R&:=&G^{IJ} R_{IJ} \nonumber \\
&=&G^{dd}R_{dd}+2 \int d\bar{\sigma} \bar{e}  G^{d \; (\mu\bar{\sigma})} R_{d \; (\mu\bar{\sigma})} +\int d\bar{\sigma} \bar{e} \int d\bar{\sigma}'\bar{e}'  G^{(\mu\bar{\sigma}) \; (\mu'\bar{\sigma}')}R_{(\mu\bar{\sigma})  \; (\mu'\bar{\sigma}')}. \nonumber 
\end{eqnarray}
The volume is  $\sqrt{G}$, where $G=det (G_{IJ})$.

By using these geometrical objects, we formulate string theory non-perturbatively as
\begin{equation}
Z=\int \mathcal{D}G \mathcal{D}Ae^{-S},
\end{equation}
where
\begin{equation}
S=\frac{1}{G_N}\int \mathcal{D} h \mathcal{D} X(\bar{\tau})\mathcal{D} \bar{\tau} 
\sqrt{G} (-R +\frac{1}{4} G_N G^{I_1 I_2} G^{J_1 J_2} F_{I_1 J_1} F_{I_2 J_2} ). \label{BosonicAction}
\end{equation}
As an example of sets of fields on the string manifolds, we consider the metric and an $u(1)$ gauge field $A_I$ whose field strength is given by $F_{IJ}$. The path integral is defined by semi-classically\footnote{It will be enough to define the path-integral by semi-classically summing classical solutions and small classical and quantum fluctuations around them, because string manifolds themselves possess quantum corrections, and loops of the fields on them do not correspond to quantum corrections as one can see in the derivation of the perturbative string theory later. The unitarity is manifest and there is also no UV divergence from loop integrals, by defining the path-integral semi-classically. }  summing over the metrics and gauge fields on $\mathcal{M}$. By definition, the theory is background independent. $\mathcal{D}h$ is the invariant measure\footnote{The invariant measure is defined implicitly by the most general invariant norm without derivatives for elements $\delta h_{mn}$ of the tangent space of the metric, $||\delta h||^2=\int d^2 \sigma \sqrt{h}(h^{mp} h^{nq}+C h^{mn}h^{pq} )\delta h_{mn} \delta h_{pq}$ with $C$ an arbitrary constant, and a normalization $\int \mathcal{D} \delta h \exp^{-\frac{1}{2}||\delta h||^2}=1.$}\footnote {$\int$ in $\int \mathcal{D} h$ includes $\sum_{\mbox{compact topologies}}$.} of the metrics $h_{mn}$ on the two-dimensional Riemannian manifolds $\Sigma$. $h_{mn}$ and $\bar{h}_{mn}$ are related to each others by the diffeomorphism and the Weyl transformations.

Under 
\begin{equation}
(\bar{\tau}, X(\bar{\tau})) \mapsto (\bar{\tau}'(\bar{\tau}, X(\bar{\tau})) , X'(\bar{\tau}')(\bar{\tau}, X(\bar{\tau}))),
\label{subdiffeo}
\end{equation} 
$G_{IJ}(\bar{h}, X(\bar{\tau}), \bar{\tau})$ and $A_I(\bar{h}, X(\bar{\tau}), \bar{\tau})$ are transformed as a symmetric tensor and a vector, respectively and the action is manifestly invariant. 

We define $G_{IJ}(\bar{h}, X(\bar{\tau}), \bar{\tau})$ and $A_I(\bar{h}, X(\bar{\tau}), \bar{\tau})$ so as to transform as scalars under $\bar{h}_{ mn}(\bar{\sigma}, \bar{\tau}) \mapsto \bar{h}_{ mn}'(\bar{\sigma}'(\bar{\sigma}), \bar{\tau})$. Under $\bar{\sigma}$ diffeomorphisms: $\bar{\sigma} \mapsto \bar{\sigma}'(\bar{\sigma})$, which are equivalent to 
\begin{eqnarray}
[\bar{h}_{ mn}(\bar{\sigma}, \bar{\tau}), X^{\mu}(\bar{\sigma}, \bar{\tau}), \bar{\tau}] 
&&\mapsto 
[\bar{h}_{ mn}'(\bar{\sigma}'(\bar{\sigma}), \bar{\tau}), X'^{\mu}(\bar{\sigma}', \bar{\tau})(X(\bar{\tau})),  \bar{\tau}], \nonumber \\
&&=[\bar{h}_{ mn}'(\bar{\sigma}'(\bar{\sigma}), \bar{\tau}), X^{\mu}(\bar{\sigma}, \bar{\tau}),  \bar{\tau}], \label{diff} 
\end{eqnarray}
$G_{d \; (\mu \bar{\sigma})}$ is transformed as a scalar;
\begin{eqnarray}
G'_{d \; (\mu \bar{\sigma}')}(\bar{h}', X'(\bar{\tau}), \bar{\tau})
&=&
G'_{d \; (\mu \bar{\sigma}')}(\bar{h}, X'(\bar{\tau}), \bar{\tau})
=
\frac{\partial X^I(\bar{\tau})}{\partial X^{'d}(\bar{\tau})}
\frac{\partial X^J(\bar{\tau})}{\partial X^{'(\mu \bar{\sigma}')}(\bar{\tau})}
G_{IJ}(\bar{h}, X(\bar{\tau}), \bar{\tau}) \nonumber \\
&=&
\frac{\partial X^I(\bar{\tau})}{\partial X^d(\bar{\tau})}
\frac{\partial X^J(\bar{\tau})}{\partial X^{(\mu \bar{\sigma})}(\bar{\tau})}
G_{IJ}(\bar{h}, X(\bar{\tau}), \bar{\tau})
=
G_{d \; (\mu \bar{\sigma})}(\bar{h}, X(\bar{\tau}), \bar{\tau}),
\end{eqnarray}
because  (\ref{subdiffeo}) and (\ref{diff}).
In the same way, the other fields are also transformed as 
\begin{eqnarray}
G'_{dd}(\bar{h}', X'(\bar{\tau}), \bar{\tau})&=&G_{dd}(\bar{h}, X(\bar{\tau}), \bar{\tau})
\nonumber \\ 
G'_{ \; (\mu \bar{\sigma}')  \; (\nu \bar{\rho}')}(\bar{h}', X'(\bar{\tau}), \bar{\tau})&=&G_{ \; (\mu \bar{\sigma})  \; (\nu \bar{\rho})}(\bar{h}, X(\bar{\tau}), \bar{\tau})
\nonumber \\
A'_d(\bar{h}', X'(\bar{\tau}), \bar{\tau})&=&A_d(\bar{h}, X(\bar{\tau}), \bar{\tau})
\nonumber \\
A'_{(\mu \bar{\sigma}')}(\bar{h}', X'(\bar{\tau}), \bar{\tau})
&=&A_{(\mu \bar{\sigma})}(\bar{h}, X(\bar{\tau}), \bar{\tau}). 
\end{eqnarray}
Thus, the action is invariant under $\bar{\sigma}$ diffeomorphisms, because $\int d\bar{\sigma}'\bar{e}'(\bar{\sigma}', \bar{\tau})=\int d\bar{\sigma}\bar{e} (\bar{\sigma}, \bar{\tau})$. Therefore, $G_{IJ}(\bar{h}, X(\bar{\tau}), \bar{\tau})$ and $A_I(\bar{h}, X(\bar{\tau}), \bar{\tau})$ are transformed covariantly and the action (\ref{BosonicAction}) is invariant under the diffeomorphisms (\ref{GeneralCoordTrans}), including the $\bar{\sigma}$ diffeomorphisms.

\section{Perturbative string amplitudes from string geometry}
\setcounter{equation}{0}

The background that represents a perturbative vacuum is given by
\begin{eqnarray}
\bar{ds}^2
&=& 2\lambda \bar{\rho}(\bar{h}) N^2(X(\bar{\tau})) (dX^d)^2 +\int d\bar{\sigma}   \bar{e} \int d\bar{\sigma}' \bar{e}' N^{\frac{2}{2-D}}(X(\bar{\tau})) \frac{\bar{e}^3(\bar{\sigma}, \bar{\tau})}{\sqrt{\bar{h}(\bar{\sigma}, \bar{\tau})}} \delta_{(\mu \bar{\sigma}) (\mu' \bar{\sigma}')}
d X^{(\mu \bar{\sigma})} d X^{(\mu' \bar{\sigma}')}, \nonumber \\
\bar{A}_d&=&i \sqrt{\frac{2-2D}{2-D}}\frac{\sqrt{2\lambda \bar{\rho}(\bar{h}) }}{\sqrt{G_N}} N(X(\bar{\tau})), \qquad
\bar{A}_{(\mu \bar{\sigma})}=0, \label{solution}
\end{eqnarray}
on $\mathcal{M}_D$ where the target metric is fixed to $\eta_{\mu \mu'}$.
$\bar{\rho}(\bar{h}):=\frac{1}{4 \pi}\int d\bar{\sigma} \sqrt{\bar{h}}\bar{R}_{\bar{h}}$, where $\bar{R}_{\bar{h}}$ is the scalar curvature of $\bar{h}_{ mn}$. $D$ is a volume of the index $(\mu \bar{\sigma})$: $D:=\int d \bar{\sigma} \bar{e} \delta_{(\mu \bar{\sigma}) (\mu \bar{\sigma})}=d 2 \pi \delta(0)$. $N(X(\bar{\tau}))=\frac{1}{1+v(X(\bar{\tau}))}$. $v(X(\bar{\tau}))= \frac{\alpha}{\sqrt{d-1}} \int d\bar{\sigma}  \epsilon_{\mu\nu}X^{\mu}(\bar{\tau}) \partial_{\bar{\sigma}} X^{\nu}(\bar{\tau})$, where 
\begin{align}
\epsilon_{\mu \nu} = -\epsilon_{\nu\mu}=
\begin{cases}
 1~\text{ for } (\mu,\nu)=(0,1),(2,3),(4,5),...,(d-2,d-1),
\\
0 ~\text{ for others}.
\end{cases}
\end{align}  The inverse of the metric is given by 
\begin{eqnarray}
\bar{G}^{dd}&=&\frac{1}{2\lambda \bar{\rho} }\frac{1}{N^2} \nonumber \\
\bar{G}^{d \; (\mu\bar{\sigma})}&=&0 \nonumber \\
\bar{G}^{(\mu\bar{\sigma}) \; (\mu'\bar{\sigma}')}&=& N^{\frac{-2}{2-D}} \frac{\sqrt{\bar{h}}}{\bar{e}^3} \delta_{(\mu \bar{\sigma}) (\mu' \bar{\sigma}')},
\end{eqnarray}
because $\int d\bar{\sigma}'' \bar{e}'' \bar{G}_{(\mu\bar{\sigma}) \; (\mu''\bar{\sigma}'')}\bar{G}^{(\mu''\bar{\sigma}'') \; (\mu'\bar{\sigma}')}=\int d\bar{\sigma}''\bar{e}'' \delta_{(\mu\bar{\sigma}) \; (\mu''\bar{\sigma}'')}\delta_{(\mu''\bar{\sigma}'') \; (\mu'\bar{\sigma}')}
= \delta_{(\mu\bar{\sigma}) \; (\mu'\bar{\sigma}')}$. 
From the metric, we obtain 
\begin{eqnarray}
&&\sqrt{\bar{G}}=N^\frac{2}{2-D}\sqrt{2\lambda \bar{\rho} \exp(\frac{D}{2\pi}\int d\bar{\sigma} \ln \frac{\bar{e}^3}{\sqrt{\bar{h}}})} 
\nonumber \\
&&\bar{R}_{dd}=-2\lambda \bar{\rho} N^{\frac{-2}{2-D}} \int d\bar{\sigma} \frac{\sqrt{\bar{h}}}{\bar{e}^2} \partial_{(\mu \bar{\sigma})}N \partial_{(\mu \bar{\sigma})}N
\nonumber \\
&&\bar{R}_{d \; (\mu \bar{\sigma})}=0 
\nonumber \\
&&\bar{R}_{(\mu\bar{\sigma}) \; (\mu'\bar{\sigma}')}
=\frac{D-1}{2-D}N^{-2}\partial_{(\mu \bar{\sigma})}N \partial_{(\mu' \bar{\sigma}')}N
\nonumber \\
&& \qquad \qquad \quad  +\frac{1}{D-2}N^{-2}
\int d\bar{\sigma}'' \frac{\sqrt{\bar{h}''}}{\bar{e}^{''2}} \partial_{(\mu'' \bar{\sigma}'')}N \partial_{(\mu'' \bar{\sigma}'')}N
\frac{\bar{e}^3}{\sqrt{\bar{h}}}
\delta_{(\mu\bar{\sigma}) \; (\mu'\bar{\sigma}')}
\nonumber \\
&&\bar{R}=\frac{D-3}{2-D} N^{\frac{2D-6}{2-D}}
\int d\bar{\sigma} \frac{\sqrt{\bar{h}}}{\bar{e}^2} \partial_{(\mu \bar{\sigma})}N \partial_{(\mu \bar{\sigma})}N.
\end{eqnarray}
By using these quantities, one can show that the background (\ref{solution}) is a classical solution\footnote{This solution is a generalization of the Majumdar-Papapetrou solution \cite{Majumdar, Papapetrou} of the Einstein-Maxwell system.} to the equations of motion of (\ref{BosonicAction}). We also need to use the fact that $v(X(\bar{\tau}))$ is a harmonic function with respect to $X^{(\mu \bar{\sigma})}(\bar{\tau})$. Actually, $\partial_{(\mu \bar{\sigma})}\partial_{(\mu \bar{\sigma})}v=0$. In these calculations, we should note that $\bar{h}_{mn}$, $X^{\mu}(\bar{\tau})$ and $\bar{\tau}$ are all independent, and thus $\frac{\partial}{\partial \bar{\tau}}$ is an explicit derivative on functions over the string manifolds, especially, $\frac{\partial}{\partial \bar{\tau}}\bar{h}_{ mn}=0$ and $\frac{\partial}{\partial \bar{\tau}}X^{\mu}(\bar{\tau})=0$. Because the equations of motion are differential equations with respect to $X^{\mu}(\bar{\tau})$ and $\bar{\tau}$, $\bar{h}_{mn}$ is a constant in the solution (\ref{solution}) to the differential equations. The dependence of $\bar{h}_{mn}$ on the background (\ref{solution}) is uniquely determined  by the consistency of the quantum theory of the fluctuations around the background. Actually, we will find that all the perturbative string amplitudes are derived. 

Let us consider fluctuations around the background (\ref{solution}), $G_{IJ}=\bar{G}_{IJ}+\tilde{G}_{IJ}$ and $A_I=\bar{A}_I+\tilde{A}_I$. The action (\ref{BosonicAction}) up to the quadratic order is given by,\begin{eqnarray}
S&=&\frac{1}{G_N} \int \mathcal{D} h \mathcal{D} X(\bar{\tau})\mathcal{D} \bar{\tau}  \sqrt{\bar{G}} 
\Bigl(-\bar{R}+\frac{1}{4}\bar{F}'_{IJ}\bar{F}'^{IJ} 
\nonumber \\
&&+\frac{1}{4}\bar{\nabla}_I \tilde{G}_{JK} \bar{\nabla}^I \tilde{G}^{JK}
-\frac{1}{4}\bar{\nabla}_I \tilde{G} \bar{\nabla}^I \tilde{G}
+\frac{1}{2}\bar{\nabla}^I \tilde{G}_{IJ} \bar{\nabla}^J \tilde{G}
-\frac{1}{2}\bar{\nabla}^I \tilde{G}_{IJ} \bar{\nabla}_K \tilde{G}^{JK}
\nonumber \\
&&-\frac{1}{4}(-\bar{R}+\frac{1}{4}\bar{F}'_{KL}\bar{F}'^{KL})
(\tilde{G}_{IJ}\tilde{G}^{IJ}-\frac{1}{2}\tilde{G}^2)
+(-\frac{1}{2}\bar{R}^{I}_{\;\; J}+\frac{1}{2}\bar{F}'^{IK}\bar{F}'_{JK})
\tilde{G}_{IL}\tilde{G}^{JL}
\nonumber \\
&&+(\frac{1}{2}\bar{R}^{IJ}-\frac{1}{4}\bar{F}'^{IK}\bar{F}'^J_{\;\;\;\; K})
\tilde{G}_{IJ}\tilde{G}
+(-\frac{1}{2}\bar{R}^{IJKL}+\frac{1}{4}\bar{F}'^{IJ}\bar{F}'^{KL})
\tilde{G}_{IK}\tilde{G}_{JL}
\nonumber \\
&&+\frac{1}{4}G_N \tilde{F}_{IJ} \tilde{F}^{IJ} 
+\sqrt{G_N} 
(\frac{1}{4} \bar{F}^{'IJ} \tilde{F}_{IJ} \tilde{G} 
-\bar{F}^{'IJ} \tilde{F}_{IK} \tilde{G}_J^{\;\; K} ) \Bigr),
\end{eqnarray}
where $\bar{F}'_{IJ}:=\sqrt{G_N}\bar{F}_{IJ}$ is independent of $G_N$. $\tilde{G}:=\bar{G}^{IJ}\tilde{G}_{IJ}$. There is no first order term because the background satisfies the equations of motion. If we take $G_N \to 0$, we obtain 
\begin{eqnarray}
S'&=&\frac{1}{G_N} \int \mathcal{D} h \mathcal{D} X(\bar{\tau})\mathcal{D} \bar{\tau}  \sqrt{\bar{G}} 
\Bigl(-\bar{R}+\frac{1}{4}\bar{F}'_{IJ}\bar{F}'^{IJ} 
\nonumber \\
&&+\frac{1}{4}\bar{\nabla}_I \tilde{G}_{JK} \bar{\nabla}^I \tilde{G}^{JK}
-\frac{1}{4}\bar{\nabla}_I \tilde{G} \bar{\nabla}^I \tilde{G}
+\frac{1}{2}\bar{\nabla}^I \tilde{G}_{IJ} \bar{\nabla}^J \tilde{G}
-\frac{1}{2}\bar{\nabla}^I \tilde{G}_{IJ} \bar{\nabla}_K \tilde{G}^{JK}
\nonumber \\
&&-\frac{1}{4}(-\bar{R}+\frac{1}{4}\bar{F}'_{KL}\bar{F}'^{KL})
(\tilde{G}_{IJ}\tilde{G}^{IJ}-\frac{1}{2}\tilde{G}^2)
+(-\frac{1}{2}\bar{R}^{I}_{\;\; J}+\frac{1}{2}\bar{F}'^{IK}\bar{F}'_{JK})
\tilde{G}_{IL}\tilde{G}^{JL}
\nonumber \\
&&+(\frac{1}{2}\bar{R}^{IJ}-\frac{1}{4}\bar{F}'^{IK}\bar{F}'^J_{\;\;\;\; K})
\tilde{G}_{IJ}\tilde{G}
+(-\frac{1}{2}\bar{R}^{IJKL}+\frac{1}{4}\bar{F}'^{IJ}\bar{F}'^{KL})
\tilde{G}_{IK}\tilde{G}_{JL} \Bigr),
\end{eqnarray}
where the fluctuation of the gauge field is suppressed. In order to fix the gauge symmetry (\ref{subdiffeo}), we take the harmonic gauge. If we add the gauge fixing term
\begin{equation}
S_{fix}=\frac{1}{G_N}\int \mathcal{D} h \mathcal{D} X(\bar{\tau})\mathcal{D} \bar{\tau}  \sqrt{\bar{G}} 
\frac{1}{2} \Bigl( \bar{\nabla}^J(\tilde{G}_{IJ}-\frac{1}{2}\bar{G}_{IJ}\tilde{G}) \Bigr)^2,\end{equation}
we obtain
\begin{eqnarray}
S'+S_{fix}&=&\frac{1}{G_N} \int \mathcal{D} h \mathcal{D} X(\bar{\tau})\mathcal{D} \bar{\tau}  \sqrt{\bar{G}} 
\Bigl(-\bar{R}+\frac{1}{4}\bar{F}'_{IJ}\bar{F}'^{IJ} 
\nonumber \\
&&+\frac{1}{4}\bar{\nabla}_I \tilde{G}_{JK} \bar{\nabla}^I \tilde{G}^{JK}
-\frac{1}{8}\bar{\nabla}_I \tilde{G} \bar{\nabla}^I \tilde{G}
\nonumber \\
&&-\frac{1}{4}(-\bar{R}+\frac{1}{4}\bar{F}'_{KL}\bar{F}'^{KL})
(\tilde{G}_{IJ}\tilde{G}^{IJ}-\frac{1}{2}\tilde{G}^2)
+(-\frac{1}{2}\bar{R}^{I}_{\;\; J}+\frac{1}{2}\bar{F}'^{IK}\bar{F}'_{JK})
\tilde{G}_{IL}\tilde{G}^{JL}
\nonumber \\
&&+(\frac{1}{2}\bar{R}^{IJ}-\frac{1}{4}\bar{F}'^{IK}\bar{F}'^J_{\;\;\;\; K})
\tilde{G}_{IJ}\tilde{G}
+(-\frac{1}{2}\bar{R}^{IJKL}+\frac{1}{4}\bar{F}'^{IJ}\bar{F}'^{KL})
\tilde{G}_{IK}\tilde{G}_{JL} \Bigr). \label{fixedaction}
\end{eqnarray}
In order to obtain perturbative string amplitudes, we perform a derivative expansion of $\tilde{G}_{IJ}$,
\begin{eqnarray}
&&\tilde{G}_{IJ} \to \frac{1}{\alpha} \tilde{G}_{IJ} \nonumber \\
&&\partial_{K}\tilde{G}_{IJ} \to \partial_{K}\tilde{G}_{IJ}\nonumber \\
&&\partial_{K}\partial_{L}\tilde{G}_{IJ} \to \alpha \partial_{K}\partial_{L}\tilde{G}_{IJ},
\end{eqnarray}
and take
\begin{equation}
\alpha \to 0,
\end{equation}
where $\alpha$ is an arbitrary constant in the solution (\ref{solution}).
We normalize the fields as $\tilde{H}_{IJ}:=Z_{IJ} \tilde{G}_{IJ}$, where 
$Z_{IJ}:=\frac{1}{\sqrt{G_N}} 
\bar{G}^{\frac{1}{4}} 
(\bar{a}_I \bar{a}_J)^{-\frac{1}{2}}$.
$\bar{a}_{I}$ represent the background metric as $\bar{G}_{IJ}=\bar{a}_I \delta_{IJ}$, where $\bar{a}_d=2\lambda\bar{\rho}$ and $\bar{a}_{(\mu \bar{\sigma})}= \frac{\bar{e}^3}{\sqrt{\bar{h}}}$. Then, (\ref{fixedaction}) with appropriate boundary conditions reduces to
\begin{equation}
S'+S_{fix} \to S_0 + S_2,
\end{equation}
where
\begin{equation}
S_0
=
\int \mathcal{D} h \mathcal{D} X(\bar{\tau})\mathcal{D} \bar{\tau}  
\left(\frac{1}{G_N} \sqrt{\bar{G}} \left(-\bar{R}+\frac{1}{4}\bar{F}'_{IJ}\bar{F}'^{IJ} \right)\right),
\end{equation}
and
\begin{equation}
S_2
=
\int \mathcal{D} h \mathcal{D} X(\bar{\tau})\mathcal{D} \bar{\tau}  
\frac{1}{8}\tilde{H}_{IJ}H_{IJ;KL}\tilde{H}_{KL}.
\end{equation}
The non-zero matrices are given by
\begin{eqnarray}
&&H_{dd;dd} \nonumber \\
&=&
-\frac{1}{2\lambda\bar{\rho}}(\frac{\partial}{\partial \bar{\tau}})^2
-\int_0^{2\pi} d \bar{\sigma} \frac{\sqrt{\bar{h}}}{\bar{e}^2} (\frac{\partial}{\partial X^{\mu}(\bar{\tau})})^2 
+\frac{10-4D}{2-D}\int_0^{2\pi} d \bar{\sigma} \frac{\sqrt{\bar{h}}}{\bar{e}^2} 
\partial_{\bar{\sigma}}X^{\mu}(\bar{\tau})\partial_{\bar{\sigma}}X_{\mu}(\bar{\tau}), \nonumber \\
&&
\end{eqnarray}

\begin{eqnarray}
&&H_{dd;(\mu, \bar{\sigma})(\mu', \bar{\sigma}')}
=
H_{(\mu, \bar{\sigma})(\mu', \bar{\sigma}');dd}\nonumber \\
&=&
\delta_{(\mu, \bar{\sigma})(\mu', \bar{\sigma}')}
\Bigl(-\frac{1}{2\lambda\bar{\rho}}(\frac{\partial}{\partial \bar{\tau}})^2
-\int_0^{2\pi} d \bar{\sigma} \frac{\sqrt{\bar{h}}}{\bar{e}^2} (\frac{\partial}{\partial X^{\mu}(\bar{\tau})})^2 
+\frac{2D-4}{(2-D)^2}\int_0^{2\pi} d \bar{\sigma} \frac{\sqrt{\bar{h}}}{\bar{e}^2} 
\partial_{\bar{\sigma}}X^{\mu}(\bar{\tau})\partial_{\bar{\sigma}}X_{\mu}(\bar{\tau}) 
\Bigr) \nonumber \\
&+& \frac{12-4D}{2-D}
\bar{h}^{\frac{1}{4}}\bar{e}^{-\frac{3}{2}}(\bar{\sigma})
\epsilon_{\mu\nu} \partial_{\bar{\sigma}} X^{\nu}(\bar{\tau})
\bar{h}^{\frac{1}{4}}\bar{e}^{-\frac{3}{2}}(\bar{\sigma}')
\epsilon_{\mu'\nu'} \partial_{\bar{\sigma}'} X^{\nu'}(\bar{\tau}),
\end{eqnarray}

\begin{eqnarray}
&&H_{d(\mu, \bar{\sigma});d(\mu', \bar{\sigma}')}\nonumber \\&=&
\delta_{(\mu, \bar{\sigma})(\mu', \bar{\sigma}')}
\Bigl(-\frac{1}{2\lambda\bar{\rho}}(\frac{\partial}{\partial \bar{\tau}})^2
-\int_0^{2\pi} d \bar{\sigma} \frac{\sqrt{\bar{h}}}{\bar{e}^2} (\frac{\partial}{\partial X^{\mu}(\bar{\tau})})^2 
+\frac{D^2-5D+8}{(2-D)^2}\int_0^{2\pi} d \bar{\sigma} \frac{\sqrt{\bar{h}}}{\bar{e}^2} 
\partial_{\bar{\sigma}}X^{\mu}(\bar{\tau})\partial_{\bar{\sigma}}X_{\mu}(\bar{\tau}) \Bigr) \nonumber \\
&+& \frac{-6+2D}{2-D}
\bar{h}^{\frac{1}{4}}\bar{e}^{-\frac{3}{2}}(\bar{\sigma})
\epsilon_{\mu\nu} \partial_{\bar{\sigma}} X^{\nu}(\bar{\tau})
\bar{h}^{\frac{1}{4}}\bar{e}^{-\frac{3}{2}}(\bar{\sigma}')
\epsilon_{\mu'\nu'} \partial_{\bar{\sigma}'} X^{\nu'}(\bar{\tau}),
\end{eqnarray}

\begin{eqnarray}
&&H_{(\mu, \bar{\sigma})(\mu', \bar{\sigma}');(\mu'', \bar{\sigma}'')(\mu''', \bar{\sigma}''')}\nonumber \\
&=&
(\delta_{(\mu, \bar{\sigma})(\mu'', \bar{\sigma}'')}
\delta_{(\mu', \bar{\sigma}')(\mu''', \bar{\sigma}''')}
+
\delta_{(\mu, \bar{\sigma})(\mu''', \bar{\sigma}''')}
\delta_{(\mu', \bar{\sigma}')(\mu'', \bar{\sigma}'')})
\nonumber \\
&&
\Bigl(-\frac{1}{2\lambda\bar{\rho}}(\frac{\partial}{\partial \bar{\tau}})^2
-\int_0^{2\pi} d \bar{\sigma} \frac{\sqrt{\bar{h}}}{\bar{e}^2} (\frac{\partial}{\partial X^{\mu}(\bar{\tau})})^2 
+\frac{4}{(2-D)^2}\int_0^{2\pi} d \bar{\sigma} \frac{\sqrt{\bar{h}}}{\bar{e}^2} 
\partial_{\bar{\sigma}}X^{\mu}(\bar{\tau})\partial_{\bar{\sigma}}X_{\mu}(\bar{\tau}) \Bigr) \nonumber \\
&-&
\delta_{(\mu, \bar{\sigma})(\mu', \bar{\sigma}')}
\delta_{(\mu'', \bar{\sigma}'')(\mu''', \bar{\sigma}''')}
\nonumber \\
&& 
\Bigl(-\frac{1}{2\lambda\bar{\rho}}(\frac{\partial}{\partial \bar{\tau}})^2
-\int_0^{2\pi} d \bar{\sigma} \frac{\sqrt{\bar{h}}}{\bar{e}^2} (\frac{\partial}{\partial X^{\mu}(\bar{\tau})})^2 \nonumber 
+\frac{2}{2-D}\int_0^{2\pi} d \bar{\sigma} \frac{\sqrt{\bar{h}}}{\bar{e}^2} 
\partial_{\bar{\sigma}}X^{\mu}(\bar{\tau})\partial_{\bar{\sigma}}X_{\mu}(\bar{\tau}) \Bigr) \nonumber \\
&+&
\delta_{(\mu, \bar{\sigma})(\mu', \bar{\sigma}')}
\frac{8}{(2-D)^2}
\bar{h}^{\frac{1}{4}}\bar{e}^{-\frac{3}{2}}(\bar{\sigma}'')
\epsilon_{\mu''\nu''} \partial_{\bar{\sigma}''} X^{\nu''}(\bar{\tau})
\bar{h}^{\frac{1}{4}}\bar{e}^{-\frac{3}{2}}(\bar{\sigma}''')
\epsilon_{\mu'''\nu'''} \partial_{\bar{\sigma}'''} X^{\nu'''}(\bar{\tau}) 
\nonumber \\
&+&
\delta_{(\mu'', \bar{\sigma}'')(\mu''', \bar{\sigma}''')}
\frac{8}{(2-D)^2}
\bar{h}^{\frac{1}{4}}\bar{e}^{-\frac{3}{2}}(\bar{\sigma})
\epsilon_{\mu\nu} \partial_{\bar{\sigma}} X^{\nu}(\bar{\tau})
\bar{h}^{\frac{1}{4}}\bar{e}^{-\frac{3}{2}}(\bar{\sigma}')
\epsilon_{\mu'\nu'} \partial_{\bar{\sigma}'} X^{\nu'}(\bar{\tau}) 
\nonumber \\
&+&
\delta_{(\mu''', \bar{\sigma}''')(\mu', \bar{\sigma}')}
\frac{8}{(2-D)^2}
\bar{h}^{\frac{1}{4}}\bar{e}^{-\frac{3}{2}}(\bar{\sigma}'')
\epsilon_{\mu''\nu''} \partial_{\bar{\sigma}''} X^{\nu''}(\bar{\tau})
\bar{h}^{\frac{1}{4}}\bar{e}^{-\frac{3}{2}}(\bar{\sigma})
\epsilon_{\mu\nu} \partial_{\bar{\sigma}} X^{\nu}(\bar{\tau}) 
\nonumber \\
&+&
\delta_{(\mu, \bar{\sigma})(\mu'', \bar{\sigma}'')}
\frac{8}{(2-D)^2}
\bar{h}^{\frac{1}{4}}\bar{e}^{-\frac{3}{2}}(\bar{\sigma}')
\epsilon_{\mu'\nu'} \partial_{\bar{\sigma}'} X^{\nu'}(\bar{\tau})
\bar{h}^{\frac{1}{4}}\bar{e}^{-\frac{3}{2}}(\bar{\sigma}''')
\epsilon_{\mu'''\nu'''} \partial_{\bar{\sigma}'''} X^{\nu'''}(\bar{\tau}) 
\nonumber \\
&+&
\delta_{(\mu, \bar{\sigma})(\mu''', \bar{\sigma}''')}
\frac{8}{(2-D)^2}
\bar{h}^{\frac{1}{4}}\bar{e}^{-\frac{3}{2}}(\bar{\sigma}'')
\epsilon_{\mu''\nu''} \partial_{\bar{\sigma}''} X^{\nu''}(\bar{\tau})
\bar{h}^{\frac{1}{4}}\bar{e}^{-\frac{3}{2}}(\bar{\sigma}')
\epsilon_{\mu'\nu'} \partial_{\bar{\sigma}'} X^{\nu'}(\bar{\tau}) 
\nonumber \\
&+&
\delta_{(\mu'', \bar{\sigma}'')(\mu', \bar{\sigma}')}
\frac{8}{(2-D)^2}
\bar{h}^{\frac{1}{4}}\bar{e}^{-\frac{3}{2}}(\bar{\sigma})
\epsilon_{\mu\nu} \partial_{\bar{\sigma}} X^{\nu}(\bar{\tau})
\bar{h}^{\frac{1}{4}}\bar{e}^{-\frac{3}{2}}(\bar{\sigma}''')
\epsilon_{\mu'''\nu'''} \partial_{\bar{\sigma}'''} X^{\nu'''}(\bar{\tau}).
\end{eqnarray}

A part of the action 
\begin{eqnarray}
\int \mathcal{D} h \mathcal{D} X(\bar{\tau})\mathcal{D} \bar{\tau} \frac{1}{8}
\int_0^{2\pi}d\bar{\sigma}\bar{e}
\int_0^{2\pi}d\bar{\sigma}'\bar{e}'
 \tilde{H}_{d(\mu \bar{\sigma})} 
H_{d(\mu, \bar{\sigma});d(\mu', \bar{\sigma}')} 
\tilde{H}_{d(\mu' \bar{\sigma}')}
\end{eqnarray}
decouples from the other modes.
Here we use Einstein notation with respect to $(\mu, \bar{\sigma})$.
By using projection of $\tilde{H}_{d(\mu \bar{\sigma})}$ on
$\bar{h}^{\frac{1}{4}}\bar{e}^{-\frac{3}{2}}(\bar{\sigma})
\epsilon_{\mu\nu} \partial_{\bar{\sigma}} X^{\nu}(\bar{\tau})$,
we obtain $\tilde{H}^0_{d(\mu \bar{\sigma})}$.
Then, $\tilde{H}_{d(\mu \bar{\sigma})}=\tilde{H}^0_{d(\mu \bar{\sigma})}+\tilde{H}^1_{d(\mu \bar{\sigma})}$
where
\begin{eqnarray}
&&\bar{h}^{\frac{1}{4}}\bar{e}^{-\frac{3}{2}}(\bar{\sigma})
\epsilon_{\mu\nu} \partial_{\bar{\sigma}} X^{\nu}(\bar{\tau})
\tilde{H}_{d(\mu \bar{\sigma})}
=
\bar{h}^{\frac{1}{4}}\bar{e}^{-\frac{3}{2}}(\bar{\sigma})
\epsilon_{\mu\nu} \partial_{\bar{\sigma}} X^{\nu}(\bar{\tau})
\tilde{H}^0_{d(\mu \bar{\sigma})} \nonumber \\
&&\tilde{H}^0_{d(\mu \bar{\sigma})}
\tilde{H}^1_{d(\mu \bar{\sigma})}=0 \label{projection}
\end{eqnarray}

There exists a decomposition
$\tilde{H}^1_{d(\mu \bar{\sigma})}=\tilde{H}^{\bot}_{d(\mu \bar{\sigma})}
+\tilde{H}'_{d(\mu \bar{\sigma})}$
such that
\begin{eqnarray}
&&\tilde{H}^{\bot}_{d(\mu \bar{\sigma})}
\tilde{H}^0_{d(\mu \bar{\sigma})}=0 \nonumber \\
&&\partial_I\tilde{H}^{\bot}_{d(\mu \bar{\sigma})}
\partial_I\tilde{H}^0_{d(\mu \bar{\sigma})}=0\nonumber \\
&&
\tilde{H}^{\bot}_{d(\mu \bar{\sigma})}
\tilde{H}'_{d(\mu \bar{\sigma})}=0\nonumber \\
&&
\partial_I\tilde{H}^{\bot}_{d(\mu \bar{\sigma})}
\partial_I\tilde{H}'_{d(\mu \bar{\sigma})}=0. \label{DecoupleEquations}
\end{eqnarray}
\begin{proof}
\begin{eqnarray}
&&\tilde{H}^{\bot}_{d(\mu \bar{\sigma})}(X^I_0)
\tilde{H}^0_{d(\mu \bar{\sigma})}(X^I_0)=0 \nonumber \\
&&
\tilde{H}^{\bot}_{d(\mu \bar{\sigma})}(X^I_0)
\tilde{H}'_{d(\mu \bar{\sigma})}(X^I_0)=0,
\label{DecomposeCondition1}
\end{eqnarray}
and
\begin{eqnarray}
&&\partial_I\tilde{H}^{\bot}_{d(\mu \bar{\sigma})}(X^I_0)
\partial_I\tilde{H}^0_{d(\mu \bar{\sigma})}(X^I_0)=0\nonumber \\
&&
\partial_I\tilde{H}^{\bot}_{d(\mu \bar{\sigma})}(X^I_0)
\partial_I\tilde{H}'_{d(\mu \bar{\sigma})}(X^I_0)=0, \label{DecomposeCondition2}
\end{eqnarray}
are necessary so that (\ref{DecoupleEquations}) are satisfied at an arbitrary point $X^I=X^I_0$. (\ref{DecomposeCondition1}) are the enough conditions for $\tilde{H}^{\bot}_{d(\mu \bar{\sigma})}(X^I_0)$, whereas there are necessary conditions for
$\partial_I\tilde{H}^{\bot}_{d(\mu \bar{\sigma})}(X^I_0)$:
\begin{eqnarray}
&&\tilde{H}^{\bot}_{d(\mu \bar{\sigma})}(X^I_0+dX^I)
\tilde{H}^0_{d(\mu \bar{\sigma})}(X^I_0+dX^I)=0 \nonumber \\
&&
\tilde{H}^{\bot}_{d(\mu \bar{\sigma})}(X^I_0+dX^I)
\tilde{H}'_{d(\mu \bar{\sigma})}(X^I_0+dX^I)=0,  \label{DecomposeCondition3}
\end{eqnarray}
in addition to (\ref{DecomposeCondition2}).  (\ref{DecomposeCondition2}) and (\ref{DecomposeCondition3}) are the enough conditions for $\partial_I\tilde{H}^{\bot}_{d(\mu \bar{\sigma})}(X^I_0)$. 

Because (\ref{DecomposeCondition1}) are 2 equations for D variables $\tilde{H}^{\bot}_{d(\mu \bar{\sigma})}(X^I_0)$, there exist solutions. Because 
(\ref{DecomposeCondition2}) and (\ref{DecomposeCondition3}) 
are 2+2(D+1) equations for D(D+1) variables $\partial_I\tilde{H}^{\bot}_{d(\mu \bar{\sigma})}(X^I_0)$, there also exist solutions. 
Thus, there exists $\tilde{H}^{\bot}_{d(\mu \bar{\sigma})}$ that satisfies (\ref{DecoupleEquations}) everywhere.
\end{proof}
By using (\ref{projection}) and (\ref{DecoupleEquations}), we obtain
\begin{eqnarray}
&&\int \mathcal{D} h \mathcal{D} X(\bar{\tau})\mathcal{D} \bar{\tau} \frac{1}{8}
\int_0^{2\pi}d\bar{\sigma}\bar{e}
\int_0^{2\pi}d\bar{\sigma}'\bar{e}'
\tilde{H}_{d(\mu \bar{\sigma})} 
H_{d(\mu, \bar{\sigma});d(\mu', \bar{\sigma}')} 
\tilde{H}_{d(\mu \bar{\sigma})} \nonumber \\
&=&
\int \mathcal{D} h \mathcal{D} X(\bar{\tau})\mathcal{D} \bar{\tau} \frac{1}{4}
\Bigl(
\int_0^{2\pi}d\bar{\sigma} \tilde{H}^{\bot}_{d(\mu \bar{\sigma})}
H
\tilde{H}^{\bot}_{d(\mu \bar{\sigma})}
+
\int_0^{2\pi}d\bar{\sigma}(\tilde{H}^0_{d(\mu \bar{\sigma})}+\tilde{H}'_{d(\mu \bar{\sigma})}) 
H
(\tilde{H}^0_{d(\mu \bar{\sigma})}+\tilde{H}'_{d(\mu \bar{\sigma})}) \nonumber \\
&-&
\frac{7-D}{2-D}
\bigl(\int_0^{2\pi}d\bar{\sigma}
\bar{h}^{\frac{1}{4}}\bar{e}^{-\frac{1}{2}}(\bar{\sigma})
\epsilon_{\mu\nu} \partial_{\bar{\sigma}} X^{\nu}(\bar{\tau})
\tilde{H}^0_{d(\mu \bar{\sigma})}\bigr)^2
\Bigr),
\end{eqnarray}
where
\begin{eqnarray}
H
&=&
-\frac{1}{2}\frac{1}{2\lambda\bar{\rho}}(\frac{\partial}{\partial \bar{\tau}})^2
-\frac{1}{2}\int_0^{2\pi} d \bar{\sigma} \frac{\sqrt{\bar{h}}}{\bar{e}^2} (\frac{\partial}{\partial X^{\mu}(\bar{\tau})})^2 
+\frac{1}{2}\frac{D^2-5D+8}{(2-D)^2}\int_0^{2\pi} d \bar{\sigma} \frac{\sqrt{\bar{h}}}{\bar{e}^2} \partial_{\bar{\sigma}}X^{\mu}(\bar{\tau})\partial_{\bar{\sigma}}X_{\mu}(\bar{\tau}). 
\nonumber \\ 
\end{eqnarray}
As a result, a part of the action 
\begin{equation}
\int \mathcal{D} h \mathcal{D} X(\bar{\tau})\mathcal{D} \bar{\tau} \frac{1}{4}
\int_0^{2\pi}d\bar{\sigma} \tilde{H}^{\bot}_{d(\mu \bar{\sigma})} 
H
\tilde{H}^{\bot}_{d(\mu \bar{\sigma})} \label{SecondOrderAction}
\end{equation}
decouples from the other modes.

By adding to (\ref{SecondOrderAction})
\begin{eqnarray}
0
&=&\int \mathcal{D} h \mathcal{D} X(\bar{\tau})\mathcal{D} \bar{\tau} 
\int_0^{2\pi}d\bar{\sigma}' 
\tilde{H}^{\bot}_{d(\mu \bar{\sigma}')} 
( \int_0^{2\pi} d \bar{\sigma}
\frac{1}{4}
\bar{n}^{\bar{\sigma}}
\partial_{\bar{\sigma}} X^{\mu}(\bar{\tau})  \frac{\partial}{\partial X^{\mu}(\bar{\tau})})
\tilde{H}^{\bot}_{d(\mu \bar{\sigma}')}, \label{zero}
\end{eqnarray}
where $\bar{n}^{\bar{\sigma}}(\bar{\sigma}, \bar{\tau})$ is the shift vector in the ADM formalism, summarized in the appendix A, 
we obtain (\ref{SecondOrderAction})
with 
\begin{eqnarray}
&&H(-i\frac{\partial}{\partial \bar{\tau}}, -i\frac{1}{\bar{e}}\frac{\partial}{\partial X(\bar{\tau})}, X(\bar{\tau}), \bar{h}) \nonumber \\
&=&
\frac{1}{2}\frac{1}{2\lambda\bar{\rho}}(-i\frac{\partial}{\partial \bar{\tau}})^2
\nonumber \\
&+&\int_0^{2\pi} d\bar{\sigma} \left( \sqrt{\bar{h}} \left(\frac{1}{2}(-i\frac{1}{\bar{e}}\frac{\partial}{\partial X^{\mu}(\bar{\tau})})^2+\frac{1}{2} \bar{e}^{-2} (\partial_{\bar{\sigma}}X^{\mu}(\bar{\tau}))^2 \right)
+i\bar{e} \bar{n}^{\bar{\sigma}} \partial_{\bar{\sigma}} X_{\mu}(\bar{\tau}) (-i\frac{1}{\bar{e}}\frac{\partial}{\partial X^{\mu}(\bar{\tau})})\right), \nonumber \\ \label{bosonicHamiltonian}
\end{eqnarray}
where we have taken $D \to \infty$.
(\ref{zero}) is true because 
\begin{eqnarray}
(r.h.s)&=&
\int \mathcal{D} h \mathcal{D} X(\bar{\tau})\mathcal{D} \bar{\tau} 
( \int_0^{2\pi} d \bar{\sigma}
\frac{1}{8}
\bar{n}^{\bar{\sigma}}
\partial_{\bar{\sigma}} X^{\mu}(\bar{\tau})  \frac{\partial}{\partial X^{\mu}(\bar{\tau})})
\int_0^{2\pi}d\bar{\sigma}'' 
\tilde{H}^{\bot}_{d(\mu \bar{\sigma}'')} 
\tilde{H}^{\bot}_{d(\mu \bar{\sigma}'')}
\nonumber \\
&=&-\int \mathcal{D} h \mathcal{D} X\mathcal{D}(\bar{\tau}) \bar{\tau}  \int_0^{2\pi} d \bar{\sigma} \lim_{\bar{\sigma}' \to \bar{\sigma}} \frac{1}{8} 
\frac{\partial}{\partial X^{\mu}(\bar{\tau}, \bar{\sigma}')}
\left(\partial_{\bar{\sigma}}X^{\mu} (\bar{\tau}, \bar{\sigma}) \right) 
\bar{n}^{\bar{\sigma}}(\bar{\sigma})
\int_0^{2\pi}d\bar{\sigma}'' 
\tilde{H}^{\bot}_{d(\mu \bar{\sigma}'')} 
\tilde{H}^{\bot}_{d(\mu \bar{\sigma}'')}\nonumber \\
&=&-\int \mathcal{D} h \mathcal{D} X(\bar{\tau})\mathcal{D} \bar{\tau}  \int_0^{2\pi} d \bar{\sigma} \lim_{\bar{\sigma}' \to \bar{\sigma}} \frac{d}{8} 
\partial_{\bar{\sigma}}\delta(\bar{\sigma}-\bar{\sigma}') 
\bar{n}^{\bar{\sigma}}(\bar{\sigma})
\int_0^{2\pi}d\bar{\sigma}'' 
\tilde{H}^{\bot}_{d(\mu \bar{\sigma}'')} 
\tilde{H}^{\bot}_{d(\mu \bar{\sigma}'')}\nonumber \\
&=&0.
\end{eqnarray}

The propagator for $\tilde{H}^{\bot}_{d(\mu \bar{\sigma})}$; 
\begin{equation}
\Delta_F(\bar{h}, X(\bar{\tau}), \bar{\tau}; \; \bar{h},'  X'(\bar{\tau}'), \bar{\tau},')=<\tilde{H}^{\bot}_{d(\mu \bar{\sigma})} (\bar{h}, X(\bar{\tau}), \bar{\tau})
\tilde{H}^{\bot}_{d(\mu \bar{\sigma})}(\bar{h},'  X'(\bar{\tau}'), \bar{\tau}')>
\end{equation}
satisfies
\begin{equation}
H(-i\frac{\partial}{\partial \bar{\tau}}, -i\frac{1}{\bar{e}}\frac{\partial}{\partial X(\bar{\tau})}, X(\bar{\tau}), \bar{h})
\Delta_F(\bar{h}, X(\bar{\tau}), \bar{\tau}; \; \bar{h},'  X'(\bar{\tau}'), \bar{\tau},')
=
\delta(\bar{h}-\bar{h}') \delta(X(\bar{\tau})-X'(\bar{\tau}'))\delta(\bar{\tau}-\bar{\tau}').
\end{equation}
In order to obtain a Schwinger representation of the propagator, we use the operator formalism $(\hat{\bar{h}}, \hat{X}(\hat{\bar{\tau}}), \hat{\bar{\tau}})$ of the first quantization, whereas the conjugate momentum is written as $(\hat{p}_{\bar{h}},  \hat{p}_{X}(\bar{\tau}), \hat{p}_{\bar{\tau}})$. The eigen state is given by $|\bar{h}, X(\bar{\tau}), \bar{\tau}>$. 
First,
\begin{eqnarray}
&&<\bar{h}, X(\bar{\tau}), \bar{\tau}| \hat{H}(\hat{p}_{\bar{\tau}}, \hat{p}_{X}(\bar{\tau}), \hat{X}(\hat{\bar{\tau}}), \hat{\bar{h}}) |\bar{h},' X'(\bar{\tau}'), \bar{\tau}' > 
\nonumber \\
&=&
H(-i\frac{\partial}{\partial \bar{\tau}}, -i\frac{1}{\bar{e}}\frac{\partial}{\partial X(\bar{\tau})}, X(\bar{\tau}), \bar{h})
\delta(\bar{h}-\bar{h}') \delta(X(\bar{\tau})-X'(\bar{\tau}'))\delta(\bar{\tau}-\bar{\tau}'),
\end{eqnarray}
because
\begin{eqnarray}
(l.h.s.)&=&\int \mathcal{D}p_{\bar{h}} \mathcal{D}p_{\bar{\tau}} \mathcal{D}p_{X}(\bar{\tau})
<\bar{h}, X(\bar{\tau}), \bar{\tau}| \hat{H}(\hat{p}_{\bar{\tau}}, \hat{p}_{X}(\bar{\tau}), \hat{X}(\hat{\bar{\tau}}), \hat{\bar{h}}) |p_{\bar{h}},  p_{X}(\bar{\tau}), p_{\bar{\tau}}>
\nonumber \\
&&<p_{\bar{h}},  p_{X}(\bar{\tau}), p_{\bar{\tau}}|\bar{h},' X'(\bar{\tau}'), \bar{\tau}' > \nonumber \\
&=&\int \mathcal{D}p_{\bar{h}} \mathcal{D}p_{\bar{\tau}} \mathcal{D}p_{X}(\bar{\tau}) 
H(p_{\bar{\tau}}, p_{X}(\bar{\tau}), X(\bar{\tau}), \bar{h})
<\bar{h}, X(\bar{\tau}), \bar{\tau} |p_{\bar{h}},  p_{X}(\bar{\tau}), p_{\bar{\tau}}>
\nonumber \\
&&<p_{\bar{h}},  p_{X}(\bar{\tau}), p_{\bar{\tau}}|\bar{h},' X'(\bar{\tau}'), \bar{\tau}' > \nonumber \\
&=&\int \mathcal{D}p_{\bar{h}} \mathcal{D}p_{\bar{\tau}} \mathcal{D}p_{X}(\bar{\tau})
H(p_{\bar{\tau}}, p_{X}(\bar{\tau}), X(\bar{\tau}), \bar{h})
e^{i p_{\bar{h}}\cdot(\bar{h}-\bar{h}')+i p_{\bar{\tau}}(\bar{\tau}-\bar{\tau}')+i p_{X}(\bar{\tau})\cdot(X(\bar{\tau})-X'(\bar{\tau}'))} \nonumber \\
&=&(r.h.s.),
\end{eqnarray}
where $p_{X}(\bar{\tau}) \cdot X(\bar{\tau}):= \int d\bar{\sigma} \bar{e} p_{X}^{\mu}(\bar{\tau}) X_{\mu}(\bar{\tau})$.
By using this, we obtain
\begin{equation}
\Delta_F(\bar{h}, X(\bar{\tau}), \bar{\tau}; \; \bar{h},'  X'(\bar{\tau}'), \bar{\tau},')
=
<\bar{h}, X(\bar{\tau}), \bar{\tau}| \hat{H}^{-1}(\hat{p}_{\bar{\tau}}, \hat{p}_{X}(\bar{\tau}), \hat{X}(\hat{\bar{\tau}}), \hat{\bar{h}}) |\bar{h},' X'(\bar{\tau}'), \bar{\tau}' >,
\label{InverseH}
\end{equation}
because
\begin{eqnarray}
&&\delta(\bar{h}-\bar{h}') \delta(X(\bar{\tau})-X'(\bar{\tau}'))\delta(\bar{\tau}-\bar{\tau}') \nonumber \\
&=&
<\bar{h}, X(\bar{\tau}), \bar{\tau}|\bar{h},' X'(\bar{\tau}'), \bar{\tau}' > \nonumber \\
&=&
<\bar{h}, X(\bar{\tau}), \bar{\tau}|\hat{H}\hat{H}^{-1}|\bar{h},' X'(\bar{\tau}'), \bar{\tau}' > \nonumber \\
&=&
\int dh'' d\bar{\tau}'' dX''(\bar{\tau}'')
<\bar{h}, X(\bar{\tau}), \bar{\tau}|\hat{H}|h'', X''(\bar{\tau}''), \bar{\tau}''><h'', X''(\bar{\tau}''), \bar{\tau}''|\hat{H}^{-1}|\bar{h},' X'(\bar{\tau}'), \bar{\tau}' > \nonumber \\
&=&
H(-i\frac{\partial}{\partial \bar{\tau}}, -i\frac{1}{\bar{e}}\frac{\partial}{\partial X(\bar{\tau})}, X(\bar{\tau}), \bar{h})
\int dh'' d\bar{\tau}'' dX''(\bar{\tau}'') 
\delta(\bar{h}-\bar{h}'')\delta(\bar{\tau}-\bar{\tau}'')\delta(X(\bar{\tau})-X''(\bar{\tau}'')) \nonumber \\
&&<h'', X''(\bar{\tau}''), \bar{\tau}''|\hat{H}^{-1}|\bar{h},' X'(\bar{\tau}'), \bar{\tau}' >
\nonumber \\
&=&
H(-i\frac{\partial}{\partial \bar{\tau}}, -i\frac{1}{\bar{e}}\frac{\partial}{\partial X(\bar{\tau})}, X(\bar{\tau}), \bar{h})
<\bar{h}, X(\bar{\tau}), \bar{\tau}|\hat{H}^{-1}|\bar{h},' X'(\bar{\tau}'), \bar{\tau}' >. 
\end{eqnarray}
On the other hand,
\begin{eqnarray}
\hat{H}^{-1}=  \int _0^{\infty} dT e^{-T\hat{H}}, \label{IntegralFormula}
\end{eqnarray}
because
\begin{equation}
\lim_{\epsilon \to 0+} \int _0^{\infty} dT e^{-T(\hat{H}+\epsilon)}
=
\lim_{\epsilon \to 0+}  \left[\frac{1}{-(\hat{H}+\epsilon)} e^{-T(\hat{H}+\epsilon)}
\right]_0^{\infty}
=\hat{H}^{-1}.
\end{equation}
This fact and (\ref{InverseH}) imply
\begin{equation}
\Delta_F(\bar{h}, X(\bar{\tau}), \bar{\tau}; \; \bar{h},'  X'(\bar{\tau}'), \bar{\tau},')
=
\int _0^{\infty} dT <\bar{h}, X(\bar{\tau}), \bar{\tau}|  e^{-T\hat{H}} |\bar{h},' X'(\bar{\tau}'), \bar{\tau}' >.
\end{equation}
In order to define two-point correlation functions that is invariant under the general coordinate transformations in the string geometry, we define in and out states as
\begin{eqnarray}
||X_i \,|\,h_f, ; h_i>_{in}&:=& \int_{h_i}^{h_f} \mathcal{D}h'|\bar{h},' X_i:=X'(\bar{\tau}'=-\infty), \bar{\tau}'=-\infty > \nonumber \\
<X_f\,|\,h_f, ; h_i||_{out}&:=& \int_{h_i}^{h_f} \mathcal{D} h <\bar{h}, X_f:=X(\bar{\tau}=\infty), \bar{\tau}=\infty|,
\end{eqnarray}
where $h_i$ and $h_f$ represent the metrics of the cylinders at $\bar{\tau}=\pm \infty$, respectively. When we insert asymptotic states later, we integrate out $X_f$, $X_i$, $h_f$ and $h_i$ in the two-point correlation function for these states;
\begin{eqnarray}
\Delta_F(X_f; X_i|h_f, ; h_i) 
&=&
\int_{h_i}^{h_f} \mathcal{D} h \int_{h_i}^{h_f} \mathcal{D}h'
<\tilde{H}^{\bot}_{d(\mu \bar{\sigma})} (\bar{h}, X_f:=X(\bar{\tau}=\infty), \bar{\tau}=\infty) \nonumber \\
&&
\tilde{H}^{\bot}_{d(\mu \bar{\sigma})}(\bar{h},' X_i:=X'(\bar{\tau}'=-\infty), \bar{\tau}'=-\infty)>. 
\end{eqnarray}
This can be written as\footnote{The correlation function is zero if $h_i$ and $h_f$ of the in state do not coincide with those of the out states, because of the delta functions in the sixth line.}
\begin{eqnarray}
&&\Delta_F(X_f; X_i|h_f, ; h_i) \nonumber \\
&:=&\int _0^{\infty} dT <X_f \,|\,h_f, ; h_i||_{out}  e^{-T\hat{H}} ||X_i \,|\,h_f, ; h_i>_{in}\nonumber \\
&=&\int _0^{\infty} dT  \lim_{N \to \infty} \int_{h_i}^{h_f} \mathcal{D} h \int_{h_i}^{h_f} \mathcal{D} h' 
\prod_{n=1}^N \int d \bar{h}_{n} dX_n(\bar{\tau}_n) d\bar{\tau}_n 
\nonumber \\
&&\prod_{m=0}^N <\bar{h}_{m+1}, X_{m+1}(\bar{\tau}_{m+1}), \bar{\tau}_{m+1}| e^{-\frac{1}{N}T \hat{H}} |\bar{h}_{m}, X_{m}(\bar{\tau}_m), \bar{\tau}_{m}> \nonumber \\
&=&\int _0^{\infty} dT_0 \lim_{N \to \infty} \int d T_{N+1} \int_{h_i}^{h_f} \mathcal{D} h \int_{h_i}^{h_f} \mathcal{D} h' \prod_{n=1}^N \int d T_n d \bar{h}_{n} dX_n(\bar{\tau}_n) d\bar{\tau}_n \nonumber \\
&&\prod_{m=0}^N  <\bar{\tau}_{m+1}, X_{m+1}(\bar{\tau}_{m+1})| e^{-\frac{1}{N}T_m \hat{H}} |\bar{\tau}_{m}, X_{m}(\bar{\tau}_m)>\delta(\bar{h}_{m}-\bar{h}_{m+1})\delta(T_{m}-T_{m+1}) \nonumber \\
&=&\int _0^{\infty} dT_0 \lim_{N \to \infty} d T_{N+1} \int_{h_i}^{h_f} \mathcal{D} h \prod_{n=1}^N \int d T_n  dX_n(\bar{\tau}_n)  d\bar{\tau}_n   \prod_{m=0}^N \int dp_{T_m}  dp_{X_m}(\bar{\tau}_m) dp_{\bar{\tau}_m} \nonumber \\
&&\exp \Biggl(-\sum_{m=0}^N \Delta t
\Bigl(-ip_{T_m} \frac{T_{m}-T_{m+1}}{\Delta t} 
-i p_{\bar{\tau}_m}\frac{\bar{\tau}_{m}-\bar{\tau}_{m+1}}{\Delta t}
-ip_{X_m}(\bar{\tau}_m)\cdot \frac{X_{m}(\bar{\tau}_m)-X_{m+1}(\bar{\tau}_{m+1})}{\Delta t} 
\nonumber \\
&&+T_m H(p_{\bar{\tau}_m}, p_{X_m}(\bar{\tau}_m), X_{m}(\bar{\tau}_m), \bar{h})\Bigr) \Biggr) \nonumber \\
&=&
 \int_{h_i X_i, -\infty}^{h_f, X_f, \infty}  \mathcal{D} h \mathcal{D} X(\bar{\tau}) \mathcal{D}\bar{\tau} 
\int \mathcal{D} T  
\int 
\mathcal{D} p_T
\mathcal{D}p_{X} (\bar{\tau})
\mathcal{D}p_{\bar{\tau}}
 \nonumber \\
&&
\exp \Biggl(- \int_{-\infty}^{\infty} dt \Bigr(
-i p_{T}(t) \frac{d}{dt} T(t) 
-i p_{\bar{\tau}}(t)\frac{d}{dt}\bar{\tau}(t)
-i p_{X}(\bar{\tau}, t)\cdot \frac{d}{dt} X(\bar{\tau}, t)\nonumber \\
&&
+T(t) H(p_{\bar{\tau}}(t), p_{X}(\bar{\tau}, t), X(\bar{\tau}, t), \bar{h})\Bigr) \Biggr),  \label{canonicalform}
\end{eqnarray}
where $\bar{h}_{ 0}=\bar{h}'$, $X_{0}(\bar{\tau}_0)=X_i$, $\bar{\tau}_0=-\infty$, $\bar{h}_{ N+1}=\bar{h}$, $X_{N+1}(\bar{\tau}_{N+1})=X_f$, $\bar{\tau}_{N+1}=\infty$, and $\Delta t:=\frac{1}{\sqrt{N}}$.
A trajectory of points $[\bar{\Sigma}, X(\bar{\tau}), \bar{\tau}]$ is necessarily continuous in $\mathcal{M}_D$ so that the kernel $<\bar{h}_{m+1}, X_{m+1}(\bar{\tau}_{m+1}), \bar{\tau}_{m+1}| e^{-\frac{1}{N}T_m \hat{H}} |\bar{h}_{m}, X_{m}(\bar{\tau}_m), \bar{\tau}_{m}>$ in the third line is non-zero when $N \to \infty$. If we integrate out $p_{\bar{\tau}}(t)$ and $p_{X}(\bar{\tau}, t)$ by using the relation of the ADM formalism in the appendix A, we obtain
\begin{eqnarray}
&&\Delta_F(X_f; X_i|h_f ; h_i) \nonumber \\
&=&
 \int_{h_i X_i, -\infty}^{h_f, X_f, \infty} 
\mathcal{D} T
\mathcal{D} h  \mathcal{D} X(\bar{\tau})\mathcal{D} \bar{\tau} 
\mathcal{D} p_T \exp \Biggl(- \int_{-\infty}^{\infty} dt \Bigl(-i p_{T}(t) \frac{d}{dt} T(t)   +\lambda\bar{\rho}\frac{1}{T(t)}(\frac{d \bar{\tau}(t)}{dt})^2\nonumber \\
&&+\int d\bar{\sigma} \sqrt{\bar{h}} ( 
\frac{1}{2}\bar{h}^{00}\frac{1}{T(t)}\partial_{t} X^{\mu}(\bar{\sigma}, \bar{\tau}, t)\partial_{t} X_{\mu}(\bar{\sigma}, \bar{\tau}, t) +\bar{h}^{01}\partial_{t} X^{\mu}(\bar{\sigma}, \bar{\tau}, t)\partial_{\bar{\sigma}} X_{\mu}(\bar{\sigma}, \bar{\tau}, t) \nonumber \\
&&+\frac{1}{2}\bar{h}^{11}T(t)\partial_{\bar{\sigma}} X^{\mu}(\bar{\sigma}, \bar{\tau}, t)\partial_{\bar{\sigma}} X_{\mu}(\bar{\sigma}, \bar{\tau}, t)
) \Bigr) \Biggr). \label{pathint1}
\end{eqnarray}
The path integral is defined over all possible trajectories $[ \bar{h}, X(t),  \bar{\tau}(t)] \in \mathcal{M}_D$ with fixed boundary values as in Fig. \ref{6Connected}. We should note that the time derivative in (\ref{pathint1}) is in terms of $t$, not $\bar{\tau}$ at this moment. In the following, we will see that $t$ can be fixed to $\bar{\tau}$ by using a reparametrization of $t$ that parametrizes a trajectory.

By inserting
$\int \mathcal{D}c \mathcal{D}b
e^{\int_0^{1} dt \left(\frac{d b(t)}{dt} \frac{d c(t)}{dt}\right)
},$
where $b(t)$ and $c(t)$ are bc ghosts, we obtain 
\begin{eqnarray}
&&\Delta_F(X_f; X_i|h_f ; h_i) \nonumber \\
&=&
Z_0 \int_{h_i X_i, -\infty}^{h_f, X_f, \infty} 
\mathcal{D} T
\mathcal{D} h  \mathcal{D} X(\bar{\tau})\mathcal{D} \bar{\tau} 
\mathcal{D} p_T
\mathcal{D}c \mathcal{D}b  \nonumber \\
&&\exp \Biggl(- \int_{-\infty}^{\infty} dt \Bigl(
-i p_{T}(t) \frac{d}{dt} T(t)  
 +\lambda\bar{\rho}\frac{1}{T(t)}(\frac{d \bar{\tau}(t)}{dt})^2 +\frac{d b(t)}{dt} \frac{d (T(t) c(t))}{dt}\nonumber \\
&&+\int d\bar{\sigma} \sqrt{\bar{h}} ( 
\frac{1}{2}\bar{h}^{00}\frac{1}{T(t)}\partial_{t} X^{\mu}(\bar{\sigma}, \bar{\tau}, t)\partial_{t} X_{\mu}(\bar{\sigma}, \bar{\tau}, t) +\bar{h}^{01}\partial_{t} X^{\mu}(\bar{\sigma}, \bar{\tau}, t)\partial_{\bar{\sigma}} X_{\mu}(\bar{\sigma}, \bar{\tau}, t) \nonumber \\
&&+\frac{1}{2}\bar{h}^{11}T(t)\partial_{\bar{\sigma}} X^{\mu}(\bar{\sigma}, \bar{\tau}, t)\partial_{\bar{\sigma}} X_{\mu}(\bar{\sigma}, \bar{\tau}, t)
) \Bigr) \Biggr), 
\end{eqnarray}
where we have redefined as $c(t) \to T(t) c(t)$. $Z_0$ represents an overall constant factor, and we will rename it $Z_1, Z_2, \cdots$ when the factor changes in the following.
This path integral is obtained if 
\begin{equation}
F_1(t):=\frac{d}{dt}T(t)=0 \label{F1gauge}
\end{equation}
 gauge is chosen in 
\begin{eqnarray}
&&\Delta_F(X_f; X_i|h_f ; h_i) \nonumber \\
&=&
Z_1 \int_{h_i X_i, -\infty}^{h_f, X_f, \infty} 
\mathcal{D} T
\mathcal{D} h  \mathcal{D} X(\bar{\tau})\mathcal{D} \bar{\tau} 
\exp \Biggl(- \int_{-\infty}^{\infty} dt \Bigl(
\lambda\bar{\rho}\frac{1}{T(t)}(\frac{d \bar{\tau}(t)}{dt})^2 \nonumber \\
&&+\int d\bar{\sigma} \sqrt{\bar{h}} ( 
\frac{1}{2}\bar{h}^{00}\frac{1}{T(t)}\partial_{t} X^{\mu}(\bar{\sigma}, \bar{\tau}, t)\partial_{t} X_{\mu}(\bar{\sigma}, \bar{\tau}, t) +\bar{h}^{01}\partial_{t} X^{\mu}(\bar{\sigma}, \bar{\tau}, t)\partial_{\bar{\sigma}} X_{\mu}(\bar{\sigma}, \bar{\tau}, t) \nonumber \\
&&+\frac{1}{2}\bar{h}^{11}T(t)\partial_{\bar{\sigma}} X^{\mu}(\bar{\sigma}, \bar{\tau}, t)\partial_{\bar{\sigma}} X_{\mu}(\bar{\sigma}, \bar{\tau}, t)
) \Bigr) \Biggr), \label{pathint2}
\end{eqnarray}
which has a manifest one-dimensional diffeomorphism symmetry with respect to $t$, where $T(t)$ is transformed as an einbein \cite{Schwinger0}. 

Under $\frac{d\bar{\tau}}{d\bar{\tau}'}=T(t)$, which implies
\begin{eqnarray}
\bar{h}^{00}&=&T^2\bar{h}^{'00} \nonumber \\
\bar{h}^{01}&=&T\bar{h}^{'01} \nonumber \\
\bar{h}^{11}&=&\bar{h}^{'11} \nonumber \\
\sqrt{\bar{h}}&=&\frac{1}{T}\sqrt{\bar{h}'} \nonumber \\
\bar{\rho}&=&\frac{1}{T}\bar{\rho}' \nonumber \\
X^{\mu}(\bar{\sigma}, \bar{\tau}, t)&=&X^{'\mu}(\bar{\sigma}, \bar{\tau}', t)
\nonumber \\
(\frac{d \bar{\tau}(t)}{dt})^2
&=&
T^2
(\frac{d \bar{\tau}'(t)}{dt})^2,
\end{eqnarray}

$T(t)$ disappears in (\ref{pathint2}) and we obtain 
\begin{eqnarray}
&&\Delta_F(X_f; X_i|h_f ; h_i) \nonumber \\
&=&
Z_2 \int_{h_i X_i, -\infty}^{h_f, X_f, \infty} 
\mathcal{D} h  \mathcal{D} X(\bar{\tau})\mathcal{D} \bar{\tau} 
\exp \Biggl(- \int_{-\infty}^{\infty} dt \Bigl(
\lambda\bar{\rho}(\frac{d \bar{\tau}(t)}{dt})^2 \nonumber \\
&&+\int d\bar{\sigma} \sqrt{\bar{h}} ( 
\frac{1}{2}\bar{h}^{00}\partial_{t} X^{\mu}(\bar{\sigma}, \bar{\tau}, t)\partial_{t} X_{\mu}(\bar{\sigma}, \bar{\tau}, t) +\bar{h}^{01}\partial_{t} X^{\mu}(\bar{\sigma}, \bar{\tau}, t)\partial_{\bar{\sigma}} X_{\mu}(\bar{\sigma}, \bar{\tau}, t) \nonumber \\
&&+\frac{1}{2}\bar{h}^{11}\partial_{\bar{\sigma}} X^{\mu}(\bar{\sigma}, \bar{\tau}, t)\partial_{\bar{\sigma}} X_{\mu}(\bar{\sigma}, \bar{\tau}, t)
) \Bigr) \Biggr). \label{pathint3}
\end{eqnarray}
This action is still invariant under the diffeomorphism with respect to t if $\bar{\tau}$ transforms in the same way as $t$.

\begin{figure}[htbp]
\begin{center}
\includegraphics[height=5cm, keepaspectratio, clip]{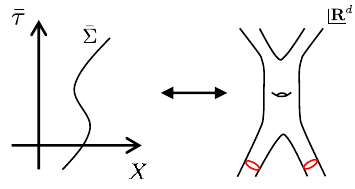}
\end{center}
\caption{A path and a Riemann surface. The line on the left is a trajectory in the path integral. The trajectory parametrized by $\bar{\tau}$ from $\bar{\tau}=-\infty$ to $\bar{\tau}=\infty$, represents a Riemann surface with fixed punctures in $\bold{R}^{d}$ on the right.} 
\label{7Pathintegral}
\end{figure}

If we choose a different gauge
\begin{equation}
F_2(t):=\bar{\tau}-t=0, \label{F2gauge}
\end{equation} 
in (\ref{pathint3}), we obtain 
\begin{eqnarray}
&&\Delta_F(X_f; X_i|h_f ; h_i) \nonumber \\
&=&
Z_3 \int_{h_i X_i, -\infty}^{h_f, X_f, \infty} 
\mathcal{D} h  \mathcal{D} X(\bar{\tau})\mathcal{D} \bar{\tau} 
\mathcal{D} \alpha \mathcal{D}c \mathcal{D}b
\nonumber \\
&&
\exp \Biggl(- \int_{-\infty}^{\infty} dt \Bigl( +\alpha(t) (\bar{\tau}-t) +b(t)c(t)(1-\frac{d \bar{\tau}(t)}{dt})  +\lambda\bar{\rho}(\frac{d \bar{\tau}(t)}{dt})^2 \nonumber \\
&&+\int d\bar{\sigma} \sqrt{\bar{h}} ( 
\frac{1}{2}\bar{h}^{00}\partial_{t} X^{\mu}(\bar{\sigma}, \bar{\tau}, t)\partial_{t} X_{\mu}(\bar{\sigma}, \bar{\tau}, t) +\bar{h}^{01}\partial_{t} X^{\mu}(\bar{\sigma}, \bar{\tau}, t)\partial_{\bar{\sigma}} X_{\mu}(\bar{\sigma}, \bar{\tau}, t) \nonumber \\
&&+\frac{1}{2}\bar{h}^{11}\partial_{\bar{\sigma}} X^{\mu}(\bar{\sigma}, \bar{\tau}, t)\partial_{\bar{\sigma}} X_{\mu}(\bar{\sigma}, \bar{\tau}, t)
) \Bigr) \Biggr) \nonumber \\
&=&
Z\int_{h_i, X_i}^{h_f, X_f} 
\mathcal{D} h  \mathcal{D} X
\exp \Biggl(- \int_{-\infty}^{\infty} d\bar{\tau} 
\int d\bar{\sigma} \sqrt{\bar{h}} ( 
\frac{\lambda }{4\pi}\bar{R}(\bar{\sigma}, \bar{\tau})
\nonumber \\ &&
+\frac{1}{2}\bar{h}^{00}\partial_{\bar{\tau}} X^{\mu}(\bar{\sigma}, \bar{\tau})\partial_{\bar{\tau}} X_{\mu}(\bar{\sigma}, \bar{\tau})
+\bar{h}^{01}\partial_{\bar{\tau}} X^{\mu}(\bar{\sigma}, \bar{\tau})\partial_{\bar{\sigma}} X_{\mu}(\bar{\sigma}, \bar{\tau}) 
 \nonumber \\
&&+\frac{1}{2}\bar{h}^{11}\partial_{\bar{\sigma}} X^{\mu}(\bar{\sigma}, \bar{\tau})\partial_{\bar{\sigma}} X_{\mu}(\bar{\sigma}, \bar{\tau})
) \Biggr). \label{prelastaction}
\end{eqnarray}
The path integral is defined over all possible two-dimensional Riemannian manifolds with fixed punctures in $\bold{R}^{d}$ as in Fig. \ref{7Pathintegral}. The diffeomorphism times Weyl invariance of the action in (\ref{prelastaction}) implies that the correlation function in the string manifold $\mathcal{M}_D$ is given by 
\begin{equation}
\Delta_F(X_f; X_i|h_f ; h_i)
=
Z
\int_{h_i, X_i}^{h_f, X_f} 
\mathcal{D} h  \mathcal{D} X
e^{-\lambda \chi}
e^{-S_{s}}, 
\label{FinalPropagator}
\end{equation}
where
\begin{equation}
S_{s}
=
\int_{-\infty}^{\infty} d\tau \int d\sigma \sqrt{h(\sigma, \tau)} \left(\frac{1}{2} h^{mn} (\sigma, \tau) \partial_m X^{\mu}(\sigma, \tau) \partial_n X_{\mu}(\sigma, \tau) \right),
\end{equation}
and $\chi$ is the Euler number of the two-dimensional Riemannian manifold.

Here, we insert asymptotic states. Punctures exist only at $\bar{\tau} = \pm \infty$. We represent as $V_{j_l}(X_i)(k_l; h_i(l))$, an incoming asymptotic state on the incoming $l$-th cylinder $\Sigma_i(l)$ in $\Sigma$ at $\bar{\tau} \simeq - \infty$, where $h_i(l)$ denotes the metric on $\Sigma_i(l)$ and $l=1, 2, \cdots, m$. Similarly, an outgoing asymptotic state is denoted by $V_{j_{l'}}(X_f)(k_{l'}; h_f(l'))$ at $\bar{\tau} \simeq \infty$, where $l'=m+1, m+2, \cdots, N$. $j_l$ and $j_{l'}$ are the levels, whereas $k^{\mu}_l=-(E_l, \bold{k}_l)$ and $k^{\mu}_{l'}=(E_{l'}, \bold{k}_{l'})$ are the momenta. By the state-operator isomorphism, these states correspond to incoming and outgoing states of vertex operators, $V_{j_l}(X)(k_l, \sigma_l)$ and $V_{j_{l'}}(X)(k_{l'}, \sigma_{l'})$. $\sigma_l$ and $\sigma_{l'}$ are points to which the cylinders $\Sigma_i(l)$ and $\Sigma_f(l')$ are conformally transformed, respectively. By inserting these asymptotic states into the propagator (\ref{FinalPropagator}), we define scattering amplitudes, 
\begin{eqnarray}
&&S_{j_1, j_2, \cdots, j_N}(k_1, k_2, \cdots, k_N) \nonumber \\
&:=&\int d h_f d h_i
d X_f d X_i
\Delta_F(X_f; X_i|h_f ; h_i)/Z V_{diff \times Weyl} \nonumber \\
&& \prod_{l,l'} V_{j_l}(X_i)(k_l; h_i(l))
V_{j_{l'}}(X_f)(k_{l'}; h_f(l'))
\nonumber \\
&=&
\int
\frac{\mathcal{D} h}{V_{diff \times Weyl} }   \mathcal{D} X(\bar{\tau})  e^{-\lambda \chi} e^{-\int d^2 \sigma \sqrt{h}(\frac{1}{2}(\partial_{m}X^{\mu})^2)
} \prod_{l=1}^N\int d^2\sigma_{l} \sqrt{h(\sigma_l)}V_{j_l}(k_l, \sigma_l)
.
\nonumber \\ \label{Smatrix}
\end{eqnarray}
For regularization, we have divided the correlation function by $Z$ and the volume of the diffeomorphism and the Weyl transformation $V_{diff \times Weyl} $, by renormalizing $\tilde{H}^{\bot}_{d(\mu \bar{\sigma})}$.  (\ref{Smatrix}) are the all-order  perturbative scattering amplitudes themselves that possess the moduli in the string theory\cite{textbook}. Especially, in string geometry, the consistency of the perturbation theory around the background (\ref{solution}) determines $d=26$ (the critical dimension).

\section{Superstring geometry}
\setcounter{equation}{0}

In this section, we will define superstring geometry and derive perturbative superstring amplitudes.

 First, let us prepare a moduli space\footnote{Strictly speaking, this should be called a parameter space of integration cycles \cite{NotesOnSupermanifolds, WittenSupermoduli} because superstring worldsheets are defined up to homology.} of type II superstring worldsheets  $\bar{\bold{\Sigma}}$ \cite{NotesOnSupermanifolds, WittenSupermoduli, SuperPeriod} with punctures $P^i$ ($i=1, \cdots, N$)\footnote{$P^i$ not necessarily represents a point, whereas the corresponding $P^i_{red}$ on a reduced space represents a point. A Ramon puncture is located over a R divisor.}. We consider two super Riemann surfaces $\bar{\bold{\Sigma}}_{L}$ and $\bar{\bold{\Sigma}}_{R}$ with Neveu-Schwarz (NS) and Ramond (R) punctures whose reduced spaces $\bar{\Sigma}_{L, red}$ and $\bar{\Sigma}_{R, red}$ are complex conjugates. A reduced space is defined by setting odd variables to zero in a super Riemann surface. The complex conjugates means that  they are complex conjugate spaces with punctures at the same points. There are four types of punctures: NS-NS, NS-R, R-NS, R-R because the punctures in $\bar{\bold{\Sigma}}_{L}$ and $\bar{\bold{\Sigma}}_{R}$ are not necessarily of the same type. A type II superstring worldsheet $\bar{\bold{\Sigma}}$ is defined by the subspace of $\bar{\bold{\Sigma}}_{L} \times \bar{\bold{\Sigma}}_{R}$ whose reduced space $\bar{\Sigma}_{L, red} \times \Sigma_{R, red}$ is restricted to its diagonal $\bar{\Sigma}_{red}$.


Next, let us define global times uniquely on $\bar{\bold{\Sigma}}$ in order to define string states by world-time constant hypersurfaces. If there are R punctures on a super Riemann surface $\bar{\bold{\Sigma}}_{R}$, the superconformal structures have singularities on the R divisors \cite{NotesOnSupermanifolds, WittenSupermoduli, SuperPeriod}. On a R divisor, a closed holomorphic 1-form takes the form, $\mu=\frac{w}{\sqrt{2\pi i}}d\theta$ (mod $z$), which is uniquely determined  by an odd period $w$.  On any other point, it takes the form, $\mu=b(z)dz+d(\theta \alpha(z))$. One can define an even period $\oint_{S} \gamma^*(\mu)$ on a cycle $S$ with dimension $1|0$, where $\gamma^*$ is a pullback by a map $\gamma: S \to \bar{\bold{\Sigma}}_R$. A- and B-periods on $\bar{\bold{\Sigma}}_{R}$ are defined by those on the reduced space because $\oint_{S} \gamma^*(\mu)=\oint_{S_{red}}b(z)dz$. The periods do not depend on a choice of reduced space because they only depend on the homology class determined by the map $\gamma$ if $\mu$ is closed. Because the period of $d(\theta \alpha(z))$ vanishes, we take the quotient of the space of 1-forms by the subspace consisting of those whose periods vanish, and thus we have $\mu=b(z)dz$. As a result, on $\bar{\bold{\Sigma}}_{R}$ except for R-divisors, a closed holomorphic 1-form is uniquely determined by even periods on the complete basis of A- and B-cycles.

 Therefore, on $\bar{\bold{\Sigma}}_{R}$, there exists an unique Abelian differential $dp$ that has simple poles with residues $f^i$ where $\sum_i f^i=0$, at $P^i$\footnote{
The odd periods do not contribute to the residues because residues are defined around $P^i$ not on $P^i$.},
if it is normalized to have purely imaginary periods with respect to all contours to fix ambiguity of adding holomorphic differentials. A global time is defined by
$\bar{w}=\bar{\tau}+i\bar{\sigma}:=\int^{P} dp$ at any point $P$ on $\bar{\bold{\Sigma}}_{R}$\footnote{We define the integral by avoiding the R punctures and define the global time on $P^i$ by a limit to $P^i$ in order that the odd periods do not contribute to the global time.}. By setting the even coordinates to $\bar{w}$ under a superconformal transformation, a reduced space $\bar{\Sigma}_{R, red}$ is canonically defined.
$\bar{\tau}$ takes the same value at the same point even if different contours are chosen in $\int^P dp$, because the real parts of the periods are zero by definition of the normalization.
 In particular, $\bar{\tau}=-\infty$ at $P^i$ with negative $f^i$ and $\bar{\tau}=\infty$ at $P^i$ with positive $f^i$. A contour integral on $\bar{\tau}$ constant line around $P^i$: $i \Delta \bar{\sigma}=\oint dp=2\pi i f^i$ indicates that the $\bar{\sigma}$ region around $P^i$ is $2\pi f^i$. This means that $\bar{\bold{\Sigma}}_{R}$ around $P^i$ represents a semi-infinite supercylinder with radius $f^i$. The condition $\sum_i f^i=0$ means that the total $\bar{\sigma}$ region of incoming supercylinders equals to that of outgoing ones if we choose the outgoing direction as positive. That is, the total $\bar{\sigma}$ region is conserved. In order to define the above global time uniquely, we fix the $\bar{\sigma}$ regions $2\pi f^i$ around $P^i$. We divide N $P^i$s to arbitrary two sets consist of $N_{-}$ and $N_{+}$ $P^i$s, respectively ($N_{-}+N_{+}=N$), then we divide  -1 to $N_{-}$ $f^i\equiv\frac{-1}{N_{-}}$ and 1 to $N_{+}$ $f^i\equiv\frac{1}{N_{+}}$ equally for all $i$.

If we give residues $-f^i$ and the same normalization on $\bar{\bold{\Sigma}}_{L}$ as on $\bar{\bold{\Sigma}}_{R}$, we can set the even coordinates on $\bar{\bold{\Sigma}}_{L}$ to the complex conjugate $\bar{\bar{w}}=\bar{\tau}-i\bar{\sigma}:=\int^{P} d\bar{p}$ by a superconformal transformation, because the Abelian differential is uniquely determined on $\bar{\bold{\Sigma}}_{L}$ and $\bar{\Sigma}_{L, red}$ is complex conjugate to $\bar{\Sigma}_{R, red}$. Therefore, we can define the global time $\bar{\tau}$ uniquely and reduced space canonically on the type II superstring worldsheet $\bar{\bold{\Sigma}}$.

Thus, under a superconformal transformation, one obtains a type II worldsheet $\bar{\bold{\Sigma}}$ that has even coordinates composed of the global time $\bar{\tau}$ and the position $\bar{\sigma}$ and $\bar{\Sigma}_{red}$ is canonically defined. Because $\bar{\bold{\Sigma}}$ can be a moduli of type II worldsheets with punctures, any two-dimensional super Riemannian manifold with punctures $\bold{\Sigma}$ can be obtained by $\bold{\Sigma}={\boldsymbol \psi}(\bar{\bold{\Sigma}})$ where ${\boldsymbol \psi}$ is a super diffeomorphism times super Weyl transformation \cite{Howe, DHokerPhong}.

Next, we will define the model space $\bold{E}$. We consider a state $(\bar{\bold{\Sigma}},  \bold{X}(\bar{\tau}_s), \bar{\tau}_s)$ determined by $\bar{\bold{\Sigma}}$, a $\bar{\tau}=\bar{\tau}_s$ constant hypersurface and an arbitrary map $\bold{X}(\bar{\tau}_s)$ from $\bar{\bold{\Sigma}}|_{\bar{\tau}_s}$ to the Euclidean space $\bold{R}^d$. $\bar{\bold{\Sigma}}$ is a union of $N_{\pm}$ supercylinders with radii $f_i$ at $\bar{\tau}\simeq \pm \infty$. Thus, we define a superstring state as an equivalence class $[\bar{\bold{\Sigma}}, \bold{X}(\bar{\tau}_s\simeq \pm \infty), \bar{\tau}_s\simeq \pm \infty]$ by a relation $(\bar{\bold{\Sigma}}, \bold{X}(\bar{\tau}_s\simeq \pm \infty), \bar{\tau}_s\simeq \pm \infty) \sim (\bar{\bold{\Sigma}}', \bold{X}'(\bar{\tau}_s\simeq \pm \infty), \bar{\tau}_s\simeq \pm \infty)$ if $N_{\pm}=N'_{\pm}$, $f_i=f'_i$, $\bold{X}(\bar{\tau}_s\simeq \pm \infty)=\bold{X}'(\bar{\tau}_s\simeq \pm \infty)$, and the corresponding supercylinders are the same type (NS-NS, NS-R, R-NS, or R-R) as in Fig. \ref{EquivalentClass}. Because the reduced space of $\bold{\Sigma}|_{\bar{\tau}_s}$ is $S^1 \cup S^1 \cup \cdots \cup S^1$ and $\bold{X}(\bar{\tau}_s): \bold{\Sigma}|_{\bar{\tau}_s} \to \bold{R}^d$, $[\bar{\bold{\Sigma}},  \bold{X}(\bar{\tau}_s), \bar{\tau}_s]$ represent many-body states of superstrings in $\bold{R}^d$ as in Fig. \ref{states}.  
In this supersymmetric case, we define the model space $\bold{E}$ such that $\bold{E}:=\cup_T\{[\bar{\bold{\Sigma}},  \bold{X}_T(\bar{\tau}_s), \bar{\tau}_s]\}$ where $T$ runs IIA and IIB. IIA and IIB GSO projections are attached for $T=$ IIA and IIB, respectively. We can define the worldsheet fermion numbers of states in a Hilbert space because the states consist of the fields over the local coordinates ${\boldsymbol X}_T^{\mu}=X^{\mu}+ \bar{\theta}^{\alpha} \psi_{\alpha}^{\mu}+\frac{1}{2} \bar{\theta}^2 F^{\mu}$, where $\psi_{\alpha}^{\mu}$ is a Majorana fermion and  $F^{\mu}$ is an auxiliary field. We abbreviate $T$ and $(\bar{\tau}_s)$ of $X^{\mu}$, $\psi_{\alpha}^{\mu}$ and $F^{\mu}$. We define the Hilbert space in these coordinates by the states only with $e^{\pi i F}=1$ and $e^{\pi i \tilde{F}}=(-1)^{\tilde{\alpha}}$ for $T=$ IIA, and $e^{\pi i F}=e^{\pi i \tilde{F}}=1$ for $T=$ IIB, where $F$ and $\tilde{F}$ are left- and right-handed fermion numbers respectively, and $\tilde{\alpha}$ is 1 or 0 when the right-handed fermion is periodic (R sector) or anti-periodic (NS sector), respectively.

Here, we will define topologies of $\bold{E}$. An $\epsilon$-open neighborhood of $[\bar{{\boldsymbol \Sigma}}, \bold{X}_{Ts}(\bar{\tau}_s), \bar{\tau}_s]$ is defined by
\begin{eqnarray}
U([\bar{{\boldsymbol \Sigma}}, \bold{X}_{Ts}(\bar{\tau}_s), \bar{\tau}_s], \epsilon)
:=
\left\{[\bar{{\boldsymbol \Sigma}}, \bold{X}_{T}(\bar{\tau}), \bar{\tau}] \bigm|
\sqrt{|\bar{\tau}-\bar{\tau}_s|^2
+\| \bold{X}_{T}(\bar{\tau}) -\bold{X}_{T s}(\bar{\tau}_s) \|^2}
<\epsilon   \right\}, \label{SuperNeighbour}
\end{eqnarray}
where
\begin{eqnarray}
&&
\| \bold{X}_{T}(\bar{\tau}) -\bold{X}_{s T}(\bar{\tau}_s) \|^2 \nonumber \\
&:=&\int_0^{2\pi}  d\bar{\sigma} 
\Bigl(|x(\bar{\tau}, \bar{\sigma})-x_s(\bar{\tau}_s, \bar{\sigma})|^2 
+(\bar{\psi}(\bar{\tau}, \bar{\sigma})-\bar{\psi}_s(\bar{\tau}_s, \bar{\sigma}))
(\psi(\bar{\tau}, \bar{\sigma})-\psi_s(\bar{\tau}_s, \bar{\sigma})) \nonumber \\
&+&|f(\bar{\tau}, \bar{\sigma})-f_s(\bar{\tau}_s, \bar{\sigma})|^2 \Bigr).
\end{eqnarray}
$U([\bar{\bold{\Sigma}}, \bold{X}_{T}(\bar{\tau}_s\simeq \pm \infty), \bar{\tau}_s\simeq \pm \infty], \epsilon)=
U([\bar{\bold{\Sigma}}', \bold{X}'_{T}(\bar{\tau}_s\simeq \pm \infty), \bar{\tau}_s\simeq \pm \infty], \epsilon)$ consistently if $N_{\pm}=N'_{\pm}$, $f_i=f'_i$, $\bold{X}_{T}(\bar{\tau}_s\simeq \pm \infty)=\bold{X}'_{T}(\bar{\tau}_s\simeq \pm \infty)$, the corresponding supercylinders are the same type (NS-NS, NS-R, R-NS, or R-R), and $\epsilon$ is small enough, because the $\bar{\tau}_s\simeq \pm \infty$ constant hypersurfaces traverses only supercylinders  overlapped by $\bar{{\boldsymbol \Sigma}}$ and $\bar{{\boldsymbol \Sigma}}'$. $U$ is defined to be an open set of $\bold{E}$ if there exists $\epsilon$ such that $U([\bar{{\boldsymbol \Sigma}}, \bold{X}_{T}(\bar{\tau}_s), \bar{\tau}_s], \epsilon) \subset U$ for an arbitrary point $[\bar{{\boldsymbol \Sigma}}, \bold{X}_{T}(\bar{\tau}_s), \bar{\tau}_s] \in U$.  In exactly the same way as in section 2, one can show that the topology of $\bold{E}$ satisfies the axiom of topology.
Although the model space is defined by using the coordinates $[\bar{\bold{\Sigma}},  \bold{X}_T(\bar{\tau}_s), \bar{\tau}_s]$, the model space does not depend on the coordinates, because the model space is a topological space.

In the following, we denote $[\bar{\bold{E}}_{M}^{\quad A}(\bar{\sigma}, \bar{\tau}, \bar{\theta}^{\alpha}), \bold{X}_T(\bar{\tau}), \bar{\tau}]$,
where  $\bar{\bold{E}}_{M}^{\quad A}(\bar{\sigma}, \bar{\tau}, \bar{\theta}^{\alpha})$ ($M=(m, \alpha)$, $A=(q, a)$, $m, q=0,1$, $\alpha, a=1,2$) is the worldsheet super vierbein on $\bar{{\boldsymbol \Sigma}}$, 
 instead of $[\bar{{\boldsymbol \Sigma}}, \bold{X}_T(\bar{\tau}), \bar{\tau}]$, because giving a super Riemann surface is equivalent to giving a super vierbein up to super diffeomorphism and super Weyl transformations.

Next, in order to define structures of superstring manifold, we consider how generally we can define general coordinate transformations between $[\bar{\bold{E}}_{M}^{\quad A}, \bold{X}_T(\bar{\tau}), \bar{\tau}]$ and $[\bar{\bold{E}}_{M}^{'\quad A}, \bold{X}'_T(\bar{\tau}'), \bar{\tau}']$ where $[\bar{\bold{E}}_{M}^{\quad A}, \bold{X}_T(\bar{\tau}), \bar{\tau}] \in U \subset \bold{E}$ and $[\bar{\bold{E}}_{M}^{'\quad A}, \bold{X}'_T(\bar{\tau}'), \bar{\tau}']\in U' \subset \bold{E}$. $\bar{\bold{E}}_{M}^{\quad A}$ does not transform to $\bar{\tau}$ and $\bold{X}_T(\bar{\tau})$ and vice versa, because $\bar{\tau}$ and $\bold{X}_T(\bar{\tau})$ are continuous variables, whereas $\bar{\bold{E}}_{M}^{\quad A}$ is a discrete variable: $\bar{\tau}$ and $\bold{X}_T(\bar{\tau})$ vary continuously, whereas $\bar{\bold{E}}_{M}^{\quad A}$ varies discretely in a trajectory on $\bold{E}$ by definition of the neighborhoods. $\bar{\tau}$ does not transform to $\bar{\sigma}$ and $\bar{\theta}$ and vice versa, because the superstring states are defined by $\bar{\tau}$ constant surfaces. Under these restrictions, the most general coordinate transformation is given by 
\begin{eqnarray}
&&[\bar{\bold{E}}_{M}^{\quad A}(\bar{\sigma}, \bar{\tau}, \bar{\theta}^{\alpha}), \bold{X}_T^{\mu}(\bar{\sigma}, \bar{\tau}, \bar{\theta}^{\alpha}), \bar{\tau}] \nonumber \\
&&\mapsto 
[\bar{\bold{E}}_{M}^{'\quad A}(\bar{\sigma}'(\bar{\sigma}, \bar{\theta}), \bar{\tau}'(\bar{\tau}, \bold{X}_T(\bar{\tau})), \bar{\theta}^{'\alpha}(\bar{\sigma}, \bar{\theta})), \bold{X}_T^{'\mu} (\bar{\sigma}', \bar{\tau}', \bar{\theta}^{'\alpha})(\bar{\tau}, \bold{X}_T(\bar{\tau})), \bar{\tau}'(\bar{\tau}, \bold{X}_T(\bar{\tau}))], \label{GeneralCoordTrans2}
\nonumber \\
&&
\end{eqnarray}
where $\bar{\bold{E}}_{M}^{\quad A} \mapsto \bar{\bold{E}}_{M}^{'\quad A}$ represents a world-sheet superdiffeomorphism transformation\footnote{
We extend the model space from $\bold{E}=\{[\bar{\bold{E}}_{M}^{\quad A}(\bar{\sigma}, \bar{\tau}, \bar{\theta}^{\alpha}), \bold{X}_T^{\mu}(\bar{\sigma}, \bar{\tau}, \bar{\theta}^{\alpha}), \bar{\tau}] \}$ to $\bold{E}=\{[\bar{\bold{E}}_{M}^{'\quad A}(\bar{\sigma}', \bar{\tau}', \bar{\theta}^{'\alpha}), \bold{X}_T^{'\mu} (\bar{\sigma}', \bar{\tau}', \bar{\theta}^{'\alpha}), \bar{\tau}'] \}$ by including the points generated by the superdiffeomorphisms $\bar{\sigma} \mapsto \bar{\sigma}'(\bar{\sigma}, \bar{\theta})$,  $\bar{\theta}^{\alpha} \mapsto \bar{\theta}^{'\alpha}(\bar{\sigma}, \bar{\theta})$, and $\bar{\tau} \mapsto \bar{\tau}'(\bar{\tau})$.}. $\bold{X}_T^{'\mu} (\bar{\tau}, \bold{X}_T(\bar{\tau}))$ and $\bar{\tau}'(\bar{\tau}, \bold{X}_T(\bar{\tau}))$ are functionals of $\bar{\tau}$ and $\bold{X}_T^{\mu}(\bar{\tau})$. Here, we consider all the manifolds which are constructed by patching open sets of the model space $\bold{E}$ by general coordinate transformations (\ref{GeneralCoordTrans2}) and call them superstring manifolds $\mathfrak{M}$.

Here, we give an example of superstring manifolds: $\mathfrak{M}_{D_T}:= \{ [\bar{\bold{\Sigma}}, \bold{x}_T(\bar{\tau}), \bar{\tau}] \}$, where $D_T$ represents a target manifold $M$ and a type of the GSO projection. $\bold{x}_T(\bar{\tau}): \bar{\bold{\Sigma}}|_{\bar{\tau}} \to M$,  where $\bold{x}_T^{\mu}=x^{\mu}+ \bar{\theta}^{\alpha} \psi_{\alpha}^{\mu}+\frac{1}{2} \bar{\theta}^2 f^{\mu}$. The image of the bosonic part of the embedding function, $x(\bar{\tau})$ has a metric: 
$ds^2= dx^{\mu}(\bar{\tau}, \bar{\sigma})
dx^{\nu}(\bar{\tau}, \bar{\sigma})
G_{\mu \nu}(x(\bar{\tau}, \bar{\sigma}))$. 

We will show that $\mathfrak{M}_{D_T}$has a structure of manifold, that is there exists a homeomorphism between the sufficiently small neighborhood around an arbitrary point $[\bar{\bold{\Sigma}}, \bold{x}_{sT}(\bar{\tau}_s), \bar{\tau}_s] \in \mathfrak{M}_{D_T}$ and an open set of $E$. There exists a general coordinate transformation $X^{\mu}(x)$ that satisfies 
$ds^2= dx^{\mu}
dx^{\nu}
G_{\mu \nu}(x)
=
dX^{\mu}
dX^{\nu}
\eta_{\mu \nu}$
on an arbitrary point $x$ in the $\epsilon_{\bar{\sigma}}$ open neighborhood around $x_s(\bar{\tau}_s, \bar{\sigma}) \in M$, if $\epsilon_{\bar{\sigma}}$ is sufficiently small. An arbitrary point $[\bar{\bold{\Sigma}}, \bold{x}_T(\bar{\tau}), \bar{\tau}]$ in the $\epsilon$ open neighborhood around  $[\bar{\bold{\Sigma}}, \bold{x}_{sT}(\bar{\tau}_s), \bar{\tau}_s] $ satisfies 
\begin{eqnarray}
&&\int_0^{2\pi}  d\bar{\sigma} 
|x(\bar{\tau}, \bar{\sigma})-x_s(\bar{\tau}_s, \bar{\sigma})|^2 \nonumber \\
&&< \epsilon^2- |\bar{\tau}-\bar{\tau}_s|^2
-\int_0^{2\pi}  d\bar{\sigma}\Bigl( (\bar{\psi}(\bar{\tau}, \bar{\sigma})-\bar{\psi}_s(\bar{\tau}_s, \bar{\sigma}))
(\psi(\bar{\tau}, \bar{\sigma})-\psi_s(\bar{\tau}_s, \bar{\sigma}))
-|f(\bar{\tau}, \bar{\sigma})-f_s(\bar{\tau}_s, \bar{\sigma})|^2 \Bigr) 
\nonumber \\
&&
 \leqq \epsilon^{2}
 \label{Inequivalent}
\end{eqnarray}
and thus 
\begin{equation}
|x(\bar{\tau}, \bar{\sigma})-x_s(\bar{\tau}_s, \bar{\sigma})|
< \epsilon_{\bar{\sigma}}'
\end{equation}
on an arbitrary $\bar{\sigma}$.
$\epsilon_{\bar{\sigma}}'< \epsilon_{\bar{\sigma}}$ is satisfied on an arbitrary $\bar{\sigma}$ if $\epsilon$ is taken to be sufficiently small. 
Then, there exists a transformation  $X^{\mu}(\bar{\tau}, \bar{\sigma}):=X^{\mu}(x(\bar{\tau}, \bar{\sigma}))$, which satisfies
\begin{equation}
ds^2= dx^{\mu}(\bar{\tau}, \bar{\sigma})
dx^{\nu}(\bar{\tau}, \bar{\sigma})
G_{\mu \nu}(x(\bar{\tau}, \bar{\sigma}))
=
dX^{\mu}(\bar{\tau}, \bar{\sigma})
dX^{\nu}(\bar{\tau}, \bar{\sigma})
\eta_{\mu \nu}. \label{SuperLocalLorents}
\end{equation}
Because the tangent vector $\bold{X}_T(\bar{\tau}, \bar{\sigma}, \bar{\theta})$ exists for each $\bold{x}_T(\bar{\tau}, \bar{\sigma}, \bar{\theta})$, there exists a vector bundle $\bold{X}_T(\bar{\tau})$ for $0 \leqq \bar{\sigma} < 2\pi$ and $\bar{\theta}$. $\bold{x}_T(\bar{\tau})$ and $\bold{X}_T(\bar{\tau})$ satisfy (\ref{SuperLocalLorents}) on each $\bar{\sigma}$, that is $\bold{X}_T(\bar{\tau}): \bar{\bold{\Sigma}}|_{\bar{\tau}} \to \bold{R}^d$. Therefore, there exists a homeomorphism between the sufficiently small neighborhood around an arbitrary point $[\bar{\bold{\Sigma}}, \bold{x}_{sT}(\bar{\tau}_s), \bar{\tau}_s] \in \mathfrak{M}_{D_T}$ and an open set of $\bold{E}$: $[\bar{\bold{\Sigma}}, \bold{x}_T(\bar{\tau}), \bar{\tau}] \mapsto [\bar{\bold{\Sigma}}, \bold{X}_T(\bar{\tau}), \bar{\tau}]$.

Actually, a map from the image $\bold{x}_T(\bar{\tau}, \bar{\sigma}, \bar{\theta})$ to the image $\bold{X}_T(\bar{\tau}, \bar{\sigma}, \bar{\theta})$ is explicitly given by an exponential map
\begin{equation}
\bold{x}_T(\bar{\tau}, \bar{\sigma}, \bar{\theta})
=
\exp_{\bold{x}_{sT}(\bar{\tau}_s, \bar{\sigma}, \bar{\theta})}
\bold{X}_T(\bar{\tau}, \bar{\sigma}, \bar{\theta})
\simeq
\bold{x}_{sT}(\bar{\tau}_s, \bar{\sigma}, \bar{\theta})
+\bold{X}_T(\bar{\tau}, \bar{\sigma}, \bar{\theta}).
\end{equation}
If we substitute this to an $\epsilon$ open neighborhood around an arbitrary point $[\bar{\bold{\Sigma}}, \bold{x}_{sT}(\bar{\tau}_s), \bar{\tau}_s] \in \mathfrak{M}_{D_T}$ (\ref{Inequivalent}),
we obtain an $\epsilon$ open neighborhood around $ [\bar{\bold{\Sigma}}, 0, \bar{\tau}] \in \bold{E}$,
\begin{equation}
\| \bold{X}_{T}(\bar{\tau}) \|^2 +  |\bar{\tau}-\bar{\tau}_s|^2 < \epsilon^{2}.
\end{equation}

By definition of the $\epsilon$-open neighborhood, arbitrary two superstring states on a connected super Riemann surface are connected continuously. Thus, there is an one-to-one correspondence between a super Riemann surface with punctures in $M$ and a curve parametrized by $\bar{\tau}$ from $\bar{\tau}=-\infty$ to $\bar{\tau}=\infty$ on $\mathfrak{M}_{D_T}$. That is, curves that represent asymptotic processes on $\mathfrak{M}_{D_T}$ reproduce the right moduli space of the super Riemann surfaces in the target manifold.

By a general curve parametrized by $t$ on $\mathfrak{M}_{D_T}$, superstring states on different super Riemann surfaces that have even different genera, can be connected continuously, for example see Fig. \ref{6Connected}, whereas different super Riemann surfaces that have different genera cannot be connected continuously in the moduli space of the super Riemann surfaces in the target space.

The tangent space is spanned by $\frac{\partial}{\partial\bar{\tau}}$ and $\frac{\partial}{\partial \bold{X}_T^{\mu}(\bar{\sigma}, \bar{\tau}, \bar{\theta})}$ as one can see from the $\epsilon$-open neighborhood (\ref{SuperNeighbour}). We should note that $\frac{\partial}{\partial \bar{\bold{E}}^{\quad A}_{M}}$  cannot be a part of basis that span the tangent space because $\bar{\bold{E}}^{\quad A}_{M}$ is just a discrete variable in $\bold{E}$. The index of $\frac{\partial}{\partial \bold{X}_T^{\mu}(\bar{\sigma}, \bar{\tau}, \bar{\theta}) }$ can be $(\mu \, \bar{\sigma} \, \bar{\theta})$. 
Then, let us define a summation over $\bar{\sigma}$ and $\bar{\theta}$ that is invariant under $(\bar{\sigma}, \bar{\theta}^{\alpha}) \mapsto (\bar{\sigma}'(\bar{\sigma}, \bar{\theta}), \bar{\theta}^{'\alpha}(\bar{\sigma}, \bar{\theta}))$ and transformed as a scalar under $\bar{\tau} \mapsto \bar{\tau}'(\bar{\tau}, \bold{X}_T(\bar{\tau}))$. First, $\int d\bar{\tau} \int d\bar{\sigma}d^2\bar{\theta} \bar{\bold{E}}(\bar{\sigma}, \bar{\tau}, \bar{\theta}^{\alpha})$ is invariant under $(\bar{\sigma}, \bar{\tau}, \bar{\theta}^{\alpha}) \mapsto (\bar{\sigma}'(\bar{\sigma}, \bar{\theta}), \bar{\tau}'(\bar{\tau}, \bold{X}_T(\bar{\tau})), \bar{\theta}^{'\alpha}(\bar{\sigma}, \bar{\theta}))$, where $\bar{\bold{E}}(\bar{\sigma}, \bar{\tau}, \bar{\theta}^{\alpha})$ is the superdeterminant of $\bar{\bold{E}}_{M}^{\quad A}(\bar{\sigma}, \bar{\tau}, \bar{\theta}^{\alpha})$. The lapse function, $\bar{n}$ transforms as an one-dimensional vector in the $\bar{\tau}$ direction: 
$\int d\bar{\tau} \bar{n}$ is invariant under $\bar{\tau} \mapsto \bar{\tau}'(\bar{\tau}, \bold{X}_T(\bar{\tau}))$ and transformed as a superscalar under $(\bar{\sigma}, \bar{\theta}^{\alpha}) \mapsto (\bar{\sigma}'(\bar{\sigma}, \bar{\theta}), \bar{\theta}^{'\alpha}(\bar{\sigma}, \bar{\theta}))$. Therefore, 
$\int d\bar{\sigma}d^2\bar{\theta} \hat{\bold{E}}(\bar{\sigma}, \bar{\tau}, \bar{\theta}^{\alpha})$,
where
$\hat{\bold{E}}(\bar{\sigma}, \bar{\tau}, \bar{\theta}^{\alpha})
:=
\frac{1}{\bar{n}}\bar{\bold{E}}(\bar{\sigma}, \bar{\tau}, \bar{\theta}^{\alpha})$,
is transformed as a scalar under $\bar{\tau} \mapsto \bar{\tau}'(\bar{\tau}, \bold{X}_T(\bar{\tau}))$ and invariant under $(\bar{\sigma}, \bar{\theta}^{\alpha}) \mapsto (\bar{\sigma}'(\bar{\sigma}, \bar{\theta}), \bar{\theta}^{'\alpha}(\bar{\sigma}, \bar{\theta}))$.

Riemannian superstring manifold is obtained by defining a metric, which is a section of an inner product on the tangent space. The general form of a metric is given by
\begin{eqnarray}
&&ds^2(\bar{\bold{E}}, \bold{X}_T(\bar{\tau}), \bar{\tau}) \nonumber \\
=&&G(\bar{\bold{E}}, \bold{X}_T(\bar{\tau}), \bar{\tau})_{dd} (d\bar{\tau})^2 \nonumber \\ 
&&+2 d\bar{\tau} \int d\bar{\sigma} d^2\bar{\theta}  \hat{\bold{E}} \sum_{\mu} G(\bar{\bold{E}}, \bold{X}_T(\bar{\tau}), \bar{\tau})_{d \; (\mu \bar{\sigma} \bar{\theta})} d \bold{X}_T^{\mu}(\bar{\sigma}, \bar{\tau}, \bar{\theta}) \nonumber \\
&&+\int d\bar{\sigma} d^2\bar{\theta} \hat{\bold{E}}  \int d\bar{\sigma}'  d^2\bar{\theta}' \hat{\bold{E}}'  \sum_{\mu, \mu'} G(\bar{\bold{E}}, \bold{X}_T(\bar{\tau}), \bar{\tau})_{ \; (\mu \bar{\sigma} \bar{\theta})  \; (\mu' \bar{\sigma}' \bar{\theta}')} d \bold{X}_T^{\mu}(\bar{\sigma}, \bar{\tau}, \bar{\theta}) d \bold{X}_T^{\mu'}(\bar{\sigma}', \bar{\tau}, \bar{\theta}'). \nonumber \\
&&
\end{eqnarray}
We summarize the vectors as $d\bold{X}_T^\bold{I}$ ($\bold{I}=d,(\mu \bar{\sigma} \bar{\theta})$), where  $d\bold{X}_T^d:=d\bar{\tau}$ and $d \bold{X}_T^{(\mu \bar{\sigma} \bar{\theta})}:=d\bold{X}_T^{\mu}(\bar{\sigma}, \bar{\tau}, \bar{\theta})$. Then, the components of the metric are summarized as $G_{\bold{I}\bold{J}}(\bar{\bold{E}}, \bold{X}_T(\bar{\tau}), \bar{\tau})$. The inverse of the metric $G^{\bold{I}\bold{J}}(\bar{\bold{E}}, \bold{X}_T(\bar{\tau}), \bar{\tau})$ is defined by $G_{\bold{I}\bold{J}}G^{\bold{J}\bold{K}}=G^{\bold{K}\bold{J}}G_{\bold{J}\bold{I}}=\delta_\bold{I}^\bold{K}$, where $\delta_d^d=1$ and $\delta_{\mu \bar{\sigma} \bar{\theta}}^{\mu' \bar{\sigma}' \bar{\theta}'}=\frac{1}{\hat{\bold{E}}}\delta_{\mu}^{\mu'}\delta(\bar{\sigma}-\bar{\sigma}') \delta^2(\bar{\theta}-\bar{\theta}')$. The components of the Riemannian curvature tensor are given by $R^\bold{I}_{\bold{J}\bold{K}\bold{L}}$ in the basis $\frac{\partial}{\partial \bold{X}_T^\bold{I}(\bar{\tau})}$. The components of the Ricci tensor are $R_{\bold{I}\bold{J}}:=R^\bold{K}_{\bold{I}\bold{K}\bold{J}}=R^d_{\bold{I}d\bold{J}}+\int d\bar{\sigma} d^2\bar{\theta} \hat{\bold{E}}  R^{(\mu \bar{\sigma} \bar{\theta})}_{\bold{I} \; (\mu \bar{\sigma} \bar{\theta}) \; \bold{J}}$. The scalar curvature is 
\begin{eqnarray}
R&:=&G^{\bold{I}\bold{J}} R_{\bold{I}\bold{J}} \nonumber \\
&=&G^{dd}R_{dd}+2 \int d\bar{\sigma} d^2\bar{\theta} \hat{\bold{E}}  G^{d \; (\mu \bar{\sigma} \bar{\theta})} R_{d \; (\mu \bar{\sigma} \bar{\theta})} 
 \nonumber \\
&&+\int d\bar{\sigma} d^2\bar{\theta} \hat{\bold{E}}  \int d\bar{\sigma}'  d^2\bar{\theta}' \hat{\bold{E}}' G^{(\mu \bar{\sigma} \bar{\theta}) \; (\mu' \bar{\sigma}' \bar{\theta}')}R_{(\mu \bar{\sigma} \bar{\theta})  \; (\mu' \bar{\sigma}' \bar{\theta}')}. \nonumber
\end{eqnarray}
The volume is  $vol=\sqrt{G}$, where $G=det (G_{\bold{I}\bold{J}})$.

By using these geometrical objects, we formulate superstring theory non-perturbatively as
\begin{equation}
Z=\int \mathcal{D}G \mathcal{D}Ae^{-S},  \label{SuperTheory}
\end{equation}
where
\begin{equation}
S=\frac{1}{G_N}\int \mathcal{D}\bold{E} \mathcal{D}\bold{X}_T(\bar{\tau}) \mathcal{D}\bar{\tau} 
\sqrt{G} (-R +\frac{1}{4} G_N G^{\bold{I}_1 \bold{I}_2} G^{\bold{J}_1 \bold{J}_2} F_{\bold{I}_1 \bold{J}_1} F_{\bold{I}_2 \bold{J}_2} ). \label{SuperAction}
\end{equation}
As an example of sets of fields on the superstring manifolds, we consider the metric and an $u(1)$ gauge field $A_\bold{I}$ whose field strength is given by $F_{\bold{I}\bold{J}}$. The path integral is defined by semi-classically\footnote{It will be enough to define the path-integral by semi-classically summing classical solutions and small classical and quantum fluctuations around them, because string manifolds themselves possess quantum corrections, and loops of the fields on them do not correspond to quantum corrections as one can see in the derivation of the perturbative string theory later. The unitarity is manifest and there is also no UV divergence from loop integrals, by defining the path-integral semi-classically. } summing over the metrics and gauge fields on $\mathfrak{M}$. By definition, the theory is background independent. $\mathcal{D}\bold{E}$ is the invariant measure of the super vierbeins $\bold{E}_{M}^{\quad A}$ on the two-dimensional super Riemannian manifolds $\bold{\Sigma}$. $\bold{E}_{M}^{\quad A}$ and $\bar{\bold{E}}_{M}^{\quad A}$ are related to each others by the super diffeomorphism and super Weyl transformations.

Under 
\begin{equation}
(\bar{\tau}, \bold{X}_T(\bar{\tau})) \mapsto (\bar{\tau}'(\bar{\tau}, \bold{X}_T(\bar{\tau})) , \bold{X}_T'(\bar{\tau}')(\bar{\tau}, \bold{X}_T(\bar{\tau}))),
\label{Supersubdiffeo}
\end{equation} 
$G_{\bold{I}\bold{J}}(\bar{\bold{E}}, \bold{X}_T(\bar{\tau}), \bar{\tau})$ and $A_{\bold{I}}(\bar{\bold{E}}, \bold{X}_T(\bar{\tau}), \bar{\tau})$ are transformed as a symmetric tensor and a vector, respectively and the action is manifestly invariant.

We define $G_{\bold{I}\bold{J}}(\bar{\bold{E}}, \bold{X}_T(\bar{\tau}), \bar{\tau})$ and $A_{\bold{I}}(\bar{\bold{E}}, \bold{X}_T(\bar{\tau}), \bar{\tau})$ so as to transform as scalars under $\bar{\bold{E}}_{M}^{\quad A}(\bar{\sigma}, \bar{\tau}, \bar{\theta}^{\alpha}) \mapsto
\bar{\bold{E}}_{M}^{'\quad A}(\bar{\sigma}'(\bar{\sigma}, \bar{\theta}), \bar{\tau}, \bar{\theta}^{'\alpha}(\bar{\sigma}, \bar{\theta}))$. Under $(\bar{\sigma}, \bar{\theta})$ superdiffeomorphisms: $(\bar{\sigma}, \bar{\theta}^{\alpha}) \mapsto (\bar{\sigma}'(\bar{\sigma}, \bar{\theta}), \bar{\theta}^{'\alpha}(\bar{\sigma}, \bar{\theta}))$, which are equivalent to 
\begin{eqnarray}
&&[\bar{\bold{E}}_{M}^{\quad A}(\bar{\sigma}, \bar{\tau}, \bar{\theta}^{\alpha}),  \bold{X}_T^{\mu}(\bar{\sigma}, \bar{\tau}, \bar{\theta}^{\alpha}), \bar{\tau}] \nonumber \\
&&\mapsto [\bar{\bold{E}}_{M}^{'\quad A}(\bar{\sigma}'(\bar{\sigma}, \bar{\theta}), \bar{\tau}, \bar{\theta}^{'\alpha}(\bar{\sigma}, \bar{\theta})), \bold{X}_T^{'\mu} (\bar{\sigma}'(\bar{\sigma}, \bar{\theta}), \bar{\tau}, \bar{\theta}^{'\alpha}(\bar{\sigma}, \bar{\theta}))(\bold{X}_T(\bar{\tau})),  \bar{\tau}] \nonumber \\
&&=[\bar{\bold{E}}_{M}^{'\quad A}(\bar{\sigma}'(\bar{\sigma}, \bar{\theta}), \bar{\tau}, \bar{\theta}^{'\alpha}(\bar{\sigma}, \bar{\theta})), \bold{X}_T^{\mu}(\bar{\sigma}, \bar{\tau}, \bar{\theta}^{\alpha}),  \bar{\tau}], \label{SuperStringGeometryTrans}
\end{eqnarray}
$G_{d \; (\mu \bar{\sigma} \bar{\theta})}$ is transformed as a superscalar;
\begin{eqnarray}
G'_{d \; (\mu \bar{\sigma}'  \bar{\theta}')}(\bar{\bold{E}}', \bold{X}_T'(\bar{\tau}),  \bar{\tau})
&=&
G'_{d \; (\mu \bar{\sigma}'  \bar{\theta}')}(\bar{\bold{E}}, \bold{X}'_T(\bar{\tau}), \bar{\tau})
=
\frac{\partial \bold{X}_T^{\bold{I}}(\bar{\tau})}{\partial \bold{X}_T^{'d}(\bar{\tau})}
\frac{\partial \bold{X}_T^{\bold{J}}(\bar{\tau})}{\partial \bold{X}_T^{'(\mu \bar{\sigma}' \bar{\theta}')}(\bar{\tau})}
G_{\bold{I} \bold{J}}(\bar{\bold{E}}, \bold{X}_T(\bar{\tau}), \bar{\tau})
\nonumber \\
&=&
\frac{\partial \bold{X}_T^{\bold{I}}(\bar{\tau})}{\partial \bold{X}_T^{d}(\bar{\tau})}
\frac{\partial \bold{X}_T^{\bold{J}}(\bar{\tau})}{\partial \bold{X}_T^{(\mu \bar{\sigma} \bar{\theta})}(\bar{\tau})}
G_{\bold{I} \bold{J}}(\bar{\bold{E}}, \bold{X}_T(\bar{\tau}), \bar{\tau})
=
G_{d \; (\mu \bar{\sigma} \bar{\theta})}(\bar{\bold{E}}, \bold{X}_T(\bar{\tau}), \bar{\tau}),  \nonumber \\
&&\label{GdmuTrans}
\end{eqnarray}
because (\ref{Supersubdiffeo}) and (\ref{SuperStringGeometryTrans}). In the same way, the other fields are also transformed as
\begin{eqnarray}
G'_{dd}(\bar{\bold{E}}', \bold{X}_T'(\bar{\tau}),  \bar{\tau})
&=&G_{dd}(\bar{\bold{E}}, \bold{X}_T(\bar{\tau}), \bar{\tau}) \nonumber \\
G'_{ \; (\mu \bar{\sigma}'\bar{\theta}')  \; (\nu \bar{\rho}'\tilde{\bar{\theta}}')}(\bar{\bold{E}}', \bold{X}_T'(\bar{\tau}),  \bar{\tau})
&=&G_{ \; (\mu \bar{\sigma} \bar{\theta})  \; (\nu \bar{\rho} \tilde{\bar{\theta}})}(\bar{\bold{E}}, \bold{X}_T(\bar{\tau}), \bar{\tau}) \nonumber \\
A'_d(\bar{\bold{E}}', \bold{X}_T'(\bar{\tau}),  \bar{\tau})
&=&A_d(\bar{\bold{E}}, \bold{X}_T(\bar{\tau}), \bar{\tau}) \nonumber \\
A'_{(\mu \bar{\sigma}'\bar{\theta}')}(\bar{\bold{E}}', \bold{X}_T'(\bar{\tau}),  \bar{\tau})&=&A_{(\mu \bar{\sigma} \bar{\theta})}(\bar{\bold{E}}, \bold{X}_T(\bar{\tau}), \bar{\tau}). \label{SuperOtherTrans}
\end{eqnarray}
Thus, the action is invariant under the $(\bar{\sigma}, \bar{\theta})$ superdiffeomorphisms, because 
\begin{equation}
\int d\bar{\sigma}' d^2\bar{\theta}'  \hat{\bold{E}}'(\bar{\sigma}', \bar{\tau}, \bar{\theta}')=\int d\bar{\sigma} d^2\bar{\theta}  \hat{\bold{E}}(\bar{\sigma}, \bar{\tau}, \bar{\theta}).
\end{equation}
Therefore, $G_{\bold{I}\bold{J}}(\bar{\bold{E}}, \bold{X}_T(\bar{\tau}), \bar{\tau})$ and $A_{\bold{I}}(\bar{\bold{E}}, \bold{X}_T(\bar{\tau}), \bar{\tau})$ are transformed covariantly and the action (\ref{SuperAction}) is invariant under the diffeomorphisms (\ref{GeneralCoordTrans2}), including the $(\bar{\sigma}, \bar{\theta})$ superdiffeomorphisms, whose infinitesimal transformations are given by 
\begin{eqnarray}
\bar{\sigma}^{\xi}&=&\bar{\sigma}+i \xi^{\alpha}(\bar{\sigma}) \gamma^1_{\alpha \beta} \bar{\theta}^{\beta} \nonumber \\
\bar{\theta}^{\xi \alpha}(\bar{\sigma})&=&\bar{\theta}^{\alpha} + \xi^{\alpha}(\bar{\sigma}). \label{SUSYtrans}
\end{eqnarray}
(\ref{SUSYtrans}) are dimensional reductions in $\bar{\tau}$ direction of the two-dimensional $\mathcal{N}=(1,1)$ local supersymmetry infinitesimal transformations. The number of supercharges
\begin{equation}
\xi^{\alpha} Q_{\alpha}
=
\xi^{\alpha}(\frac{\partial}{\partial \bar{\theta}^{\alpha}} + i \gamma^1_{\alpha \beta}\bar{\theta}^{\beta} \frac{\partial}{\partial \bar{\sigma}})
\end{equation}
 of the transformations is the same as of the two-dimensional ones. The supersymmetry algebra closes in a field-independent sense as in ordinary supergravities.

The background that represents a perturbative vacuum is given by 
\begin{eqnarray}
\bar{ds}^2
&=& 2\lambda \bar{\rho}(\bar{h}) N^2(\bold{X}_T(\bar{\tau})) (d\bold{X}_T^d)^2 \nonumber \\ 
&&+\int d\bar{\sigma} d^2\bar{\theta}  \hat{\bold{E}} \int d\bar{\sigma}' d^2\bar{\theta}'  \hat{\bold{E}}'N^{\frac{2}{2-\bold{D}}}(\bold{X}_T(\bar{\tau})) 
\frac{\hat{\bold{E}}^3(\bar{\sigma}, \bar{\tau}, \bar{\theta})}
{\bold{E}(\bar{\sigma}, \bar{\tau}, \bar{\theta})} 
\delta_{(\mu \bar{\sigma} \bar{\theta}) (\mu' \bar{\sigma}' \bar{\theta}')}
d \bold{X}_T^{(\mu \bar{\sigma} \bar{\theta})} d \bold{X}_T^{(\mu' \bar{\sigma}' \bar{\theta}')}, \nonumber \\
\bar{A}_d&=&i \sqrt{\frac{2-2\bold{D}}{2-\bold{D}}}\frac{\sqrt{2\lambda \bar{\rho}(\bar{h}) }}{\sqrt{G_N}} N(\bold{X}_T(\bar{\tau})), \qquad
\bar{A}_{(\mu \bar{\sigma} \bar{\theta})}=0, \label{Supersolution}
\end{eqnarray}
on $\mathfrak{M}_{D_T}$ where the target metric is fixed to $\eta_{\mu \mu'}$. 
$\bar{\rho}(\bar{h}):=\frac{1}{4 \pi}\int d\bar{\sigma} \sqrt{\bar{h}}\bar{R}_{\bar{h}}$, where $\bar{R}_{\bar{h}}$ is the scalar curvature of $\bar{h}_{ mn}$. $\bold{D}$ is a volume of the index $(\mu \bar{\sigma} \bar{\theta})$: $\bold{D}:=\int d\bar{\sigma} d^2\bar{\theta}  \hat{\bold{E}} \delta_{(\mu \bar{\sigma} \bar{\theta}) (\mu \bar{\sigma} \bar{\theta})}=d \int d\bar{\sigma} d^2\bar{\theta} \delta(\bar{\sigma}-\bar{\sigma}) \delta^2(\bar{\theta}-\bar{\theta})$. $N(\bold{X}_T(\bar{\tau}))=\frac{1}{1+v(\bold{X}_T(\bar{\tau}))}$, where 
\begin{equation}
v(\bold{X}_T(\bar{\tau}))= \frac{\alpha}{\sqrt{d-1}} \int d\bar{\sigma} d^2\bar{\theta}
\hat{\bold{E}} \epsilon_{\mu\nu}\bold{X}_T^{\mu}(\bar{\tau}) \sqrt{\tilde{\bold{D}}_{\alpha}^2} \bold{X}_T^{\nu}(\bar{\tau}).
\end{equation} 
$\tilde{\bold{D} }_{\alpha}$ is a $\bar{\tau}$ independent super derivative that satisfies 
\begin{eqnarray}
&&\int d\bar{\tau} d\bar{\sigma} d^2\bar{\theta} \bar{\bold{E}}  \frac{1}{2}(\tilde{\bold{D}}_{\alpha}\bold{X}_{T\mu}(\bar{\tau}))^2\nonumber \\
&=&
\int d\bar{\tau} d\bar{\sigma} \sqrt{\bar{h}} \frac{1}{2}\biggl(-(-\frac{\bar{n}^{\bar{\sigma}}}{\bar{n}} \partial_{\bar{\sigma}} X^{\mu}
+\frac{1}{2}\bar{n} \bar{\chi}_{m} \bar{E}^{0}_{r} \gamma^{r} \bar{E}^{m}_{q} \gamma^q \psi_{\mu})^2 \nonumber \\
&&
+\bar{h}^{11}\partial_{\bar{\sigma}} X^{\mu} \partial_{\bar{\sigma}} X_{\mu} -\bar{\psi}^{\mu} \bar{E}^{1}_q \gamma^{q} \partial_{\bar{\sigma}} \psi_{\mu}
-(F^{\mu})^2 \nonumber \\
&& + \bar{\chi}_{m} \bar{E}^{1}_{r} \gamma^{r} \bar{E}^{m}_{q} \gamma^q \psi_{\mu} \partial_{\bar{\sigma}} X^{\mu}
-\frac{1}{8}\bar{\psi}^{\mu}\psi_{\mu} \bar{\chi}_{m}\bar{E}^{n}_q\gamma^{q}\bar{E}^{m}_r \gamma^r \chi_{n}\biggr),
\label{quad2}
\end{eqnarray}
where $\bar{E}^{m}_{\; q}$, $\bar{\chi}_m$, and $\gamma^q$ are a vierbein, a gravitino, and gamma matrices in the two dimensions, respectively. On the other hand, the ordinary super covariant derivative $\bar{\bold{D}}_{\alpha}$ satisfies \cite{Howe, SuperStringAction}
\begin{eqnarray}
&&\int d\bar{\tau} d\bar{\sigma} d^2\bar{\theta} \bar{\bold{E}}  \frac{1}{2}(\bar{\bold{D}}_{\alpha}\bold{X}_{T\mu}(\bar{\tau}))^2 \nonumber \\
&=&
\int d\bar{\tau} d\bar{\sigma} \sqrt{\bar{h}} \frac{1}{2}(\bar{h}^{mn}\bar{\partial}_{m} X^{\mu} \bar{\partial}_{n} X_{\mu} -\bar{\psi}^{\mu} \bar{E}^{m}_q \gamma^{q} \bar{\partial}_{m} \psi_{\mu}
-(F^{\mu})^2 \nonumber \\
&& + \bar{\chi}_{m} \bar{E}^{n}_{r} \gamma^{r} \bar{E}^{m}_{q} \gamma^q \psi_{\mu} \bar{\partial}_{n} X^{\mu}
-\frac{1}{8}\bar{\psi}^{\mu}\psi_{\mu} \bar{\chi}_{m}\bar{E}^{n}_q\gamma^{q}\bar{E}^{m}_r \gamma^r \chi_{n}). \label{quad1}
\end{eqnarray}
The inverse of the metric is given by 
\begin{eqnarray}
\bar{G}^{dd}&=&\frac{1}{2\lambda \bar{\rho} }\frac{1}{N^2} \nonumber \\
\bar{G}^{d \; (\mu\bar{\sigma} \bar{\theta})}&=&0 \nonumber \\
\bar{G}^{(\mu\bar{\sigma} \bar{\theta}) \; (\mu'\bar{\sigma}' \bar{\theta}')}&=& N^{\frac{-2}{2-\bold{D}}} 
\frac{\bold{E}} 
{\hat{\bold{E}}^3}
 \delta_{(\mu \bar{\sigma} \bar{\theta}) (\mu' \bar{\sigma}' \bar{\theta}')},
\end{eqnarray}
because $\int d\bar{\sigma}'' d^2\bar{\theta}''  \hat{\bold{E}}''  \bar{G}_{(\mu\bar{\sigma}\bar{\theta}) \; (\mu''\bar{\sigma}'' \bar{\theta}'')}\bar{G}^{(\mu''\bar{\sigma}'' \bar{\theta}'') \; (\mu'\bar{\sigma}' \bar{\theta}')}=\int d\bar{\sigma}'' d^2\bar{\theta}''  \hat{\bold{E}}''  \delta_{(\mu\bar{\sigma}\bar{\theta}) \; (\mu''\bar{\sigma}''\bar{\theta}'')}\delta_{(\mu''\bar{\sigma}''\bar{\theta}'') \; (\mu'\bar{\sigma}'\bar{\theta}')}
= \delta_{(\mu\bar{\sigma}\bar{\theta}) \; (\mu'\bar{\sigma}'\bar{\theta}')}$. 
From the metric, we obtain 
\begin{eqnarray}
&&\sqrt{\bar{G}}=N^\frac{2}{2-\bold{D}}\sqrt{2\lambda \bar{\rho} \exp(\int d\bar{\sigma}d^2\bar{\theta} \hat{\bold{E}} \delta_{(\mu\bar{\sigma}\bar{\theta}) \; (\mu \bar{\sigma} \bar{\theta} )} \ln\frac{\hat{\bold{E}}^3}
{\bold{E}} )} 
\nonumber \\
&&\bar{R}_{dd}=-2\lambda \bar{\rho} N^{\frac{-2}{2-\bold{D}}} \int d\bar{\sigma} d^2\bar{\theta}\frac{\bold{E}}
{\hat{\bold{E}}^2}  \partial_{(\mu \bar{\sigma} \bar{\theta})}N \partial_{(\mu \bar{\sigma} \bar{\theta})}N
\nonumber \\
&&\bar{R}_{d \; (\mu \bar{\sigma} \bar{\theta})}=0 
\nonumber \\
&&\bar{R}_{(\mu\bar{\sigma}\bar{\theta}) \; (\mu' \bar{\sigma}' \bar{\theta}')}
=\frac{\bold{D}-1}{2-\bold{D}}N^{-2}\partial_{(\mu\bar{\sigma}\bar{\theta})}N \partial_{(\mu' \bar{\sigma}' \bar{\theta}')}N
\nonumber \\
&& \qquad \qquad \qquad +\frac{1}{\bold{D}-2}N^{-2}
\int d\bar{\sigma}'' d^2\bar{\theta}''\frac{\bold{E}''}
{\hat{\bold{E}}^{''2}}\partial_{(\mu'' \bar{\sigma}'' \bar{\theta}'')}N \partial_{(\mu'' \bar{\sigma}'' \bar{\theta}'')}N
\frac{\hat{\bold{E}}^3}
{\bold{E}} 
\delta_{(\mu\bar{\sigma}\bar{\theta}) \; (\mu' \bar{\sigma}' \bar{\theta}')}
\nonumber \\
&&\bar{R}=\frac{\bold{D}-3}{2-\bold{D}} N^{\frac{2\bold{D}-6}{2-\bold{D}}}
\int d\bar{\sigma} d^2\bar{\theta} \frac{\bold{E}}
{\hat{\bold{E}}^2} \partial_{(\mu\bar{\sigma}\bar{\theta})}N \partial_{(\mu\bar{\sigma}\bar{\theta})}N.
\end{eqnarray}
By using these quantities, one can show that the background (\ref{Supersolution}) is a classical solution to the equations of motion of (\ref{SuperAction}). We also need to use the fact that $v(\bold{X}_T(\bar{\tau}))$ is a harmonic function with respect to $\bold{X}_T^{(\mu\bar{\sigma}\bar{\theta})}(\bar{\tau})$. Actually, $\partial_{(\mu\bar{\sigma}\bar{\theta})}\partial_{(\mu\bar{\sigma}\bar{\theta})}v=0$. In these calculations, we should note that $\bar{\bold{E}}_{M}^{\quad A}$, $\bold{X}_T^{\mu}(\bar{\tau})$ and $\bar{\tau}$ are all independent, and thus $\frac{\partial}{\partial \bar{\tau}}$ is an explicit derivative on functions over the superstring manifolds, especially, $\frac{\partial}{\partial \bar{\tau}}\bar{\bold{E}}_{M}^{\quad A}=0$ and $\frac{\partial}{\partial \bar{\tau}}\bold{X}_T^{\mu}(\bar{\tau})=0$. Because the equations of motion are differential equations with respect to $\bold{X}_T^{\mu}(\bar{\tau})$ and $\bar{\tau}$, $\bar{\bold{E}}_{M}^{\quad A}$ is a constant in the solution (\ref{Supersolution}) to the differential equations. The dependence of $\bar{\bold{E}}_{M}^{\quad A}$ on the background (\ref{Supersolution}) is uniquely determined  by the consistency of the quantum theory of the fluctuations around the background. Actually, we will find that all the perturbative superstring amplitudes are derived.

 Let us consider fluctuations around the background (\ref{Supersolution}), $G_{\bold{I}\bold{J}}=\bar{G}_{\bold{I}\bold{J}}+\tilde{G}_{\bold{I}\bold{J}}$ and $A_\bold{I}=\bar{A}_\bold{I}+\tilde{A}_\bold{I}$. Here we  fix the charts, where we choose $T$=IIA or IIB. The action (\ref{SuperAction}) up to the quadratic order is given by,\begin{eqnarray}
S&=&\frac{1}{G_N} \int \mathcal{D}\bold{E} \mathcal{D}\bold{X}_T(\bar{\tau}) \mathcal{D}\bar{\tau}  \sqrt{\bar{G}} 
\Bigl(-\bar{R}+\frac{1}{4}\bar{F}'_{\bold{I}\bold{J}}\bar{F}'^{\bold{I}\bold{J}} 
\nonumber \\
&&+\frac{1}{4}\bar{\nabla}_\bold{I} \tilde{G}_{\bold{J}\bold{K}} \bar{\nabla}^\bold{I} \tilde{G}^{\bold{J}\bold{K}}
-\frac{1}{4}\bar{\nabla}_\bold{I} \tilde{G} \bar{\nabla}^\bold{I} \tilde{G}
+\frac{1}{2}\bar{\nabla}^\bold{I} \tilde{G}_{\bold{I}\bold{J}} \bar{\nabla}^\bold{J} \tilde{G}
-\frac{1}{2}\bar{\nabla}^\bold{I} \tilde{G}_{\bold{I}\bold{J}} \bar{\nabla}_\bold{K} \tilde{G}^{\bold{J}\bold{K}}
\nonumber \\
&&-\frac{1}{4}(-\bar{R}+\frac{1}{4}\bar{F}'_{\bold{K}\bold{L}}\bar{F}'^{\bold{K}\bold{L}})
(\tilde{G}_{\bold{I}\bold{J}}\tilde{G}^{\bold{I}\bold{J}}-\frac{1}{2}\tilde{G}^2)
+(-\frac{1}{2}\bar{R}^{\bold{I}}_{\;\; \bold{J}}+\frac{1}{2}\bar{F}'^{\bold{I}\bold{K}}\bar{F}'_{\bold{J}\bold{K}})
\tilde{G}_{\bold{I}\bold{L}}\tilde{G}^{\bold{J}\bold{L}}
\nonumber \\
&&+(\frac{1}{2}\bar{R}^{\bold{I}\bold{J}}-\frac{1}{4}\bar{F}'^{\bold{I}\bold{K}}\bar{F}'^{\bold{J}}_{\;\;\;\; \bold{K}})
\tilde{G}_{\bold{I}\bold{J}}\tilde{G}
+(-\frac{1}{2}\bar{R}^{\bold{I}\bold{J}\bold{K}\bold{L}}+\frac{1}{4}\bar{F}'^{\bold{I}\bold{J}}\bar{F}'^{\bold{K}\bold{L}})
\tilde{G}_{\bold{I}\bold{K}}\tilde{G}_{\bold{J}\bold{L}}
\nonumber \\
&&+\frac{1}{4}G_N \tilde{F}_{\bold{I}\bold{J}} \tilde{F}^{\bold{I}\bold{J}} 
+\sqrt{G_N} 
(\frac{1}{4} \bar{F}^{'\bold{I}\bold{J}} \tilde{F}_{\bold{I}\bold{J}} \tilde{G} 
-\bar{F}^{'\bold{I}\bold{J}} \tilde{F}_{\bold{I}\bold{K}} \tilde{G}_\bold{J}^{\;\; \bold{K}} ) \Bigr), \label{fluctuation}
\end{eqnarray}
where $\bar{F}'_{\bold{I}\bold{J}}:=\sqrt{G_N}\bar{F}_{\bold{I}\bold{J}}$ is independent of $G_N$. $\tilde{G}:=\bar{G}^{\bold{I}\bold{J}}\tilde{G}_{\bold{I}\bold{J}}$. There is no first order term because the background satisfies the equations of motion. If we take $G_N \to 0$, we obtain 
\begin{eqnarray}
S'&=&\frac{1}{G_N} \int \mathcal{D}\bold{E} \mathcal{D}\bold{X}_T(\bar{\tau}) \mathcal{D}\bar{\tau}  \sqrt{\bar{G}} 
\Bigl(-\bar{R}+\frac{1}{4}\bar{F}'_{\bold{I}\bold{J}}\bar{F}'^{\bold{I}\bold{J}} 
\nonumber \\
&&+\frac{1}{4}\bar{\nabla}_\bold{I} \tilde{G}_{\bold{J}\bold{K}} \bar{\nabla}^\bold{I} \tilde{G}^{\bold{J}\bold{K}}
-\frac{1}{4}\bar{\nabla}_\bold{I} \tilde{G} \bar{\nabla}^\bold{I} \tilde{G}
+\frac{1}{2}\bar{\nabla}^\bold{I} \tilde{G}_{\bold{I}\bold{J}} \bar{\nabla}^\bold{J} \tilde{G}
-\frac{1}{2}\bar{\nabla}^\bold{I} \tilde{G}_{\bold{I}\bold{J}} \bar{\nabla}_\bold{K} \tilde{G}^{\bold{J}\bold{K}}
\nonumber \\
&&-\frac{1}{4}(-\bar{R}+\frac{1}{4}\bar{F}'_{\bold{K}\bold{L}}\bar{F}'^{\bold{K}\bold{L}})
(\tilde{G}_{\bold{I}\bold{J}}\tilde{G}^{\bold{I}\bold{J}}-\frac{1}{2}\tilde{G}^2)
+(-\frac{1}{2}\bar{R}^{\bold{I}}_{\;\; \bold{J}}+\frac{1}{2}\bar{F}'^{\bold{I}\bold{K}}\bar{F}'_{\bold{J}\bold{K}})
\tilde{G}_{\bold{I}\bold{L}}\tilde{G}^{\bold{J}\bold{L}}
\nonumber \\
&&+(\frac{1}{2}\bar{R}^{\bold{I}\bold{J}}-\frac{1}{4}\bar{F}'^{\bold{I}\bold{K}}\bar{F}'^{\bold{J}}_{\;\;\;\; \bold{K}})
\tilde{G}_{\bold{I}\bold{J}}\tilde{G}
+(-\frac{1}{2}\bar{R}^{\bold{I}\bold{J}\bold{K}\bold{L}}+\frac{1}{4}\bar{F}'^{\bold{I}\bold{J}}\bar{F}'^{\bold{K}\bold{L}})
\tilde{G}_{\bold{I}\bold{K}}\tilde{G}_{\bold{J}\bold{L}} \Bigr),
\end{eqnarray}
where the fluctuation of the gauge field is suppressed. In order to fix the gauge symmetry (\ref{Supersubdiffeo}), we take the harmonic gauge. If we add the gauge fixing term
\begin{equation}
S_{fix}=\frac{1}{G_N}\int \mathcal{D}\bold{E} \mathcal{D}\bold{X}_T(\bar{\tau}) \mathcal{D}\bar{\tau}  \sqrt{\bar{G}} 
\frac{1}{2} \Bigl( \bar{\nabla}^\bold{J}(\tilde{G}_{\bold{I}\bold{J}}-\frac{1}{2}\bar{G}_{\bold{I}\bold{J}}\tilde{G}) \Bigr)^2,\end{equation}
we obtain
\begin{eqnarray}
S'+S_{fix}&=&\frac{1}{G_N} \int \mathcal{D}\bold{E} \mathcal{D}\bold{X}_T(\bar{\tau}) \mathcal{D}\bar{\tau}  \sqrt{\bar{G}} 
\Bigl(-\bar{R}+\frac{1}{4}\bar{F}'_{\bold{I}\bold{J}}\bar{F}'^{\bold{I}\bold{J}} 
\nonumber \\
&&+\frac{1}{4}\bar{\nabla}_\bold{I} \tilde{G}_{\bold{J}\bold{K}} \bar{\nabla}^\bold{I} \tilde{G}^{\bold{J}\bold{K}}
-\frac{1}{8}\bar{\nabla}_\bold{I} \tilde{G} \bar{\nabla}^\bold{I} \tilde{G}
\nonumber \\
&&-\frac{1}{4}(-\bar{R}+\frac{1}{4}\bar{F}'_{\bold{K}\bold{L}}\bar{F}'^{\bold{K}\bold{L}})
(\tilde{G}_{\bold{I}\bold{J}}\tilde{G}^{\bold{I}\bold{J}}-\frac{1}{2}\tilde{G}^2)
+(-\frac{1}{2}\bar{R}^{\bold{I}}_{\;\; \bold{J}}+\frac{1}{2}\bar{F}'^{\bold{I}\bold{K}}\bar{F}'_{\bold{J}\bold{K}})
\tilde{G}_{\bold{I}\bold{L}}\tilde{G}^{\bold{J}\bold{L}}
\nonumber \\
&&+(\frac{1}{2}\bar{R}^{\bold{I}\bold{J}}-\frac{1}{4}\bar{F}'^{\bold{I}\bold{K}}\bar{F}'^{\bold{J}}_{\;\;\;\; \bold{K}})
\tilde{G}_{\bold{I}\bold{J}}\tilde{G}
+(-\frac{1}{2}\bar{R}^{\bold{I}\bold{J}\bold{K}\bold{L}}+\frac{1}{4}\bar{F}'^{\bold{I}\bold{J}}\bar{F}'^{\bold{K}\bold{L}})
\tilde{G}_{\bold{I}\bold{K}}\tilde{G}_{\bold{J}\bold{L}} \Bigr). \label{Sfixedaction}\end{eqnarray}

In order to obtain perturbative string amplitudes, we perform a derivative expansion of $\tilde{G}_{\bold{I}\bold{J}}$,
\begin{eqnarray}
&&\tilde{G}_{\bold{I}\bold{J}} \to \frac{1}{\alpha} \tilde{G}_{\bold{I}\bold{J}} \nonumber \\
&&\partial_{\bold{K}}\tilde{G}_{\bold{I}\bold{J}} \to \partial_{\bold{K}}\tilde{G}_{\bold{I}\bold{J}}\nonumber \\
&&\partial_{\bold{K}}\partial_{\bold{L}}\tilde{G}_{\bold{I}\bold{J}} \to \alpha \partial_{\bold{K}}\partial_{\bold{L}}\tilde{G}_{\bold{I}\bold{J}},
\end{eqnarray}
and take
\begin{equation}
\alpha \to 0,
\end{equation}
where $\alpha$ is an arbitrary constant in the solution (\ref{Supersolution}). We normalize the fields as $\tilde{H}_{\bold{I}\bold{J}}:=Z_{\bold{I}\bold{J}} \tilde{G}_{\bold{I}\bold{J}}$, where $Z_{\bold{I}\bold{J}}:=\frac{1}{\sqrt{G_N}} 
\bar{G}^{\frac{1}{4}} 
(\bar{a}_\bold{I} \bar{a}_\bold{J})^{-\frac{1}{2}}$. $\bar{a}_{\bold{I}}$ represent the background metric as $\bar{G}_{\bold{I}\bold{J}}=\bar{a}_\bold{I} \delta_{\bold{I}\bold{J}}$, where $\bar{a}_d=2\lambda\bar{\rho}$ and $\bar{a}_{(\mu \bar{\sigma} \bar{\theta})}=\frac{\hat{\bold{E}}^3}
{\bold{E}}$. Then, (\ref{Sfixedaction}) with appropriate boundary conditions  reduces to
\begin{equation}
S'+S_{fix} \to S_0 + S_2,
\end{equation}
where
\begin{equation}
S_0
=
\frac{1}{G_N} \int \mathcal{D}\bold{E} \mathcal{D}\bold{X}_T(\bar{\tau}) \mathcal{D}\bar{\tau}  \sqrt{\bar{G}} 
\Bigl(-\bar{R}+\frac{1}{4}\bar{F}'_{\bold{I}\bold{J}}\bar{F}'^{\bold{I}\bold{J}} \Bigr),
\end{equation}
and
\begin{equation}
S_2
=
\int \mathcal{D}\bold{E} \mathcal{D}\bold{X}_T(\bar{\tau}) \mathcal{D}\bar{\tau}
\frac{1}{8}\tilde{H}_{\bold{I}\bold{J}}H_{{\bold{I}\bold{J}};\bold{K}\bold{L}}\tilde{H}_{\bold{K}\bold{L}}.
\end{equation}

In the same way as in the previous section, a part of the action 
\begin{equation}
\int \mathcal{D}\bold{E} \mathcal{D}\bold{X}_T(\bar{\tau}) \mathcal{D}\bar{\tau} \frac{1}{4}
\int_0^{2\pi}d\bar{\sigma}d^2\bar{\theta} \tilde{H}^{\bot}_{d(\mu \bar{\sigma} \bar{\theta})} 
H
\tilde{H}^{\bot}_{d(\mu \bar{\sigma} \bar{\theta})} \label{SuperSecondOrderAction}
\end{equation}
with 
\begin{eqnarray}
H
&=&
-\frac{1}{2}\frac{1}{2\lambda\bar{\rho}}(\frac{\partial}{\partial \bar{\tau}})^2
-\frac{1}{2}\int_0^{2\pi} d \bar{\sigma} \int d^2\bar{\theta} \frac{\bold{E}} {\hat{\bold{E}}^2}
 (\frac{\partial}{\partial \bold{X}_T^{\mu}(\bar{\tau})})^2  \nonumber \\
&&+\frac{1}{2}\frac{\bold{D}^2-9\bold{D}+20}{(2-\bold{D})^2}\int_0^{2\pi} d \bar{\sigma} \int d^2\bar{\theta} \bar{\bold{E}} (\tilde{\bold{\bold{D}} }_{\alpha}\bold{X}_T^{\mu}(\bar{\tau}))^2,
\end{eqnarray}
decouples from the other modes.

In the following, we consider a sector that consists of local fluctuations in a sense of strings as 
\begin{equation}
\tilde{H}_{\bold{I}\bold{J}}=\int_0^{2\pi}  d \bar{\sigma} d^2\bar{\theta} \hat{\bold{E}}
\tilde{h}_{\bold{I}\bold{J}}
(\bar{\bold{E}}(\bar{\sigma},  \bar{\tau}, \bar{\theta}), \bold{X}_T(\bar{\sigma},  \bar{\tau}, \bar{\theta}), \bar{\tau})).
\label{LocalSector}
\end{equation}
Because $\bold{X}_T^{\mu}(\bar{\tau})$ can be expanded as $\bold{X}_T^{\mu}(\bar{\tau})= X^{\mu}+ \bar{\theta}^{\alpha} \psi_{\alpha}^{\mu}+\frac{1}{2}\bar{\theta}^2 F^{\mu}$, we have 
\begin{eqnarray}
\left(\frac{\partial}{\partial X^{\mu}(\bar{\sigma},  \bar{\tau})} \right)^2 \tilde{H}_{\bold{I}\bold{J}}
&=&
\left(\frac{\partial}{\partial X^{\mu}(\bar{\sigma},  \bar{\tau})} \right)
\int_0^{2\pi}  d \bar{\sigma}' d^2\bar{\theta}' 
\frac{\partial \bold{X}_T^{\nu}(\bar{\sigma}',  \bar{\tau}, \bar{\theta}')}
{\partial X^{\mu}(\bar{\sigma},  \bar{\tau})}
\frac{\partial}{\partial \bold{X}_T^{\nu}(\bar{\sigma}',  \bar{\tau}, \bar{\theta}')} 
 \tilde{H}_{\bold{I}\bold{J}} \nonumber \\
&=&
\left(\frac{\partial}{\partial X^{\mu}(\bar{\sigma},  \bar{\tau})} \right)
\int_0^{2\pi}   d^2\bar{\theta}' 
\frac{\partial}{\partial \bold{X}_T^{\nu}(\bar{\sigma},  \bar{\tau}, \bar{\theta}')} 
 \tilde{H}_{\bold{I}\bold{J}} \nonumber \\
&=&
\int_0^{2\pi}   d^2\bar{\theta}
\int_0^{2\pi}   d^2\bar{\theta}' 
\frac{\partial}{\partial \bold{X}_T^{\nu}(\bar{\sigma},  \bar{\tau}, \bar{\theta})} 
\frac{\partial}{\partial \bold{X}_T^{\nu}(\bar{\sigma},  \bar{\tau}, \bar{\theta}')} 
 \tilde{H}_{\bold{I}\bold{J}} \nonumber \\
&=&
\int_0^{2\pi}   d^2\bar{\theta}
\frac{\bar{e}}{\hat{\bold{E}}}
(\frac{\partial}{\partial \bold{X}_T^{\nu}(\bar{\sigma},  \bar{\tau}, \bar{\theta})})^2 
 \tilde{H}_{\bold{I}\bold{J}}, 
\end{eqnarray}
where we have used the following facts in the last equality.  (\ref{LocalSector}) implies that
\begin{equation}
\frac{\partial}{\partial \bold{X}_T^{\nu}(\bar{\sigma},  \bar{\tau}, \bar{\theta})} 
\frac{\partial}{\partial \bold{X}_T^{\nu}(\bar{\sigma},  \bar{\tau}, \bar{\theta}')} 
 \tilde{H}_{\bold{I}\bold{J}}
 \sim
\delta^2(\bar{\theta}-\bar{\theta}')
 \frac{\partial}{\partial \bold{X}_T^{\nu}(\bar{\sigma},  \bar{\tau}, \bar{\theta})} 
\frac{\partial}{\partial \bold{X}_T^{\nu}(\bar{\sigma},  \bar{\tau}, \bar{\theta})} 
 \tilde{H}_{\bold{I}\bold{J}}.
\end{equation}
Under the diffeomorphism transformation of  $\bar{\sigma}$ and $\bar{\theta}$, 
$\int  d \bar{\sigma} \bar{e}$
and
$\int  d \bar{\sigma} d^2\bar{\theta} \hat{\bold{E}}$
are transformed as scalars, then 
$\frac{1}{\bar{e}}\delta(\bar{\sigma}-\bar{\sigma}')$
and
$\frac{1}{\hat{\bold{E}}}\delta(\bar{\sigma}-\bar{\sigma}') \delta^2(\bar{\theta}-\bar{\theta}')$
are scalars, thus
$\frac{\bar{e}}{\hat{\bold{E}}}\delta^2(\bar{\theta}-\bar{\theta}')$
is a scalar. Therefore,
\begin{equation}
\frac{\partial}{\partial \bold{X}_T^{\nu}(\bar{\sigma},  \bar{\tau}, \bar{\theta})} 
\frac{\partial}{\partial \bold{X}_T^{\nu}(\bar{\sigma},  \bar{\tau}, \bar{\theta}')} 
 \tilde{H}_{\bold{I}\bold{J}}
 =
 \frac{\bar{e}}{\hat{\bold{E}}}\delta^2(\bar{\theta}-\bar{\theta}')
 \frac{\partial}{\partial \bold{X}_T^{\nu}(\bar{\sigma},  \bar{\tau}, \bar{\theta})} 
\frac{\partial}{\partial \bold{X}_T^{\nu}(\bar{\sigma},  \bar{\tau}, \bar{\theta})} 
 \tilde{H}_{\bold{I}\bold{J}}.
\end{equation}

Then, (\ref{SuperSecondOrderAction}) can be simplified where
\begin{eqnarray}
H
&=&
-\frac{1}{2}\frac{1}{2\lambda\bar{\rho}}(\frac{\partial}{\partial \bar{\tau}})^2
-\frac{1}{2}\int_0^{2\pi} d \bar{\sigma} \frac{\sqrt{\bar{h}}}{\bar{e}^2} (\frac{\partial}{\partial X^{\mu}})^2 
+\frac{1}{2}\frac{\bold{D}^2-9\bold{D}+20}{(2-\bold{D})^2}\int_0^{2\pi} d \bar{\sigma} \int d^2\bar{\theta} \bar{\bold{E}} (\tilde{\bold{D} }_{\alpha}\bold{X}_T^{\mu}(\bar{\tau}))^2.
\nonumber \\
&&
\end{eqnarray}
By adding to (\ref{SuperSecondOrderAction}), a formula similar to the bosonic case

\begin{eqnarray}
0
&=&
\int \mathcal{D}\bold{E} \mathcal{D}\bold{X}_T(\bar{\tau}) \mathcal{D}\bar{\tau} \frac{1}{4}
\int_0^{2\pi}d\bar{\sigma}'d^2\bar{\theta}' \tilde{H}^{\bot}_{d(\mu \bar{\sigma}' \bar{\theta}')} ( \int_0^{2\pi} d \bar{\sigma}
\bar{n}^{\bar{\sigma}}
\partial_{\bar{\sigma}} X^{\mu}  \frac{\partial}{\partial X^{\mu}})
\tilde{H}^{\bot}_{d(\mu \bar{\sigma}' \bar{\theta}')} 
\end{eqnarray}
and
\begin{equation}
0=\int \mathcal{D}\bold{E} \mathcal{D}\bold{X}_T(\bar{\tau}) \mathcal{D}\bar{\tau} \frac{1}{4}
\int_0^{2\pi}d\bar{\sigma}'d^2\bar{\theta}'
\tilde{H}^{\bot}_{d(\mu \bar{\sigma}' \bar{\theta}')}
 \int_0^{2\pi} d \bar{\sigma} \bar{E} 
\frac{-i}{2} \bar{n} \bar{\chi}_{m} \bar{E}^0_{\; r} \gamma^{r}
\bar{E}^{m}_{\; q} \gamma^q \psi^{\mu}(-i \frac{1}{\bar{e}}\frac{\partial}{\partial X^{\mu}})
\tilde{H}^{\bot}_{d(\mu \bar{\sigma}' \bar{\theta}')}, 
\label{SuperZero}
\end{equation}
we obtain (\ref{SuperSecondOrderAction}) 
with
\begin{eqnarray}
&&H(-i\frac{\partial}{\partial \bar{\tau}}, -i\frac{1}{\bar{e}}\frac{\partial}{\partial X}, \bold{X}_T(\bar{\tau}), \bar{\bold{E}}) \nonumber \\
&=&
\frac{1}{2}\frac{1}{2\lambda\bar{\rho}}(-i\frac{\partial}{\partial \bar{\tau}})^2
+\int d\bar{\sigma} \Biggl( \sqrt{\bar{h}} \left(\frac{1}{2}(-i\frac{1}{\bar{e}}\frac{\partial}{\partial X^{\mu}})^2  -\frac{i}{2} \bar{n} \bar{\chi}_{m} \bar{E}^0_{\; r} \gamma^{r} \bar{E}^{m}_{\; q} \gamma^q \psi^{\mu}(-i\frac{1}{\bar{e}}\frac{\partial}{\partial X^{\mu}})\right)
\nonumber \\
&&+i \bar{e} \bar{n}^{\bar{\sigma}} \partial_{\bar{\sigma}} X^{\mu} (-i\frac{1}{\bar{e}}\frac{\partial}{\partial X^{\mu}})\Biggr)
+\int d\bar{\sigma} d^2\bar{\theta} \bar{\bold{E}}  \frac{1}{2}(\tilde{\bold{D} }_{\alpha} \bold{X}_{T\mu}(\bar{\tau}))^2, \nonumber \\ \label{SuperbosonicHamiltonian}
\end{eqnarray}
where we have taken $\bold{D} \to \infty$.
(\ref{SuperZero}) is true because the integrand of the right hand side is a total derivative with respect to $X^{\mu}$.

The propagator for $\tilde{H}^{\bot}_{d(\mu \bar{\sigma} \bar{\theta})}$;\begin{equation}
\Delta_F(\bar{\bold{E}}, \bold{X}_T(\bar{\tau}), \bar{\tau}; \; \bar{\bold{E}}', \bold{X}'_T(\bar{\tau}'), \bar{\tau}')
=<\tilde{H}^{\bot}_{d(\mu \bar{\sigma} \bar{\theta})}
(\bar{\bold{E}}, \bold{X}_T(\bar{\tau}), \bar{\tau})
\tilde{H}^{\bot}_{d(\mu \bar{\sigma} \bar{\theta})}
(\bar{\bold{E}}', \bold{X}'_T(\bar{\tau}'), \bar{\tau}')>
\end{equation}
satisfies
\begin{equation}
H(-i\frac{\partial}{\partial \bar{\tau}}, -i\frac{1}{\bar{e}}\frac{\partial}{\partial X}, \bold{X}_T(\bar{\tau}), \bar{\bold{E}})
\Delta_F(\bar{\bold{E}}, \bold{X}_T(\bar{\tau}), \bar{\tau}; \; \bar{\bold{E}}', \bold{X}'_T(\bar{\tau}'), \bar{\tau}')=
\delta(\bar{\bold{E}}-\bar{\bold{E}}')\delta(\bold{X}_T(\bar{\tau})-\bold{X}_T'(\bar{\tau}'))\delta(\bar{\tau}-\bar{\tau}'). \label{SuperPropagatorDef}
\end{equation}

In order to obtain a Schwinger representation of the propagator, we use the operator formalism $(\hat{\bar{\bold{E}}},  \hat{\bold{X}}_T(\hat{\bar{\tau}}), \hat{\bar{\tau}})$ of the first quantization. 
The eigen state for $(\hat{\bar{\bold{E}}},  \hat{X}, \hat{\bar{\tau}})$ is given by $|\bar{\bold{E}}, X, \bar{\tau}>$. The conjugate momentum is written as $(\hat{\bold{p}}_{\bar{\bold{E}}}, \hat{p}_{X}, \hat{p}_{\bar{\tau}})$. There is no conjugate momentum for the auxiliary field $F^{\mu}$, whereas the Majorana fermion $\psi^{\mu}_{\alpha}$ is self-conjugate. The renormalized operators $\hat{\tilde{\psi}}^{\mu}_{\alpha}$ satisfy $\{ \hat{\tilde{\psi}}_{\alpha}^{\mu}(\bar{\sigma}), \hat{\tilde{\psi}}_{\beta}^{\nu}(\bar{\sigma}') \}= \frac{1}{\bar{E}}\delta_{\alpha \beta} \eta^{\mu \nu} \delta(\bar{\sigma}-\bar{\sigma}')$ as summarized in the appendix B.  By defining creation and annihilation operators for $\tilde{\psi}^{\mu}_{\alpha}$ as $\hat{\tilde{\psi}}^{\mu \dagger}:= \frac{1}{\sqrt{2}}(\hat{\tilde{\psi}}^{\mu}_1-i\hat{\tilde{\psi}}^{\mu}_2)$ and $\hat{\tilde{\psi}}^{\mu}:= \frac{1}{\sqrt{2}}(\hat{\tilde{\psi}}^{\mu}_1+i\hat{\tilde{\psi}}^{\mu}_2)$,
one obtains an algebra $\{ \hat{\tilde{\psi}}^{\mu}(\bar{\sigma}), \hat{\tilde{\psi}}^{\nu \dagger}(\bar{\sigma}') \}= \frac{1}{\bar{E}} \eta^{\mu \nu} \delta(\bar{\sigma}-\bar{\sigma}')$, $\{ \hat{\tilde{\psi}}^{\mu}(\bar{\sigma}), \hat{\tilde{\psi}}^{\nu}(\bar{\sigma}') \}= 0$, and $\{ \hat{\tilde{\psi}}^{\mu \dagger}(\bar{\sigma}), \hat{\tilde{\psi}}^{\nu \dagger}(\bar{\sigma}') \}=0$. The vacuum $|0>$ for this algebra is defined by $\hat{\tilde{\psi}}^{\mu}(\bar{\sigma}) |0>=0$. The eigen state $|\tilde{\psi}>$, which satisfies $\hat{\tilde{\psi}}^{\mu}(\bar{\sigma}) |\tilde{\psi}>= \tilde{\psi}^{\mu}(\bar{\sigma}) |\tilde{\psi}>$, is given by 
$
e^{-\tilde{\psi} \cdot \hat{\tilde{\psi}}^{\dagger}} |0>
=
e^{- \int d\bar{\sigma} \bar{E} \tilde{\psi}_{\mu}(\bar{\sigma}) \hat{\tilde{\psi}}^{\mu \dagger}(\bar{\sigma})} |0>$. Then, the inner product is given by $<\tilde{\psi} | \tilde{\psi}'>=e^{\tilde{\psi}^{\dagger} \cdot \tilde{\psi}'}$, whereas the completeness relation is
$\int \mathcal{D}\tilde{\psi}^{\dagger} \mathcal{D}\tilde{\psi} |\tilde{\psi}>e^{-\tilde{\psi}^{\dagger} \cdot \tilde{\psi}}<\tilde{\psi}|=1$.

Because (\ref{SuperPropagatorDef}) means that $\Delta_F$ is an inverse of $H$, $\Delta_F$ can be expressed by a matrix element of the operator $\hat{H}^{-1}$ as
\begin{equation}
\Delta_F(\bar{\bold{E}}, \bold{X}_T(\bar{\tau}), \bar{\tau}; \; \bar{\bold{E}}',  \bold{X}_T'(\bar{\tau}'), \bar{\tau}')
=
<\bar{\bold{E}}, \bold{X}_T(\bar{\tau}), \bar{\tau}| \hat{H}^{-1}(\hat{p}_{\bar{\tau}}, \hat{p}_{X},  \hat{\bold{X}}_T(\hat{\bar{\tau}}), \hat{\bar{\bold{E}}}) |\bar{\bold{E}}',  \bold{X}_T'(\bar{\tau}'), \bar{\tau}'>.
\label{SuperInverseH2}
\end{equation}
(\ref{IntegralFormula}) implies that
\begin{equation}
\Delta_F(\bar{\bold{E}}, \bold{X}_T(\bar{\tau}), \bar{\tau}; \; \bar{\bold{E}}',  \bold{X}_T'(\bar{\tau}'), \bar{\tau}')
=
\int _0^{\infty} dT <\bar{\bold{E}}, \bold{X}_T(\bar{\tau}), \bar{\tau}|  e^{-T\hat{H}} |\bar{\bold{E}}',  \bold{X}_T'(\bar{\tau}'), \bar{\tau}'>.
\end{equation}
In order to define two-point correlation functions that is invariant under the general coordinate transformations in the superstring geometry, we define in and out states as
\begin{eqnarray}
||\bold{X}_{T i} \,|\,\bold{E}_f, ; \bold{E}_i>_{in}&:=& \int_{\bold{E}_i}^{\bold{E}_f} \mathcal{D}\bold{E}'|\bar{\bold{E}},' \bold{X}_{T i}:=\bold{X}'_T(\bar{\tau}'=-\infty), \bar{\tau}=-\infty > \nonumber \\
<\bold{X}_{T f}\,|\,\bold{E}_f, ; \bold{E}_i||_{out}&:=& \int_{\bold{E}_i}^{\bold{E}_f} \mathcal{D} \bold{E} <\bar{\bold{E}}, \bold{X}_{T f}:=\bold{X}_T(\bar{\tau}=\infty), \bar{\tau}=\infty|,
\end{eqnarray}
where $\bold{E}_i$ and $\bold{E}_f$ represent the super vierbeins of the supercylinders at $\bar{\tau}=\pm \infty$, respectively. When we insert asymptotic states, we integrate out $\bold{X}_{T f}$, $\bold{X}_{T i}$, $\bold{E}_f$ and $\bold{E}_i$ in the two-point correlation function for these states;
\begin{eqnarray}
\Delta_F(\bold{X}_{T f}; \bold{X}_{T i}|\bold{E}_f, ; \bold{E}_i) 
&=&
\int_{\bold{E}_i}^{\bold{E}_f} \mathcal{D}\bold{E}
\int_{\bold{E}_i}^{\bold{E}_f} \mathcal{D} \bold{E}' 
<\tilde{H}^{\bot}_{d(\mu \bar{\sigma} \bar{\theta})}
(\bar{\bold{E}}, \bold{X}_{T f}:=\bold{X}_T(\bar{\tau}=\infty), \bar{\tau}=\infty )
\nonumber \\
&&\tilde{H}^{\bot}_{d(\mu \bar{\sigma} \bar{\theta})}
(\bar{\bold{E}}', \bold{X}_{T i}:=\bold{X}'_T(\bar{\tau}'=-\infty), \bar{\tau}'=-\infty)>.
\end{eqnarray}
By inserting 
\begin{eqnarray}
1&=&
\int d\bar{\bold{E}}_{m} d\bar{\tau}_m d\bold{X}_{Tm} (\bar{\tau}_m ) 
|\bar{\bold{E}}_{m}, \bar{\tau}_m, \bold{X}_{Tm}(\bar{\tau}_m) > 
e^{-\tilde{\psi}^{\dagger}_m \cdot \tilde{\psi}_m}
<\bar{\bold{E}}_{m}, \bar{\tau}_m, \bold{X}_{Tm}(\bar{\tau}_m) |
\nonumber \\ 
1&=&
\int  dp_{\bar{\tau}}^i dp_{X}^i
|p_{\bar{\tau}}^i, p_{X}^i>
<p_{\bar{\tau}}^i, p_{X}^i|,
\end{eqnarray}
This can be written as\footnote{The correlation function is zero if $\bold{E}_i$ and $\bold{E}_f$ of the in state do not coincide with those of the out states, because of the delta functions in the sixth line.}
\begin{eqnarray}
&&\Delta_F(\bold{X}_{T f}; \bold{X}_{T i}|\bold{E}_f, ; \bold{E}_i) \nonumber \\
&:=&\int _0^{\infty} dT <\bold{X}_{T f} \,|\,\bold{E}_f, ; \bold{E}_i||_{out}  e^{-T\hat{H}} ||\bold{X}_{T i} \,|\,\bold{E}_f, ; \bold{E}_i>_{in}\nonumber \\
&=&
\int _0^{\infty} dT \lim_{N \to \infty}
\int_{\bold{E}_i}^{\bold{E}_f}\mathcal{D} \bold{E}
\int_{\bold{E}_i}^{\bold{E}_f}\mathcal{D} \bold{E}'
\prod_{m=1}^N \prod_{i=0}^N
\int d\bar{\bold{E}}_{m} d\bar{\tau}_m d\bold{X}_{Tm}(\bar{\tau}_m)  
e^{-\tilde{\psi}^{\dagger}_m \cdot \tilde{\psi}_m} \nonumber \\
&&
<\bar{\bold{E}}_{i+1}, \bar{\tau}_{i+1}, \bold{X}_{Ti+1}(\bar{\tau}_{i+1})| e^{-\frac{1}{N}T \hat{H}} |\bar{\bold{E}}_{i}, \bar{\tau}_i, \bold{X}_{Ti}(\bar{\tau}_i)>
\nonumber \\
&=&
\int _0^{\infty} dT_0 \lim_{N \to \infty} \int d T_{N+1}
\int_{\bold{E}_i}^{\bold{E}_f}\mathcal{D} \bold{E}
\int_{\bold{E}_i}^{\bold{E}_f}\mathcal{D} \bold{E}'
\prod_{m=1}^N \prod_{i=0}^N
\int d T_m  d\bar{\bold{E}}_{m} d\bar{\tau}_m d\bold{X}_{Tm}(\bar{\tau}_m)  e^{-\tilde{\psi}^{\dagger}_m \cdot \tilde{\psi}_m}
\nonumber \\
&&
<\bar{\tau}_{i+1}, \bold{X}_{Ti+1}(\bar{\tau}_{i+1})| e^{-\frac{1}{N}T_i \hat{H}}  |\bar{\tau}_i, \bold{X}_{Ti}(\bar{\tau}_{i})> \delta(T_{i}-T_{i+1}) \delta(\bar{\bold{E}}_{i+1}-\bar{\bold{E}}_{i})
\nonumber \\
&=&\int _0^{\infty} dT_0 \lim_{N \to \infty} \int d T_{N+1}
\int_{\bold{E}_i}^{\bold{E}_f}\mathcal{D} \bold{E}  
\prod_{m=1}^N \prod_{i=0}^N
\int d T_m   d\bar{\tau}_m d\bold{X}_{Tm}(\bar{\tau}_m)  e^{-\tilde{\psi}^{\dagger}_m \cdot \tilde{\psi}_m}
\nonumber \\
&&
\int dp_{\bar{\tau}}^i dp_{X}^i
<\bar{\tau}_{i+1}, X_{i+1}| 
p_{\bar{\tau}}^i, p_{X}^i>
\nonumber \\
&&
<p_{\bar{\tau}}^i, p_{X}^i|
<\tilde{\psi}_{i+1}, F_{i+1}|
e^{-\frac{1}{N}T_i \hat{H}}
|\tilde{\psi}_{i}, F_{i}> |\bar{\tau}_i, X_i> \delta(T_{i}-T_{i+1}) 
\nonumber \\
&=&
\int _0^{\infty} dT_0 \lim_{N \to \infty} \int d T_{N+1} 
\int_{\bold{E}_i}^{\bold{E}_f}\mathcal{D} \bold{E}
\prod_{m=1}^N \prod_{i=0}^N
\int d T_m d\bar{\tau}_m d\bold{X}_{Tm}(\bar{\tau}_m)  e^{-\tilde{\psi}^{\dagger}_m \cdot \tilde{\psi}_m}
\nonumber \\
&&
\int dp_{\bar{\tau}}^i dp_{X}^i
 e^{-\frac{1}{N}T_i H (p_{\bar{\tau}}^i, p_{X}^i, \bold{X}_{Ti}(\bar{\tau}_{i}), \bar{\bold{E}})}
e^{\tilde{\psi}^{\dagger}_{i+1} \cdot \tilde{\psi}_i}
\delta(F_{i}-F_{i+1}) \delta(T_{i}-T_{i+1}) 
\nonumber \\
&&
e^{i(p_{\bar{\tau}}^i(\bar{\tau}_{i+1}-\bar{\tau}_{i})+p_{X}^i\cdot(X_{i+1}-X_{i}))}
\nonumber \\
&=&
\int _0^{\infty} dT_0 \lim_{N \to \infty} \int d T_{N+1} 
\int_{\bold{E}_i}^{\bold{E}_f}\mathcal{D} \bold{E}
\prod_{m=1}^N \prod_{i=0}^N
\int d T_m  d\bar{\tau}_m d\bold{X}_{Tm}(\bar{\tau}_m)  
\int dp_{T_{i}}  dp_{\bar{\tau}}^i dp_{X}^i dp_{F_{i}}
\nonumber \\
&&
\exp \Biggl(- \sum_{i=0}^{N} \triangle t \Bigl(-ip_{T_{i}} \frac{T_{i+1}-T_{i}}{\Delta t} 
-ip_{F_{i}} \frac{F_{i+1}-F_{i}}{\Delta t} 
+\tilde{\psi}^{\dagger}_{i+1} \cdot \frac{\tilde{\psi}_{i+1}-\tilde{\psi}_{i}}{\triangle t}\nonumber \\
&&
-ip_{\bar{\tau}}^{i} \frac{\bar{\tau}_{i+1}-\bar{\tau}_{i}}{\triangle t} -ip_{X}^{i}\cdot \frac{X_{i+1}-X_{i}}{\triangle t}
+T_i H(p_{\bar{\tau}}^{i}, p_{X}^{i}, \bold{X}_{Ti}(\bar{\tau}_{i}), \bar{\bold{E}}) \Bigr)\Biggr)
 \nonumber \\
&&
e^{\tilde{\psi}^{\dagger}_{N+1} \cdot \tilde{\psi}_{N+1}}
\nonumber \\
&=&
\int^{\bold{E}_f, \bold{X}_{T f}, \infty }_{\bold{E}_i, \bold{X}_{T i}, -\infty} 
\mathcal{D} T \mathcal{D}\bold{E} \mathcal{D}\bar{\tau} \mathcal{D}\bold{X}_T(\bar{\tau}) 
\int 
\mathcal{D} p_T
\mathcal{D}p_{\bar{\tau}}  \mathcal{D}p_{X} \mathcal{D}p_{F} 
\nonumber \\
&&e^{- \int_{-\infty}^{\infty} dt (-ip_{T} \cdot \frac{d}{dt} T  -ip_{F} \cdot \frac{d}{dt} F +\tilde{\psi}^{\dagger} \cdot \frac{d}{dt}\tilde{\psi}-ip_{\bar{\tau}} \frac{d}{dt} \bar{\tau}-ip_{X} \cdot \frac{d}{dt} X+TH(p_{\bar{\tau}}, p_{X}, \bold{X}_T(\bar{\tau}), \bar{\bold{E}}))},
\end{eqnarray}
where $\bar{\bold{E}}_{0}=\bar{\bold{E}}'$, $\bar{\tau}_0=-\infty$, $\bold{X}_{T 0}=\bold{X}_{T i}$, $\bar{\bold{E}}_{N+1}=\bar{\bold{E}}$, $\bar{\tau}_{N+1}=\infty$, and $\bold{X}_{T N+1}=\bold{X}_{T f}$. $p_{X} \cdot \frac{d}{dt}X=\int d\bar{\sigma} \bar{e} p_{X}^{\mu} \frac{d}{dt} X_{\mu}$ and $\Delta t=\frac{1}{\sqrt{N}}$ as in the bosonic case. 
A trajectory of points $[\bar{{\boldsymbol \Sigma}}, \bold{X}_T(\bar{\tau}), \bar{\tau}]$ is necessarily continuous in $\mathfrak{M}_{D_T}$ so that the kernel $<\bar{\bold{E}}_{i+1}, \bar{\tau}_{i+1}, \bold{X}_{T i+1}(\bar{\tau}_{i+1})| e^{-\frac{1}{N}T_i \hat{H}}  |\bar{\bold{E}}_{i}, \bar{\tau}_i, \bold{X}_{T i}(\bar{\tau}_i)>$ in the fourth line is non-zero when $N \to \infty$. If we integrate out $p_{\bar{\tau}}(t)$, $p_{X}(t)$ and $p_{F}(t)$ by using the relation of the ADM formalism and the relation between  $\tilde{\psi}^{\mu}$ and $\psi^{\mu}$ in the appendix A and B, we obtain
\begin{eqnarray}
&&\Delta_F(\bold{X}_{T f}; \bold{X}_{T i}|\bold{E}_f, ; \bold{E}_i) \nonumber \\
&=&
\int^{\bold{E}_f, \bold{X}_{T f}, \infty }_{\bold{E}_i, \bold{X}_{T i}, -\infty}  
\mathcal{D} T
\mathcal{D}\bold{E} \mathcal{D}\bar{\tau} \mathcal{D}\bold{X}_T(\bar{\tau})\int 
\mathcal{D} p_T
\nonumber \\
&& \exp \Biggl(- \int_{-\infty}^{\infty} dt \Bigl(-i p_{T}(t) \frac{d}{dt} T(t)   +\lambda\bar{\rho}\frac{1}{T(t)}(\frac{d \bar{\tau}(t)}{dt})^2\nonumber \\
&&+\int d\bar{\sigma} \sqrt{\bar{h}} T(t) \biggl( \frac{1}{2 \bar{n}^2}(\frac{1}{T(t)}\frac{\partial}{\partial t} X^{\mu} -\bar{n}^{\bar{\sigma}} \partial_{\bar{\sigma}} X^{\mu}
+\frac{1}{2}\bar{n}^2 \bar{\chi}_{m} \bar{E}^{0}_{r} \gamma^{r} \bar{E}^{m}_{q} \gamma^q \psi_{\mu})^2 
\nonumber \\
&&
 -\frac{1}{2}\frac{1}{T(t)} \bar{\psi}^{\mu} \bar{E}^{0}_{q}\gamma^{q}\frac{\partial}{\partial t} \psi_{\mu} \biggr)  +\int d\bar{\sigma} d^2\bar{\theta}(\bar{\bold{E}}  \frac{1}{2}T(t)(\tilde{\bold{D} }_{\alpha} \bold{X}_{T\mu}(\bar{\tau}))^2)\Bigr) \Biggr)
\nonumber \\
&=&
\int^{\bold{E}_f, \bold{X}_{T f}, \infty }_{\bold{E}_i, \bold{X}_{T i}, -\infty}  
\mathcal{D} T 
\mathcal{D}\bold{E} \mathcal{D}\bar{\tau} \mathcal{D}\bold{X}_T(\bar{\tau})\int 
\mathcal{D} p_T
\exp \Biggl(- \int_{-\infty}^{\infty} dt \Bigl(-i p_{T}(t) \frac{d}{dt} T(t) 
\nonumber \\
&&  +\lambda\bar{\rho}\frac{1}{T(t)}(\frac{d \bar{\tau}(t)}{dt})^2
+\int d\bar{\sigma} d^2\bar{\theta}(\bar{\bold{E}}  \frac{1}{2}T(t)(\bar{\bold{D}}'_{\alpha} \bold{X}_{T\mu}(\bar{\tau}))^2)\Bigr) \Biggr).
\end{eqnarray}
When the last equality is obtained, we use (\ref{quad1}) and (\ref{quad2}).
In the last line, $F^{\mu}$ is constant with respect to $t$, and $\bar{\bold{D}} '_{\alpha}$ is given by replacing $\frac{\partial}{\partial \bar{\tau}}$ with $\frac{1}{T(t)}\frac{\partial}{\partial t}$ in $\bar{\bold{D}}_{\alpha}$. 
The path integral is defined over all possible trajectories with fixed boundary values, on the superstring manifold $\mathfrak{M}_{D_T}$.

By inserting
$\int \mathcal{D}c \mathcal{D}b
e^{\int_0^{1} dt \left(\frac{d b(t)}{dt} \frac{d c(t)}{dt}\right)
},$
where $b(t)$ and $c(t)$ are bc ghosts, we obtain 
\begin{eqnarray}
&&\Delta_F(\bold{X}_{T f}; \bold{X}_{T i}|\bold{E}_f, ; \bold{E}_i)
\nonumber \\
&=&
\bold{Z}_0\int^{\bold{E}_f, \bold{X}_{T f}, \infty }_{\bold{E}_i, \bold{X}_{T i}, -\infty}  
\mathcal{D} T 
\mathcal{D}\bold{E} \mathcal{D}\bar{\tau} \mathcal{D}\bold{X}_T(\bar{\tau})
\mathcal{D}c \mathcal{D}b
\int 
\mathcal{D} p_T
\exp \Biggl(- \int_{-\infty}^{\infty} dt \Bigl(-i p_{T}(t) \frac{d}{dt} T(t) 
\nonumber \\
&&
 +\frac{d b(t)}{dt} \frac{d (T(t) c(t))}{dt}
+\lambda\bar{\rho}\frac{1}{T(t)}(\frac{d \bar{\tau}(t)}{dt})^2
+\int d\bar{\sigma} d^2\bar{\theta}(\bar{\bold{E}}  \frac{1}{2}T(t)(\bar{\bold{D}}'_{\alpha} \bold{X}_{T\mu}(\bar{\tau}))^2)\Bigr) \Biggr). \nonumber \\
&&
\end{eqnarray}
where we have redefined as $c(t) \to T(t) c(t)$. $\bold{Z}_0$ represents an overall constant factor, and we will rename it $\bold{Z}_1, \bold{Z}_2, \cdots$ when the factor changes in the following.
This path integral is obtained if 
\begin{equation}
F_1(t):=\frac{d}{dt}T(t)=0 \label{superF1gauge}
\end{equation}
 gauge is chosen in 
\begin{eqnarray}
&&\Delta_F(\bold{X}_{T f}; \bold{X}_{T i}|\bold{E}_f, ; \bold{E}_i)
\nonumber \\
&=&
\bold{Z}_1\int^{\bold{E}_f, \bold{X}_{T f}, \infty }_{\bold{E}_i, \bold{X}_{T i}, -\infty}  
\mathcal{D} T 
\mathcal{D}\bold{E} \mathcal{D}\bar{\tau} \mathcal{D}\bold{X}_T(\bar{\tau})
\int 
\exp \Biggl(- \int_{-\infty}^{\infty} dt \Bigl(
\nonumber \\
&&  
+\lambda\bar{\rho}\frac{1}{T(t)}(\frac{d \bar{\tau}(t)}{dt})^2
+\int d\bar{\sigma} d^2\bar{\theta}(\bar{\bold{E}}  \frac{1}{2}T(t)(\bar{\bold{D}}'_{\alpha} \bold{X}_{T\mu}(\bar{\tau}))^2)\Bigr) \Biggr),
\label{superpathint}
\end{eqnarray}
which has a manifest one-dimensional diffeomorphism symmetry with respect to $t$, where $T(t)$ is transformed as an einbein \cite{Schwinger0}. 

Under $\frac{d\bar{\tau}}{d\bar{\tau}'}=T(t)$, $T(t)$ disappears in (\ref{superpathint}) as in the bosonic case, and we obtain 
\begin{eqnarray}
&&\Delta_F(\bold{X}_{T f}; \bold{X}_{T i}|\bold{E}_f, ; \bold{E}_i)
\nonumber \\
&=&
\bold{Z}_2\int^{\bold{E}_f, \bold{X}_{T f}, \infty }_{\bold{E}_i, \bold{X}_{T i}, -\infty}  
\mathcal{D}\bold{E} \mathcal{D}\bar{\tau} \mathcal{D}\bold{X}_T(\bar{\tau})
\int 
\exp \Biggl(- \int_{-\infty}^{\infty} dt \Bigl(
\nonumber \\
&&  
+\lambda\bar{\rho}(\frac{d \bar{\tau}(t)}{dt})^2
+\int d\bar{\sigma} d^2\bar{\theta}(\bar{\bold{E}}  \frac{1}{2}(\bar{\bold{D}}''_{\alpha} \bold{X}_{T\mu}(\bar{\tau}))^2)\Bigr) \Biggr),
\label{superpathint2}
\end{eqnarray}
where $\bar{\bold{D}} ''_{\alpha}$ is given by replacing $\frac{\partial}{\partial \bar{\tau}}$ with $\frac{\partial}{\partial t}$ in $\bar{\bold{D}}_{\alpha}$. This action is still invariant under the diffeomorphism with respect to t if $\bar{\tau}$ transforms in the same way as $t$.

If we choose a different gauge
\begin{equation}
F_2(t):=\bar{\tau}-t=0, \label{superF2gauge}
\end{equation} 
in (\ref{superpathint2}), we obtain 
\begin{eqnarray}
&&\Delta_F(\bold{X}_{T f}; \bold{X}_{T i}|\bold{E}_f, ; \bold{E}_i)
\nonumber \\
&=&
\bold{Z}_3\int^{\bold{E}_f, \bold{X}_{T f}, \infty }_{\bold{E}_i, \bold{X}_{T i}, -\infty}  
\mathcal{D}\bold{E} \mathcal{D}\bar{\tau} \mathcal{D}\bold{X}_T(\bar{\tau})
\int 
\mathcal{D} \alpha \mathcal{D}c \mathcal{D}b
\nonumber \\
&&
\exp \Biggl(- \int_{-\infty}^{\infty} dt \Bigl(\alpha(t) (\bar{\tau}-t) +b(t)c(t)(1-\frac{d \bar{\tau}(t)}{dt}) +\lambda \bar{\rho}(\frac{d \bar{\tau}(t)}{dt})^2 \nonumber \\
&&+\int d\bar{\sigma} d^2\bar{\theta}(\bar{\bold{E}}  \frac{1}{2}(\bar{\bold{D}}''_{\alpha} \bold{X}_{T\mu}(\bar{\tau}))^2)\Bigr) \Biggr)\nonumber \\
&=&
\bold{Z}
\int^{\bold{E}_f, \bold{X}_{T f}}_{\bold{E}_i, \bold{X}_{T i}}
\mathcal{D}\bold{E}  \mathcal{D}\bold{X}_T
\nonumber \\
&&
\exp \Biggl(- \int_{-\infty}^{\infty} d\bar{\tau} 
\Bigl(
\frac{1}{4 \pi}\int d\bar{\sigma} \sqrt{\bar{h}} 
\lambda \bar{R}(\bar{\sigma}, \bar{\tau})+\int d\bar{\sigma} d^2\bar{\theta}(\bar{\bold{E}}  \frac{1}{2}(\bar{\bold{D}}_{\alpha} \bold{X}_{T\mu})^2)\Bigr) \Biggr). \nonumber \\
\end{eqnarray}
In the second equality, we have redefined as $c(t)(1-\frac{d \bar{\tau}(t)}{dt}) \to c(t)$ and integrated out the ghosts. The path integral is defined over all possible two-dimensional super Riemannian manifolds with fixed punctures in $\bold{R}^{d}$ as in Fig. \ref{7Pathintegral}. By using the two-dimensional superdiffeomorphism and super Weyl invariance of the action, we obtain
\begin{equation}
\Delta_F(\bold{X}_{T f}; \bold{X}_{T i}|\bold{E}_f, ; \bold{E}_i)
=
\bold{Z}
\int^{\bold{E}_f, \bold{X}_{T f}}_{\bold{E}_i, \bold{X}_{T i}}
\mathcal{D}\bold{E} \mathcal{D}\bold{X}_T
e^{-\lambda \chi} e^{-\int d^2\sigma d^2\theta \bold{E}  \frac{1}{2}(\bold{D}_{\alpha} \bold{X}_{T\mu})^2}, \label{cSuperLast}
\end{equation}
where $\chi$ is the Euler number of the two-dimensional super Riemannian manifold. By inserting asymptotic states to (\ref{cSuperLast}) and renormalizing the metric, we obtain the all-order  perturbative scattering amplitudes that possess the supermoduli in the type IIA and IIB superstring theory for $T=$ IIA and IIB, respectively\cite{textbook}. Especially, in superstring geometry, the consistency of the perturbation theory around the background (\ref{solution}) determines $d=10$ (the critical dimension).

\section{Including open strings}
\setcounter{equation}{0}

Let us define unique global times on oriented open-closed string worldsheets $\bar{\Sigma}$ with open and closed punctures in order to define string states by world-time constant lines. $\bar{\Sigma}$ can be given by $\bar{\Sigma}=\bar{\Sigma}_c/Z_2$ where $Z_2$ is generated by an anti-holomorphic involution $\rho$ and $\bar{\Sigma}_c$ is an oriented Riemann surface with closed punctures that satisfies $\rho(\bar{\Sigma}_c)=\bar{\Sigma}_c^* \cong \bar{\Sigma}_c$. That is, $\bar{\Sigma}_c$ is an oriented closed double cover of $\bar{\Sigma}$ \cite{WittenSupermoduli}. First of all, we define global coordinates on $\bar{\Sigma}_c$ in the same way as in section 2. The real part of the global coordinates $\bar{\tau}$ remains on $\bar{\Sigma}$ because $\rho$ is an anti-holomorphic involution. If $\rho$ maps a puncture to another puncture on $\bar{\Sigma}_c$, the discs around the punctures are identified and give a disk $D^i$ around a closed puncture $P^i$ on $\bar{\Sigma}$. On the other hand, if $\rho$ maps a puncture to itself on $\bar{\Sigma}_c$, the disc around the puncture is identified with itself and gives an upper half disk $\tilde{D}^j$ around an open puncture $\tilde{P}^j$ on $\bar{\Sigma}$. The $\bar{\sigma}$ regions around $P^i$ and $\tilde{P}^j$ are $2\pi f^i$ and $\pi \tilde{f}^j$, respectively where $\sum_{i=1}^N 2f^i+ \sum_{j=1}^M \tilde{f}^j=0$. This means that $2\pi f^i$ is the circumference of a cylinder from $P^i$, whereas $\pi \tilde{f}^j$ is the width of a strip from $\tilde{P}^j$. $\bar{\tau}=-\infty$ at $P^i$ and $\tilde{P}^j$ with negative $f^i$ and $\tilde{f}^j$, respectively, whereas $\bar{\tau}=\infty$ at $P^i$ and $\tilde{P}^j$ with positive $f^i$ and $\tilde{f}^j$, respectively. The condition $\sum_{i=1}^N f^i+ \sum_{j=1}^M \frac{1}{2} \tilde{f}^j=0$ means that the total $\bar{\sigma}$ region of incoming cylinders and strips equals to that of outgoing ones if we choose the outgoing direction as positive. That is, the total $\bar{\sigma}$ region is conserved. In order to define the above global time uniquely, we need to fix the $\bar{\sigma}$ regions $2\pi f^i$ and $\pi \tilde{f}^j$ around $P^i$ and $\tilde{P}^j$, respectively. We divide ($N \, P^i, M \, \tilde{P}^j$) to arbitrary two sets consist of ($N_- \, P^i, M_- \, \tilde{P}^j$) for incoming punctures and ($N_+ \, P^i, M_+ \, \tilde{P}^j$) for outgoing punctures ($N_{-}+N_{+}=N$, $M_{-}+M_{+}=M$), then we divide  $-1$ to $N_{-}$ $f^i\equiv-\frac{2}{2 N_- + M_-}$ and $M_-$ $\tilde{f}^j\equiv-\frac{1}{2 N_- + M_-}$, and $1$ to $N_{+}$ $f^i\equiv\frac{2}{2 N_+ + M_+}$ and $M_+$ $\tilde{f}^j\equiv\frac{1}{2 N_+ + M_+}$, equally for all $i$ and $j$.

Thus, under a conformal transformation, one obtains $\bar{\Sigma}$ that has coordinates composed of the global time $\bar{\tau}$ and the position $\bar{\sigma}$. Because $\bar{\Sigma}$ can be a moduli of oriented open-closed string worldsheets with open and closed punctures, any two-dimensional oriented Riemannian manifold with open and closed punctures and with or without boundaries $\Sigma$ can be obtained by $\Sigma=\psi(\bar{\Sigma})$ where $\psi$ is a diffeomorphism times Weyl transformation.

Next, we will define  the model space $E$. Here we fix not only the Euclidean space $\bold{R}^d$ but also all the backgrounds except for the metric, that consist of a NS-NS B-field, a dilaton, a set of submanifolds $L$ of $\bold{R}^d$,  and $N$ dimensional vector bundles $E$ with gauge connections on them, which we call D-submanifolds and D-bundles, respectively. We may also fix orientifold planes on $\bold{R}^d$ consistently.  We consider a state $(\bar{\Sigma}, X_{\hat{D}}(\bar{\tau}_s), \bold{I}(\bar{\tau}_s), \bar{\tau}_s)$ determined by a $\bar{\tau}=\bar{\tau}_s$ constant line. $X_{\hat{D}}(\bar{\tau}_s): \bar{\Sigma}|_{\bar{\tau}_s} \to \bold{R}^d$ is an arbitrary map that maps a boundary component into a D-submanifold, where $\hat{D}$ represents the above fixed background.  
$\bold{I}({\bar{\tau}_s})=(i|_{\bar{\tau}_s \, 1}, \cdots, i|_{\bar{\tau}_s \, k}, \cdots, i|_{\bar{\tau}_s \, b|_{\bar{\tau}_s}})$ represents a set of the Chan-Paton indices where $i|_{\bar{\tau}_s \, k}$ represents a Chan-Paton index on the $k$-th intersection between $\Sigma|_{\bar{\tau}_s}$ and boundary components on $\Sigma$.  $i|_{\bar{\tau}_s \, k}$ runs from $1$ to $N_k$ that represents the dimension of the D-bundle where the $k$-th intersection maps.  An open string that has Chan-Paton indices $(i|_{\bar{\tau}_s \, k-1}, i|_{\bar{\tau}_s \, k})$ is represented by a part of $\Sigma|_{\bar{\tau}_s}$ that is surrounded by the $k-1$- and $k$-th intersections, whose $\bar{\sigma}$ coordinates are represented by $\bar{\sigma}_{k-1}$ and $\bar{\sigma}_{k}$, respectively. The zero mode and the boundary conditions on $\bar{\sigma}_{k}$ of $X_{\hat{D}}(\bar{\tau}_s)$ are determined by the background including D-submanifolds $L$ and the gauge connections.  $\bar{\Sigma}$ is a union of $N_{\pm}$ cylinders with radii $f_i$ and $M_{\pm}$ strips with width $\pi \tilde{f}^j$ at $\bar{\tau}\simeq \pm \infty$. Thus, we define a string state as an equivalence class $[\bar{\Sigma}, X_{\hat{D}}(\bar{\tau}_s\simeq \pm \infty), \bold{I}(\bar{\tau}_s\simeq \pm \infty), \bar{\tau}_s\simeq \pm \infty]$ by a relation $(\bar{\Sigma}, X_{\hat{D}}(\bar{\tau}_s\simeq \pm \infty), \bold{I}(\bar{\tau}_s\simeq \pm \infty), \bar{\tau}_s\simeq \pm \infty) \sim (\bar{\Sigma}', X'_{\hat{D}}(\bar{\tau}_s\simeq \pm \infty), \bold{I}'(\bar{\tau}_s\simeq \pm \infty), \bar{\tau}_s\simeq \pm \infty)$ if $N_{\pm}=N'_{\pm}$, $M_{\pm}=M'_{\pm}$, $f^i=f'^i$, $\tilde{f}^j=\tilde{f}'^j$, $\bold{I}(\bar{\tau}_s\simeq \pm \infty)=\bold{I}'(\bar{\tau}_s\simeq \pm \infty)$ and $X_{\hat{D}}(\bar{\tau}_s\simeq \pm \infty)=X'_{\hat{D}}(\bar{\tau}_s \simeq \pm \infty)$. $[\bar{\Sigma}, X_{\hat{D}}(\bar{\tau}_s), \bold{I}(\bar{\tau}_s), \bar{\tau}_s]$ represent many-body states of open and closed strings in $M$ because $\bar{\Sigma}|_{\bar{\tau}_s} \simeq S^1 \times \cdots \times S^1 \times I^1 \times \cdots \times I^1$ where $I^1$ represents a line segment, and $X_{\hat{D}}(\bar{\tau}_s): \bar{\Sigma}|_{\bar{\tau}_s} \to \bold{R}^d$. The model space $E$ is defined as $E:= \bigcup_{\hat{D}}\{[\bar{\Sigma}, X_{\hat{D}}(\bar{\tau}_s), \bold{I}(\bar{\tau}_s), \bar{\tau}_s]\}$ where disjoint unions are taken over all the backgrounds $\hat{D}$ except for the metric.

\begin{figure}[htbp]
\begin{center}
\includegraphics[height=7cm, keepaspectratio, clip]{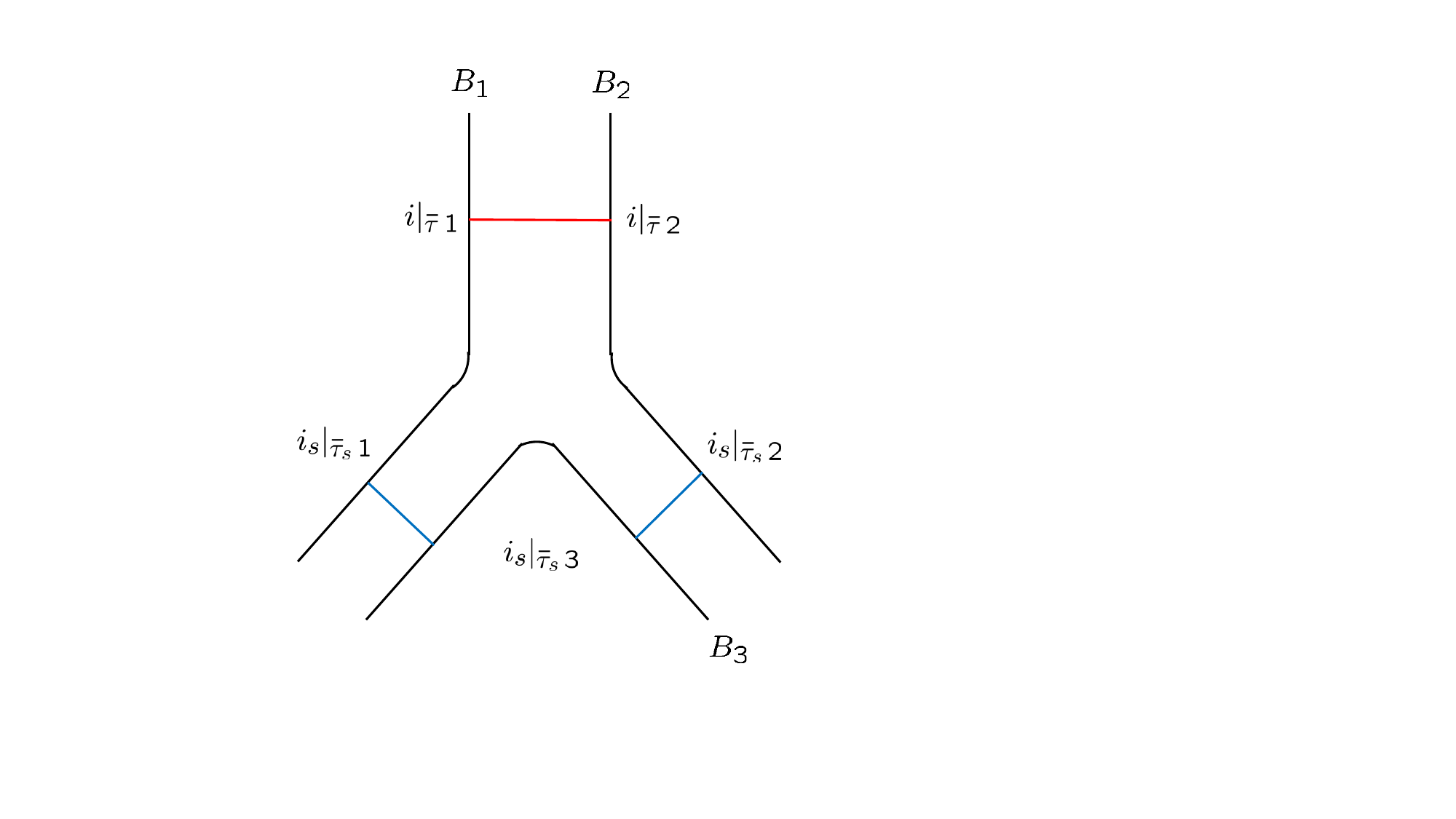}
\end{center}
\caption{In this figure, $\bold{I}_s(\bar{\tau}_s) \cong \bold{I}(\bar{\tau})$ implies $i_s|_{\bar{\tau}_s \, i}=i|_{\bar{\tau}\, i}$ ($i=1,2$), where $i_s|_{\bar{\tau}_s \, i}$ ($i|_{\bar{\tau}\, i}$) represents a Chan-Paton index defined on the intersection between a boundary component $B_i$ and $\Sigma|_{\bar{\tau}_s}$(the blue lines) ($\Sigma|_{\bar{\tau}}$(the red line)). There is no restriction on $i_s|_{\bar{\tau}_s \, 3}$, which is a Chan-Paton index defined on the intersection between a boundary component $B_3$ and $\Sigma|_{\bar{\tau}_s}$ because there is no intersection between $B_3$ and $\Sigma|_{\bar{\tau}}$.
}
\label{9open}
\end{figure}

Here, we will define topologies of $E$. An $\epsilon$-open neighborhood of $[\bar{\Sigma}, X_{s \, \hat{D}}(\bar{\tau}_s), \bold{I}(\bar{\tau}_s), \bar{\tau}_s]$
is defined by 
\begin{eqnarray}
&&U([\bar{\Sigma}, X_{s \, \hat{D}}(\bar{\tau}_s), \bold{I}(\bar{\tau}_s), \bar{\tau}_s], \epsilon) \nonumber \\
&:=&
\left\{[\bar{\Sigma}, X_{\hat{D}}(\bar{\tau}), \bold{I}(\bar{\tau}), \bar{\tau}]
 \bigm|
 \sqrt{|\bar{\tau}-\bar{\tau}_s|^2
+ \| X_{\hat{D}} (\bar{\tau})-X_{s \, \hat{D}}(\bar{\tau}_s) \|^2}
<\epsilon, \bold{I}_s(\bar{\tau}_s) \cong \bold{I}(\bar{\tau})
\right\}, \label{OpenNeighbour}
\end{eqnarray}
where
\begin{equation}
\| X(\bar{\tau})-X_s(\bar{\tau}_s) \|^2:=\int_0^{2\pi}  d\bar{\sigma} 
|X(\bar{\tau}, \bar{\sigma})-X_s(\bar{\tau}_s, \bar{\sigma})|^2.
\label{OpenNorm}
\end{equation}
$\bold{I}_s(\bar{\tau}_s) \cong \bold{I}(\bar{\tau})$ means that a Chan-Paton index on the intersection between $\Sigma|_{\bar{\tau}_s}$ and a boundary component on $\Sigma$ equals a Chan-Paton index on the intersection between $\Sigma|_{\bar{\tau}}$ and the boundary component, excepting that the corresponding intersection does not exist. For example, see Fig. \ref{9open}. 
$U([\bar{\Sigma}, X_{\hat{D}}(\bar{\tau}_s\simeq \pm \infty), \bold{I}(\bar{\tau}_s\simeq \pm \infty), \bar{\tau}_s\simeq \pm \infty], \epsilon)
=
U([\bar{\Sigma}', X'_{\hat{D}}(\bar{\tau}_s\simeq \pm \infty), \bold{I}'(\bar{\tau}_s\simeq \pm \infty), \bar{\tau}_s\simeq \pm \infty], \epsilon)$ consistently  if $N_{\pm}=N'_{\pm}$, $M_{\pm}=M'_{\pm}$, $f^i=f'^i$, $\tilde{f}^j=\tilde{f}'^j$, $\bold{I}(\bar{\tau}_s\simeq \pm \infty)=\bold{I}'(\bar{\tau}_s\simeq \pm \infty)$, $X_{\hat{D}}(\bar{\tau}_s\simeq \pm \infty)=X'_{\hat{D}}(\bar{\tau}_s\simeq \pm \infty)$, and $\epsilon$ is small enough, because the $\bar{\tau}=\bar{\tau}_s$ constant line traverses only propagators overlapped by $\bar{\Sigma}$ and $\bar{\Sigma}'$. $U$ is defined to be an open set of $E$ if there exists $\epsilon$ such that $U([\bar{\Sigma}, X_{\hat{D}}(\bar{\tau}_s), \bold{I}(\bar{\tau}_s), \bar{\tau}_s], \epsilon) \subset U$ for an arbitrary point $[\bar{\Sigma}, X_{\hat{D}}(\bar{\tau}_s), \bold{I}(\bar{\tau}_s), \bar{\tau}_s] \in U$. In exactly the same way as in section 2, one can show that the topology of $E$ satisfies the axiom of topology.
Although the model space is defined by using the coordinates $[\bar{\Sigma},  X_D(\bar{\tau}_s), \bar{\tau}_s]$, the model space does not depend on the coordinates, because the model space is a topological space.

In the following, we denote $[\bar{h}_{ mn}, X_{\hat{D}}(\bar{\tau}), \bold{I}(\bar{\tau}), \bar{\tau}]$, where $\bar{h}_{ mn} (\bar{\sigma}, \bar{\tau})$ is the worldsheet metric of $\bar{\Sigma}$,  instead of $[\bar{\Sigma}, X_{\hat{D}}(\bar{\tau}), \bold{I}(\bar{\tau}), \bar{\tau}]$ because giving a Riemann surface is equivalent to giving a  metric up to diffeomorphism and Weyl transformations.

In order to define structures of manifold, let us consider how generally we can define general coordinate transformations between $[\bar{h}_{ mn}, X_{\hat{D}}(\bar{\tau}), \bold{I}(\bar{\tau}), \bar{\tau}]$ and $[\bar{h}'_{ mn}, X'_{\hat{D}}(\bar{\tau}'), \bold{I}'(\bar{\tau}'), \bar{\tau}']$ where $[\bar{h}_{ mn}, X_{\hat{D}}(\bar{\tau}), \bold{I}(\bar{\tau}), \bar{\tau}] \in U \subset E$ and $[\bar{h}'_{ mn}, X'_{\hat{D}}(\bar{\tau}'), \bold{I}'(\bar{\tau}'), \bar{\tau}'] \in U'\subset E$. 
 $\bar{h}_{ mn}$ and $\bold{I}(\bar{\tau})$ do not transform to $\bar{\tau}$ and $X_{\hat{D}}(\bar{\tau})$ and vice versa, because $\bar{\tau}$ and $X_{\hat{D}}(\bar{\tau})$ are continuous variables, whereas $\bar{h}_{ mn}$ and $\bold{I}(\bar{\tau})$ are discrete variables: $\bar{\tau}$ and $X_{\hat{D}}(\bar{\tau})$ vary continuously, whereas $\bar{h}_{ mn}$ and $\bold{I}(\bar{\tau})$ vary discretely in a trajectory on $E$ by definition of the neighborhoods. $\bar{\tau}$ and $\bar{\sigma}$ do not transform to each other because the string states are defined by $\bar{\tau}$ constant lines. 
Under these restrictions, the most general coordinate transformation is given by 
\begin{eqnarray}
&&[\bar{h}_{ mn}(\bar{\sigma}, \bar{\tau}), X_{\hat{D}}^{\mu}(\bar{\sigma}, \bar{\tau}), \bold{I}(\bar{\tau}), \bar{\tau}] 
\nonumber \\
&&\mapsto [h'_{ mn}(\bar{\sigma}'(\bar{\sigma}), \bar{\tau}'(\bar{\tau}, X_{\hat{D}})), X_{\hat{D}}'^{\mu}(\bar{\sigma}', \bar{\tau}')(\bar{\tau}, X_{\hat{D}}), \bold{I}(\bar{\tau}'), \bar{\tau}'(\bar{\tau}, X_{\hat{D}}))], \nonumber \\
&&\label{OGeneralCoordTrans}
\end{eqnarray}
where $\bar{h}_{ mn} \mapsto \bar{h}_{ mn}'$ represents a world-sheet diffeomorphism transformation\footnote{
We extend the model space from  $E=\{[\bar{h}_{ mn}(\bar{\sigma}, \bar{\tau}), X_{\hat{D}}^{\mu}(\bar{\sigma}, \bar{\tau}), \bold{I}(\bar{\tau}), \bar{\tau}]\}$ to $E=\{[h'_{ mn}(\bar{\sigma}', \bar{\tau}'), X_{\hat{D}}'^{\mu}(\bar{\sigma}', \bar{\tau}'), \bold{I}(\bar{\tau}'), \bar{\tau}']\}$ by including the points generated by the diffeomorphisms $\bar{\sigma} \mapsto \bar{\sigma}'(\bar{\sigma})$ and $\bar{\tau} \mapsto \bar{\tau}'(\bar{\tau})$.}. $X_{\hat{D}}^{'\mu}(\bar{\tau}, X_{\hat{D}}(\bar{\tau}))$ and $\bar{\tau}'(\bar{\tau}, X_{\hat{D}}(\bar{\tau}))$ are functionals of $\bar{\tau}$ and $X_{\hat{D}}(\bar{\tau})$. $\mu=0, 1, \cdots d-1$. 
Here, we consider all the manifolds which are constructed by patching open sets of the model space $E$ by the general coordinate transformations (\ref{OGeneralCoordTrans}) and call them string manifolds $\mathcal{M}$.


Here, we give an example of string manifolds: $\mathcal{M}_D:= \{ [\bar{\Sigma}, x_{D}(\bar{\tau}), \bold{I}(\bar{\tau}), \bar{\tau}] \}$, where $D$ represents a target metric $G_{\mu \nu}$, O-planes and D-bundles with gauge connections where all the other backgrounds are turned off except for $G_{\mu \nu}$. $x_D(\bar{\tau}): \bar{\Sigma}|_{\bar{\tau}} \to M$,  where the image of the embedding function $x_D(\bar{\tau})$ has a metric: 
$ds^2= dx_D^{\mu}(\bar{\tau}, \bar{\sigma})
dx_D^{\nu}(\bar{\tau}, \bar{\sigma})
G_{\mu \nu}(x_D(\bar{\tau}, \bar{\sigma}))$.

We will show that $ \mathcal{M}_D$ has a structure of manifold, that is there exists a homeomorphism between the sufficiently small neighborhood around an arbitrary point $ [\bar{\Sigma}, x_{s D}(\bar{\tau}_s), \bold{I}_s(\bar{\tau}_s), \bar{\tau}_s]\in \mathcal{M}_D$ and an open set of $E$. There exists a general coordinate transformation  $X_{\hat{D}}(x_D)$ that satisfies 
$ds^2= dx_D^{\mu}
dx_D^{\nu}
G_{\mu \nu}(x_D)
=
dX_{\hat{D}}^{\mu}
dX_{\hat{D}}^{\nu}
\eta_{\mu \nu}$
on an arbitrary point $x_D$ in the $\epsilon_{\bar{\sigma}}$ open neighborhood around $x_{s D}(\bar{\tau}_s, \bar{\sigma}) \in M$, if $\epsilon_{\bar{\sigma}}$ is sufficiently small. An arbitrary point $ [\bar{\Sigma}, x_{s D}(\bar{\tau}_s), \bold{I}_s(\bar{\tau}_s), \bar{\tau}_s]\in \mathcal{M}_D$ in the $\epsilon$ open neighborhood around $[\bar{\Sigma}, x_{s D}(\bar{\tau}_s), \bold{I}_s(\bar{\tau}_s), \bar{\tau}_s]$ satisfies 
\begin{equation}
\int_0^{2\pi}  d\bar{\sigma} 
|x_D(\bar{\tau}, \bar{\sigma}) -x_{s D}(\bar{\tau}_s, \bar{\sigma})|^2
< \epsilon^2- |\bar{\tau}-\bar{\tau}_s|^2 \leqq \epsilon^{2}, \label{xxneighopen}
\end{equation} 
and thus 
\begin{equation}
|x_D(\bar{\tau}, \bar{\sigma})-x_{s D}(\bar{\tau}_s, \bar{\sigma})|
< \epsilon_{\bar{\sigma}}'
\end{equation}
on an arbitrary $\bar{\sigma}$.
$\epsilon_{\bar{\sigma}}'< \epsilon_{\bar{\sigma}}$ is satisfied on an arbitrary $\bar{\sigma}$ if $\epsilon$ is taken to be sufficiently small. 
Then, there exists a transformation  $X^{\mu}_{\hat{D}}(\bar{\tau}, \bar{\sigma}):=X_{\hat{D}}^{\mu}(x_D(\bar{\tau}, \bar{\sigma}))$, which satisfies
\begin{equation}
ds^2= dx_D^{\mu}(\bar{\tau}, \bar{\sigma})
dx_D^{\nu}(\bar{\tau}, \bar{\sigma})
G_{\mu \nu}(x_D(\bar{\tau}, \bar{\sigma}))
=
dX_{\hat{D}}^{\mu}(\bar{\tau}, \bar{\sigma})
dX_{\hat{D}}^{\nu}(\bar{\tau}, \bar{\sigma}) \eta_{\mu \nu}. \label{oLocalLorents}
\end{equation}

Because the tangent vector $X_{\hat{D}}(\bar{\tau}, \bar{\sigma})$ exists for each $x_D(\bar{\tau}, \bar{\sigma})$, there exists a vector bundle $X_{\hat{D}}(\bar{\tau})$ for $0 \leqq \bar{\sigma} < 2\pi$. $x_D(\bar{\tau})$ and $X_{\hat{D}}(\bar{\tau})$ satisfy (\ref{oLocalLorents}) on each $\bar{\sigma}$, that is $X_{\hat{D}}(\bar{\tau}): \bar{\Sigma}|_{\bar{\tau}} \to \bold{R}^d$. Therefore, there exists a homeomorphism between the sufficiently small neighborhood around an arbitrary point $[\bar{\Sigma}, x_{s D}(\bar{\tau}_s), \bold{I}_s(\bar{\tau}_s), \bar{\tau}_s]\in \mathcal{M}_D$ and an open set of $E$: $[\bar{\Sigma}, x_{D}(\bar{\tau}), \bold{I}(\bar{\tau}), \bar{\tau}]\mapsto[\bar{\Sigma}, X_{\hat{D}}(\bar{\tau}), \bold{I}(\bar{\tau}), \bar{\tau}]$. 
Actually, a map from the image $x_{D}(\bar{\tau}, \bar{\sigma})$ to the image $X_{\hat{D}}(\bar{\tau}, \bar{\sigma})$ is explicitly given by an exponential map
\begin{equation}
x_{D}(\bar{\tau}, \bar{\sigma})
=
\exp_{x_{sD}(\bar{\tau}_s, \bar{\sigma})}
X_{\hat{D}}(\bar{\tau}, \bar{\sigma})
\simeq
x_{sD}(\bar{\tau}_s, \bar{\sigma})
+X_{\hat{D}}(\bar{\tau}, \bar{\sigma}).
\end{equation}
If we substitute this to an $\epsilon$ open neighborhood around an arbitrary point $[\bar{\Sigma}, x_{s D}(\bar{\tau}_s), \bold{I}_s(\bar{\tau}_s), \bar{\tau}_s]\in \mathcal{M}_D$ (\ref{xxneighopen}), we obtain an $\epsilon$ open neighborhood around $ [\bar{\Sigma}, 0, \bold{I}_s(\bar{\tau}_s), \bar{\tau}_s]\in E$,
\begin{equation}
\int_0^{2\pi}  d\bar{\sigma} 
|X_{\hat{D}}(\bar{\tau}, \bar{\sigma})|^2 +  |\bar{\tau}-\bar{\tau}_s|^2 < \epsilon^{2}.
\end{equation}

By definition of the $\epsilon$-open neighborhood, on a connected Riemann surface with open and closed punctures and with or without boundaries in $M$, arbitrary two string states with the same Chan-Paton indices are connected continuously. Thus, there is an one-to-one correspondence between such a Riemann surface that have Chan-Paton indices and a curve  parametrized by $\bar{\tau}$ from $\bar{\tau}=-\infty$ to $\bar{\tau}=\infty$ on $\mathcal{M}_D$. That is, curves that represent asymptotic processes on $\mathcal{M}_D$ reproduce the right moduli space of the Riemann surfaces.

By a general curve parametrized by $t$ on $\mathcal{M}_D$, string states on the different Riemann surfaces that have even different genera, can be connected continuously, whereas the different Riemann surfaces that have different genera cannot be connected continuously in the moduli space of the Riemann surfaces in the target space.

The tangent space is spanned by $\frac{\partial}{\partial\bar{\tau}}$ and $\frac{\partial}{\partial X_{\hat{D}}^{\mu}(\bar{\sigma}, \bar{\tau})}$ as one can see from the definition of the neighborhood (\ref{OpenNeighbour}). We should note that $\frac{\partial}{\partial \bar{h}_{ mn}}$  and $\frac{\partial}{\partial\bold{I}(\bar{\tau})}$ cannot be a part of basis that span the tangent space, because $\bar{h}_{ mn}$ and $\bold{I}(\bar{\tau})$ are just discrete variables in $E$. The index of $\frac{\partial}{\partial X_{\hat{D}}^{\mu}(\bar{\sigma}, \bar{\tau}) }$ can be $(\mu \, \bar{\sigma})$.
We define a summation over $ \bar{\sigma}$ by $\int d\bar{\sigma}\bar{e} (\bar{\sigma}, \bar{\tau})$, where $\bar{e}:=\sqrt{\bar{h}_{ \bar{\sigma} \bar{\sigma}}}$
This summation is invariant under $\bar{\sigma} \mapsto \bar{\sigma}'(\bar{\sigma})$ and transformed as a scalar under $\bar{\tau} \mapsto \bar{\tau}'(\bar{\tau}, X_{\hat{D}}(\bar{\tau}))$.

Riemannian string manifold is obtained by defining a metric, which is a section of an inner product on the tangent space. The general form of a metric is given by
\begin{eqnarray}
&&ds^2(\bar{h}, X_{\hat{D}}(\bar{\tau}), \bold{I}(\bar{\tau}), \bar{\tau}) \nonumber \\
=&&G(\bar{h}, X_{\hat{D}}(\bar{\tau}), \bold{I}(\bar{\tau}), \bar{\tau})_{dd} (d\bar{\tau})^2 +2 d\bar{\tau} \int d\bar{\sigma}  \bar{e} (\bar{\sigma}, \bar{\tau})  \sum_{\mu}  G(\bar{h}, X_{\hat{D}}(\bar{\tau}), \bold{I}(\bar{\tau}), \bar{\tau})_{d \; (\mu \bar{\sigma})} d X_{\hat{D}}^{\mu}(\bar{\sigma}, \bar{\tau}) \nonumber \\
&&+\int d\bar{\sigma}   \bar{e} (\bar{\sigma}, \bar{\tau}) \int d\bar{\sigma}' \bar{e} (\bar{\sigma}', \bar{\tau})  \sum_{\mu, \mu'} G(\bar{h}, X_{\hat{D}}(\bar{\tau}), \bold{I}(\bar{\tau}), \bar{\tau})_{ \; (\mu \bar{\sigma})  \; (\mu' \bar{\sigma}')} d X_{\hat{D}}^{\mu}(\bar{\sigma}, \bar{\tau}) d X_{\hat{D}}^{\mu'}(\bar{\sigma}', \bar{\tau}). \nonumber \\
\end{eqnarray}
We summarize the vectors as $dX_{\hat{D}}^I$ ($I=d,(\mu\bar{\sigma})$), where  $dX_{\hat{D}}^d:=d\bar{\tau}$ and $d X_{\hat{D}}^{(\mu\bar{\sigma})}:=dX_{\hat{D}}^{\mu}(\bar{\sigma}, \bar{\tau})$. Then, the components of the metric are summarized as $G_{IJ}(\bar{h}, X_{\hat{D}}(\bar{\tau}), \bold{I}(\bar{\tau}), \bar{\tau})$. The inverse of the metric $G^{IJ}(\bar{h}, X_{\hat{D}}(\bar{\tau}), \bold{I}(\bar{\tau}), \bar{\tau})$ is defined by $G_{IJ}G^{JK}=G^{KJ}G_{JI}=\delta_I^K$, where $\delta_d^d=1$ and $\delta_{\mu\bar{\sigma}}^{\mu'\bar{\sigma}'}=\frac{1}{\bar{e} (\bar{\sigma}, \bar{\tau}) } \delta_{\mu}^{\mu'}\delta(\bar{\sigma}-\bar{\sigma}')$. The components of the Riemannian curvature tensor are given by $R^I_{JKL}$ in the basis $\frac{\partial}{\partial X_{\hat{D}}^I(\bar{\tau})}$. The components of the Ricci tensor are $R_{IJ}:=R^K_{IKJ}=R^d_{IdJ}+\int d\bar{\sigma} \bar{e}  R^{(\mu \bar{\sigma})}_{I \; (\mu \bar{\sigma}) \; J}$. The scalar curvature is 
\begin{eqnarray}
R&:=&G^{IJ} R_{IJ} \nonumber \\
&=&G^{dd}R_{dd}+2 \int d\bar{\sigma} \bar{e}  G^{d \; (\mu\bar{\sigma})} R_{d \; (\mu\bar{\sigma})} +\int d\bar{\sigma} \bar{e} \int d\bar{\sigma}'\bar{e}'  G^{(\mu\bar{\sigma}) \; (\mu'\bar{\sigma}')}R_{(\mu\bar{\sigma})  \; (\mu'\bar{\sigma}')}. \nonumber 
\end{eqnarray}
The volume is  $\sqrt{G}$, where $G=det (G_{IJ})$.

By using these geometrical objects, we formulate string theory non-perturbatively as
\begin{equation}
Z=\int \mathcal{D}G \mathcal{D}Ae^{-S}, \label{OBosonicTheory}
\end{equation}
where
\begin{equation}
S=\frac{1}{G_N}\int \mathcal{D} h \mathcal{D} X_{\hat{D}}(\bar{\tau}) \mathcal{D} \bar{\tau} \mathcal{D} \bold{I}(\bar{\tau})
\sqrt{G} (-R +\frac{1}{4} G_N G^{I_1 I_2} G^{J_1 J_2} F_{I_1 J_1} F_{I_2 J_2} ). \label{OBosonicAction}
\end{equation}
As an example of sets of fields on the string manifolds, we consider the metric and an $u(1)$ gauge field $A_I$ whose field strength is given by $F_{IJ}$. The path integral is defined by semi-classically\footnote{It will be enough to define the path-integral by semi-classically summing classical solutions and small classical and quantum fluctuations around them, because string manifolds themselves possess quantum corrections, and loops of the fields on them do not correspond to quantum corrections as one can see in the derivation of the perturbative string theory later. The unitarity is manifest and there is also no UV divergence from loop integrals, by defining the path-integral semi-classically. } summing over the metrics and gauge fields on $\mathcal{M}$. By definition, the theory is background independent. $\mathcal{D}h$ is the invariant measure of the metrics $h_{mn}$ on the two-dimensional Riemannian manifolds $\Sigma$. $h_{mn}$ and $\bar{h}_{mn}$ are related to each others by the diffeomorphism and the Weyl transformations.

Under 
\begin{equation}
(\bar{\tau}, X_{\hat{D}}(\bar{\tau})) \mapsto (\bar{\tau}'(\bar{\tau}, X_{\hat{D}}(\bar{\tau})) , X_{\hat{D}}'(\bar{\tau}')(\bar{\tau}, X_{\hat{D}}(\bar{\tau}))),
\label{Osubdiffeo}
\end{equation} 
$G_{IJ}(\bar{h}, X_{\hat{D}}(\bar{\tau}), \bold{I}(\bar{\tau}), \bar{\tau})$ and $A_I(\bar{h}, X_{\hat{D}}(\bar{\tau}), \bold{I}(\bar{\tau}), \bar{\tau})$ are transformed as a symmetric tensor and a vector, respectively and the action is manifestly invariant. 

We define $G_{IJ}(\bar{h}, X_{\hat{D}}(\bar{\tau}), \bold{I}(\bar{\tau}), \bar{\tau})$ and $A_I(\bar{h}, X_{\hat{D}}(\bar{\tau}), \bold{I}(\bar{\tau}), \bar{\tau})$ so as to transform as scalars under $\bar{h}_{ mn}(\bar{\sigma}, \bar{\tau}) \mapsto \bar{h}_{ mn}'(\bar{\sigma}'(\bar{\sigma}), \bar{\tau})$. 
Under $\bar{\sigma}$ diffeomorphisms: $\bar{\sigma} \mapsto \bar{\sigma}'(\bar{\sigma})$, which are equivalent to 
\begin{eqnarray}
[\bar{h}_{ mn}(\bar{\sigma}, \bar{\tau}), X^{\mu}_{\hat{D}}(\bar{\sigma}, \bar{\tau}), \bold{I}(\bar{\tau}), \bar{\tau}] 
&&\mapsto 
[\bar{h}_{ mn}'(\bar{\sigma}'(\bar{\sigma}), \bar{\tau}), X'^{\mu}_{\hat{D}}(\bar{\sigma}', \bar{\tau})(X_{\hat{D}}(\bar{\tau})), \bold{I}(\bar{\tau}), \bar{\tau}], \nonumber \\
&&=[\bar{h}_{ mn}'(\bar{\sigma}'(\bar{\sigma}), \bar{\tau}), X^{\mu}_{\hat{D}}(\bar{\sigma}, \bar{\tau}), \bold{I}(\bar{\tau}), \bar{\tau}], \label{Odiff}
\end{eqnarray}
$G_{d \; (\mu \bar{\sigma})}$ is transformed as a scalar;
\begin{eqnarray}
&&G'_{d \; (\mu \bar{\sigma}')}(\bar{h}', X'_{\hat{D}}(\bar{\tau}), \bold{I}(\bar{\tau}), \bar{\tau}) \nonumber \\
&=&
G'_{d \; (\mu \bar{\sigma}')}(\bar{h}, X'_{\hat{D}}(\bar{\tau}), \bold{I}(\bar{\tau}), \bar{\tau})
=
\frac{\partial X^I_{\hat{D}}(\bar{\tau})}{\partial X^{'d}_{\hat{D}}(\bar{\tau})}
\frac{\partial X^J_{\hat{D}}(\bar{\tau})}{\partial X^{'(\mu \bar{\sigma}')}_{\hat{D}}(\bar{\tau})}
G_{IJ}(\bar{h}, X_{\hat{D}}(\bar{\tau}), \bold{I}(\bar{\tau}), \bar{\tau})
\nonumber \\
&=&
\frac{\partial X^I_{\hat{D}}(\bar{\tau})}{\partial X^d_{\hat{D}}(\bar{\tau})}
\frac{\partial X^J_{\hat{D}}(\bar{\tau})}{\partial X^{(\mu \bar{\sigma})}_{\hat{D}}(\bar{\tau})}
G_{IJ}(\bar{h}, X_{\hat{D}}(\bar{\tau}), \bold{I}(\bar{\tau}), \bar{\tau})
=
G_{d \; (\mu \bar{\sigma})}(\bar{h}, X_{\hat{D}}(\bar{\tau}), \bold{I}(\bar{\tau}), \bar{\tau}), \end{eqnarray}
because (\ref{Osubdiffeo}) and (\ref{Odiff}).
In the same way, the other fields are also transformed as 
\begin{eqnarray}
G'_{dd}(\bar{h}', X'_{\hat{D}}(\bar{\tau}), \bold{I}(\bar{\tau}), \bar{\tau})&=&G_{dd}(\bar{h}, X_{\hat{D}}(\bar{\tau}), \bold{I}(\bar{\tau}), \bar{\tau})
\nonumber \\
G'_{ \; (\mu \bar{\sigma}')  \; (\nu \bar{\rho}')}(\bar{h}', X'_{\hat{D}}(\bar{\tau}), \bold{I}(\bar{\tau}), \bar{\tau})&=&G_{ \; (\mu \bar{\sigma})  \; (\nu \bar{\rho})}(\bar{h}, X_{\hat{D}}(\bar{\tau}), \bold{I}(\bar{\tau}), \bar{\tau})
\nonumber \\
A'_d(\bar{h}', X'_{\hat{D}}(\bar{\tau}), \bold{I}(\bar{\tau}), \bar{\tau})&=&A_d(\bar{h}, X_{\hat{D}}(\bar{\tau}), \bold{I}(\bar{\tau}), \bar{\tau})
\nonumber \\
A'_{(\mu \bar{\sigma}')}(\bar{h}', X'_{\hat{D}}(\bar{\tau}), \bold{I}(\bar{\tau}), \bar{\tau})
&=&A_{(\mu \bar{\sigma})}(\bar{h}, X_{\hat{D}}(\bar{\tau}), \bold{I}(\bar{\tau}), \bar{\tau}).
\end{eqnarray}
Thus, the action is invariant under $\bar{\sigma}$ diffeomorphisms, because $\int d\bar{\sigma}'\bar{e}'(\bar{\sigma}', \bar{\tau})=\int d\bar{\sigma}\bar{e} (\bar{\sigma}, \bar{\tau})$. 
Therefore, $G_{IJ}(\bar{h}, X_{\hat{D}}(\bar{\tau}), \bold{I}(\bar{\tau}), \bar{\tau})$ and $A_I(\bar{h}, X_{\hat{D}}(\bar{\tau}), \bold{I}(\bar{\tau}), \bar{\tau})$ are transformed covariantly and the action (\ref{OBosonicAction}) is invariant under the diffeomorphisms (\ref{OGeneralCoordTrans}) including the $\bar{\sigma}$ diffeomorphisms.

The background that represents a perturbative vacuum is given by 
\begin{eqnarray}
\bar{ds}^2
&=& 2\lambda \bar{\rho}(\bar{h}) N^2(X_{\hat{D}}(\bar{\tau})) (dX^d_{\hat{D}})^2 
\nonumber \\
&&+\int d\bar{\sigma}   \bar{e} \int d\bar{\sigma}' \bar{e}' N^{\frac{2}{2-D}}(X_{\hat{D}}(\bar{\tau})) \frac{\bar{e}^3(\bar{\sigma}, \bar{\tau})}{\sqrt{\bar{h}(\bar{\sigma}, \bar{\tau})}} \delta_{(\mu \bar{\sigma}) (\mu' \bar{\sigma}')}
d X_{\hat{D}}^{(\mu \bar{\sigma})} d X_{\hat{D}}^{(\mu' \bar{\sigma}')}, \nonumber \\
\bar{A}_d&=&i \sqrt{\frac{2-2D}{2-D}}\frac{\sqrt{2\lambda \bar{\rho}(\bar{h}) }}{\sqrt{G_N}} N(X_{\hat{D}}(\bar{\tau})), \qquad
\bar{A}_{(\mu \bar{\sigma})}=0, \label{Osolution}
\end{eqnarray}
on $\mathcal{M}_D$ where we fix the target metric to $\eta_{\mu \mu'}$, a set of D-submanifolds to arbitrary one, and the gauge connections to zero, respectively. $\bar{\rho}(\bar{h}):=\frac{1}{4\pi}\int d\bar{\sigma} \sqrt{\bar{h}}\bar{R}_{\bar{h}}+\frac{1}{2\pi}\bar{k}_{\bar{h}}$, where $\bar{R}_{\bar{h}}$ is the scalar curvature of $\bar{h}_{ mn}$ and $\bar{k}_{\bar{h}}$ is the geodesic curvature of  $\bar{h}_{ mn}$. $D$ is a volume of the index $(\mu \bar{\sigma})$: $D:=\int d \bar{\sigma} \bar{e} \delta_{(\mu \bar{\sigma}) (\mu \bar{\sigma})}=d 2 \pi \delta(0)$. $N(X_{\hat{D}}(\bar{\tau}))=\frac{1}{1+v(X_{\hat{D}}(\bar{\tau}))}$, where $v(X_{\hat{D}}(\bar{\tau}))= \frac{\alpha}{\sqrt{d-1}} \int d\bar{\sigma}  \epsilon_{\mu\nu}X_{\hat{D}}^{\mu}(\bar{\tau}) \partial_{\bar{\sigma}} X_{\hat{D}}^{\nu}(\bar{\tau})$. One can show that the background (\ref{Osolution}) is a classical solution\footnote{This solution is a generalization of the Majumdar-Papapetrou solution \cite{Majumdar, Papapetrou} of the Einstein-Maxwell system.} to the equations of motion of (\ref{OBosonicAction}) as in section 3. The dependence of $\bar{h}_{mn}$ on the background (\ref{Osolution}) is uniquely determined  by the consistency of the quantum theory of the fluctuations around the background. Actually, we will find that all the perturbative string amplitudes are derived as follows.

Let us consider fluctuations around the background (\ref{Osolution}), $G_{IJ}=\bar{G}_{IJ}+\tilde{G}_{IJ}$ and $A_I=\bar{A}_I+\tilde{A}_I$. The action (\ref{OBosonicAction}) up to the quadratic order is given by,\begin{eqnarray}
S&=&\frac{1}{G_N} \int \mathcal{D}h  \mathcal{D}X_{\hat{D}}(\bar{\tau}) \mathcal{D}\bar{\tau} 
\sqrt{\bar{G}} 
\Bigl(-\bar{R}+\frac{1}{4}\bar{F}'_{IJ}\bar{F}'^{IJ} 
\nonumber \\
&&+\frac{1}{4}\bar{\nabla}_I \tilde{G}_{JK} \bar{\nabla}^I \tilde{G}^{JK}
-\frac{1}{4}\bar{\nabla}_I \tilde{G} \bar{\nabla}^I \tilde{G}
+\frac{1}{2}\bar{\nabla}^I \tilde{G}_{IJ} \bar{\nabla}^J \tilde{G}
-\frac{1}{2}\bar{\nabla}^I \tilde{G}_{IJ} \bar{\nabla}_K \tilde{G}^{JK}
\nonumber \\
&&-\frac{1}{4}(-\bar{R}+\frac{1}{4}\bar{F}'_{KL}\bar{F}'^{KL})
(\tilde{G}_{IJ}\tilde{G}^{IJ}-\frac{1}{2}\tilde{G}^2)
+(-\frac{1}{2}\bar{R}^{I}_{\;\; J}+\frac{1}{2}\bar{F}'^{IK}\bar{F}'_{JK})
\tilde{G}_{IL}\tilde{G}^{JL}
\nonumber \\
&&+(\frac{1}{2}\bar{R}^{IJ}-\frac{1}{4}\bar{F}'^{IK}\bar{F}'^J_{\;\;\;\; K})
\tilde{G}_{IJ}\tilde{G}
+(-\frac{1}{2}\bar{R}^{IJKL}+\frac{1}{4}\bar{F}'^{IJ}\bar{F}'^{KL})
\tilde{G}_{IK}\tilde{G}_{JL}
\nonumber \\
&&+\frac{1}{4}G_N \tilde{F}_{IJ} \tilde{F}^{IJ} 
+\sqrt{G_N} 
(\frac{1}{4} \bar{F}^{'IJ} \tilde{F}_{IJ} \tilde{G} 
-\bar{F}^{'IJ} \tilde{F}_{IK} \tilde{G}_J^{\;\; K} ) \Bigr).
\end{eqnarray}
The Lagrangian is independent of Chan-Paton indices because the background (\ref{Osolution}) is independent of them. $\bar{F}'_{IJ}:=\sqrt{G_N}\bar{F}_{IJ}$ is independent of $G_N$. $\tilde{G}:=\bar{G}^{IJ}\tilde{G}_{IJ}$. There is no first order term because the background satisfies the equations of motion. 

From these fluctuations,  we obtain the correlation function in the string manifold $\mathcal{M}_D$ in exactly the same way as in section 3,
\begin{equation}
\Delta_F(X_{\hat{D} f}; X_{\hat{D} i}|h_f ; h_i)
=
Z
\int_{h_i, X_{\hat{D} i}}^{h_f, X_{\hat{D} f}} 
\mathcal{D} h  \mathcal{D} X_{\hat{D}}
e^{-\lambda \chi}
e^{-S_{s}}, 
\label{OFinalPropagator}
\end{equation}
where
\begin{equation}
S_{s}
=
\int_{-\infty}^{\infty} d\tau \int d\sigma \sqrt{h(\sigma, \tau)} \left(\frac{1}{2} h^{mn} (\sigma, \tau) \partial_m X_{\hat{D}}^{\mu}(\sigma, \tau) \partial_n X_{\hat{D} \mu}(\sigma, \tau) \right),
\end{equation}
and $\chi$ is the Euler number of the two-dimensional Riemannian manifold with boundaries. By inserting asymptotic states with Chan-Paton matrices to (\ref{OFinalPropagator}) and renormalizing the metric, we obtain the all-order  perturbative scattering amplitudes that possess the moduli in the open-closed string theory with Dirichlet and Neumann boundary conditions in the normal and tangential directions to the D-submanifolds, respectively \cite{textbook}. Therefore, a set of D-submanifolds represents a D-brane background where back reactions from the D-branes are ignored. The consistency of the perturbation theory around the background (\ref{Osolution}) determines $d=26$ (the critical dimension).

\section{Non-perturbative formulation of superstring theory}
\setcounter{equation}{0}

In this section, we complete superstring geometry including open superstrings. Let us define unique global times on oriented open-closed superstring worldsheets with punctures $\bar{\bold{\Sigma}}$ \cite{NotesOnSupermanifolds, WittenSupermoduli, SuperPeriod} in order to define string states by world-time constant hypersurfaces. The worldsheets with boundaries can be constructed as orientifolds of ordinary type II superstring worldsheets with punctures $\bar{\bold{\Sigma}}_{c}$ as follows. $\bar{\bold{\Sigma}}_{c}$ are embedded in $\bar{\bold{\Sigma}}_{L} \times \bar{\bold{\Sigma}}_{R}$, which have reduced spaces $\rho(\Sigma_{L, red}) \times \Sigma_{R, red}$. We restrict $\bar{\bold{\Sigma}}_{c}$ such that $\rho(\Sigma_{L, red})=\Sigma_{R, red}$. Then, $\rho(\Sigma_{L, red})=\Sigma_{L, red}$, where $\rho$ is the anti-holomorphic involution. An orbifold action $Z_2$ on $\bar{\bold{\Sigma}}_{c}$ is defined in local coordinates by $(z, \theta_1, \rho(z),  \theta_2) \mapsto (\rho(z), \theta_1, z,  -\theta_2)$. $\bar{\bold{\Sigma}}_{c}/Z_2$ are topologically classified into open and/or unoriented super Riemann surfaces. By restricting them, we obtain oriented open-closed superstring worldsheets with punctures $\bar{\bold{\Sigma}}$. Thus, for each $\bar{\bold{\Sigma}}$, there exists a closed double cover $\bar{\bold{\Sigma}}_{c}$.  Because $\bar{\Sigma}_{red}/Z_2 \cong \Sigma_{L, red}/Z_2 =\Sigma_{R, red}/Z_2 $, reduced spaces $\bar{\Sigma}_{red}$ of $\bar{\bold{\Sigma}}$ are oriented open-closed worldsheets with punctures. First of all, we define global coordinates on the double covers $\bar{\bold{\Sigma}}_{c}$ of $\bar{\bold{\Sigma}}$ in the same way as in section 4. The real part of the global coordinates $\bar{\tau}$ remains on $\bar{\bold{\Sigma}}$ because $\rho$ is an anti-holomorphic involution. If $\rho$ maps a puncture to another puncture on $\bar{\bold{\Sigma}}_{c}$, the superdiscs around the punctures are identified and give a superdisk $\bold{D}^i$ around a closed puncture $P^i$ on $\bar{\bold{\Sigma}}$. On the other hand, if $\rho$ maps a puncture to itself on $\bar{\bold{\Sigma}}_{c}$, the superdisk around the puncture is identified with itself and gives an upper half superdisk $\tilde{\bold{D}}^j$ around an open puncture $\tilde{P}^j$ on $\bar{\bold{\Sigma}}$. The $\bar{\sigma}$ regions around $P^i$ and $\tilde{P}^j$ are $2\pi f^i$ and $\pi \tilde{f}^j$, respectively where $\sum_{i=1}^n 2f^i+ \sum_{j=1}^m \tilde{f}^j=0$.  $\bar{\tau}=-\infty$ at $P^i$ and $\tilde{P}^j$ with negative $f^i$ and $\tilde{f}^j$, respectively, whereas $\bar{\tau}=\infty$ at $P^i$ and $\tilde{P}^j$ with positive $f^i$ and $\tilde{f}^j$, respectively. In order to define the above global time uniquely, we fix the $\bar{\sigma}$ regions $2\pi f^i$ and $\pi \tilde{f}^j$ around $P^i$ and $\tilde{P}^j$, respectively in exactly the same way as in section 5. 
 
Thus, under a superconformal transformation, one obtains $\bar{\bold{\Sigma}}$ that has even coordinates composed of the global time $\bar{\tau}$ and the position $\bar{\sigma}$, and $\bar{\Sigma}_{red}$ is canonically defined. Because $\bar{\bold{\Sigma}}$ can be a moduli of oriented open-closed superstring worldsheets with open and closed punctures, any two-dimensional oriented open-closed super Riemannian manifold with open and closed punctures $\bold{\Sigma}$ can be obtained by $\bold{\Sigma}={\boldsymbol \psi}(\bar{\bold{\Sigma}})$ where ${\boldsymbol \psi}$ is a superdiffeomorphism times super Weyl transformation.

Next, we will define the model space $\bold{E}$. Here we fix not only a d-dimensional Euclidean space $\bold{R}^d$ but also  backgrounds except for the metric, that consist of a NS-NS B-field, a dilaton, R-R fields
\footnote{$\bold{X}_{\hat{D}_T}$ do not depend on the R-R backgrounds because strings do not couple with them. However, open sets of the model space need to possess the R-R backgrounds (We may write $\bold{E}:=\bigcup_{\hat{D}_T}\{[\bar{{\boldsymbol \Sigma}}, \bold{X}_{\hat{D}_T}, \bold{I}(\bar{\tau}), \bar{\tau}]_{\hat{D}_T}\}$.) in order that D-brane states in the Hilbert space defined on the open sets couple with the R-R backgrounds.},  
a set of  submanifolds $L$ in $\bold{R}^d$, $N$ dimensional vector bundles $E$ with gauge connections on them, which we call D-submanifolds and D-bundles, respectively, and consistent configurations of O-planes.

We consider a state $(\bar{\bold{\Sigma}}, \bold{X}_{\hat{D}_T}(\bar{\tau}_s), \bold{I}(\bar{\tau}_s), \bar{\tau}_s)$ determined by a $\bar{\tau}=\bar{\tau}_s$ constant hypersurface. $\bold{X}_{\hat{D}_T}(\bar{\tau}_s): \bold{\Sigma}|_{\bar{\tau}_s} \to \bold{R}^d$ is an arbitrary map that maps a boundary component of the reduced space into a D-submanifold. $\hat{D}_T$ represents the above fixed quantities, where $T$ runs IIA, IIB and I. The IIA GSO projection is attached for $T=$ IIA, and the IIB GSO projection is attached for $T=$ IIB and I. $\Omega$ projection is imposed and 32 D9-submanifolds are fixed for $T=I$.  We can define the worldsheet fermion numbers of states in a Hilbert space because the states consist of the fields over the local coordinates ${\boldsymbol X}^{\mu}_{\hat{D}_T}(\bar{\tau}_s)=X^{\mu}+ \bar{\theta}^{\alpha} \psi_{\alpha}^{\mu}+\frac{1}{2} \bar{\theta}^2 F^{\mu}$, where $\mu=0, 1, \cdots d-1$, $\psi_{\alpha}^{\mu}$ is a Majorana fermion and  $F^{\mu}$ is an auxiliary field. We abbreviate $\hat{D}_T$ and $(\bar{\tau}_s)$ of $X^{\mu}$, $\psi_{\alpha}^{\mu}$ and $F^{\mu}$. We define the Hilbert space in these coordinates by the states only with $e^{\pi i F}=1$ and $e^{\pi i \tilde{F}}=(-1)^{\tilde{\alpha}}$ for $T=$ IIA and $e^{\pi i F}=e^{\pi i \tilde{F}}=1$ for $T=$ IIB and I, where $F$ and $\tilde{F}$ are left- and right-handed fermion numbers respectively, and $\tilde{\alpha}$ is 1 or 0 when the right-handed fermion is periodic (R sector) or anti-periodic (NS sector), respectively. 

$\bold{I}({\bar{\tau}_s})=(i|_{\bar{\tau}_s \, 1}, \cdots, i|_{\bar{\tau}_s \, k}, \cdots, i|_{\bar{\tau}_s \, b|_{\bar{\tau}_s}})$ represents a set of the Chan-Paton indices where $i|_{\bar{\tau}_s \, k}$ represents a Chan-Paton index on the $k$-th intersection between $\Sigma|_{\bar{\tau}_s}$ and boundary components on $\Sigma$.  $i|_{\bar{\tau}_s \, k}$ runs from $1$ to $N_k$ that represents the dimension of the D-bundle where the $k$-th intersection maps.  An open string that has Chan-Paton indices $(i|_{\bar{\tau}_s \, k-1}, i|_{\bar{\tau}_s \, k})$ is represented by a part of $\Sigma|_{\bar{\tau}_s}$ that is surrounded by the $k-1$- and $k$-th intersections, whose $\bar{\sigma}$ coordinates are represented by $\bar{\sigma}_{k-1}$ and $\bar{\sigma}_{k}$, respectively. 
 The zero mode and the boundary conditions on  $\bar{\sigma}_{k}$ of $X^{\mu}$ are determined by the background including the D-submanifolds $L$ and the gauge connections.

$\bar{\bold{\Sigma}}$ is a union of $N_{\pm}$ supercylinders with radii $f_i$ and $M_{\pm}$ superstrips with width $\pi \tilde{f}^j$ at $\bar{\tau}\simeq \pm \infty$. Thus, we define a superstring state as an equivalence class $[\bar{\bold{\Sigma}}, \bold{X}_{\hat{D}_T}(\bar{\tau}_s\simeq \pm \infty), \bold{I}(\bar{\tau}_s\simeq \pm \infty), \bar{\tau}_s\simeq \pm \infty]$ by a relation $(\bar{\bold{\Sigma}}, \bold{X}_{\hat{D}_T}(\bar{\tau}_s\simeq \pm \infty), \bold{I}(\bar{\tau}_s\simeq \pm \infty), \bar{\tau}_s\simeq \pm \infty) \sim (\bar{\bold{\Sigma}}', \bold{X}'_{\hat{D}_T}(\bar{\tau}_s\simeq \pm \infty), \bold{I}'(\bar{\tau}_s\simeq \pm \infty), \bar{\tau}_s\simeq \pm \infty)$ if $N_{\pm}=N'_{\pm}$, $M_{\pm}=M'_{\pm}$, $f^i=f'^i$, $\tilde{f}^j=\tilde{f}'^j$, $\bold{I}(\bar{\tau}_s\simeq \pm \infty)=\bold{I}'(\bar{\tau}_s\simeq \pm \infty)$, $\bold{X}_{\hat{D}_T}(\bar{\tau}_s\simeq \pm \infty)=\bold{X}'_{\hat{D}_T}(\bar{\tau}_s\simeq \pm \infty)$, and the corresponding supercylinders and superstrips are the same type (NS-NS, NS-R, R-NS, or R-R) and (NS or R), respectively as in Fig. \ref{EquivalentClass}.
 $[\bar{\bold{\Sigma}}, \bold{X}_{\hat{D}_T}(\bar{\tau}_s), \bold{I}(\bar{\tau}_s), \bar{\tau}_s]$ represent many-body states of open and closed superstrings in $\bold{R}^d$ as in Fig. \ref{states}, because the reduced space of $\bold{\Sigma}|_{\bar{\tau}_s}$ is $S^1 \times \cdots \times S^1 \times I^1 \times \cdots \times I^1$ where $I^1$ represents a line segment, and $\bold{X}_{\hat{D}_T}(\bar{\tau}_s): \bold{\Sigma}|_{\bar{\tau}_s} \to \bold{R}^d$.  We define the model space $\bold{E}$ such that $\bold{E}:=\bigcup_{\hat{D}_T}\{[\bar{{\boldsymbol \Sigma}}, \bold{X}_{\hat{D}_T}(\bar{\tau}), \bold{I}(\bar{\tau}), \bar{\tau}]\}$ where  disjoint unions are taken over all the backgrounds $\hat{D}_T$ except for the metric.

Here, we will define topologies of $\bold{E}$. An $\epsilon$-open neighborhood of $[\bar{\bold{\Sigma}}, \bold{X}_{s \, \hat{D}_T}(\bar{\tau}_s), \bold{I}_s(\bar{\tau}_s), \bar{\tau}_s]$
is defined by 
\begin{eqnarray}
&&U([\bar{\bold{\Sigma}}, \bold{X}_{s \, \hat{D}_T}(\bar{\tau}_s), \bold{I}_s(\bar{\tau}_s), \bar{\tau}_s], \epsilon) \nonumber \\
&:=&
\left\{[\bar{\bold{\Sigma}}, \bold{X}_{\hat{D}_T}(\bar{\tau}), \bold{I}(\bar{\tau}), \bar{\tau}]
 \bigm|
 \sqrt{|\bar{\tau}-\bar{\tau}_s|^2
+ \| \bold{X}_{\hat{D}_T} (\bar{\tau})-\bold{X}_{s \, \hat{D}_T}(\bar{\tau}_s) \|^2}
<\epsilon, \bold{I}_s(\bar{\tau}_s) \cong \bold{I}(\bar{\tau})
\right\}, \nonumber \\
&& \label{SuperOpenNeighbour}
\end{eqnarray}
where
\begin{eqnarray}
&&
\| \bold{X}_{\hat{D}_T} (\bar{\tau})-\bold{X}_{s \, \hat{D}_T}(\bar{\tau}_s) \|^2 \nonumber \\
&:=&\int_0^{2\pi}  d\bar{\sigma} 
\Bigl(|x(\bar{\tau}, \bar{\sigma})-x_s(\bar{\tau}_s, \bar{\sigma})|^2 
+(\bar{\psi}(\bar{\tau}, \bar{\sigma})-\bar{\psi}_s(\bar{\tau}_s, \bar{\sigma}))
(\psi(\bar{\tau}, \bar{\sigma})-\psi_s(\bar{\tau}_s, \bar{\sigma})) \nonumber \\
&+&|f(\bar{\tau}, \bar{\sigma})-f_s(\bar{\tau}_s, \bar{\sigma})|^2 \Bigr).
\end{eqnarray}
$\bold{I}_s(\bar{\tau}_s) \cong \bold{I}(\bar{\tau})$ means that a Chan-Paton index on the intersection between $\Sigma|_{\bar{\tau}_s}$ and a boundary component on $\Sigma$ equals a Chan-Paton index on the intersection between $\Sigma|_{\bar{\tau}}$ and the boundary component, excepting that the corresponding intersection does not exist. For example, see Fig. \ref{9open}.
$U([\bar{\bold{\Sigma}}, \bold{X}_{\hat{D}_T}(\bar{\tau}_s\simeq \pm \infty), \bold{I}(\bar{\tau}_s\simeq \pm \infty), \bar{\tau}_s\simeq \pm \infty], \epsilon)
=
U([\bar{\bold{\Sigma}}', \bold{X}'_{\hat{D}_T}(\bar{\tau}_s\simeq \pm \infty), \bold{I}'(\bar{\tau}_s\simeq \pm \infty), \bar{\tau}_s\simeq \pm \infty], \epsilon)$ consistently  if $N_{\pm}=N'_{\pm}$, $M_{\pm}=M'_{\pm}$, $f^i=f'^i$, $\tilde{f}^j=\tilde{f}'^j$, the corresponding supercylinders and superstrips are the same type (NS-NS, NS-R, R-NS, or R-R) and (NS or R), respectively, $\bold{I}(\bar{\tau}_s\simeq \pm \infty)=\bold{I}'(\bar{\tau}_s\simeq \pm \infty)$, $\bold{X}_{\hat{D}_T}(\bar{\tau}_s\simeq \pm \infty)=\bold{X}'_{\hat{D}_T}(\bar{\tau}_s\simeq \pm \infty)$,  and $\epsilon$ is small enough, because the $\bar{\tau}=\bar{\tau}_s$ constant line traverses only propagators overlapped by $\bar{\bold{\Sigma}}$ and $\bar{\bold{\Sigma}}'$. $U$ is defined to be an open set of $\bold{E}$ if there exists $\epsilon$ such that $U([\bar{\bold{\Sigma}}, \bold{X}_{\hat{D}_T}(\bar{\tau}_s), \bold{I}(\bar{\tau}_s), \bar{\tau}_s], \epsilon) \subset U$ for an arbitrary point $[\bar{\bold{\Sigma}}, \bold{X}_{\hat{D}_T}(\bar{\tau}_s), \bold{I}(\bar{\tau}_s), \bar{\tau}_s] \in U$. In exactly the same way as in section 2, one can show that the topology of $\bold{E}$ satisfies the axiom of topology.
Although the model space is defined by using the coordinates $[\bar{{\boldsymbol \Sigma}}, \bold{X}_{\hat{D}_T}(\bar{\tau}), \bold{I}(\bar{\tau}), \bar{\tau}]$, the model space does not depend on the coordinates, because the model space is a topological space.

In the following, we denote $[\bar{\bold{E}}_{M}^{\quad A}(\bar{\sigma}, \bar{\tau}, \bar{\theta}^{\alpha}), \bold{X}_{\hat{D}_T}(\bar{\tau}), \bold{I}(\bar{\tau}), \bar{\tau}]$, where  $\bar{\bold{E}}_{M}^{\quad A}(\bar{\sigma}, \bar{\tau}, \bar{\theta}^{\alpha})$ ($M=(m, \alpha)$, $A=(q, a)$, $m, q=0,1$, $\alpha, a=1,2$) is the worldsheet super vierbein on $\bar{{\boldsymbol \Sigma}}$,  instead of $[\bar{{\boldsymbol \Sigma}}, \bold{X}_{\hat{D}_T}(\bar{\tau}), \bold{I}(\bar{\tau}), \bar{\tau}]$ because giving a super Riemann surface is equivalent to giving a super vierbein up to super diffeomorphism and super Weyl transformations.

In order to define structures of manifold, let us consider how generally we can define general coordinate transformations between $[\bar{\bold{E}}_{M}^{\quad A}, \bold{X}_{\hat{D}_T}(\bar{\tau}), \bold{I}(\bar{\tau}), \bar{\tau}]$ and $[\bar{\bold{E}}_{M}^{'\quad A}, \bold{X}'_{\hat{D}_T}(\bar{\tau}'), \bold{I}'(\bar{\tau}'), \bar{\tau}']$ where $[\bar{\bold{E}}_{M}^{\quad A}, \bold{X}_{\hat{D}_T}(\bar{\tau}), \bold{I}(\bar{\tau}), \bar{\tau}] \in U \subset \bold{E}$ and $[\bar{\bold{E}}_{M}^{'\quad A}, \bold{X}'_{\hat{D}_T}(\bar{\tau}'), \bold{I}'(\bar{\tau}'), \bar{\tau}']\in U' \subset \bold{E}$.  $\bar{\bold{E}}_{M}^{\quad A}$ and $\bold{I}(\bar{\tau})$ do not transform to $\bar{\tau}$ and $\bold{X}_{\hat{D}_T}(\bar{\tau})$ and vice versa, because $\bar{\tau}$ and $\bold{X}_{\hat{D}_T}(\bar{\tau})$ are continuous variables, whereas $\bar{\bold{E}}_{M}^{\quad A}$ and $\bold{I}(\bar{\tau})$ are discrete variables: $\bar{\tau}$ and $\bold{X}_{\hat{D}_T}(\bar{\tau})$ vary continuously, whereas $\bar{\bold{E}}_{M}^{\quad A}$ and $\bold{I}(\bar{\tau})$ vary discretely in a trajectory on $\bold{E}$ by definition of the neighborhoods. $\bar{\tau}$ does not transform to $\bar{\sigma}$ and $\bar{\theta}$ and vice versa, because the superstring states are defined by $\bar{\tau}$ constant hypersurfaces. Under these restrictions, the most general coordinate transformation is given by 
\begin{eqnarray}
&&[\bar{\bold{E}}_{M}^{\quad A}(\bar{\sigma}, \bar{\tau}, \bar{\theta}^{\alpha}), \bold{X}_{\hat{D}_T}^{\mu}(\bar{\sigma}, \bar{\tau}, \bar{\theta}^{\alpha}), \bold{I}(\bar{\tau}), \bar{\tau}] \nonumber \\
&\mapsto& [\bold{E}_{M}^{' \quad A}(\bar{\sigma}'(\bar{\sigma}, \bar{\theta}), \bar{\tau}', \bar{\theta}^{'\alpha}(\bar{\sigma}, \bar{\theta})), \bold{X}_{\hat{D}_T}^{'\mu} (\bar{\sigma}', \bar{\tau}', \bar{\theta}^{'\alpha})(\bar{\tau}, \bold{X}_{\hat{D}_T}(\bar{\tau})), \bold{I}(\bar{\tau}'), \bar{\tau}'(\bar{\tau}, \bold{X}_{\hat{D}_T}(\bar{\tau}))], \nonumber \\
\label{SuperGeneralCoordTrans2}
\end{eqnarray}
where $\bar{\bold{E}}_{M}^{\quad A} \mapsto \bar{\bold{E}}_{M}^{'\quad A}$ represents a world-sheet superdiffeomorphism transformation\footnote{
We extend the model space from $\bold{E}=\{[\bar{\bold{E}}_{M}^{\quad A}(\bar{\sigma}, \bar{\tau}, \bar{\theta}^{\alpha}), \bold{X}_{\hat{D}_T}^{\mu}(\bar{\sigma}, \bar{\tau}, \bar{\theta}^{\alpha}), \bold{I}(\bar{\tau}), \bar{\tau}]\}$ to $\bold{E}=\{[\bold{E}_{M}^{' \quad A}(\bar{\sigma}', \bar{\tau}', \bar{\theta}^{'\alpha}), \bold{X}_{\hat{D}_T}^{'\mu} (\bar{\sigma}', \bar{\tau}', \bar{\theta}^{'\alpha}), \bold{I}(\bar{\tau}'), \bar{\tau}'] \}$ by including the points generated by the superdiffeomorphisms $\bar{\sigma} \mapsto \bar{\sigma}'(\bar{\sigma}, \bar{\theta})$,  $\bar{\theta}^{\alpha} \mapsto \bar{\theta}^{'\alpha}(\bar{\sigma}, \bar{\theta})$, and $\bar{\tau} \mapsto \bar{\tau}'(\bar{\tau})$.}. $\bold{X}_{\hat{D}_T}^{'\mu} (\bar{\tau}, \bold{X}_{\hat{D}_T}(\bar{\tau}))$ and $\bar{\tau}'(\bar{\tau}, \bold{X}_{\hat{D}_T}(\bar{\tau}))$ are functionals of $\bar{\tau}$ and $\bold{X}_{\hat{D}_T}(\bar{\tau})$. 
Here, we consider all the manifolds which are constructed by patching open sets of the model space $\bold{E}$ by general coordinate transformations (\ref{SuperGeneralCoordTrans2}) and call them superstring manifolds $\mathfrak{M}$.


Here, we give an example of superstring manifolds: $\mathfrak{M}_{D_T}:= \{[\bar{{\boldsymbol \Sigma}}, \bold{x}_{D_T}(\bar{\tau}), \bold{I}(\bar{\tau}), \bar{\tau}]\}$, where $D_T$ represents a target metric $G_{\mu \nu}$, O-planes and D-bundles with gauge connections where all the other backgrounds are turned off except for $G_{\mu \nu}$,  and a type of the GSO projection. $\bold{x}_{D_T}(\bar{\tau}): \bar{\bold{\Sigma}}|_{\bar{\tau}} \to M$,  where $\bold{x}_{D_T}^{\mu}(\bar{\tau})=x^{\mu}+ \bar{\theta}^{\alpha} \psi_{\alpha}^{\mu}+\frac{1}{2} \bar{\theta}^2 f^{\mu}$. The image of the bosonic part of the embedding function, $x(\bar{\tau})$ has a metric: 
$ds^2= dx^{\mu}(\bar{\tau}, \bar{\sigma})
dx^{\nu}(\bar{\tau}, \bar{\sigma})
G_{\mu \nu}(x(\bar{\tau}, \bar{\sigma}))$. 

We will show that $\mathfrak{M}_{D_T}$has a structure of manifold, that is there exists a homeomorphism between the sufficiently small neighborhood around an arbitrary point $[\bar{\bold{\Sigma}}, \bold{x}_{s D_T}(\bar{\tau}_s), \bold{I}_s(\bar{\tau}_s), \bar{\tau}_s] \in \mathfrak{M}_{D_T}$ and an open set of $E$. There exists a general coordinate transformation $X^{\mu}(x)$ that satisfies 
$ds^2= dx^{\mu}
dx^{\nu}
G_{\mu \nu}(x)
=
dX^{\mu}
dX^{\nu}
\eta_{\mu \nu}$
on an arbitrary point $x$ in the $\epsilon_{\bar{\sigma}}$ open neighborhood around $x_s(\bar{\tau}_s, \bar{\sigma}) \in M$, if $\epsilon_{\bar{\sigma}}$ is sufficiently small. An arbitrary point $[\bar{\bold{\Sigma}}, \bold{x}_{D_T}(\bar{\tau}), \bold{I}(\bar{\tau}), \bar{\tau}]$ in the $\epsilon$ open neighborhood around  $[\bar{\bold{\Sigma}}, \bold{x}_{s  D_T}(\bar{\tau}_s), \bold{I}_s(\bar{\tau}_s), \bar{\tau}_s] $ satisfies 
\begin{eqnarray}
&&\int_0^{2\pi}  d\bar{\sigma} 
|x(\bar{\tau}, \bar{\sigma})-x_s(\bar{\tau}_s, \bar{\sigma})|^2 \nonumber \\
&&< \epsilon^2- |\bar{\tau}-\bar{\tau}_s|^2
-\int_0^{2\pi}  d\bar{\sigma}\Bigl( (\bar{\psi}(\bar{\tau}, \bar{\sigma})-\bar{\psi}_s(\bar{\tau}_s, \bar{\sigma}))
(\psi(\bar{\tau}, \bar{\sigma})-\psi_s(\bar{\tau}_s, \bar{\sigma}))
-|f(\bar{\tau}, \bar{\sigma})-f_s(\bar{\tau}_s, \bar{\sigma})|^2 \Bigr)\nonumber \\
&&
 \leqq \epsilon^{2} \label{neighcomplete}
\end{eqnarray}
and thus 
\begin{equation}
|x(\bar{\tau}, \bar{\sigma})-x_s(\bar{\tau}_s, \bar{\sigma})|
< \epsilon_{\bar{\sigma}}'
\end{equation}
on an arbitrary $\bar{\sigma}$.
$\epsilon_{\bar{\sigma}}'< \epsilon_{\bar{\sigma}}$ is satisfied on an arbitrary $\bar{\sigma}$ if $\epsilon$ is taken to be sufficiently small. 
Then, there exists a transformation  $X^{\mu}(\bar{\tau}, \bar{\sigma}):=X^{\mu}(x(\bar{\tau}, \bar{\sigma}))$, which satisfies
\begin{equation}
ds^2= dx^{\mu}(\bar{\tau}, \bar{\sigma})
dx^{\nu}(\bar{\tau}, \bar{\sigma})
G_{\mu \nu}(x(\bar{\tau}, \bar{\sigma}))
=
dX^{\mu}(\bar{\tau}, \bar{\sigma})
dX^{\nu}(\bar{\tau}, \bar{\sigma})
\eta_{\mu \nu}. \label{oSuperLocalLorents}
\end{equation}
Because the tangent vector $\bold{X}_{\hat{D}_T}(\bar{\tau}, \bar{\sigma}, \bar{\theta})$ exists for each $\bold{x}_{D_T}(\bar{\tau}, \bar{\sigma}, \bar{\theta})$, there exists a vector bundle $\bold{X}_{\hat{D}_T}(\bar{\tau})$ for $0 \leqq \bar{\sigma} < 2\pi$ and $\bar{\theta}$. $\bold{x}_{D_T}(\bar{\tau})$ and $\bold{X}_{\hat{D}_T}(\bar{\tau})$ satisfy (\ref{oSuperLocalLorents}) on each $\bar{\sigma}$, that is $\bold{X}_{\hat{D}_T}(\bar{\tau}): \bar{\bold{\Sigma}}|_{\bar{\tau}} \to \bold{R}^d$. Therefore, there exists a homeomorphism between the sufficiently small neighborhood around an arbitrary point $[\bar{\bold{\Sigma}}, \bold{x}_{s D_T}(\bar{\tau}_s), \bold{I}_s(\bar{\tau}_s), \bar{\tau}_s] \in \mathfrak{M}_{D_T}$ and an open set of $\bold{E}$: $[\bar{\bold{\Sigma}}, \bold{x}_{D_T}(\bar{\tau}), \bold{I}(\bar{\tau}), \bar{\tau}] \mapsto [\bar{\bold{\Sigma}}, \bold{X}_{\hat{D}_T}(\bar{\tau}), \bold{I}(\bar{\tau}), \bar{\tau}]$.
Actually, a map from the image $\bold{x}_{D_T}(\bar{\tau}, \bar{\sigma}, \bar{\theta})$ to the image $\bold{X}_{\hat{D}_T}(\bar{\tau}, \bar{\sigma}, \bar{\theta})$ is explicitly given by an exponential map
\begin{equation}
\bold{x}_{D_T}(\bar{\tau}, \bar{\sigma}, \bar{\theta})
=
\exp_{\bold{x}_{sD_T}(\bar{\tau}_s, \bar{\sigma}, \bar{\theta})}
\bold{X}_{\hat{D}_T}(\bar{\tau}, \bar{\sigma}, \bar{\theta})
\simeq
\bold{x}_{sD_T}(\bar{\tau}_s, \bar{\sigma}, \bar{\theta})
+\bold{X}_{\hat{D}_T}(\bar{\tau}, \bar{\sigma}, \bar{\theta}).
\end{equation}
If we substitute this to an $\epsilon$ open neighborhood around an arbitrary point  $[\bar{\bold{\Sigma}}, \bold{x}_{s D_T}(\bar{\tau}_s), \bold{I}_s(\bar{\tau}_s), \bar{\tau}_s] \in \mathfrak{M}_{D_T}$ (\ref{neighcomplete}),
we obtain an $\epsilon$ open neighborhood around $ [\bar{\bold{\Sigma}}, 0, \bold{I}_s(\bar{\tau}_s), \bar{\tau}_s] \in \bold{E}$,
\begin{equation}
\| \bold{X}_{\hat{D}_T}(\bar{\tau}) \|^2 +  |\bar{\tau}-\bar{\tau}_s|^2 < \epsilon^{2}.
\end{equation}

By definition of the $\epsilon$-open neighborhood, on a connected super Riemann surface with open and closed punctures and with or without boundaries in $M$, arbitrary two superstring states with the same Chan-Paton indices are connected continuously. Thus, there is an one-to-one correspondence between such a  super Riemann surface that have Chan-Paton indices and a curve  parametrized by $\bar{\tau}$ from $\bar{\tau}=-\infty$ to $\bar{\tau}=\infty$ on $\mathfrak{M}_{D_T}$. That is, curves that represent asymptotic processes on $\mathfrak{M}_{D_T}$ reproduce the right moduli space of the super Riemann surfaces.

By a general curve parametrized by $t$ on $\mathfrak{M}_{D_T}$, superstring states on the different super Riemann surfaces that have even different genera, can be connected continuously, whereas the different super Riemann surfaces that have different genera cannot be connected continuously in the moduli space of the super Riemann surfaces in the target space.

The tangent space is spanned by $\frac{\partial}{\partial\bar{\tau}}$ and $\frac{\partial}{\partial \bold{X}_{\hat{D}_T}^{\mu}(\bar{\sigma}, \bar{\tau}, \bar{\theta})}$ as one can see from the  $\epsilon$-open neighborhood  (\ref{SuperOpenNeighbour}). We should note that $\frac{\partial}{\partial \bar{\bold{E}}^{\quad A}_{\, M}}$ and $\frac{\partial}{\partial \bold{I}(\bar{\tau})}$ cannot be a part of basis that span the tangent space because $\bar{\bold{E}}^{\quad A}_{\, M}$ and $\bold{I}(\bar{\tau})$ are just discrete variables in $\bold{E}$. The index of $\frac{\partial}{\partial \bold{X}_{\hat{D}_T}^{\mu}(\bar{\sigma}, \bar{\tau}, \bar{\theta}) }$ can be $(\mu \, \bar{\sigma} \, \bar{\theta})$. 
We define a summation over $\bar{\sigma}$ and $\bar{\theta}$ by $\int d\bar{\sigma}d^2\bar{\theta} \hat{\bold{E}}(\bar{\sigma}, \bar{\tau}, \bar{\theta}^{\alpha})$,
where
$\hat{\bold{E}}(\bar{\sigma}, \bar{\tau}, \bar{\theta}^{\alpha})
:=
\frac{1}{\bar{n}}\bar{\bold{E}}(\bar{\sigma}, \bar{\tau}, \bar{\theta}^{\alpha})$, because it is transformed as a scalar under $\bar{\tau} \mapsto \bar{\tau}'(\bar{\tau}, \bold{X}_{\hat{D}_T}(\bar{\tau}))$ and invariant under $(\bar{\sigma}, \bar{\theta}^{\alpha}) \mapsto (\bar{\sigma}'(\bar{\sigma}, \bar{\theta}), \bar{\theta}^{'\alpha}(\bar{\sigma}, \bar{\theta}))$ as one can see in section 4.

Riemannian superstring manifold is obtained by defining a metric, which is a section of an inner product on the tangent space. The general form of a metric is given by
\begin{eqnarray}
&&ds^2(\bar{\bold{E}}, \bold{X}_{\hat{D}_T}(\bar{\tau}), \bold{I}(\bar{\tau}), \bar{\tau}) \nonumber \\
=&&G(\bar{\bold{E}}, \bold{X}_{\hat{D}_T}(\bar{\tau}), \bold{I}(\bar{\tau}), \bar{\tau})_{dd} (d\bar{\tau})^2 \nonumber \\ 
&&+2 d\bar{\tau} \int d\bar{\sigma} d^2\bar{\theta}  \hat{\bold{E}} \sum_{\mu} G(\bar{\bold{E}}, \bold{X}_{\hat{D}_T}(\bar{\tau}), \bold{I}(\bar{\tau}), \bar{\tau})_{d \; (\mu \bar{\sigma} \bar{\theta})} d \bold{X}_{\hat{D}_T}^{\mu}(\bar{\sigma}, \bar{\tau}, \bar{\theta}) \nonumber \\
&&+\int d\bar{\sigma} d^2\bar{\theta} \hat{\bold{E}}  \int d\bar{\sigma}'  d^2\bar{\theta}' \hat{\bold{E}}'\sum_{\mu, \mu'} G(\bar{\bold{E}}, \bold{X}_{\hat{D}_T}(\bar{\tau}), \bold{I}(\bar{\tau}), \bar{\tau})_{ \; (\mu \bar{\sigma} \bar{\theta})  \; (\mu' \bar{\sigma}' \bar{\theta}')} d \bold{X}_{\hat{D}_T}^{\mu}(\bar{\sigma}, \bar{\tau}, \bar{\theta}) d \bold{X}_{\hat{D}_T}^{\mu'}(\bar{\sigma}', \bar{\tau}, \bar{\theta}'). \nonumber \\
&&
\end{eqnarray}
We summarize the vectors as $d\bold{X}_{\hat{D}_T}^\bold{I}$ ($\bold{I}=d,(\mu \bar{\sigma} \bar{\theta})$), where  $d\bold{X}_{\hat{D}_T}^d:=d\bar{\tau}$ and $d \bold{X}_{\hat{D}_T}^{(\mu \bar{\sigma} \bar{\theta})}:=d\bold{X}_{\hat{D}_T}^{\mu}(\bar{\sigma}, \bar{\tau}, \bar{\theta})$. Then, the components of the metric are summarized as $G_{\bold{I}\bold{J}}(\bar{\bold{E}}, \bold{X}_{\hat{D}_T}(\bar{\tau}), \bold{I}(\bar{\tau}), \bar{\tau})$. The inverse of the metric $G^{\bold{I}\bold{J}}(\bar{\bold{E}}, \bold{X}_{\hat{D}_T}(\bar{\tau}), \bold{I}(\bar{\tau}), \bar{\tau})$ is defined by $G_{\bold{I}\bold{J}}G^{\bold{J}\bold{K}}=G^{\bold{K}\bold{J}}G_{\bold{J}\bold{I}}=\delta_\bold{I}^\bold{K}$, where $\delta_d^d=1$ and $\delta_{\mu \bar{\sigma} \bar{\theta}}^{\mu' \bar{\sigma}' \bar{\theta}'}=\frac{1}{\hat{\bold{E}}}\delta_{\mu}^{\mu'}\delta(\bar{\sigma}-\bar{\sigma}') \delta^2(\bar{\theta}-\bar{\theta}')$. The components of the Riemannian curvature tensor are given by $R^\bold{I}_{\bold{J}\bold{K}\bold{L}}$ in the basis $\frac{\partial}{\partial \bold{X}_{\hat{D}_T}^\bold{I}(\bar{\tau})}$. The components of the Ricci tensor are $R_{\bold{I}\bold{J}}:=R^\bold{K}_{\bold{I}\bold{K}\bold{J}}=R^d_{\bold{I}d\bold{J}}+\int d\bar{\sigma} d^2\bar{\theta} \hat{\bold{E}}  R^{(\mu \bar{\sigma} \bar{\theta})}_{\bold{I} \; (\mu \bar{\sigma} \bar{\theta}) \; \bold{J}}$. The scalar curvature is 
\begin{eqnarray}
R&:=&G^{\bold{I}\bold{J}} R_{\bold{I}\bold{J}} \nonumber \\
&=&G^{dd}R_{dd}+2 \int d\bar{\sigma} d^2\bar{\theta} \hat{\bold{E}}  G^{d \; (\mu \bar{\sigma} \bar{\theta})} R_{d \; (\mu \bar{\sigma} \bar{\theta})} 
\nonumber \\
&&+\int d\bar{\sigma} d^2\bar{\theta} \hat{\bold{E}}  \int d\bar{\sigma}'  d^2\bar{\theta}' \hat{\bold{E}}' G^{(\mu \bar{\sigma} \bar{\theta}) \; (\mu' \bar{\sigma}' \bar{\theta}')}R_{(\mu \bar{\sigma} \bar{\theta})  \; (\mu' \bar{\sigma}' \bar{\theta}')}. \nonumber
\end{eqnarray}
The volume is  $vol=\sqrt{G}$, where $G=det (G_{\bold{I}\bold{J}})$.

By using these geometrical objects, we formulate superstring theory non-perturbatively as
\begin{equation}
Z=\int \mathcal{D}G \mathcal{D}Ae^{-S}, \label{OSSuperTheory}
\end{equation}
where
\begin{equation}
S=\frac{1}{G_N}\int \mathcal{D}\bold{E} \mathcal{D}\bar{\tau} \mathcal{D}\bold{X}_{\hat{D}_T}(\bar{\tau}) \mathcal{D} \bold{I}(\bar{\tau})
\sqrt{G} (-R +\frac{1}{4} G_N G^{\bold{I}_1 \bold{I}_2} G^{\bold{J}_1 \bold{J}_2} F_{\bold{I}_1 \bold{J}_1} F_{\bold{I}_2 \bold{J}_2} ). \label{OSSuperAction}
\end{equation}
As an example of sets of fields on the superstring manifolds, we consider the metric and an $u(1)$ gauge field $A_\bold{I}$ whose field strength is given by $F_{\bold{I}\bold{J}}$. The path integral is defined by semi-classically\footnote{It will be enough to define the path-integral by semi-classically summing classical solutions and small classical and quantum fluctuations around them, because string manifolds themselves possess quantum corrections, and loops of the fields on them do not correspond to quantum corrections as one can see in the derivation of the perturbative string theory later. The unitarity is manifest and there is also no UV divergence from loop integrals, by defining the path-integral semi-classically. } summing over the metrics and gauge fields on $\mathfrak{M}$. By definition, the theory is background independent. $\mathcal{D}\bold{E}$ is the invariant measure of the super vierbeins $\bold{E}_{M}^{\quad A}$ on the two-dimensional super Riemannian manifolds $\bold{\Sigma}$. $\bold{E}_{M}^{\quad A}$ and $\bar{\bold{E}}_{M}^{\quad A}$ are related to each others by the super diffeomorphism and super Weyl transformations.

Under 
\begin{equation}
(\bar{\tau}, \bold{X}_{\hat{D}_T}(\bar{\tau})) \mapsto (\bar{\tau}'(\bar{\tau}, \bold{X}_{\hat{D}_T}(\bar{\tau})) , \bold{X}_{\hat{D}_T}'(\bar{\tau}')(\bar{\tau}, \bold{X}_{\hat{D}_T}(\bar{\tau}))),
\label{OSSupersubdiffeo}
\end{equation} 
$G_{\bold{I}\bold{J}}(\bar{\bold{E}}, \bold{X}_{\hat{D}_T}(\bar{\tau}), \bold{I}(\bar{\tau}), \bar{\tau})$ and $A_{\bold{I}}(\bar{\bold{E}}, \bold{X}_{\hat{D}_T}(\bar{\tau}), \bold{I}(\bar{\tau}), \bar{\tau})$ are transformed as a symmetric tensor and a vector, respectively and the action is manifestly invariant.

We define $G_{\bold{I}\bold{J}}(\bar{\bold{E}}, \bold{X}_{\hat{D}_T}(\bar{\tau}), \bold{I}(\bar{\tau}), \bar{\tau})$ and $A_{\bold{I}}(\bar{\bold{E}}, \bold{X}_{\hat{D}_T}(\bar{\tau}), \bold{I}(\bar{\tau}), \bar{\tau})$ so as to transform as scalars under $\bar{\bold{E}}_{M}^{\quad A}(\bar{\sigma}, \bar{\tau}, \bar{\theta}^{\alpha}) \mapsto
\bar{\bold{E}}_{M}^{'\quad A}(\bar{\sigma}'(\bar{\sigma}, \bar{\theta}), \bar{\tau}, \bar{\theta}^{'\alpha}(\bar{\sigma}, \bar{\theta}))$. Under $(\bar{\sigma}, \bar{\theta})$ superdiffeomorphisms: $(\bar{\sigma}, \bar{\theta}^{\alpha}) \mapsto (\bar{\sigma}'(\bar{\sigma}, \bar{\theta}), \bar{\theta}^{'\alpha}(\bar{\sigma}, \bar{\theta}))$, which are equivalent to 
\begin{eqnarray}
&&[\bar{\bold{E}}_{M}^{\quad A}(\bar{\sigma}, \bar{\tau}, \bar{\theta}^{\alpha}),  \bold{X}_{\hat{D}_T}^{\mu}(\bar{\sigma}, \bar{\tau}, \bar{\theta}^{\alpha}), \bold{I}(\bar{\tau}), \bar{\tau}] \nonumber \\
&&\mapsto [\bar{\bold{E}}_{M}^{'\quad A}(\bar{\sigma}'(\bar{\sigma}, \bar{\theta}), \bar{\tau}, \bar{\theta}^{'\alpha}(\bar{\sigma}, \bar{\theta})), \bold{X}_{\hat{D}_T}^{'\mu} (\bar{\sigma}'(\bar{\sigma}, \bar{\theta}), \bar{\tau}, \bar{\theta}^{'\alpha}(\bar{\sigma}, \bar{\theta}))(\bold{X}_{\hat{D}_T}(\bar{\tau})),  \bold{I}(\bar{\tau}), \bar{\tau}] \nonumber \\
&&=[\bar{\bold{E}}_{M}^{'\quad A}(\bar{\sigma}'(\bar{\sigma}, \bar{\theta}), \bar{\tau}, \bar{\theta}^{'\alpha}(\bar{\sigma}, \bar{\theta})), \bold{X}_{\hat{D}_T}^{\mu}(\bar{\sigma}, \bar{\tau}, \bar{\theta}^{\alpha}),  \bold{I}(\bar{\tau}), \bar{\tau}], \label{OSSuperStringGeometryTrans}
\end{eqnarray}
$G_{d \; (\mu \bar{\sigma} \bar{\theta})}$ is transformed as a superscalar;
\begin{eqnarray}
&&G'_{d \; (\mu \bar{\sigma}'  \bar{\theta}')}(\bar{\bold{E}}', \bold{X}_{\hat{D}_T}'(\bar{\tau}),  \bold{I}(\bar{\tau}), \bar{\tau}) \nonumber \\
&=&
G'_{d \; (\mu \bar{\sigma}'  \bar{\theta}')}(\bar{\bold{E}}, \bold{X}'_T(\bar{\tau}), \bold{I}(\bar{\tau}), \bar{\tau})
=
\frac{\partial \bold{X}_{\hat{D}_T}^{\bold{I}}(\bar{\tau})}{\partial \bold{X}_{\hat{D}_T}^{'d}(\bar{\tau})}
\frac{\partial \bold{X}_{\hat{D}_T}^{\bold{J}}(\bar{\tau})}{\partial \bold{X}_{\hat{D}_T}^{'(\mu \bar{\sigma}' \bar{\theta}')}(\bar{\tau})}
G_{\bold{I} \bold{J}}(\bar{\bold{E}}, \bold{X}_{\hat{D}_T}(\bar{\tau}), \bold{I}(\bar{\tau}), \bar{\tau})
\nonumber \\
&=&
\frac{\partial \bold{X}_{\hat{D}_T}^{\bold{I}}(\bar{\tau})}{\partial \bold{X}_{\hat{D}_T}^{d}(\bar{\tau})}
\frac{\partial \bold{X}_{\hat{D}_T}^{\bold{J}}(\bar{\tau})}{\partial \bold{X}_{\hat{D}_T}^{(\mu \bar{\sigma} \bar{\theta})}(\bar{\tau})}
G_{\bold{I} \bold{J}}(\bar{\bold{E}}, \bold{X}_{\hat{D}_T}(\bar{\tau}), \bold{I}(\bar{\tau}), \bar{\tau})
=
G_{d \; (\mu \bar{\sigma} \bar{\theta})}(\bar{\bold{E}}, \bold{X}_{\hat{D}_T}(\bar{\tau}), \bold{I}(\bar{\tau}), \bar{\tau}), \label{GdmuTrans} \nonumber \\
&&
\end{eqnarray}
because (\ref{OSSupersubdiffeo}) and (\ref{OSSuperStringGeometryTrans}). In the same way, the other fields are also transformed as
\begin{eqnarray}
G'_{dd}(\bar{\bold{E}}', \bold{X}_{\hat{D}_T}'(\bar{\tau}),  \bold{I}(\bar{\tau}), \bar{\tau})
&=&G_{dd}(\bar{\bold{E}}, \bold{X}_{\hat{D}_T}(\bar{\tau}), \bold{I}(\bar{\tau}), \bar{\tau}) \nonumber \\
G'_{ \; (\mu \bar{\sigma}'\bar{\theta}')  \; (\nu \bar{\rho}'\tilde{\bar{\theta}}')}(\bar{\bold{E}}', \bold{X}_{\hat{D}_T}'(\bar{\tau}),  \bold{I}(\bar{\tau}), \bar{\tau})
&=&G_{ \; (\mu \bar{\sigma} \bar{\theta})  \; (\nu \bar{\rho} \tilde{\bar{\theta}})}(\bar{\bold{E}}, \bold{X}_{\hat{D}_T}(\bar{\tau}), \bold{I}(\bar{\tau}), \bar{\tau}) \nonumber \\
A'_d(\bar{\bold{E}}', \bold{X}_{\hat{D}_T}'(\bar{\tau}),  \bold{I}(\bar{\tau}), \bar{\tau})
&=&A_d(\bar{\bold{E}}, \bold{X}_{\hat{D}_T}(\bar{\tau}), \bold{I}(\bar{\tau}), \bar{\tau}) \nonumber \\
A'_{(\mu \bar{\sigma}'\bar{\theta}')}(\bar{\bold{E}}', \bold{X}_{\hat{D}_T}'(\bar{\tau}),  \bold{I}(\bar{\tau}), \bar{\tau})&=&A_{(\mu \bar{\sigma} \bar{\theta})}(\bar{\bold{E}}, \bold{X}_{\hat{D}_T}(\bar{\tau}), \bold{I}(\bar{\tau}), \bar{\tau}). \label{SuperOtherTrans}
\end{eqnarray}
Thus, the action is invariant under the $(\bar{\sigma}, \bar{\theta})$ superdiffeomorphisms, because 
\begin{equation}
\int d\bar{\sigma}' d^2\bar{\theta}'  \hat{\bold{E}}'(\bar{\sigma}', \bar{\tau}, \bar{\theta}')=\int d\bar{\sigma} d^2\bar{\theta}  \hat{\bold{E}}(\bar{\sigma}, \bar{\tau}, \bar{\theta}).
\end{equation}
Therefore, $G_{\bold{I}\bold{J}}(\bar{\bold{E}}, \bold{X}_{\hat{D}_T}(\bar{\tau}), \bold{I}(\bar{\tau}), \bar{\tau})$ and $A_{\bold{I}}(\bar{\bold{E}}, \bold{X}_{\hat{D}_T}(\bar{\tau}), \bold{I}(\bar{\tau}), \bar{\tau})$ are transformed covariantly and the action (\ref{OSSuperAction}) is invariant under the diffeomorphisms (\ref{SuperGeneralCoordTrans2}) including the $(\bar{\sigma}, \bar{\theta})$ superdiffeomorphisms, whose
infinitesimal transformations are given by 
\begin{eqnarray}
\bar{\sigma}^{\xi}&=&\bar{\sigma}+i \xi^{\alpha}(\bar{\sigma}) \gamma^1_{\alpha \beta} \bar{\theta}^{\beta} \nonumber \\
\bar{\theta}^{\xi \alpha}(\bar{\sigma})&=&\bar{\theta}^{\alpha} + \xi^{\alpha}(\bar{\sigma}). \label{OSUSYtrans}
\end{eqnarray}
(\ref{OSUSYtrans}) are dimensional reductions in $\bar{\tau}$ direction of the two-dimensional $\mathcal{N}=(1,1)$ local supersymmetry infinitesimal transformations. The number of supercharges
\begin{equation}
\xi^{\alpha} Q_{\alpha}
=
\xi^{\alpha}(\frac{\partial}{\partial \bar{\theta}^{\alpha}} + i \gamma^1_{\alpha \beta}\bar{\theta}^{\beta} \frac{\partial}{\partial \bar{\sigma}})
\end{equation}
 of the transformations is the same as of the two-dimensional ones. The supersymmetry algebra closes in a field-independent sense as in ordinary supergravities. 

The background that represents a perturbative vacuum is given by
\begin{eqnarray}
\bar{ds}^2
&=& 2\lambda \bar{\rho}(\bar{h}) N^2(\bold{X}_{\hat{D}_T}(\bar{\tau})) (d\bold{X}_{\hat{D}_T}^d)^2 \nonumber \\ 
&&+\int d\bar{\sigma} d^2\bar{\theta}  \hat{\bold{E}} \int d\bar{\sigma}' d^2\bar{\theta}'  \hat{\bold{E}}' N^{\frac{2}{2-\bold{D}}}(\bold{X}_{\hat{D}_T}(\bar{\tau}))
\frac{\hat{\bold{E}}^3(\bar{\sigma}, \bar{\tau}, \bar{\theta})}
{\bold{E}(\bar{\sigma}, \bar{\tau}, \bar{\theta})} 
 \delta_{(\mu \bar{\sigma} \bar{\theta}) (\mu' \bar{\sigma}' \bar{\theta}')}
d \bold{X}_{\hat{D}_T}^{(\mu \bar{\sigma} \bar{\theta})} d \bold{X}_{\hat{D}_T}^{(\mu' \bar{\sigma}' \bar{\theta}')}, \nonumber \\
\bar{A}_d&=&i \sqrt{\frac{2-2\bold{D}}{2-\bold{D}}}\frac{\sqrt{2\lambda \bar{\rho}(\bar{h}) }}{\sqrt{G_N}} N(\bold{X}_{\hat{D}_T}(\bar{\tau})), \qquad
\bar{A}_{(\mu \bar{\sigma} \bar{\theta})}=0, \label{OSupersolution}
\end{eqnarray}
on  $\mathfrak{M}_{D_T}$ where we fix the target metric to $\eta_{\mu \mu'}$, a set of D-submanifolds to arbitrary one, and the gauge connections to zero, respectively. $\bar{\rho}(\bar{h}):=\frac{1}{4\pi}\int d\bar{\sigma} \sqrt{\bar{h}}\bar{R}_{\bar{h}}+\frac{1}{2\pi}\bar{k}_{\bar{h}}$, where $\bar{R}_{\bar{h}}$ is the scalar curvature of $\bar{h}_{ mn}$ and $\bar{k}_{\bar{h}}$ is the geodesic curvature of  $\bar{h}_{ mn}$. $\bold{D}$ is a volume of the index $(\mu \bar{\sigma} \bar{\theta})$: $\bold{D}:=\int d\bar{\sigma} d^2\bar{\theta}  \hat{\bold{E}} \delta_{(\mu \bar{\sigma} \bar{\theta}) (\mu \bar{\sigma} \bar{\theta})}=d \int d\bar{\sigma} d^2\bar{\theta} \delta(\bar{\sigma}-\bar{\sigma}) \delta^2(\bar{\theta}-\bar{\theta})$. $N(\bold{X}_{\hat{D}_T}(\bar{\tau}))=\frac{1}{1+v(\bold{X}_{\hat{D}_T}(\bar{\tau}))}$, where $v(\bold{X}_{\hat{D}_T}(\bar{\tau}))= \frac{\alpha}{\sqrt{d-1}} \int d\bar{\sigma} d^2\bar{\theta}\hat{\bold{E}}  \epsilon_{\mu\nu}\bold{X}_{\hat{D}_T}^{\mu}(\bar{\tau}) \sqrt{\tilde{\bold{D}}_{\alpha}^2} \bold{X}_{\hat{D}_T}^{\nu}(\bar{\tau})$. 
$\tilde{\bold{D} }_{\alpha}$ is a $\bar{\tau}$ independent super derivative that satisfies 
\begin{eqnarray}
&&\int d\bar{\tau} d\bar{\sigma} d^2\bar{\theta} \bar{\bold{E}}  \frac{1}{2}(\tilde{\bold{D}}_{\alpha}\bold{X}_{\hat{D}_T \, \mu}(\bar{\tau}))^2 \nonumber \\
&=&
\int d\bar{\tau} d\bar{\sigma} \sqrt{\bar{h}} \frac{1}{2}\biggl(-(-\frac{\bar{n}^{\bar{\sigma}}}{\bar{n}} \partial_{\bar{\sigma}} X^{\mu}
+\frac{1}{2}\bar{n} \bar{\chi}_{m} \bar{E}^{0}_{r} \gamma^{r} \bar{E}^{m}_{q} \gamma^q \psi_{\mu})^2 \nonumber \\
&&
+\bar{h}^{11}\partial_{\bar{\sigma}} X^{\mu} \partial_{\bar{\sigma}} X_{\mu} -\bar{\psi}^{\mu} \bar{E}^{1}_q \gamma^{q} \partial_{\bar{\sigma}} \psi_{\mu}
-(F^{\mu})^2 \nonumber \\
&& + \bar{\chi}_{m} \bar{E}^{1}_{r} \gamma^{r} \bar{E}^{m}_{q} \gamma^q \psi_{\mu} \partial_{\bar{\sigma}} X^{\mu}
-\frac{1}{8}\bar{\psi}^{\mu}\psi_{\mu} \bar{\chi}_{m}\bar{E}^{n}_q\gamma^{q}\bar{E}^{m}_r \gamma^r \chi_{n}\biggr), 
\label{Oquad2}
\end{eqnarray}
where $\bar{E}^{m}_{\; q}$, $\bar{\chi}_m$, and $\gamma^q$ are a vierbein, a gravitino, and gamma matrices in the two dimensions, respectively.  
On the other hand, the ordinary super covariant derivative $\bar{\bold{D}}_{\alpha}$ satisfies \cite{Howe, SuperStringAction}
\begin{eqnarray}
&&\int d\bar{\tau} d\bar{\sigma} d^2\bar{\theta} \bar{\bold{E}}  \frac{1}{2}(\bar{\bold{D}}_{\alpha}\bold{X}_{\hat{D}_T \, \mu}(\bar{\tau}))^2 \nonumber \\
&=&
\int d\bar{\tau} d\bar{\sigma} \sqrt{\bar{h}} \frac{1}{2}(\bar{h}^{mn}\bar{\partial}_{m} X^{\mu} \bar{\partial}_{n} X_{\mu} -\bar{\psi}^{\mu} \bar{E}^{m}_q \gamma^{q} \bar{\partial}_{m} \psi_{\mu}
-(F^{\mu})^2 \nonumber \\
&& + \bar{\chi}_{m} \bar{E}^{n}_{r} \gamma^{r} \bar{E}^{m}_{q} \gamma^q \psi_{\mu} \bar{\partial}_{n} X^{\mu}
-\frac{1}{8}\bar{\psi}^{\mu}\psi_{\mu} \bar{\chi}_{m}\bar{E}^{n}_q\gamma^{q}\bar{E}^{m}_r \gamma^r \chi_{n}). \label{Oquad1}
\end{eqnarray}
One can show that the background (\ref{OSupersolution}) is a classical solution\footnote{This solution is a generalization of the Majumdar-Papapetrou solution \cite{Majumdar, Papapetrou} of the Einstein-Maxwell system.} to the equations of motion of (\ref{OSSuperAction}) as in section 4. The dependence of $\bar{\bold{E}}_{M}^{\quad A}$ on the background (\ref{OSupersolution}) is uniquely determined  by the consistency of the quantum theory of the fluctuations around the background. Actually, we will find that all the perturbative superstring amplitudes are derived as follows.

 Let us consider fluctuations around the background (\ref{OSupersolution}), $G_{\bold{I}\bold{J}}=\bar{G}_{\bold{I}\bold{J}}+\tilde{G}_{\bold{I}\bold{J}}$ and $A_\bold{I}=\bar{A}_\bold{I}+\tilde{A}_\bold{I}$.  Here we  fix the charts, where we choose $T$=IIA, IIB or I. The action (\ref{OSSuperAction}) up to the quadratic order is given by,\begin{eqnarray}
S&=&\frac{1}{G_N} \int \mathcal{D}\bold{E} \mathcal{D}\bold{X}_{\hat{D}_T}(\bar{\tau}) \mathcal{D}\bar{\tau} 
\sqrt{\bar{G}} 
\Bigl(-\bar{R}+\frac{1}{4}\bar{F}'_{\bold{I}\bold{J}}\bar{F}'^{\bold{I}\bold{J}} \nonumber \\
&&+\frac{1}{4}\bar{\nabla}_\bold{I} \tilde{G}_{\bold{J}\bold{K}} \bar{\nabla}^\bold{I} \tilde{G}^{\bold{J}\bold{K}}
-\frac{1}{4}\bar{\nabla}_\bold{I} \tilde{G} \bar{\nabla}^\bold{I} \tilde{G}
+\frac{1}{2}\bar{\nabla}^\bold{I} \tilde{G}_{\bold{I}\bold{J}} \bar{\nabla}^\bold{J} \tilde{G}
-\frac{1}{2}\bar{\nabla}^\bold{I} \tilde{G}_{\bold{I}\bold{J}} \bar{\nabla}_\bold{K} \tilde{G}^{\bold{J}\bold{K}}
\nonumber \\
&&-\frac{1}{4}(-\bar{R}+\frac{1}{4}\bar{F}'_{\bold{K}\bold{L}}\bar{F}'^{\bold{K}\bold{L}})
(\tilde{G}_{\bold{I}\bold{J}}\tilde{G}^{\bold{I}\bold{J}}-\frac{1}{2}\tilde{G}^2)
+(-\frac{1}{2}\bar{R}^{\bold{I}}_{\;\; \bold{J}}+\frac{1}{2}\bar{F}'^{\bold{I}\bold{K}}\bar{F}'_{\bold{J}\bold{K}})
\tilde{G}_{\bold{I}\bold{L}}\tilde{G}^{\bold{J}\bold{L}}
\nonumber \\
&&+(\frac{1}{2}\bar{R}^{\bold{I}\bold{J}}-\frac{1}{4}\bar{F}'^{\bold{I}\bold{K}}\bar{F}'^{\bold{J}}_{\;\;\;\; \bold{K}})
\tilde{G}_{\bold{I}\bold{J}}\tilde{G}
+(-\frac{1}{2}\bar{R}^{\bold{I}\bold{J}\bold{K}\bold{L}}+\frac{1}{4}\bar{F}'^{\bold{I}\bold{J}}\bar{F}'^{\bold{K}\bold{L}})
\tilde{G}_{\bold{I}\bold{K}}\tilde{G}_{\bold{J}\bold{L}}
\nonumber \\
&&+\frac{1}{4}G_N \tilde{F}_{\bold{I}\bold{J}} \tilde{F}^{\bold{I}\bold{J}} 
+\sqrt{G_N} 
(\frac{1}{4} \bar{F}^{'\bold{I}\bold{J}} \tilde{F}_{\bold{I}\bold{J}} \tilde{G} 
-\bar{F}^{'\bold{I}\bold{J}} \tilde{F}_{\bold{I}\bold{K}} \tilde{G}_\bold{J}^{\;\; \bold{K}} ) \Bigr). \label{Sopenfluctuation}
\end{eqnarray}
The Lagrangian is independent of Chan-Paton indices because the background (\ref{OSupersolution}) is independent of them. $\bar{F}'_{\bold{I}\bold{J}}:=\sqrt{G_N}\bar{F}_{\bold{I}\bold{J}}$ is independent of $G_N$. $\tilde{G}:=\bar{G}^{\bold{I}\bold{J}}\tilde{G}_{\bold{I}\bold{J}}$. There is no first order term because the background satisfies the equations of motion.

From these fluctuations,  we obtain the correlation function in the string manifold $\mathfrak{M}_{D_T}$ in exactly the same way as in section 4,
\begin{equation}
\Delta_F(\bold{X}_{\hat{D}_T f}; \bold{X}_{\hat{D}_T i}|\bold{E}_f, ; \bold{E}_i)
=
\bold{Z}
\int^{\bold{E}_f, \bold{X}_{\hat{D}_T f}}_{\bold{E}_i, \bold{X}_{\hat{D}_T i}}
\mathcal{D}\bold{E} \mathcal{D}\bold{X}_{\hat{D}_T} 
e^{-\lambda \chi} e^{-\int d^2\sigma d^2\theta \bold{E}  \frac{1}{2}(\bold{D}_{\alpha} \bold{X}_{\hat{D}_T \mu})^2}, \label{OSuperLast}\end{equation}
where $\chi$ is the Euler number of the reduced space. By inserting asymptotic states with Chan-Paton matrices to (\ref{OSuperLast}) and renormalizing the metric as in section 3, we obtain the all-order perturbative scattering amplitudes that possess the supermoduli in the type IIA, type IIB and SO(32) type I  superstring theory for $T=$ IIA, IIB and I, respectively in the presence of D-branes with arbitrary configuration\footnote{This includes the case there is no D-brane.}. The open superstrings possess Dirichlet and Neumann boundary conditions in the normal and tangential directions to the D-submanifolds, respectively \cite{textbook}. Therefore, a set of D-submanifolds represents a D-brane background where back reactions from the D-branes are ignored. Especially, in superstring geometry, the consistency of the perturbation theory around the background (\ref{OSupersolution}) determines $d=10$ (the critical dimension).

\section{Matrix models for superstring geometry}
\setcounter{equation}{0}

There are some types of supersymmetric matrix models such as, models that consist of super matrices \cite{Smolin, AzumaIsoKawaiOhwashi, OkudaTakayanagi, DijkgraafHeidenreichJeffersonVafa}, and dimensional reductions of supersymmetric Yang-Mills for examples, IKKT matrix model \cite{IKKT}, BFSS matrix model \cite{BFSS} and matrix string \cite{Motl, BS, DVV, DM}. In this section, we propose a new type of supersymmetric matrix models (\ref{matrixmodel1}) and (\ref{matrixmodel2}), namely, manifestly supersymmetric models that consist of matrices whose infinite dimensional indices include super coordinates. 

It is shown in \cite{HanadaKawaiKimura} that the equations of motions of the 10-dimensional gravity theory coupled with a $u(1)$ gauge field 
\begin{equation}
S_e= \frac{1}{G_N}\int d^{10} x \sqrt{g}(-R + \frac{1}{4}G_N F_{\mu \nu}F^{\mu \nu}),
\end{equation}
are equivalent to the equations of motions of a matrix model
\begin{equation}
S_m=\tr(-[A_{\mu}, A_{\nu}][A^{\mu}, A^{\nu}]),
\end{equation}
if the matrices are mapped to the covariant derivatives on the manifolds.

In the same way, one can show that the equations of motions of (\ref{OSSuperAction}) are equivalent to the equations of motions of a supersymmetric matrix model\footnote{$\bold{I}=(d,(\mu \bar{\sigma} \bar{\theta}))$, whereas $\tilde{\bold{I}}$ represent the Chan-Paton indices.} 
\begin{equation}
S_M=
\int \mathcal{D} \bold{E} \mathcal{D} \tilde{\bold{I}} \tr(-[A_{\hat{D}_T \; \bold{I}}(\bar{\bold{E}}, \tilde{\bold{I}}), A_{\hat{D}_T \; \bold{J}}(\bar{\bold{E}}, \tilde{\bold{I}})][A_{\hat{D}_T}^{\bold{I}}(\bar{\bold{E}}, \tilde{\bold{I}}), A_{\hat{D}_T}^{\bold{J}}(\bar{\bold{E}}, \tilde{\bold{I}})]), \label{matrixmodel1}
\end{equation}
which is decomposed as
\begin{eqnarray}
S_M&=&
\int \mathcal{D} \bold{E}  \mathcal{D} \tilde{\bold{I}}
\tr\biggl(- 2\int d\bar{\sigma} d^2\bar{\theta} \hat{\bold{E}}
[A_{\hat{D}_T \; d}(\bar{\bold{E}}, \tilde{\bold{I}}), A_{\hat{D}_T \; (\mu \bar{\sigma} \bar{\theta})}(\bar{\bold{E}}, \tilde{\bold{I}})][A_{\hat{D}_T \; d}(\bar{\bold{E}}, \tilde{\bold{I}}), A_{\hat{D}_T \; (\mu \bar{\sigma} \bar{\theta})}(\bar{\bold{E}}, \tilde{\bold{I}})] \nonumber \\
&&\qquad \qquad- \int d\bar{\sigma} d^2\bar{\theta} \hat{\bold{E}} 
\int d\bar{\sigma}' d^2\bar{\theta}' \hat{\bold{E}}'
[A_{\hat{D}_T \; (\mu \bar{\sigma} \bar{\theta})}(\bar{\bold{E}}, \tilde{\bold{I}}), A_{\hat{D}_T \; (\mu' \bar{\sigma}' \bar{\theta}')}(\bar{\bold{E}}, \tilde{\bold{I}})]
\nonumber \\
&&\qquad \qquad[A_{\hat{D}_T \; (\mu \bar{\sigma} \bar{\theta})}(\bar{\bold{E}}, \tilde{\bold{I}}), A_{\hat{D}_T \; (\mu' \bar{\sigma}' \bar{\theta}')}(\bar{\bold{E}}, \tilde{\bold{I}})]\biggr), \nonumber \\
&&
\end{eqnarray}
if the matrices are mapped to the covariant derivatives on the superstring manifolds.

Moreover, it is interesting to study relations between the superstring geometry and a more simple supersymmetric matrix model
\begin{equation}
S_{M_0}=
\tr(-[A_{\bold{I}}, A_{\bold{J}}][A^{\bold{I}}, A^{\bold{J}}]), \label{matrixmodel2}
\end{equation}
decomposed as
\begin{eqnarray}
S_{M_0}&=&
\tr\biggl(- 2\int d\bar{\sigma} d^2\bar{\theta} 
[A_{d}, A_{(\mu \bar{\sigma} \bar{\theta})}][A_{d}, A_{(\mu \bar{\sigma} \bar{\theta})}] \nonumber \\
&&\qquad \qquad- \int d\bar{\sigma} d^2\bar{\theta}
\int d\bar{\sigma}' d^2\bar{\theta}' 
[A_{(\mu \bar{\sigma} \bar{\theta})}, A_{(\mu' \bar{\sigma}' \bar{\theta}')}][A_{(\mu \bar{\sigma} \bar{\theta})}, A_{(\mu' \bar{\sigma}' \bar{\theta}')}]\biggr), \nonumber \\
&&
\end{eqnarray}
because topological expansions of worldsheets can be derived in general by perturbations of matrix models \cite{Hooft, BrezinItzyksonParisiZuber, KazakovMigdal, DistlerKawai, DasJevicki, GrossMiljkovic, GinspargZinn-Justin, Itoyama, GinspargMoore}. (\ref{matrixmodel2}) may correspond to (\ref{matrixmodel1}) by an extension of the large N reduction \cite{RM}.

\section{Heterotic construction}
\setcounter{equation}{0}

In this section, based on superstring geometry, we formulate and study a theory that manifestly possesses the $SO(32)$ and $E_8 \times E_8$ heterotic perturbative vacua. We expect that this theory is equivalent to the theory in section 6, which manifestly possesses the type IIA, type IIB and $SO(32)$ type I perturbative vacua, because of the S-duality.

 First, let us prepare a moduli space\footnote{Strictly speaking, this should be called a parameter space of integration cycles \cite{NotesOnSupermanifolds, WittenSupermoduli} because superstring worldsheets are defined up to homology.} of heterotic superstring worldsheets $\bar{\bold{\Sigma}}$ \cite{NotesOnSupermanifolds, WittenSupermoduli, SuperPeriod} with punctures $P^i$ ($i=1, \cdots, N$)\footnote{$P^i$ not necessarily represents a point, whereas the corresponding $P^i_{red}$ on a reduced space represents a point. A Ramond puncture is located over a R divisor.}.  We consider a super Riemann surface $\bar{\bold{\Sigma}}_{R}$ with Neveu-Schwarz (NS) and Ramond (R) punctures whose reduced space $\bar{\Sigma}_{R, red}$ is complex conjugate to a Riemann surface $\bar{\Sigma}_L$. A reduced space is defined by setting odd variables to zero in a super Riemann surface. The complex conjugates means that  they are complex conjugate spaces with punctures at the same points. A heterotic superstring worldsheet $\bar{\bold{\Sigma}}$ is defined by the subspace of $\bar{\Sigma}_{L} \times \bar{\bold{\Sigma}}_{R}$ whose reduced space $\bar{\Sigma}_L \times \bar{\Sigma}_{R, red}$ is restricted to its diagonal $\bar{\Sigma}_{red}$.

Next, we define global times uniquely on $\bar{\bold{\Sigma}}_{R}$ in exactly the same way as in section 4. If we give residues $-f^i$ and the same normalization on $\bar{\Sigma}_L$ as on $\bar{\Sigma}_{R, red}$, we can set the coordinates on $\bar{\Sigma}_{L}$ to the complex conjugate $\bar{\bar{w}}=\bar{\tau}-i\bar{\sigma}:=\int^{P} d\bar{p}$ by a conformal transformation, because the Abelian differential is uniquely determined on $\bar{\Sigma}_{L}$ and $\bar{\Sigma}_L$ is complex conjugate to $\bar{\Sigma}_{R, red}$. Therefore, we can define the global time $\bar{\tau}$ uniquely and reduced space canonically on $\bar{\bold{\Sigma}}$.

Thus, under a superconformal transformation, one obtains a heterotic worldsheet $\bar{\bold{\Sigma}}$ that has even coordinates composed of the global time $\bar{\tau}$ and the position $\bar{\sigma}$ and $\bar{\Sigma}_{red}$ is canonically defined. Because $\bar{\bold{\Sigma}}$ can be a moduli of heterotic worldsheets with punctures, any two-dimensional heterotic super Riemannian manifold with punctures $\bold{\Sigma}$ can be obtained by $\bold{\Sigma}={\boldsymbol \psi}(\bar{\bold{\Sigma}})$ where ${\boldsymbol \psi}$ is a superdiffeomorphism times super Weyl transformation.

Next, we will define the heterotic model space $\bold{E}$.  We consider a state $(\bar{{\boldsymbol \Sigma}}, \bold{X}_{\hat{D}_T}(\bar{\tau}_s), \lambda_{\hat{D}_T}(\bar{\tau}_s), \bar{\tau}_s)$ determined by a $\bar{\tau}=\bar{\tau}_s$ constant hypersurface, an arbitrary map $\bold{X}_{\hat{D}_T}(\bar{\tau}_s)$ from $\bold{\Sigma}|_{\bar{\tau}_s}$ to the d-dimensional Euclidean space $\bold{R}^d$, and an arbitrary left-handed fermionic map $\lambda_{\hat{D}_T}(\bar{\tau}_s)$ from $\bar{\Sigma}_{red}|_{\bar{\tau}_s}$ to a 32-dimensional internal vector bundle $V$ on $\bold{R}^d$. In order to define $V$ globally, we need to patch local vector spaces. The transition functions among them require gauge connections.  $\hat{D}_T$ represents all the backgrounds except for a target metric $G_{\mu\nu}$, that consist of a NS-NS B-field, a dilaton, and a gauge field, where the $SO(32)$ and $E_8 \times E_8$ heterotic GSO projections are attached for $T=$ $SO(32)$ and $E_8 \times E_8$, respectively.  We can define the worldsheet fermion numbers of states in a Hilbert space because the states consist of the fields over the local coordinates ${\boldsymbol X}_{\hat{D}_T}^{\mu}(\bar{\tau}_s)=X^{\mu}+ \bar{\theta} \psi^{\mu}$ and $\lambda^A_{\hat{D}_T}(\bar{\tau}_s)$, where $\mu=0, 1, \cdots d-1$ and $\psi^{\mu}$ is a Majorana fermion.  We abbreviate $\hat{D}_T$ of $X^{\mu}$ and $\psi^{\mu}$.
For $T=$ $SO(32)$, we take periodicities 
\begin{equation}
\lambda_{\hat{D}_{SO(32)}}^A(\bar{\tau}, \bar{\sigma}+2\pi)=\pm \lambda_{\hat{D}_{SO(32)}}^A(\bar{\tau}, \bar{\sigma}) \quad (A=1, \cdots 32)
\end{equation}
with the same sign on all 32 components. 
We define the Hilbert space in these coordinates by the states only with $e^{\pi i F}=1$ and $e^{\pi i \tilde{F}}=1$, where $F$ and $\tilde{F}$ are the numbers of left- and right- handed fermions $\lambda^A_{\hat{D}_{SO(32)}}$ and $\psi^{\mu}$, respectively. 
For $T=$ $E_8 \times E_8$, the periodicity is given by 
\begin{eqnarray}
\lambda_{\hat{D}_{E_8 \times E_8}}^A(\bar{\tau}, \bar{\sigma}+2\pi)= 
\left\{
\begin{array}{c}
\eta \lambda_{\hat{D}_{E_8 \times E_8}}^A(\bar{\tau}, \bar{\sigma}) \quad (1 \leqq A \leqq 16) \\
\eta' \lambda_{\hat{D}_{E_8 \times E_8}}^A(\bar{\tau}, \bar{\sigma}) \quad (17 \leqq A \leqq 32), 
\end{array}
\right.
\end{eqnarray}
with the same sign $\eta(= \pm 1)$ and $\eta'(= \pm 1)$ on each 16 components.  
The GSO projection is given by $e^{\pi i F_1}=1$, $e^{\pi i F_2}=1$ and $e^{\pi i \tilde{F}}=1$, where $F_1$, $F_2$ and $\tilde{F}$ are the numbers of $\lambda_{\hat{D}_{E_8 \times E_8}}^{A_1}$ ($A_1=1, \cdots, 16$), $\lambda_{\hat{D}_{E_8 \times E_8}}^{A_2}$ ($A_2=17, \cdots, 32$) and $\psi^{\mu}$, respectively.

$\bar{\bold{\Sigma}}$ is a union of $N_{\pm}$ supercylinders with radii $f_i$ at $\bar{\tau}\simeq \pm \infty$. Thus, we define a superstring state as an equivalence class $[\bar{\bold{\Sigma}},  \bold{X}_{\hat{D}_T}(\bar{\tau}_s\simeq \pm \infty), \lambda_{\hat{D}_T}(\bar{\tau}_s\simeq \pm \infty), \bar{\tau}_s\simeq \pm \infty]$ by a relation $(\bar{\bold{\Sigma}},  \bold{X}_{\hat{D}_T}(\bar{\tau}_s\simeq \pm \infty), \lambda_{\hat{D}_T}(\bar{\tau}_s\simeq \pm \infty), \bar{\tau}_s\simeq \pm \infty) \sim (\bar{\bold{\Sigma}}',  \bold{X}'_{\hat{D}_T}(\bar{\tau}_s\simeq \pm \infty),  \lambda'_{\hat{D}_T}(\bar{\tau}_s\simeq \pm \infty), \bar{\tau}_s\simeq \pm \infty)$ if $N_{\pm}=N'_{\pm}$, $f_i=f'_i$, $\bold{X}_{\hat{D}_T}(\bar{\tau}_s\simeq \pm \infty)=\bold{X}'_{\hat{D}_T}(\bar{\tau}_s\simeq \pm \infty)$, 
$\lambda_{\hat{D}_T}(\bar{\tau}_s \simeq \pm \infty)
=
\lambda'_{\hat{D}_T}(\bar{\tau}_s \simeq \pm \infty)$  and the corresponding supercylinders are the same type (NS or R). Because the reduced space of $\bold{\Sigma}|_{\bar{\tau}_s}$ is $S^1 \times S^1 \times \cdots \times S^1$ and $\bold{X}_{\hat{D}_T}(\bar{\tau}_s): \bold{\Sigma}|_{\bar{\tau}_s} \to \bold{R}^d$, $[\bar{\bold{\Sigma}},  \bold{X}_{\hat{D}_T}(\bar{\tau}_s), \lambda_{\hat{D}_T}(\bar{\tau}_s), \bar{\tau}_s]$ represent many-body states of superstrings in $\bold{R}^d$. A heterotic model space $\bold{E}$ is defined by $\bigcup_{\hat{D}_T}\{[\bar{\bold{\Sigma}},  \bold{X}_{\hat{D}_T}(\bar{\tau}_s), \lambda_{\hat{D}_T}(\bar{\tau}_s), \bar{\tau}_s]\}$, where  disjoint unions are taken over all the backgrounds $\hat{D}_T$ except for the metric.

Here, we will define topologies of $\bold{E}$. We define an $\epsilon$-open neighborhood of 
\linebreak
$[\bar{\bold{\Sigma}},  \bold{X}_{s \hat{D}_T}(\bar{\tau}_s), \lambda_{s \hat{D}_T}(\bar{\tau}_s), \bar{\tau}_s]$ by
\begin{eqnarray}
&&U([\bar{\bold{\Sigma}},  \bold{X}_{s \hat{D}_T}(\bar{\tau}_s), \lambda_{s \hat{D}_T}(\bar{\tau}_s), \bar{\tau}_s], \epsilon) \nonumber \\
&:=&
\biggl\{[\bar{\bold{\Sigma}},  \bold{X}_{\hat{D}_T}(\bar{\tau}), \lambda_{\hat{D}_T}(\bar{\tau}), \bar{\tau}]
\bigm| \nonumber \\
&&\sqrt{|\bar{\tau}-\bar{\tau}_s|^2
+\| \bold{X}_{\hat{D}_T}(\bar{\tau}) -\bold{X}_{s \hat{D}_T}(\bar{\tau}_s) \|^2
+\| \lambda_{\hat{D}_T}(\bar{\tau})
- \lambda_{s \hat{D}_T}(\bar{\tau}_s)\|^2}
<\epsilon  \biggr\}, \label{HeteroNeighbour}
\end{eqnarray}
where 
\begin{eqnarray}
&&
\| \bold{X}_{\hat{D}_T}(\bar{\tau}) -\bold{X}_{s \hat{D}_T}(\bar{\tau}_s) \|^2 \nonumber \\
&:=&\int_0^{2\pi}  d\bar{\sigma} 
\Bigl(|x(\bar{\tau}, \bar{\sigma})-x_s(\bar{\tau}_s, \bar{\sigma})|^2 
+(\bar{\psi}(\bar{\tau}, \bar{\sigma})-\bar{\psi}_s(\bar{\tau}_s, \bar{\sigma}))
(\psi(\bar{\tau}, \bar{\sigma})-\psi_s(\bar{\tau}_s, \bar{\sigma})) \nonumber \\
&+&|f(\bar{\tau}, \bar{\sigma})-f_s(\bar{\tau}_s, \bar{\sigma})|^2 \Bigr) \nonumber \\&&  \nonumber \\
&&\| \lambda_{\hat{D}_T}(\bar{\tau})
- \lambda_{s \hat{D}_T}(\bar{\tau}_s)\|^2 
\nonumber \\
&:=&\int_0^{2\pi}  d\bar{\sigma} 
( \bar{\lambda}_{\hat{D}_T}(\bar{\tau}, \bar{\sigma})-\bar{\lambda}_{s \hat{D}_T}(\bar{\tau}_s, \bar{\sigma}))
(\lambda_{\hat{D}_T}(\bar{\tau}, \bar{\sigma})-\lambda_{s \hat{D}_T}(\bar{\tau}_s, \bar{\sigma})) .
\end{eqnarray}
$U([\bar{\bold{\Sigma}},  \bold{X}_{\hat{D}_T}(\bar{\tau}_s\simeq \pm \infty), \lambda_{\hat{D}_T}(\bar{\tau}_s\simeq \pm \infty), \bar{\tau}_s\simeq \pm \infty], \epsilon)
=
U([\bar{\bold{\Sigma}}',  \bold{X}'_{\hat{D}_T}(\bar{\tau}_s\simeq \pm \infty), \lambda'_{\hat{D}_T}(\bar{\tau}_s\simeq \pm \infty), \bar{\tau}_s\simeq \pm \infty], \epsilon)$ consistently if $N_{\pm}=N'_{\pm}$, $f_i=f'_i$, $\bold{X}_{\hat{D}_T}(\bar{\tau}_s\simeq \pm \infty)=\bold{X}'(\bar{\tau}_s\simeq \pm \infty)$, 
$\lambda_{\hat{D}_T}(\bar{\tau}_s\simeq \pm \infty)
=
\lambda'_{\hat{D}_T}(\bar{\tau}_s\simeq \pm \infty)$,
the corresponding supercylinders are the same type (NS or R), and $\epsilon$ is small enough, because the $\bar{\tau}_s\simeq \pm \infty$ constant hypersurfaces traverses only supercylinders  overlapped by $\bar{{\boldsymbol \Sigma}}$ and $\bar{{\boldsymbol \Sigma}}'$. 
$U$ is defined to be an open set of $\mathfrak{M}_{D_T}$ if there exists $\epsilon$ such that $U([\bar{\bold{\Sigma}},  \bold{X}_{\hat{D}_T}(\bar{\tau}_s), \lambda_{\hat{D}_T}(\bar{\tau}_s), \bar{\tau}_s], \epsilon) \subset U$ for an arbitrary point $[\bar{\bold{\Sigma}},  \bold{X}_{\hat{D}_T}(\bar{\tau}_s), \lambda_{\hat{D}_T}(\bar{\tau}_s), \bar{\tau}_s] \in U$. In exactly the same way as in section 2, one can show that the topology of $\mathfrak{M}_{D_T}$ satisfies the axiom of topology.
Although the model space is defined by using the coordinates $[\bar{\bold{\Sigma}},  \bold{X}_{\hat{D}_T}(\bar{\tau}_s), \lambda_{\hat{D}_T}(\bar{\tau}_s), \bar{\tau}_s]$, the model space does not depend on the coordinates, because the model space is a topological space.

In the following,  instead of $[\bar{{\boldsymbol \Sigma}}, \bold{X}_{\hat{D}_T}(\bar{\tau}), \lambda_{\hat{D}_T}(\bar{\tau}), \bar{\tau}]$,  we denote $[\bar{\bold{E}}_{M}^{\quad A}(\bar{\sigma}, \bar{\tau}, \bar{\theta}), \bold{X}_{\hat{D}_T}(\bar{\tau}), \lambda_{\hat{D}_T}(\bar{\tau}), \bar{\tau}]$, 
where $\bar{\bold{E}}_{M}^{\quad A}(\bar{\sigma}, \bar{\tau}, \bar{\theta}^+)$ ($M=(m, +)$, $A=(q, +)$, $m, q=0,1$, $\theta^+ :=\theta$) is the worldsheet super vierbein on $\bar{{\boldsymbol \Sigma}}$ \cite{BrooksMuhammadGates}, 
because giving a super Riemann surface is equivalent to giving a super vierbein up to super diffeomorphism and super Weyl transformations.
$\bar{h}_{ mn} (\bar{\sigma}, \bar{\tau})$ ($m, n =0,1$) and $\bar{E}^{\quad z}_{m}$ ($\bar{E}^{\quad \bar{z}}_{m}$) are the worldsheet metric and its vierbein of $\bar{\Sigma}_{red}$, respectively,

In order to define structures of manifold, let us consider how generally we can define general coordinate transformations between 
$[\bar{\bold{E}}_{M}^{\quad A}, \bold{X}_{\hat{D}_T}(\bar{\tau}), \lambda_{\hat{D}_T}(\bar{\tau}), \bar{\tau}]$ 
and
$[\bold{E}_{M}^{'\quad A}, \bold{X}'_{\hat{D}_T}(\bar{\tau}'), \lambda'_{\hat{D}_T}(\bar{\tau}'), \bar{\tau}']$. where $[\bar{\bold{E}}_{M}^{\quad A}, \bold{X}_{\hat{D}_T}(\bar{\tau}), \lambda_{\hat{D}_T}(\bar{\tau}), \bar{\tau}] \in U \subset \bold{E}$ and $[\bold{E}_{M}^{'\quad A}, \bold{X}'_{\hat{D}_T}(\bar{\tau}'), \lambda'_{\hat{D}_T}(\bar{\tau}'), \bar{\tau}']\in U' \subset \bold{E}$. $\bar{\bold{E}}_{M}^{\quad A}$ does not transform to $\bar{\tau}$, $\bold{X}_{\hat{D}_T}(\bar{\tau})$ and $\lambda_{\hat{D}_T}(\bar{\tau})$, and vice versa, because $\bar{\tau}$, $\bold{X}_{\hat{D}_T}(\bar{\tau})$ and $\lambda_{\hat{D}_T}(\bar{\tau})$ are continuous variables, whereas $\bar{\bold{E}}_{M}^{\quad A}$ is a discrete variable: $\bar{\tau}$, $\bold{X}_{\hat{D}_T}(\bar{\tau})$ and $\lambda_{\hat{D}_T}(\bar{\tau})$ vary continuously, whereas $\bar{\bold{E}}_{M}^{\quad A}$ varies discretely in a trajectory on $\bold{E}$ by definition of the neighborhoods. $\bar{\tau}$ does not transform to $\bar{\sigma}$ and $\bar{\theta}$ and vice versa, because the superstring states are defined by $\bar{\tau}$ constant surfaces. Under these restrictions, the most general coordinate transformation is given by 
\begin{eqnarray}
&&[\bar{\bold{E}}_{M}^{\quad A}(\bar{\sigma}, \bar{\tau}, \bar{\theta}), \bold{X}_{\hat{D}_T}^{\mu}(\bar{\tau}), 
\lambda_{\hat{D}_T}^A(\bar{\tau}), \bar{\tau}] \nonumber \\
&&\mapsto [\bold{E}_{M}^{'\quad A}(\bar{\sigma}'(\bar{\sigma}, \bar{\theta}), \bar{\tau}'(\bar{\tau}, \bold{X}_{\hat{D}_T}(\bar{\tau}), \lambda_{\hat{D}_T}(\bar{\tau})), \bar{\theta}'(\bar{\sigma}, \bar{\theta})), \bold{X}_{\hat{D}_T}^{'\mu}(\bar{\sigma}', \bar{\tau}', \bar{\theta}')(\bar{\tau}, \bold{X}_{\hat{D}_T}(\bar{\tau}), \lambda_{\hat{D}_T}(\bar{\tau})), 
\nonumber \\
&&
\qquad 
\lambda_{\hat{D}_T}^{'A}(\bar{\sigma}', \bar{\tau}')(\bar{\tau}, \bold{X}_{\hat{D}_T}(\bar{\tau}), \lambda_{\hat{D}_T}(\bar{\tau})), 
\bar{\tau}'(\bar{\tau}, \bold{X}_{\hat{D}_T}(\bar{\tau}), \lambda_{\hat{D}_T}(\bar{\tau}))], 
\label{HGeneralCoordTrans2}
\end{eqnarray}
where $\bar{\bold{E}}_{M}^{\quad A} \mapsto \bar{\bold{E}}_{M}^{'\quad A}$ represents a world-sheet superdiffeomorphism transformation\footnote{
We extend the model space from $\bold{E}=\{[\bar{\bold{E}}_{M}^{\quad A}(\bar{\sigma}, \bar{\tau}, \bar{\theta}), \bold{X}_{\hat{D}_T}^{\mu}(\bar{\sigma}, \bar{\tau}, \bar{\theta}), 
\lambda_{\hat{D}_T}^A(\bar{\sigma}, \bar{\tau}), \bar{\tau}] \}$ to $\bold{E}=\{[\bold{E}_{M}^{'\quad A}(\bar{\sigma}', \bar{\tau}', \bar{\theta}'), \bold{X}_{\hat{D}_T}^{'\mu}(\bar{\sigma}', \bar{\tau}', \bar{\theta}'), 
\lambda_{\hat{D}_T}^{'A}(\bar{\sigma}', \bar{\tau}'), 
\bar{\tau}'] \}$ by including the points generated by the superdiffeomorphisms $\bar{\sigma} \mapsto \bar{\sigma}'(\bar{\sigma}, \bar{\theta})$,  $\bar{\theta} \mapsto \bar{\theta}'(\bar{\sigma}, \bar{\theta})$, and $\bar{\tau} \mapsto \bar{\tau}'(\bar{\tau})$.}.  \linebreak
$\bold{X}_{\hat{D}_T}^{'\mu}(\bar{\tau}, \bold{X}_{\hat{D}_T}(\bar{\tau}), \lambda_{\hat{D}_T}(\bar{\tau}))$, 
$\lambda_{\hat{D}_T}^{'A}(\bar{\tau}, \bold{X}_{\hat{D}_T}(\bar{\tau}), \lambda_{\hat{D}_T}(\bar{\tau}))$ and 
$\bar{\tau}'(\bar{\tau}, \bold{X}_{\hat{D}_T}(\bar{\tau}), \lambda_{\hat{D}_T}(\bar{\tau}))$
are functionals of 
$\bar{\tau}$, $\bold{X}_{\hat{D}_T}(\bar{\tau})$ and  $\lambda_{\hat{D}_T}(\bar{\tau})$.
Here, we consider all the manifolds which are constructed by patching open sets of the model space $\bold{E}$ by general coordinate transformations (\ref{HGeneralCoordTrans2}) and call them heterotic superstring manifolds $\mathfrak{M}$.

Here, we give an example of heterotic superstring manifolds: $\mathfrak{M}_{D_T}:= \{[\bar{\bold{\Sigma}},  \bold{x}_{D_T}(\bar{\tau}), \lambda_{D_T}(\bar{\tau}), \bar{\tau}]\}$, where $D_T$ represents a type of the GSO projection and a target metric $G_{\mu \nu}$ where all the other backgrounds are turned off.  
$\bold{x}_{D_T}(\bar{\tau}): \bar{\bold{\Sigma}}|_{\bar{\tau}} \to M$,  where $\bold{x}_{D_T}^{\mu}(\bar{\tau})=x^{\mu}+ \bar{\theta}^{\alpha} \psi_{\alpha}^{\mu}+\frac{1}{2} \bar{\theta}^2 f^{\mu}$. The image of the bosonic part of the embedding function, $x(\bar{\tau})$ has a metric: 
$ds^2= dx^{\mu}(\bar{\tau}, \bar{\sigma})
dx^{\nu}(\bar{\tau}, \bar{\sigma})
G_{\mu \nu}(x(\bar{\tau}, \bar{\sigma}))$. 

We will show that $\mathfrak{M}_{D_T}$has a structure of manifold, that is there exists a homeomorphism between the sufficiently small neighborhood around an arbitrary point $[\bar{\bold{\Sigma}},  \bold{x}_{s D_T}(\bar{\tau}_s), \lambda_{s D_T}(\bar{\tau}_s), \bar{\tau}_s] \in \mathfrak{M}_{D_T}$ and an open set of $E$. There exists a general coordinate transformation $X^{\mu}(x)$ that satisfies 
$ds^2= dx^{\mu}
dx^{\nu}
G_{\mu \nu}(x)
=
dX^{\mu}
dX^{\nu}
\eta_{\mu \nu}$
on an arbitrary point $x$ in the $\epsilon_{\bar{\sigma}}$ open neighborhood around $x_s(\bar{\tau}_s, \bar{\sigma}) \in M$, if $\epsilon_{\bar{\sigma}}$ is sufficiently small. An arbitrary point $[\bar{\bold{\Sigma}},  \bold{x}_{D_T}(\bar{\tau}), \lambda_{D_T}(\bar{\tau}), \bar{\tau}]$ in the $\epsilon$ open neighborhood around  $[\bar{\bold{\Sigma}},  \bold{x}_{s D_T}(\bar{\tau}_s), \lambda_{s D_T}(\bar{\tau}_s), \bar{\tau}_s] $ satisfies 
\begin{eqnarray}
&&\int_0^{2\pi}  d\bar{\sigma} 
|x(\bar{\tau}, \bar{\sigma})-x_s(\bar{\tau}_s, \bar{\sigma})|^2 \nonumber \\
&&< \epsilon^2
-\int_0^{2\pi}  d\bar{\sigma}(\bar{\psi}(\bar{\tau}, \bar{\sigma})-\bar{\psi}_s(\bar{\tau}_s, \bar{\sigma}))
(\psi(\bar{\tau}, \bar{\sigma})-\psi_s(\bar{\tau}_s, \bar{\sigma}))
-\int_0^{2\pi}  d\bar{\sigma}|f(\bar{\tau}, \bar{\sigma})-f_s(\bar{\tau}_s, \bar{\sigma})|^2 \nonumber \\
&&
-\int_0^{2\pi}  d\bar{\sigma} 
( \bar{\lambda}_{\hat{D}_T}(\bar{\tau}, \bar{\sigma})-\bar{\lambda}_{s \hat{D}_T}(\bar{\tau}_s, \bar{\sigma}))
(\lambda_{\hat{D}_T}(\bar{\tau}, \bar{\sigma})-\lambda_{s \hat{D}_T}(\bar{\tau}_s, \bar{\sigma})) 
- |\bar{\tau}-\bar{\tau}_s|^2
\nonumber \\
&& \leqq \epsilon^{2} \label{heteroneigh}
\end{eqnarray}
and thus 
\begin{equation}
|x(\bar{\tau}, \bar{\sigma})-x_s(\bar{\tau}_s, \bar{\sigma})|
< \epsilon_{\bar{\sigma}}'
\end{equation}
on an arbitrary $\bar{\sigma}$.
$\epsilon_{\bar{\sigma}}'< \epsilon_{\bar{\sigma}}$ is satisfied on an arbitrary $\bar{\sigma}$ if $\epsilon$ is taken to be sufficiently small. 
Then, there exists a transformation  $X^{\mu}(\bar{\tau}, \bar{\sigma}):=X^{\mu}(x(\bar{\tau}, \bar{\sigma}))$, which satisfies
\begin{equation}
ds^2= dx^{\mu}(\bar{\tau}, \bar{\sigma})
dx^{\nu}(\bar{\tau}, \bar{\sigma})
G_{\mu \nu}(x(\bar{\tau}, \bar{\sigma}))
=
dX^{\mu}(\bar{\tau}, \bar{\sigma})
dX^{\nu}(\bar{\tau}, \bar{\sigma})
\eta_{\mu \nu}. \label{HeteroSuperLocalLorents}
\end{equation}
Because the tangent vectors $\bold{X}_{\hat{D}_T}(\bar{\tau}, \bar{\sigma}, \bar{\theta})$ and $\lambda_{\hat{D}_T}(\bar{\tau}, \bar{\sigma})$ exist for each $\bold{x}_{D_T}(\bar{\tau}, \bar{\sigma}, \bar{\theta})$ and $\lambda_{D_T}(\bar{\tau}, \bar{\sigma})$, there exist vector bundles  $\bold{X}_{\hat{D}_T}(\bar{\tau})$ and $\lambda_{\hat{D}_T}(\bar{\tau})$ for $0 \leqq \bar{\sigma} < 2\pi$ and $\bar{\theta}$. $\bold{x}_{D_T}(\bar{\tau})$ and $\bold{X}_{\hat{D}_T}(\bar{\tau})$ satisfy (\ref{HeteroSuperLocalLorents}) on each $\bar{\sigma}$, that is $\bold{X}_{\hat{D}_T}(\bar{\tau}): \bar{\bold{\Sigma}}|_{\bar{\tau}} \to \bold{R}^d$. Therefore, there exists a homeomorphism between the sufficiently small neighborhood around an arbitrary point $[\bar{\bold{\Sigma}}, \bold{x}_{s D_T}(\bar{\tau}_s), \lambda_{s D_T}(\bar{\tau}_s), \bar{\tau}_s] \in \mathfrak{M}_{D_T}$ and an open set of $\bold{E}$: $[\bar{\bold{\Sigma}},  \bold{x}_{D_T}(\bar{\tau}), \lambda_{D_T}(\bar{\tau}), \bar{\tau}] \mapsto [\bar{\bold{\Sigma}}, \bold{X}_{\hat{D}_T}(\bar{\tau}), \lambda_{\hat{D}_T}(\bar{\tau}), \bar{\tau}]$.
Actually, a map from the image $\bold{x}_{D_T}(\bar{\tau}, \bar{\sigma}, \bar{\theta})$ and $\lambda_{D_T}(\bar{\tau}, \bar{\sigma}, \bar{\theta})$ to the images $\bold{X}_{\hat{D}_T}(\bar{\tau}, \bar{\sigma}, \bar{\theta})$ and $\lambda_{\hat{D}_T}(\bar{\tau}, \bar{\sigma}, \bar{\theta})$ are explicitly given by exponential maps,
\begin{eqnarray}
\bold{x}_{D_T}(\bar{\tau}, \bar{\sigma}, \bar{\theta})
&=&
\exp_{\bold{x}_{sD_T}(\bar{\tau}, \bar{\sigma}, \bar{\theta})}
\bold{X}_{\hat{D}_T}(\bar{\tau}, \bar{\sigma}, \bar{\theta})
\simeq
\bold{x}_{sD_T}(\bar{\tau}, \bar{\sigma}, \bar{\theta})
+\bold{X}_{\hat{D}_T}(\bar{\tau}, \bar{\sigma}, \bar{\theta})\nonumber\\
\lambda_{D_T}(\bar{\tau}, \bar{\sigma}, \bar{\theta})
&=&
\exp_{\lambda_{sD_T}(\bar{\tau}, \bar{\sigma}, \bar{\theta})}
\lambda_{\hat{D}_T}(\bar{\tau}, \bar{\sigma}, \bar{\theta})
\simeq
\lambda_{sD_T}(\bar{\tau}, \bar{\sigma}, \bar{\theta})
+\lambda_{\hat{D}_T}(\bar{\tau}, \bar{\sigma}, \bar{\theta}).
\end{eqnarray}
If we substitute these to an $\epsilon$ open neighborhood around an arbitrary point $[\bar{\bold{\Sigma}}, \bold{x}_{s D_T}(\bar{\tau}_s), \lambda_{s D_T}(\bar{\tau}_s), \bar{\tau}_s] \in \mathfrak{M}_{D_T}$ (\ref{heteroneigh}),
we obtain an $\epsilon$ open neighborhood around $[\bar{\bold{\Sigma}}, 0, 0, \bar{\tau}_s] \in \bold{E}$,
\begin{equation}
\| \bold{X}_{\hat{D}_T}(\bar{\tau}) \|^2
+\| \lambda_{\hat{D}_T}(\bar{\tau})
)\|^2 +  |\bar{\tau}-\bar{\tau}_s|^2 < \epsilon^{2}.
\end{equation}

By definition of the $\epsilon$-open neighborhood, arbitrary two superstring states on a connected heterotic super Riemann surface are connected continuously. Thus, there is an one-to-one correspondence between a heterotic super Riemann surface with punctures in $M$ and a curve parametrized by $\bar{\tau}$ from $\bar{\tau}=-\infty$ to $\bar{\tau}=\infty$ on $\mathfrak{M}_{D_T}$. That is, curves that represent asymptotic processes on $\mathfrak{M}_{D_T}$ reproduce the right moduli space of the heterotic super Riemann surfaces in the target manifold.

By a general curve parametrized by $t$ on $\mathfrak{M}_{D_T}$, superstring states on different heterotic super Riemann surfaces that have even different genera, can be connected continuously, for example see Fig. \ref{6Connected}, whereas different super Riemann surfaces that have different genera cannot be connected continuously in the moduli space of the heterotic super Riemann surfaces in the target space.

 In the following, instead of the fermionic coordinate $\lambda_{\hat{D}_T}^A(\bar{\sigma}, \bar{\tau})$, we use a bosonic coordinate $\bold{X}_{L\hat{D}_T}^A(\bar{\sigma}, \bar{\tau}, \bar{\theta}^-):= \bar{\theta}^- \lambda_{\hat{D}_T}^A(\bar{\sigma}, \bar{\tau})$ where $\bar{\theta}^-$ has the opposite chirality to $\bar{\theta}^+$.

The tangent space is spanned by $\frac{\partial}{\partial \bold{X}_{\hat{D}_T}^{\mu}(\bar{\sigma}, \bar{\tau}, \bar{\theta})}$, $\frac{\partial}{\partial \bold{X}_{L\hat{D}_T}^A(\bar{\sigma}, \bar{\tau}, \bar{\theta}^-)}$ and  $\frac{\partial}{\partial\bar{\tau}}$ as one can see from the  $\epsilon$-open neighborhood (\ref{HeteroNeighbour}). We should note that $\frac{\partial}{\partial \bar{\bold{E}}^{\quad A}_{\, M}}$  cannot be a part of basis that span the tangent space because $\bar{\bold{E}}^{\quad A}_{\, M}$ is just a discrete variable in $\bold{E}$. The indices of $\frac{\partial}{\partial \bold{X}_{\hat{D}_T}^{\mu}(\bar{\sigma}, \bar{\tau}, \bar{\theta}) }$ and $\frac{\partial}{\partial \bold{X}_{L\hat{D}_T}^A(\bar{\sigma}, \bar{\tau}, \bar{\theta}^-)}$ can be $(\mu \, \bar{\sigma} \, \bar{\theta})$ and $(A \, \bar{\sigma} \, \bar{\theta}^-)$, where $\mu=0, 1, \cdots, d-1$ and $A=1, \cdots 32$, respectively. 
Then, let us define a summation over $\bar{\sigma}$ and $\bar{\theta}$ that is invariant under $(\bar{\sigma}, \bar{\theta}) \mapsto (\bar{\sigma}'(\bar{\sigma}, \bar{\theta}), \bar{\theta}'(\bar{\sigma}, \bar{\theta}))$ and transformed as a scalar under $\bar{\tau} \mapsto \bar{\tau}'(\bar{\tau}, \bold{X}_{\hat{D}_T}(\bar{\tau}),  \bold{X}_{L\hat{D}_T}(\bar{\tau}))$. First, $\int d\bar{\tau} \int d\bar{\sigma}d\bar{\theta} \bar{\bold{E}}(\bar{\sigma}, \bar{\tau}, \bar{\theta})$ is invariant under $(\bar{\sigma}, \bar{\tau}, \bar{\theta}) \mapsto (\bar{\sigma}'(\bar{\sigma}, \bar{\theta}), \bar{\tau}'(\bar{\tau}, \bold{X}_{\hat{D}_T}(\bar{\tau}), \bold{X}_{L\hat{D}_T}(\bar{\tau})), \bar{\theta}'(\bar{\sigma}, \bar{\theta}))$, where $\bar{\bold{E}}(\bar{\sigma}, \bar{\tau}, \bar{\theta})$ is the superdeterminant of $\bar{\bold{E}}_{M}^{\quad A}(\bar{\sigma}, \bar{\tau}, \bar{\theta})$. 
The lapse function, $\bar{n}$ transforms as an one-dimensional vector in the $\bar{\tau}$ direction: 
$\int d\bar{\tau}\bar{n}$ is invariant under $\bar{\tau} \mapsto \bar{\tau}'(\bar{\tau}, \bold{X}_{\hat{D}_T}(\bar{\tau}),  \bold{X}_{L\hat{D}_T}(\bar{\tau}))$ and transformed as a superscalar under $(\bar{\sigma}, \bar{\theta}) \mapsto (\bar{\sigma}'(\bar{\sigma}, \bar{\theta}), \bar{\theta}'(\bar{\sigma}, \bar{\theta}))$. Therefore, 
$\int d\bar{\sigma}d\bar{\theta} \hat{\bold{E}}(\bar{\sigma}, \bar{\tau}, \bar{\theta})$,
where
$\hat{\bold{E}}(\bar{\sigma}, \bar{\tau}, \bar{\theta})
:=
\frac{1}{\bar{n}}\bar{\bold{E}}(\bar{\sigma}, \bar{\tau}, \bar{\theta})$,
is transformed as a scalar under $\bar{\tau} \mapsto \bar{\tau}'(\bar{\tau}, \bold{X}_{\hat{D}_T}(\bar{\tau}),  \bold{X}_{L\hat{D}_T}(\bar{\tau}))$ and invariant under $(\bar{\sigma}, \bar{\theta}) \mapsto (\bar{\sigma}'(\bar{\sigma}, \bar{\theta}), \bar{\theta}'(\bar{\sigma}, \bar{\theta}))$. The summation over $\bar{\sigma}$ and $\bar{\theta}^-$ is defined by $\int d\bar{\sigma} d\bar{\theta}^- \bar{e} (\bar{\sigma}, \bar{\tau})$, where $\bar{e}:=\sqrt{\bar{h}_{ \bar{\sigma} \bar{\sigma}}}$. This summation is also invariant under $(\bar{\sigma}, \bar{\theta}) \mapsto (\bar{\sigma}'(\bar{\sigma}, \bar{\theta}), \bar{\theta}'(\bar{\sigma}, \bar{\theta}))$, where $\bar{\theta}^-$ is not transformed, and transformed as a scalar under $\bar{\tau} \mapsto \bar{\tau}'(\bar{\tau}, \bold{X}_{\hat{D}_T}(\bar{\tau}),  \bold{X}_{L\hat{D}_T}(\bar{\tau}))$.

Riemannian heterotic superstring manifold is obtained by defining a metric, which is a section of an inner product on the tangent space. The general form of a metric is given by
\begin{eqnarray}
&&ds^2(\bar{\bold{E}}, \bold{X}_{\hat{D}_T}(\bar{\tau}),  \bold{X}_{L\hat{D}_T}(\bar{\tau}), \bar{\tau}) \nonumber \\
=&&G(\bar{\bold{E}}, \bold{X}_{\hat{D}_T}(\bar{\tau}),  \bold{X}_{L\hat{D}_T}(\bar{\tau}), \bar{\tau})_{dd} (d\bar{\tau})^2 \nonumber \\ 
&&+2 d\bar{\tau} \int d\bar{\sigma} d \bar{\theta}   \hat{\bold{E}} \sum_{\mu} G(\bar{\bold{E}}, \bold{X}_{\hat{D}_T}(\bar{\tau}),  \bold{X}_{L\hat{D}_T}(\bar{\tau}), \bar{\tau})_{d \; (\mu \bar{\sigma} \bar{\theta})} d \bold{X}_{\hat{D}_T}^{\mu}(\bar{\sigma}, \bar{\tau}, \bar{\theta}) \nonumber \\ 
&&+2 d\bar{\tau} \int d\bar{\sigma} d \bar{\theta}^- \bar{e} \sum_{A} G(\bar{\bold{E}}, \bold{X}_{\hat{D}_T}(\bar{\tau}),  \bold{X}_{L\hat{D}_T}(\bar{\tau}), \bar{\tau})_{d \; (A \bar{\sigma} \bar{\theta}^-)} d \bold{X}_{L\hat{D}_T}^A(\bar{\sigma}, \bar{\tau}, \bar{\theta}^-) \nonumber \\
&&+\int d\bar{\sigma} d \bar{\theta}  \hat{\bold{E}}  \int d\bar{\sigma}'  d \bar{\theta} ' \hat{\bold{E}}'  \sum_{\mu, \mu'} G(\bar{\bold{E}}, \bold{X}_{\hat{D}_T}(\bar{\tau}),  \bold{X}_{L\hat{D}_T}(\bar{\tau}), \bar{\tau})_{ \; (\mu \bar{\sigma} \bar{\theta})  \; (\mu' \bar{\sigma}' \bar{\theta}')} d \bold{X}_{\hat{D}_T}^{\mu}(\bar{\sigma}, \bar{\tau}, \bar{\theta}) d \bold{X}_{\hat{D}_T}^{\mu'}(\bar{\sigma}', \bar{\tau}, \bar{\theta}') \nonumber \\
&&+\int d\bar{\sigma} d \bar{\theta}  \hat{\bold{E}}  \int d\bar{\sigma}'  d \bar{\theta}^- \bar{e}' \sum_{\mu, A} G(\bar{\bold{E}}, \bold{X}_{\hat{D}_T}(\bar{\tau}),  \bold{X}_{L\hat{D}_T}(\bar{\tau}), \bar{\tau})_{ \; (\mu \bar{\sigma} \bar{\theta})  \; (A \bar{\sigma}' \bar{\theta}^-)} d \bold{X}_{\hat{D}_T}^{\mu}(\bar{\sigma}, \bar{\tau}, \bar{\theta}) d \bold{X}_{L\hat{D}_T}^A(\bar{\sigma}', \bar{\tau}, \bar{\theta}^-) \nonumber \\
&&+\int d\bar{\sigma} d \bar{\theta}^-  \bar{e}  \int d\bar{\sigma}'  d \bar{\theta}^{'-} \bar{e}'  \sum_{A, A'} G(\bar{\bold{E}}, \bold{X}_{\hat{D}_T}(\bar{\tau}),  \bold{X}_{L\hat{D}_T}(\bar{\tau}), \bar{\tau})_{ \; (A \bar{\sigma} \bar{\theta}^-)  \; (A' \bar{\sigma}' \bar{\theta}^{'-})} 
\nonumber \\
&&d \bold{X}_{L\hat{D}_T}^{A}(\bar{\sigma}, \bar{\tau}, \bar{\theta}^-) d \bold{X}_{L\hat{D}_T}^{A'}(\bar{\sigma}', \bar{\tau}, \bar{\theta}^{'-}). \nonumber \\ &&
\end{eqnarray}
We summarize the vectors as $d\bold{X}_{\hat{D}_T}^\bold{I}$ ($\bold{I}=d,(\mu \bar{\sigma} \bar{\theta}), (A \bar{\sigma} \bar{\theta}^-)$), where  $d\bold{X}_{\hat{D}_T}^d:=d\bar{\tau}$, $d \bold{X}_{\hat{D}_T}^{(\mu \bar{\sigma} \bar{\theta})}:=d\bold{X}_{\hat{D}_T}^{\mu}(\bar{\sigma}, \bar{\tau}, \bar{\theta})$ and $d \bold{X}_{\hat{D}_T}^{(A \bar{\sigma} \bar{\theta}^-)}:=d\bold{X}_{L\hat{D}_T}^{A}(\bar{\sigma}, \bar{\tau}, \bar{\theta}^-)$. Then, the components of the metric are summarized as $G_{\bold{I}\bold{J}}(\bar{\bold{E}}, \bold{X}_{\hat{D}_T}(\bar{\tau}),  \bold{X}_{L\hat{D}_T}(\bar{\tau}), \bar{\tau})$. The inverse of the metric $G^{\bold{I}\bold{J}}(\bar{\bold{E}}, \bold{X}_{\hat{D}_T}(\bar{\tau}),  \bold{X}_{L\hat{D}_T}(\bar{\tau}), \bar{\tau})$ is defined by $G_{\bold{I}\bold{J}}G^{\bold{J}\bold{K}}=G^{\bold{K}\bold{J}}G_{\bold{J}\bold{I}}=\delta_\bold{I}^\bold{K}$, where $\delta_d^d=1$, $\delta_{\mu \bar{\sigma} \bar{\theta}}^{\mu' \bar{\sigma}' \bar{\theta}'}=\frac{1}{\hat{\bold{E}}}\delta_{\mu}^{\mu'}\delta(\bar{\sigma}-\bar{\sigma}') \delta(\bar{\theta}-\bar{\theta}')$ and $\delta_{A \bar{\sigma} \bar{\theta}^-}^{A' \bar{\sigma}' \bar{\theta}^{'-}}=\frac{1}{\bar{e}}\delta_{A}^{A'}\delta(\bar{\sigma}-\bar{\sigma}') \delta(\bar{\theta}^--\bar{\theta}^{'-})$. The components of the Riemannian curvature tensor are given by $R^\bold{I}_{\bold{J}\bold{K}\bold{L}}$ in the basis $\frac{\partial}{\partial \bold{X}_{\hat{D}_T}^\bold{I}}$. The Ricci tensor is $R_{\bold{I}\bold{J}}:=R^\bold{K}_{\bold{I}\bold{K}\bold{J}}$ and the scalar curvature is $R:=G^{\bold{I}\bold{J}} R_{\bold{I}\bold{J}}$. The volume is  $vol=\sqrt{G}$, where $G=det (G_{\bold{I}\bold{J}})$.

By using these geometrical objects, we define a superstring theory non-perturbatively as
\begin{equation}
Z=\int \mathcal{D}G \mathcal{D}Ae^{-S}, \label{HSuperTheory}
\end{equation}
where
\begin{equation}
S=\frac{1}{G_N}\int \mathcal{D}\bold{E} \mathcal{D}\bold{X}_{\hat{D}_T}(\bar{\tau}) \mathcal{D}\bold{X}_{L\hat{D}_T}(\bar{\tau})\mathcal{D}\bar{\tau} 
\sqrt{G} (-R +\frac{1}{4} G_N G^{\bold{I}_1 \bold{I}_2} G^{\bold{J}_1 \bold{J}_2} F_{\bold{I}_1 \bold{J}_1} F_{\bold{I}_2 \bold{J}_2} ). \label{HSuperAction}
\end{equation}
As an example of sets of fields on the superstring manifolds, we consider the metric and an $u(1)$ gauge field $A_\bold{I}$ whose field strength is given by $F_{\bold{I}\bold{J}}$. The path integral is defined by semi-classically\footnote{It will be enough to define the path-integral by semi-classically summing classical solutions and small classical and quantum fluctuations around them, because string manifolds themselves possess quantum corrections, and loops of the fields on them do not correspond to quantum corrections as one can see in the derivation of the perturbative string theory later. The unitarity is manifest and there is also no UV divergence from loop integrals, by defining the path-integral semi-classically. } summing over the metrics and gauge fields on $\mathfrak{M}$. By definition, the theory is background independent. $\mathcal{D}\bold{E}$ is the invariant measure of the super vierbeins $\bold{E}_{M}^{\quad A}$ on the two-dimensional super Riemannian manifolds $\bold{\Sigma}$. $\bold{E}_{M}^{\quad A}$ and $\bar{\bold{E}}_{M}^{\quad A}$ are related to each others by the super diffeomorphism and super Weyl transformations.

Under 
\begin{eqnarray}
&&(\bar{\tau}, \bold{X}_{\hat{D}_T}(\bar{\tau}), \bold{X}_{L\hat{D}_T}(\bar{\tau}))  \nonumber \\
&\mapsto& (\bar{\tau}'(\bar{\tau}, \bold{X}_{\hat{D}_T}(\bar{\tau}), \bold{X}_{L\hat{D}_T}(\bar{\tau})) , \bold{X}_{\hat{D}_T}'(\bar{\tau}') (\bar{\tau}, \bold{X}_{\hat{D}_T}(\bar{\tau}) , \bold{X}_{L\hat{D}_T}(\bar{\tau})), \bold{X}_{L\hat{D}_T}'(\bar{\tau}') (\bar{\tau}, \bold{X}_{\hat{D}_T}(\bar{\tau}), \bold{X}_{L\hat{D}_T}(\bar{\tau}))),\label{HSupersubdiffeo} \nonumber \\
&&
\end{eqnarray} 
$G_{\bold{I}\bold{J}}(\bar{\bold{E}}, \bold{X}_{\hat{D}_T}(\bar{\tau}),  \bold{X}_{L\hat{D}_T}(\bar{\tau}), \bar{\tau})$ and $A_{\bold{I}}(\bar{\bold{E}}, \bold{X}_{\hat{D}_T}(\bar{\tau}),  \bold{X}_{L\hat{D}_T}(\bar{\tau}), \bar{\tau})$ are transformed as a symmetric tensor and a vector, respectively and the action is manifestly invariant. 

We define $G_{\bold{I}\bold{J}}(\bar{\bold{E}}, \bold{X}_{\hat{D}_T}(\bar{\tau}),  \bold{X}_{L\hat{D}_T}(\bar{\tau}), \bar{\tau})$ and $A_{\bold{I}}(\bar{\bold{E}}, \bold{X}_{\hat{D}_T}(\bar{\tau}),  \bold{X}_{L\hat{D}_T}(\bar{\tau}), \bar{\tau})$ so as to transform as scalars under $\bar{\bold{E}}_{M}^{\quad A}(\bar{\sigma}, \bar{\tau}, \bar{\theta}) \mapsto
\bar{\bold{E}}_{M}^{'\quad A}(\bar{\sigma}'(\bar{\sigma}, \bar{\theta}), \bar{\tau}, \bar{\theta}^{'}(\bar{\sigma}, \bar{\theta}))$. Under $(\bar{\sigma}, \bar{\theta})$ superdiffeomorphisms: $(\bar{\sigma}, \bar{\theta}) \mapsto (\bar{\sigma}'(\bar{\sigma}, \bar{\theta}), \bar{\theta}'(\bar{\sigma}, \bar{\theta}))$, which are equivalent to
\begin{eqnarray}
&&[\bar{\bold{E}}_{M}^{\quad A}(\bar{\sigma}, \bar{\tau}, \bar{\theta}), \bold{X}_{\hat{D}_T}^{\mu}(\bar{\sigma}, \bar{\tau}, \bar{\theta}), 
\bold{X}_{L\hat{D}_T}^A(\bar{\sigma}, \bar{\tau}, \bar{\theta}^-), \bar{\tau}] \nonumber \\
&\mapsto& [\bar{\bold{E}}_{M}^{'\quad A}(\bar{\sigma}'(\bar{\sigma}, \bar{\theta}), \bar{\tau}, \bar{\theta}'(\bar{\sigma}, \bar{\theta})),
\bold{X}_{\hat{D}_T}^{'\mu} (\bar{\sigma}'(\bar{\sigma}, \bar{\theta}), \bar{\tau}, \bar{\theta}'(\bar{\sigma}, \bar{\theta})) (\bold{X}_{\hat{D}_T}(\bar{\tau})),
\bold{X}_{L\hat{D}_T}^{'A}(\bar{\sigma}'(\bar{\sigma}, \bar{\theta}), \bar{\tau}, \bar{\theta}^{-})(\bold{X}_{L\hat{D}_T}(\bar{\tau})), \bar{\tau}] \nonumber \\
&=&
[\bar{\bold{E}}_{M}^{'\quad A}(\bar{\sigma}'(\bar{\sigma}, \bar{\theta}), \bar{\tau}, \bar{\theta}'(\bar{\sigma}, \bar{\theta})), \bold{X}_{\hat{D}_T}^{\mu}(\bar{\sigma}, \bar{\tau}, \bar{\theta}), 
\bold{X}_{L\hat{D}_T}^A(\bar{\sigma}, \bar{\tau}, \bar{\theta}^-), \bar{\tau}],
\label{HsuperTrans}
\end{eqnarray}
$G_{d \; (A \bar{\sigma} \bar{\theta}^-)}$ is transformed as a scalar;
\begin{eqnarray}
G'_{d \; (A \bar{\sigma}' \bar{\theta}^{-})}(\bold{E}', \bold{X}_{\hat{D}_T}'(\bar{\tau}) , \bold{X}_{L\hat{D}_T}'(\bar{\tau}) , \bar{\tau})
&=&
G'_{d \; (A \bar{\sigma}' \bar{\theta}^{-})}(\bar{\bold{E}}, \bold{X}_{\hat{D}_T}'(\bar{\tau}) , \bold{X}_{L\hat{D}_T}'(\bar{\tau}) , \bar{\tau})
\nonumber \\
&=&
\frac{\partial \bold{X}_{\hat{D}_T}^{\bold{I}}(\bar{\tau}) }{\partial \bold{X}_{\hat{D}_T}^{'d}(\bar{\tau}) }
\frac{\partial \bold{X}_{\hat{D}_T}^{\bold{J}}(\bar{\tau}) }{\partial \bold{X}_{\hat{D}_T}^{'(A \bar{\sigma}' \bar{\theta}^{-})}(\bar{\tau}) }
G_{\bold{I} \bold{J}}(\bar{\bold{E}}, \bold{X}_{\hat{D}_T}(\bar{\tau}),  \bold{X}_{L\hat{D}_T}(\bar{\tau}), \bar{\tau})
\nonumber \\
&=&
\frac{\partial \bold{X}_{\hat{D}_T}^{\bold{I}}(\bar{\tau}) }{\partial \bold{X}_{\hat{D}_T}^{d}(\bar{\tau}) }
\frac{\partial \bold{X}_{\hat{D}_T}^{\bold{J}}(\bar{\tau}) }{\partial \bold{X}_{\hat{D}_T}^{(A \bar{\sigma} \bar{\theta}^{-})}(\bar{\tau}) }
G_{\bold{I} \bold{J}}(\bar{\bold{E}}, \bold{X}_{\hat{D}_T}(\bar{\tau}),  \bold{X}_{L\hat{D}_T}(\bar{\tau}), \bar{\tau})
\nonumber \\
&=&
G_{d \; (A \bar{\sigma} \bar{\theta}^-)}(\bar{\bold{E}}, \bold{X}_{\hat{D}_T}(\bar{\tau}),  \bold{X}_{L\hat{D}_T}(\bar{\tau}), \bar{\tau}),
\label{HGdmuTrans}
\end{eqnarray}
because (\ref{HSupersubdiffeo}) and (\ref{HsuperTrans}).
In the same way, the other fields are also transformed as 
\begin{eqnarray}
G'_{dd}(\bold{E}', \bold{X}_{\hat{D}_T}'(\bar{\tau}) , \bold{X}_{L\hat{D}_T}'(\bar{\tau}) , \bar{\tau})
&=&G_{dd}(\bar{\bold{E}}, \bold{X}_{\hat{D}_T}(\bar{\tau}),  \bold{X}_{L\hat{D}_T}(\bar{\tau}), \bar{\tau})
\nonumber \\
G'_{d \; (\mu \bar{\sigma}'\bar{\theta}')}(\bold{E}', \bold{X}_{\hat{D}_T}'(\bar{\tau}), \bold{X}_{L\hat{D}_T}'(\bar{\tau}), \bar{\tau})
&=&G_{d \; (\mu \bar{\sigma} \bar{\theta})}(\bar{\bold{E}}, \bold{X}_{\hat{D}_T}(\bar{\tau}),  \bold{X}_{L\hat{D}_T}(\bar{\tau}), \bar{\tau})
\nonumber \\
G'_{ \; (\mu \bar{\sigma}'\bar{\theta}')  \; (\nu \bar{\rho}'\tilde{\bar{\theta}}')}(\bold{E}', \bold{X}_{\hat{D}_T}'(\bar{\tau}), \bold{X}_{L\hat{D}_T}'(\bar{\tau}), \bar{\tau})
&=&G_{ \; (\mu \bar{\sigma} \bar{\theta})  \; (\nu \bar{\rho} \tilde{\bar{\theta}})}(\bar{\bold{E}}, \bold{X}_{\hat{D}_T}(\bar{\tau}),  \bold{X}_{L\hat{D}_T}(\bar{\tau}), \bar{\tau})
\nonumber \\
G'_{ \; (\mu \bar{\sigma}'\bar{\theta}')  \; (A \bar{\rho}'\tilde{\bar{\theta}}^{-})}(\bold{E}', \bold{X}_{\hat{D}_T}'(\bar{\tau}), \bold{X}_{L\hat{D}_T}'(\bar{\tau}), \bar{\tau})
&=&G_{ \; (\mu \bar{\sigma} \bar{\theta})  \; (A \bar{\rho} \tilde{\bar{\theta}}^-)}(\bar{\bold{E}}, \bold{X}_{\hat{D}_T}(\bar{\tau}),  \bold{X}_{L\hat{D}_T}(\bar{\tau}), \bar{\tau})
\nonumber \\
G'_{ \; (A \bar{\sigma}' \bar{\theta}^{-})  \; (B \bar{\rho}' \tilde{\bar{\theta}}^{-})}(\bold{E}', \bold{X}_{\hat{D}_T}'(\bar{\tau}), \bold{X}_{L\hat{D}_T}'(\bar{\tau}), \bar{\tau})
&=&G_{ \; (A \bar{\sigma} \bar{\theta}^-)  \; (B \bar{\rho} \tilde{\bar{\theta}}^-)}(\bar{\bold{E}}, \bold{X}_{\hat{D}_T}(\bar{\tau}),  \bold{X}_{L\hat{D}_T}(\bar{\tau}), \bar{\tau})
\nonumber \\
A'_d(\bold{E}', \bold{X}_{\hat{D}_T}'(\bar{\tau}), \bold{X}_{L\hat{D}_T}'(\bar{\tau}), \bar{\tau})
&=&A_d(\bar{\bold{E}}, \bold{X}_{\hat{D}_T}(\bar{\tau}),  \bold{X}_{L\hat{D}_T}(\bar{\tau}), \bar{\tau})
\nonumber \\
A'_{(\mu \bar{\sigma}'\bar{\theta}')}(\bold{E}', \bold{X}_{\hat{D}_T}'(\bar{\tau}), \bold{X}_{L\hat{D}_T}'(\bar{\tau}), \bar{\tau})
&=&A_{(\mu \bar{\sigma} \bar{\theta})}(\bar{\bold{E}}, \bold{X}_{\hat{D}_T}(\bar{\tau}),  \bold{X}_{L\hat{D}_T}(\bar{\tau}), \bar{\tau})
\nonumber \\
A'_{(A \bar{\sigma}' \bar{\theta}^{-})}(\bold{E}', \bold{X}_{\hat{D}_T}'(\bar{\tau}), \bold{X}_{L\hat{D}_T}'(\bar{\tau}), \bar{\tau})
&=&A_{(A \bar{\sigma} \bar{\theta}^-)}(\bar{\bold{E}}, \bold{X}_{\hat{D}_T}(\bar{\tau}),  \bold{X}_{L\hat{D}_T}(\bar{\tau}), \bar{\tau}). 
\label{HOtherSuperTrans}
\end{eqnarray} 
Thus, the action is invariant under the $(\bar{\sigma}, \bar{\theta})$ superdiffeomorphisms, because 
\begin{eqnarray}
\int d\bar{\sigma}' d \bar{\theta} '  \hat{\bold{E}}'(\bar{\sigma}', \bar{\tau}, \bar{\theta}')&=&\int d\bar{\sigma} d \bar{\theta}   \hat{\bold{E}}(\bar{\sigma}, \bar{\tau}, \bar{\theta}) \nonumber \\
\int d\bar{\sigma}' d\bar{\theta}^- \bar{e}' (\bar{\sigma}', \bar{\tau})
&=&
\int d\bar{\sigma} d\bar{\theta}^- \bar{e} (\bar{\sigma}, \bar{\tau})
\end{eqnarray}
Therefore, $G_{\bold{I}\bold{J}}(\bar{\bold{E}}, \bold{X}_{\hat{D}_T}(\bar{\tau}),  \bold{X}_{L\hat{D}_T}(\bar{\tau}), \bar{\tau})$ and $A_{\bold{I}}(\bar{\bold{E}}, \bold{X}_{\hat{D}_T}(\bar{\tau}),  \bold{X}_{L\hat{D}_T}(\bar{\tau}), \bar{\tau})$ are transformed covariantly and the action (\ref{HSuperAction}) is invariant under the diffeomorphisms (\ref{HGeneralCoordTrans2}) including the $(\bar{\sigma}, \bar{\theta})$ superdiffeomorphisms,
whose infinitesimal transformations are given by 
\begin{eqnarray}
\bar{\sigma}^{\xi}&=&\bar{\sigma}-\frac{i}{2} \xi \bar{\theta} \nonumber \\
\bar{\theta}^{\xi}(\bar{\sigma})&=&\bar{\theta} + \xi(\bar{\sigma}), \label{HSUSYtrans}
\end{eqnarray}
(\ref{HSUSYtrans}) are dimensional reductions in $\bar{\tau}$ direction of the two-dimensional $\mathcal{N}=(0,1)$ local supersymmetry infinitesimal transformations. The number of supercharges
\begin{equation}
\xi Q
=
\xi (\frac{\partial}{\partial \bar{\theta}} - \frac{i}{2} \bar{\theta} \frac{\partial}{\partial \bar{\sigma}})
\end{equation}
 of the transformations is the same as of the two-dimensional ones. The supersymmetry algebra closes in a field-independent sense as in ordinary supergravities.

The background that represents a perturbative vacuum is given by
\begin{eqnarray}
\bar{ds}^2
&=& 2\zeta \bar{\rho}(\bar{h}) N^2(\bold{X}_{\hat{D}_T}(\bar{\tau}), \bold{X}_{L\hat{D}_T}(\bar{\tau})) (d\bold{X}_{\hat{D}_T}^d)^2 \nonumber \\ 
&&+\int d\bar{\sigma} d \bar{\theta}   \hat{\bold{E}} \int d\bar{\sigma}' d \bar{\theta} '  \hat{\bold{E}}' N^{\frac{2}{2-\bold{D}}}(\bold{X}_{\hat{D}_T}(\bar{\tau}), \bold{X}_{L\hat{D}_T}(\bar{\tau}))
\frac{\hat{\bold{E}}^3(\bar{\sigma}, \bar{\tau}, \bar{\theta})}
{\bold{E}(\bar{\sigma}, \bar{\tau}, \bar{\theta})}  \delta_{(\mu \bar{\sigma} \bar{\theta}) (\mu' \bar{\sigma}' \bar{\theta}')}
d \bold{X}_{\hat{D}_T}^{(\mu \bar{\sigma} \bar{\theta})} d \bold{X}_{\hat{D}_T}^{(\mu' \bar{\sigma}' \bar{\theta}')} \nonumber \\
&&+\int d\bar{\sigma} d \bar{\theta}^-   \bar{e} \int d\bar{\sigma}' d \bar{\theta}^{'-}  \bar{e}' N^{\frac{2}{2-\bold{D}}}(\bold{X}_{\hat{D}_T}(\bar{\tau}), \bold{X}_{L\hat{D}_T}(\bar{\tau}))
\frac{\bar{e}^3(\bar{\sigma}, \bar{\tau})}{\sqrt{\bar{h}
(\bar{\sigma}, \bar{\tau})}} 
\delta_{(A \bar{\sigma} \bar{\theta}^-) (A' \bar{\sigma}' \bar{\theta}^{'-})}
d \bold{X}_{L\hat{D}_T}^{(A \bar{\sigma} \bar{\theta}^-)} d \bold{X}_{L\hat{D}_T}^{(A' \bar{\sigma}' \bar{\theta}^{'-})}, \nonumber \\
\bar{A}_d&=&i \sqrt{\frac{2-2\bold{D}}{2-\bold{D}}}\frac{\sqrt{2\zeta \bar{\rho}(\bar{h}) }}{\sqrt{G_N}} N(\bold{X}_{\hat{D}_T}(\bar{\tau}), \bold{X}_{L\hat{D}_T}(\bar{\tau})), \qquad
\bar{A}_{(\mu \bar{\sigma} \bar{\theta})}=0, \qquad
\bar{A}_{(A \bar{\sigma} \bar{\theta}^-)}=0, \label{HSupersolution}
\end{eqnarray}
on  $\mathfrak{M}_{D_T}$ where we fix the target metric to $\eta_{\mu \mu'}$.
$\bar{\rho}(\bar{h}):=\frac{1}{4 \pi}\int d\bar{\sigma} \sqrt{\bar{h}}\bar{R}_{\bar{h}}$, where $\bar{R}_{\bar{h}}$ is the scalar curvature of $\bar{h}_{ mn}$. $\bold{D}$ is a volume of the index $(\mu \bar{\sigma} \bar{\theta})$ and $(A \bar{\sigma} \bar{\theta}^-)$: $\bold{D}:=\int d\bar{\sigma} d \bar{\theta}   \hat{\bold{E}} \delta_{(\mu \bar{\sigma} \bar{\theta}) (\mu \bar{\sigma} \bar{\theta})}
+\int d\bar{\sigma} d \bar{\theta}^-   \bar{e} \delta_{(A \bar{\sigma} \bar{\theta}^-) (A \bar{\sigma} \bar{\theta}^-)}
=(d+32) \int d\bar{\sigma} d \bar{\theta}  \delta(\bar{\sigma}-\bar{\sigma}) \delta(\bar{\theta}-\bar{\theta})$. $N(\bold{X}_{\hat{D}_T}(\bar{\tau}), \bold{X}_{L\hat{D}_T}(\bar{\tau}))=\frac{1}{1+v(\bold{X}_{\hat{D}_T}(\bar{\tau}), \bold{X}_{L\hat{D}_T}(\bar{\tau}))}$, where 
\begin{eqnarray}
v(\bold{X}_{\hat{D}_T}(\bar{\tau}), \bold{X}_{L\hat{D}_T}(\bar{\tau}))&=& \frac{\alpha}{\sqrt{d-1}} \int d\bar{\sigma} d \bar{\theta} \hat{\bold{E}} \epsilon_{\mu\nu}\bold{X}_{\hat{D}_T}^{\mu}(\bar{\tau}) \sqrt{\tilde{\partial}_{z} \tilde{\bold{D}}_{\theta}} \bold{X}_{\hat{D}_T}^{\nu}(\bar{\tau})  \nonumber \\
&&+\frac{\beta}{\sqrt{31}} \int d\bar{\sigma} d \bar{\theta}^- \bar{e}  \frac{\sqrt{\bar{\bold{E}}^L}}{(\bar{h})^{\frac{1}{4}}} \epsilon_{AB}\bold{X}_{L\hat{D}_T}^A(\bar{\tau}) \sqrt{\tilde{\partial}_{\bar{z}} \tilde{\bold{D}}_{\theta^-}} \bold{X}_{L\hat{D}_T}^B(\bar{\tau}) .
\end{eqnarray} 
$\tilde{\partial}_{z}  \tilde{\bold{D} }_{\theta}$ is a $\bar{\tau}$ independent operator that satisfies 
\begin{eqnarray}
&&\int d\bar{\tau} d\bar{\sigma} d \bar{\theta}  \bar{\bold{E}}  \frac{-1}{2} \bold{X}_{\hat{D}_T}^{\mu}(\bar{\tau})  \tilde{\partial}_{z}  \tilde{\bold{D}}_{\theta}\bold{X}_{\hat{D}_T \mu} (\bar{\tau})  \nonumber \\
&=&
\int d\bar{\tau} d\bar{\sigma} \sqrt{\bar{h}} \frac{1}{2}\biggl(
\bar{e}^{-2}({\partial}_{\bar{\sigma}} X^{\mu})^2
+\bar{E}_{\bar{z}}^1 \psi_{\mu} \bar{\chi}_z \partial_{\bar{\sigma}} X^{\mu}
+\psi^{\mu} \bar{E}_{z}^1 \partial_{\bar{\sigma}} \psi_{\mu}
\nonumber \\
&&-\frac{1}{4}\bar{n}^2 \bar{E}_{\bar{z}}^0 \psi^{\mu} \bar{\chi}_z \bar{E}_{\bar{z}}^0 \psi_{\mu} \bar{\chi}_z 
+\bar{n}^{\bar{\sigma}} \partial_{\bar{\sigma}} X_{\mu} \bar{E}_{\bar{z}}^0 \psi^{\mu} \bar{\chi}_z \biggr),
\label{Hquad2}
\end{eqnarray}
where $\bar{E}^{m}_{\; q}$ and $\bar{\chi}_z$ are a vierbein and a gravitino in the two dimensions, respectively. On the other hand, the ordinary super covariant derivative $\bar{\bold{D}}_{\theta}=\partial_{\bar{\theta}}+\bar{\theta} \partial_{\bar{z}}$ satisfies \cite{BrooksMuhammadGates, HullWitten, BergshoeffSezginNishino}
\begin{eqnarray}
&&\int d\bar{\tau} d\bar{\sigma} d \bar{\theta}  \bar{\bold{E}}  \frac{-1}{2}  \bold{X}_{\hat{D}_T}^{\mu}(\bar{\tau})  \partial_{z} \bar{\bold{D}}_{\theta}\bold{X}_{\hat{D}_T \mu}(\bar{\tau})  \nonumber \\
&=&
\int d\bar{\tau} d\bar{\sigma} \sqrt{\bar{h}} \frac{1}{2}(\bar{h}^{mn}\bar{\partial}_{m} X^{\mu} \bar{\partial}_{n} X_{\mu}
+\psi^{\mu} \bar{E}_z^m \bar{\partial}_m \psi_{\mu}
+\bar{E}_{\bar{z}}^m \bar{\partial}_m X^{\mu} \psi_{\mu} \bar{\chi}_z). \label{Hquad1}
\end{eqnarray}
$\bar{\bold{E}}^L$ is the chiral conjugate of $\bar{\bold{E}}$. $\tilde{\partial}_{\bar{z}}:=\bar{E}_{\bar{z}}^1\partial_{\bar{\sigma}}$ and $\tilde{\bold{D} }_{\theta^-}:=\partial_{\bar{\theta}^-}+\bar{\theta}^- \bar{E}_{z}^1\partial_{\bar{\sigma}}$ satisfy
\begin{eqnarray}
\int d\bar{\tau} d\bar{\sigma} d \bar{\theta}^-  \bar{\bold{E}}^L  \frac{-1}{2} \bold{X}_{L\hat{D}_T}^A(\bar{\tau})  \tilde{\partial}_{\bar{z}}  \tilde{\bold{D}}_{\theta^-}\bold{X}_{L\hat{D}_T \, A}(\bar{\tau})   =
\int d\bar{\tau} d\bar{\sigma} \sqrt{\bar{h}} \frac{1}{2}
\lambda^A \bar{E}_{\bar{z}}^1 \partial_{\bar{\sigma}} \lambda_A,
\label{Hquad4}
\end{eqnarray}
whereas the ordinary super covariant derivative $\bar{\bold{D}}_{\theta^-}=\partial_{\bar{\theta}^-}+\bar{\theta}^- \partial_{z}$ satisfies 
\begin{eqnarray}
\int d\bar{\tau} d\bar{\sigma} d \bar{\theta}^-  \bar{\bold{E}}^L  \frac{-1}{2}  \bold{X}_{L\hat{D}_T}^A(\bar{\tau})  \partial_{\bar{z}} \bar{\bold{D}}_{\theta^-}\bold{X}_{L\hat{D}_T\, A}(\bar{\tau}) =
\int d\bar{\tau} d\bar{\sigma} \sqrt{\bar{h}} \frac{1}{2}\lambda^A \bar{E}_{\bar{z}}^m \bar{\partial}_m \lambda_A. \label{Hquad3}
\end{eqnarray}
The inverse of the metric is given by 
\begin{eqnarray}
\bar{G}^{dd}&=&\frac{1}{2\zeta \bar{\rho} }\frac{1}{N^2} \nonumber \\
\bar{G}^{(\mu\bar{\sigma} \bar{\theta}) \; (\mu'\bar{\sigma}' \bar{\theta}')}&=& N^{\frac{-2}{2-\bold{D}}} \frac{\bold{E}} 
{\hat{\bold{E}}^3} \delta_{(\mu \bar{\sigma} \bar{\theta}) (\mu' \bar{\sigma}' \bar{\theta}')} \nonumber \\
\bar{G}^{(A \bar{\sigma} \bar{\theta}^-) \; (A'\bar{\sigma}' \bar{\theta}^{'-})}&=& N^{\frac{-2}{2-\bold{D}}} \frac{\sqrt{\bar{h}}}{\bar{e}^3} \delta_{(A \bar{\sigma} \bar{\theta}^-) (A' \bar{\sigma}' \bar{\theta}^{'-})},
\end{eqnarray}
where the other components are zero.
From the metric, we obtain 
\begin{eqnarray}
&&\sqrt{\bar{G}}=N^\frac{2}{2-\bold{D}}\sqrt{2\zeta \bar{\rho} \exp(\int d\bar{\sigma}d \bar{\theta}  \hat{\bold{E}} \delta_{(\mu\bar{\sigma}\bar{\theta}) \; (\mu \bar{\sigma} \bar{\theta} )} \ln \frac{\hat{\bold{E}}^3}
{\bold{E}}
+\int d\bar{\sigma}d \bar{\theta}^-  \bar{e} \delta_{(A \bar{\sigma} \bar{\theta}^-) \; (A \bar{\sigma} \bar{\theta}^- )} \ln \frac{\bar{e}^3}{\sqrt{\bar{h}}})} 
\nonumber \\
&&\bar{R}_{dd}=-2\zeta \bar{\rho} N^{\frac{-2}{2-\bold{D}}} \left(\int d\bar{\sigma} d \bar{\theta} \frac{\bold{E}}
{\hat{\bold{E}}^2} \partial_{(\mu \bar{\sigma} \bar{\theta})}N \partial_{(\mu \bar{\sigma} \bar{\theta})}N
+
\int d\bar{\sigma} d \bar{\theta}^-  \frac{\sqrt{\bar{h}}}{\bar{e}^2} \partial_{(A \bar{\sigma} \bar{\theta}^-)}N \partial_{(A \bar{\sigma} \bar{\theta}^-)}N
\right)
\nonumber \\
&&\bar{R}_{(\mu\bar{\sigma}\bar{\theta}) \; (\mu' \bar{\sigma}' \bar{\theta}')}
=\frac{\bold{D}-1}{2-\bold{D}}N^{-2}\partial_{(\mu\bar{\sigma}\bar{\theta})}N \partial_{(\mu' \bar{\sigma}' \bar{\theta}')}N
\nonumber \\
&& \qquad \qquad \qquad +\frac{1}{\bold{D}-2}N^{-2}
\Biggl(\int d\bar{\sigma}'' d \bar{\theta} ''\frac{\bold{E}''}
{\hat{\bold{E}}^{''2}}
\partial_{(\mu'' \bar{\sigma}'' \bar{\theta}'')}N \partial_{(\mu'' \bar{\sigma}'' \bar{\theta}'')}N \nonumber \\
&& \qquad \qquad \qquad \qquad \qquad \qquad
+\int d\bar{\sigma}'' d \bar{\theta}^{''-} \frac{\sqrt{\bar{h}''}}{\bar{e}^{''2}}\partial_{(A'' \bar{\sigma}'' \bar{\theta}^{''-})}N \partial_{(A'' \bar{\sigma}'' \bar{\theta}^{''-})}N \Biggr)
\frac{\hat{\bold{E}}^3}
{\bold{E}} 
\delta_{(\mu\bar{\sigma}\bar{\theta}) \; (\mu' \bar{\sigma}' \bar{\theta}')}
\nonumber \\
&&\bar{R}_{(A\bar{\sigma}\bar{\theta}^-) \; (A' \bar{\sigma}' \bar{\theta}^{'-})}
=\frac{\bold{D}-1}{2-\bold{D}}N^{-2}\partial_{(A \bar{\sigma} \bar{\theta}^-)}N \partial_{(A' \bar{\sigma}' \bar{\theta}^{'-})}N
\nonumber \\
&& \qquad \qquad \qquad +\frac{1}{\bold{D}-2}N^{-2}
\Biggl(\int d\bar{\sigma}'' d \bar{\theta} ''\frac{\bold{E}''}
{\hat{\bold{E}}^{''2}}\partial_{(\mu'' \bar{\sigma}'' \bar{\theta}'')}N \partial_{(\mu'' \bar{\sigma}'' \bar{\theta}'')}N \nonumber \\
&& \qquad \qquad \qquad \qquad \qquad \qquad
+\int d\bar{\sigma}'' d \bar{\theta}^{''-} \frac{\sqrt{\bar{h}''}}{\bar{e}^{''2}}\partial_{(A'' \bar{\sigma}'' \bar{\theta}^{''-})}N \partial_{(A'' \bar{\sigma}'' \bar{\theta}^{''-})}N \Biggr)
\frac{\hat{\bold{E}}^3}
{\bold{E}} 
\delta_{(A \bar{\sigma} \bar{\theta}^-) \; (A' \bar{\sigma}' \bar{\theta}^{'-})}
\nonumber \\
&&\bar{R}=\frac{\bold{D}-3}{2-\bold{D}} N^{\frac{2\bold{D}-6}{2-\bold{D}}}
\left(\int d\bar{\sigma} d \bar{\theta}  \frac{\bold{E}}
{\hat{\bold{E}}^2} \partial_{(\mu \bar{\sigma} \bar{\theta})}N \partial_{(\mu \bar{\sigma} \bar{\theta})}N
+
\int d\bar{\sigma} d \bar{\theta}^-  \frac{\sqrt{\bar{h}}}{\bar{e}^2} \partial_{(A \bar{\sigma} \bar{\theta}^-)}N \partial_{(A \bar{\sigma} \bar{\theta}^-)}N
\right). \nonumber \\
&&
\end{eqnarray}
By using these quantities, one can show that the background (\ref{HSupersolution}) is a classical solution to the equations of motion of (\ref{HSuperAction}). We also need to use the fact that $v(\bold{X}_{\hat{D}_T}(\bar{\tau}), \bold{X}_{L\hat{D}_T}(\bar{\tau}))$ is a harmonic function with respect to $\bold{X}_{\hat{D}_T}^{(\mu\bar{\sigma}\bar{\theta})}(\bar{\tau}) $ and $\bold{X}_{L\hat{D}_T}^{(A \bar{\sigma} \bar{\theta}^-)}(\bar{\tau}) $. Actually, 
$\partial_{(\mu\bar{\sigma}\bar{\theta})}\partial_{(\mu\bar{\sigma}\bar{\theta})}v
=
\partial_{(A \bar{\sigma} \bar{\theta}^-)}\partial_{(A \bar{\sigma} \bar{\theta}^-)}v
=0$. In these calculations, we should note that $\bar{\bold{E}}_{M}^{\quad A}$, $\bar{\tau}$, $\bold{X}_{\hat{D}_T}^{\mu}(\bar{\tau}) $ and $\bold{X}_{L\hat{D}_T}^A(\bar{\tau})  $ are all independent, and thus $\frac{\partial}{\partial \bar{\tau}}$ is an explicit derivative on functions over the superstring manifolds, especially, $\frac{\partial}{\partial \bar{\tau}}\bar{\bold{E}}_{M}^{\quad A}=0$,
$\frac{\partial}{\partial \bar{\tau}}\bold{X}_{\hat{D}_T}^{\mu}(\bar{\tau}) =0$
and $\frac{\partial}{\partial \bar{\tau}}\bold{X}_{L\hat{D}_T}^A(\bar{\tau}) =0$. 
Because the equations of motion are differential equations with respect to $\bar{\tau}$, $\bold{X}_{\hat{D}_T}^{\mu}(\bar{\tau}) $ and $\bold{X}_{L\hat{D}_T}^A(\bar{\tau})  $, $\bar{\bold{E}}_{M}^{\quad A}$ is a constant in the solution (\ref{HSupersolution}) to the differential equations. The dependence of $\bar{\bold{E}}_{M}^{\quad A}$ on the background (\ref{HSupersolution}) is uniquely determined  by the consistency of the quantum theory of the fluctuations around the background. Actually, we will find that all the perturbative superstring amplitudes are derived.

 Let us consider fluctuations around the background (\ref{HSupersolution}), $G_{\bold{I}\bold{J}}=\bar{G}_{\bold{I}\bold{J}}+\tilde{G}_{\bold{I}\bold{J}}$ and $A_\bold{I}=\bar{A}_\bold{I}+\tilde{A}_\bold{I}$. Here we  fix the charts, where we choose $T$=$SO(32)$ or $E_8 \times E_8$. The action (\ref{HSuperAction}) up to the quadratic order is given by,\begin{eqnarray}
S&=&\frac{1}{G_N} \int \mathcal{D}\bold{E} \mathcal{D}\bold{X}_{\hat{D}_T}(\bar{\tau}) \mathcal{D}\bold{X}_{L\hat{D}_T}(\bar{\tau})\mathcal{D}\bar{\tau}  \sqrt{\bar{G}} 
\Bigl(-\bar{R}+\frac{1}{4}\bar{F}'_{\bold{I}\bold{J}}\bar{F}'^{\bold{I}\bold{J}} 
\nonumber \\
&&+\frac{1}{4}\bar{\nabla}_\bold{I} \tilde{G}_{\bold{J}\bold{K}} \bar{\nabla}^\bold{I} \tilde{G}^{\bold{J}\bold{K}}
-\frac{1}{4}\bar{\nabla}_\bold{I} \tilde{G} \bar{\nabla}^\bold{I} \tilde{G}
+\frac{1}{2}\bar{\nabla}^\bold{I} \tilde{G}_{\bold{I}\bold{J}} \bar{\nabla}^\bold{J} \tilde{G}
-\frac{1}{2}\bar{\nabla}^\bold{I} \tilde{G}_{\bold{I}\bold{J}} \bar{\nabla}_\bold{K} \tilde{G}^{\bold{J}\bold{K}}
\nonumber \\
&&-\frac{1}{4}(-\bar{R}+\frac{1}{4}\bar{F}'_{\bold{K}\bold{L}}\bar{F}'^{\bold{K}\bold{L}})
(\tilde{G}_{\bold{I}\bold{J}}\tilde{G}^{\bold{I}\bold{J}}-\frac{1}{2}\tilde{G}^2)
+(-\frac{1}{2}\bar{R}^{\bold{I}}_{\;\; \bold{J}}+\frac{1}{2}\bar{F}'^{\bold{I}\bold{K}}\bar{F}'_{\bold{J}\bold{K}})
\tilde{G}_{\bold{I}\bold{L}}\tilde{G}^{\bold{J}\bold{L}}
\nonumber \\
&&+(\frac{1}{2}\bar{R}^{\bold{I}\bold{J}}-\frac{1}{4}\bar{F}'^{\bold{I}\bold{K}}\bar{F}'^{\bold{J}}_{\;\;\;\; \bold{K}})
\tilde{G}_{\bold{I}\bold{J}}\tilde{G}
+(-\frac{1}{2}\bar{R}^{\bold{I}\bold{J}\bold{K}\bold{L}}+\frac{1}{4}\bar{F}'^{\bold{I}\bold{J}}\bar{F}'^{\bold{K}\bold{L}})
\tilde{G}_{\bold{I}\bold{K}}\tilde{G}_{\bold{J}\bold{L}}
\nonumber \\
&&+\frac{1}{4}G_N \tilde{F}_{\bold{I}\bold{J}} \tilde{F}^{\bold{I}\bold{J}} 
+\sqrt{G_N} 
(\frac{1}{4} \bar{F}^{'\bold{I}\bold{J}} \tilde{F}_{\bold{I}\bold{J}} \tilde{G} 
-\bar{F}^{'\bold{I}\bold{J}} \tilde{F}_{\bold{I}\bold{K}} \tilde{G}_\bold{J}^{\;\; \bold{K}} ) \Bigr), \label{Hfluctuation}
\end{eqnarray}
where $\bar{F}'_{\bold{I}\bold{J}}:=\sqrt{G_N}\bar{F}_{\bold{I}\bold{J}}$ is independent of $G_N$. $\tilde{G}:=\bar{G}^{\bold{I}\bold{J}}\tilde{G}_{\bold{I}\bold{J}}$. There is no first order term because the background satisfies the equations of motion. If we take $G_N \to 0$, we obtain 
\begin{eqnarray}
S'&=&\frac{1}{G_N} \int \mathcal{D}\bold{E} \mathcal{D}\bold{X}_{\hat{D}_T}(\bar{\tau}) \mathcal{D}\bold{X}_{L\hat{D}_T}(\bar{\tau})\mathcal{D}\bar{\tau}  \sqrt{\bar{G}} 
\Bigl(-\bar{R}+\frac{1}{4}\bar{F}'_{\bold{I}\bold{J}}\bar{F}'^{\bold{I}\bold{J}} 
\nonumber \\
&&+\frac{1}{4}\bar{\nabla}_\bold{I} \tilde{G}_{\bold{J}\bold{K}} \bar{\nabla}^\bold{I} \tilde{G}^{\bold{J}\bold{K}}
-\frac{1}{4}\bar{\nabla}_\bold{I} \tilde{G} \bar{\nabla}^\bold{I} \tilde{G}
+\frac{1}{2}\bar{\nabla}^\bold{I} \tilde{G}_{\bold{I}\bold{J}} \bar{\nabla}^\bold{J} \tilde{G}
-\frac{1}{2}\bar{\nabla}^\bold{I} \tilde{G}_{\bold{I}\bold{J}} \bar{\nabla}_\bold{K} \tilde{G}^{\bold{J}\bold{K}}
\nonumber \\
&&-\frac{1}{4}(-\bar{R}+\frac{1}{4}\bar{F}'_{\bold{K}\bold{L}}\bar{F}'^{\bold{K}\bold{L}})
(\tilde{G}_{\bold{I}\bold{J}}\tilde{G}^{\bold{I}\bold{J}}-\frac{1}{2}\tilde{G}^2)
+(-\frac{1}{2}\bar{R}^{\bold{I}}_{\;\; \bold{J}}+\frac{1}{2}\bar{F}'^{\bold{I}\bold{K}}\bar{F}'_{\bold{J}\bold{K}})
\tilde{G}_{\bold{I}\bold{L}}\tilde{G}^{\bold{J}\bold{L}}
\nonumber \\
&&+(\frac{1}{2}\bar{R}^{\bold{I}\bold{J}}-\frac{1}{4}\bar{F}'^{\bold{I}\bold{K}}\bar{F}'^{\bold{J}}_{\;\;\;\; \bold{K}})
\tilde{G}_{\bold{I}\bold{J}}\tilde{G}
+(-\frac{1}{2}\bar{R}^{\bold{I}\bold{J}\bold{K}\bold{L}}+\frac{1}{4}\bar{F}'^{\bold{I}\bold{J}}\bar{F}'^{\bold{K}\bold{L}})
\tilde{G}_{\bold{I}\bold{K}}\tilde{G}_{\bold{J}\bold{L}} \Bigr),
\end{eqnarray}
where the fluctuation of the gauge field is suppressed. In order to fix the gauge symmetry (\ref{HSupersubdiffeo}), we take the harmonic gauge. If we add the gauge fixing term
\begin{equation}
S_{fix}=\frac{1}{G_N}\int \mathcal{D}\bold{E} \mathcal{D}\bold{X}_{\hat{D}_T}(\bar{\tau}) \mathcal{D}\bold{X}_{L\hat{D}_T}(\bar{\tau})\mathcal{D}\bar{\tau}  \sqrt{\bar{G}} 
\frac{1}{2} \Bigl( \bar{\nabla}^\bold{J}(\tilde{G}_{\bold{I}\bold{J}}-\frac{1}{2}\bar{G}_{\bold{I}\bold{J}}\tilde{G}) \Bigr)^2,\end{equation}
we obtain
\begin{eqnarray}
S'+S_{fix}&=&\frac{1}{G_N} \int \mathcal{D}\bold{E} \mathcal{D}\bold{X}_{\hat{D}_T}(\bar{\tau}) \mathcal{D}\bold{X}_{L\hat{D}_T}(\bar{\tau})\mathcal{D}\bar{\tau}  \sqrt{\bar{G}} 
\Bigl(-\bar{R}+\frac{1}{4}\bar{F}'_{\bold{I}\bold{J}}\bar{F}'^{\bold{I}\bold{J}} 
\nonumber \\
&&+\frac{1}{4}\bar{\nabla}_\bold{I} \tilde{G}_{\bold{J}\bold{K}} \bar{\nabla}^\bold{I} \tilde{G}^{\bold{J}\bold{K}}
-\frac{1}{8}\bar{\nabla}_\bold{I} \tilde{G} \bar{\nabla}^\bold{I} \tilde{G}
\nonumber \\
&&-\frac{1}{4}(-\bar{R}+\frac{1}{4}\bar{F}'_{\bold{K}\bold{L}}\bar{F}'^{\bold{K}\bold{L}})
(\tilde{G}_{\bold{I}\bold{J}}\tilde{G}^{\bold{I}\bold{J}}-\frac{1}{2}\tilde{G}^2)
+(-\frac{1}{2}\bar{R}^{\bold{I}}_{\;\; \bold{J}}+\frac{1}{2}\bar{F}'^{\bold{I}\bold{K}}\bar{F}'_{\bold{J}\bold{K}})
\tilde{G}_{\bold{I}\bold{L}}\tilde{G}^{\bold{J}\bold{L}}
\nonumber \\
&&+(\frac{1}{2}\bar{R}^{\bold{I}\bold{J}}-\frac{1}{4}\bar{F}'^{\bold{I}\bold{K}}\bar{F}'^{\bold{J}}_{\;\;\;\; \bold{K}})
\tilde{G}_{\bold{I}\bold{J}}\tilde{G}
+(-\frac{1}{2}\bar{R}^{\bold{I}\bold{J}\bold{K}\bold{L}}+\frac{1}{4}\bar{F}'^{\bold{I}\bold{J}}\bar{F}'^{\bold{K}\bold{L}})
\tilde{G}_{\bold{I}\bold{K}}\tilde{G}_{\bold{J}\bold{L}} \Bigr). \label{Hfixedaction}
\end{eqnarray}

In order to obtain perturbative string amplitudes, we perform a derivative expansion of $\tilde{G}_{\bold{I}\bold{J}}$,
\begin{eqnarray}
&&\tilde{G}_{\bold{I}\bold{J}} \to \frac{1}{\alpha} \tilde{G}_{\bold{I}\bold{J}} \nonumber \\
&&\partial_{\bold{K}}\tilde{G}_{\bold{I}\bold{J}} \to \partial_{\bold{K}}\tilde{G}_{\bold{I}\bold{J}}\nonumber \\
&&\partial_{\bold{K}}\partial_{\bold{L}}\tilde{G}_{\bold{I}\bold{J}} \to \alpha \partial_{\bold{K}}\partial_{\bold{L}}\tilde{G}_{\bold{I}\bold{J}},
\end{eqnarray}
and take
\begin{equation}
\alpha =\beta \to 0,
\end{equation}
where $\alpha$ and $\beta$ are arbitrary constants in the solution (\ref{HSupersolution}). We normalize the fields as $\tilde{H}_{\bold{I}\bold{J}}:=Z_{\bold{I}\bold{J}} \tilde{G}_{\bold{I}\bold{J}}$, where $Z_{\bold{I}\bold{J}}:=\frac{1}{\sqrt{G_N}} 
\bar{G}^{\frac{1}{4}} 
(\bar{a}_\bold{I} \bar{a}_\bold{J})^{-\frac{1}{2}}$. $\bar{a}_{\bold{I}}$ represent the background metric as $\bar{G}_{\bold{I}\bold{J}}=\bar{a}_\bold{I} \delta_{\bold{I}\bold{J}}$, where $\bar{a}_d=2\zeta \bar{\rho}$, $\bar{a}_{(\mu \bar{\sigma} \bar{\theta})}=\frac{\hat{\bold{E}}^3}
{\bold{E}}$ and $\bar{a}_{(A \bar{\sigma} \bar{\theta}^-)}=\frac{\bar{e}^3}{\sqrt{\bar{h}}}$. Then, (\ref{Hfixedaction}) with appropriate boundary conditions  reduces to
\begin{equation}
S'+S_{fix} \to S_0 + S_2,
\end{equation}
where
\begin{equation}
S_0
=
\frac{1}{G_N} \int \mathcal{D}\bold{E} \mathcal{D}\bold{X}_{\hat{D}_T}(\bar{\tau}) \mathcal{D}\bold{X}_{L\hat{D}_T}(\bar{\tau})\mathcal{D}\bar{\tau}  \sqrt{\bar{G}} 
\Bigl(-\bar{R}+\frac{1}{4}\bar{F}'_{\bold{I}\bold{J}}\bar{F}'^{\bold{I}\bold{J}} \Bigr),
\end{equation}
and
\begin{equation}
S_2
=
\int \mathcal{D}\bold{E} \mathcal{D}\bold{X}_{\hat{D}_T}(\bar{\tau}) \mathcal{D}\bold{X}_{L\hat{D}_T}(\bar{\tau})\mathcal{D}\bar{\tau}
\frac{1}{8}\tilde{H}_{\bold{I}\bold{J}}H_{{\bold{I}\bold{J}};\bold{K}\bold{L}}\tilde{H}_{\bold{K}\bold{L}}.
\end{equation}

In the same way as in section 3, a part of the action 
\begin{equation}
\int \mathcal{D}\bold{E} \mathcal{D}\bold{X}_{\hat{D}_T}(\bar{\tau}) \mathcal{D}\bold{X}_{L\hat{D}_T}(\bar{\tau})\mathcal{D}\bar{\tau} \frac{1}{4}
\int_0^{2\pi}d\bar{\sigma}d\bar{\theta} \tilde{H}^{\bot}_{d(\mu \bar{\sigma} \bar{\theta})} 
H
\tilde{H}^{\bot}_{d(\mu \bar{\sigma} \bar{\theta})} \label{HSuperSecondOrderAction}
\end{equation}
with 

\begin{eqnarray}
H
&=&
-\frac{1}{2}\frac{1}{2\zeta \bar{\rho}}(\frac{\partial}{\partial \bar{\tau}})^2
-\frac{1}{2}\int_0^{2\pi} d \bar{\sigma} \int d \bar{\theta}  \frac{\bold{E}} {\hat{\bold{E}}^2} (\frac{\partial}{\partial \bold{X}_{\hat{D}_T}^{\mu}(\bar{\tau})})^2 
\nonumber \\
&&- \frac{1}{2}\frac{\bold{D}^2-9\bold{D}+20}{(2-\bold{D})^2}
\biggl( \int_0^{2\pi} d \bar{\sigma} \int d \bar{\theta}  \bar{\bold{E}}  \bold{X}_{\hat{D}_T}^{\mu}(\bar{\tau}) \tilde{\partial}_{z} \tilde{\bold{D}}_{\theta} \bold{X}_{\hat{D}_T \mu}(\bar{\tau})  \nonumber \\
&+& \int_0^{2\pi} d \bar{\sigma} \int d \bar{\theta}^-  \bar{\bold{E}}^L  \bold{X}_{L\hat{D}_T}^A(\bar{\tau}) \tilde{\partial}_{\bar{z}} \tilde{\bold{D}}_{\theta^-} \bold{X}_{L\hat{D}_T \, A}(\bar{\tau})\biggr) \label{HeteroH}
\nonumber \\
&&
\end{eqnarray}
decouples from the other modes. In (\ref{HSuperSecondOrderAction}), the term including $(\frac{\partial}{\partial \bold{X}_{L\hat{D}_T}^A(\bar{\tau})})^2 $ vanishes because  $\tilde{H}^{\bot}_{d(\mu \bar{\sigma} \bar{\theta})}$ needs to be proportional to $(\bold{X}_{L\hat{D}_T}^A(\bar{\tau}))^2= (\bar{\theta}^- \lambda_{\hat{D}_T}^A(\bar{\tau}))^2=0$ so as not to vanish.

In the following, we consider a sector that consists of local fluctuations in a sense of strings as 
\begin{equation}
\tilde{H}_{\bold{I}\bold{J}}=\int_0^{2\pi}  d \bar{\sigma} d\bar{\theta} \hat{\bold{E}}
 \tilde{h}_{\bold{I}\bold{J}}
(\bar{\bold{E}}( \bar{\tau}, \bar{\sigma}, \bar{\theta}), \bold{X}_{\hat{D}_T}( \bar{\tau}, \bar{\sigma}, \bar{\theta}),  \bold{X}_{L\hat{D}_T}( \bar{\tau}, \bar{\sigma}, \bar{\theta}), \bar{\tau})).
\end{equation}
Because we have
\begin{equation}
\int d\bar{\theta}' \frac{\bold{E}'} {\hat{\bold{E}'}^2}\left(\frac{\partial}{\partial \bold{X}_{\hat{D}_T}^{\mu}(\bar{\tau}, \bar{\sigma}', \bar{\theta}')} \right)^2 \tilde{H}^{\bot}_{d(\mu \bar{\sigma} \bar{\theta})}
=
\frac{\sqrt{\bar{h'}}}{\bar{e'}^{2}}
\left(\frac{\partial}{\partial X^{\mu}(\bar{\tau}, \bar{\sigma}')} \right)^2 \tilde{H}^{\bot}_{d(\mu \bar{\sigma} \bar{\theta})}, 
\end{equation}
 as in section 4, 
(\ref{HSuperSecondOrderAction}) can be simplified 
with
\begin{eqnarray}
&&H(-i\frac{\partial}{\partial \bar{\tau}}, -i\frac{1}{\bar{e}}\frac{\partial}{\partial X}, \bold{X}_{\hat{D}_T}(\bar{\tau}), \bold{X}_{L\hat{D}_T}(\bar{\tau}), \bar{\bold{E}}) \nonumber \\
&=&
\frac{1}{2}\frac{1}{2\zeta \bar{\rho}}(-i\frac{\partial}{\partial \bar{\tau}})^2 +\int d\bar{\sigma} \sqrt{\bar{h}} \frac{1}{2}(-i\frac{1}{\bar{e}}\frac{\partial}{\partial X})^2   -\int d\bar{\sigma} d \bar{\theta}  \bar{\bold{E}}  \frac{1}{2}\bold{X}_{\hat{D}_T}^{\mu}(\bar{\tau}) \tilde{\partial}_{z} \tilde{\bold{D}}_{\theta} \bold{X}_{\hat{D}_T \mu}(\bar{\tau})
\nonumber \\
&&-\int d\bar{\sigma} d \bar{\theta}^-  \bar{\bold{E}}^L  \frac{1}{2}\bold{X}_{L\hat{D}_T}^A(\bar{\tau}) \tilde{\partial}_{\bar{z}} \tilde{\bold{D}}_{\theta^-} \bold{X}_{L\hat{D}_T \, A}(\bar{\tau}),
\end{eqnarray}
where we have taken $\bold{D} \to \infty$.
By adding to (\ref{HSuperSecondOrderAction}), a formula similar to the bosonic case

\begin{eqnarray}
0
&=&
\int \mathcal{D}\bold{E} \mathcal{D}\bold{X}_{\hat{D}_T}(\bar{\tau}) \mathcal{D}\bold{X}_{L\hat{D}_T}(\bar{\tau})\mathcal{D}\bar{\tau} \frac{1}{4}
\int_0^{2\pi}d\bar{\sigma}'d\bar{\theta}' \tilde{H}^{\bot}_{d(\mu \bar{\sigma}' \bar{\theta}')} ( \int_0^{2\pi} d \bar{\sigma}
\bar{n}^{\bar{\sigma}}
\partial_{\bar{\sigma}} X^{\mu}  \frac{\partial}{\partial X^{\mu}})
\tilde{H}^{\bot}_{d(\mu \bar{\sigma}' \bar{\theta}')},  \nonumber \\
&&
\end{eqnarray}
and
\begin{equation}
0=\int \mathcal{D}\bold{E}    \mathcal{D}\bold{X}_{\hat{D}_T}(\bar{\tau}) \mathcal{D}\bold{X}_{L\hat{D}_T}(\bar{\tau}) \mathcal{D}\bar{\tau} 
\frac{1}{4}
\int_0^{2\pi}d\bar{\sigma}'d\bar{\theta}' \tilde{H}^{\bot}_{d(\mu \bar{\sigma}' \bar{\theta}')} 
\int_0^{2\pi} d \bar{\sigma} \bar{E}  \frac{-i}{2} \bar{n} \bar{\chi}_z \bar{E}_{\; \bar{z}}^0 \psi^{\mu}(-i \frac{1}{\bar{e}}\frac{\partial}{\partial X^{\mu}})
\tilde{H}^{\bot}_{d(\mu \bar{\sigma}' \bar{\theta}')}, 
\label{HSuperZero}
\end{equation}
we obtain (\ref{HSuperSecondOrderAction}) 
with
\begin{eqnarray}
&&H(-i\frac{\partial}{\partial \bar{\tau}}, -i\frac{1}{\bar{e}}\frac{\partial}{\partial X}, \bold{X}_{\hat{D}_T}(\bar{\tau}), \lambda_{\hat{D}_T}(\bar{\tau}), \bar{\bold{E}}) \nonumber \\
&=&
\frac{1}{2}\frac{1}{2\zeta \bar{\rho}}(-i\frac{\partial}{\partial \bar{\tau}})^2
+\int d\bar{\sigma} \Biggl( \sqrt{\bar{h}} \left(\frac{1}{2}( -i\frac{1}{\bar{e}}\frac{\partial}{\partial X})^2  -\frac{i}{2} \bar{n} \bar{\chi}_z \bar{E}_{\; \bar{z}}^0  \psi_{\mu}( -i\frac{1}{\bar{e}}\frac{\partial}{\partial X})\right)
+i \bar{e} \bar{n}^{\bar{\sigma}} \partial_{\bar{\sigma}} X_{\mu} ( -i\frac{1}{\bar{e}}\frac{\partial}{\partial X})\Biggr) \nonumber \\
&&-\int d\bar{\sigma} d \bar{\theta}  \bar{\bold{E}}  \frac{1}{2}\bold{X}_{\hat{D}_T}^{\mu}(\bar{\tau}) \tilde{\partial}_{z} \tilde{\bold{D}}_{\theta} \bold{X}_{\hat{D}_T \mu}(\bar{\tau})
+\int d\bar{\sigma} \sqrt{\bar{h}} \frac{1}{2}
\lambda_{\hat{D}_T}^A(\bar{\tau}) \bar{E}_{\bar{z}}^1 \partial_{\bar{\sigma}} \lambda_{\hat{D}_T A}(\bar{\tau}),
\label{HSuperbosonicHamiltonian}
\end{eqnarray}
where we have used (\ref{Hquad4}). (\ref{HSuperZero}) is true because the integrand of the right hand side is a total derivative with respect to $X^{\mu}$.

The propagator for $\tilde{H}^{\bot}_{d(\mu \bar{\sigma} \bar{\theta})}$;
\begin{eqnarray}
&&\Delta_F(\bar{\bold{E}}, \bar{\tau}, \bold{X}_{\hat{D}_T}(\bar{\tau}), \lambda_{\hat{D}_T}(\bar{\tau}); \; \bar{\bold{E}},' \bar{\tau},'  \bold{X}'_{\hat{D}_T}(\bar{\tau}'), \lambda'_{\hat{D}_T}(\bar{\tau}')) \nonumber \\
&=&<\tilde{H}^{\bot}_{d(\mu \bar{\sigma} \bar{\theta})}
(\bar{\bold{E}}, \bar{\tau}, \bold{X}_{\hat{D}_T}(\bar{\tau}), \lambda_{\hat{D}_T}(\bar{\tau}))
\tilde{H}^{\bot}_{d(\mu \bar{\sigma} \bar{\theta})}
(\bar{\bold{E}},' \bar{\tau},'  \bold{X}'_{\hat{D}_T}(\bar{\tau}'), \lambda'_{\hat{D}_T}(\bar{\tau}'))>
\end{eqnarray}
satisfies
\begin{eqnarray}
&&
H(-i\frac{\partial}{\partial \bar{\tau}}, -i\frac{1}{\bar{e}}\frac{\partial}{\partial X}, \bold{X}_{\hat{D}_T}(\bar{\tau}), \lambda_{\hat{D}_T}(\bar{\tau}), \bar{\bold{E}})
\Delta_F(\bar{\bold{E}}, \bar{\tau}, \bold{X}_{\hat{D}_T}(\bar{\tau}), \lambda_{\hat{D}_T}(\bar{\tau}); \; \bar{\bold{E}},' \bar{\tau}',  \bold{X}'_{\hat{D}_T}(\bar{\tau}'), \lambda'_{\hat{D}_T}(\bar{\tau}')) \nonumber \\
&=&
\delta(\bar{\bold{E}}-\bar{\bold{E}}')\delta(\bar{\tau}-\bar{\tau}')\delta(\bold{X}_{\hat{D}_T}(\bar{\tau})-\bold{X}'_{\hat{D}_T}(\bar{\tau}')) \delta(\lambda_{\hat{D}_T}(\bar{\tau})-\lambda'_{\hat{D}_T}(\bar{\tau}')). 
\label{HPropagatorDefinition}
\end{eqnarray}

In order to obtain a Schwinger representation of the propagator, we use the operator formalism $(\hat{\bar{\bold{E}}}, \hat{\bar{\tau}}, \hat{\bold{X}}_{\hat{D}_T}(\hat{\bar{\tau}}), \hat{\lambda}_{\hat{D}_T}(\hat{\bar{\tau}}))$ of the first quantization. 
The eigen state for $(\hat{\bar{\bold{E}}}, \hat{\bar{\tau}}, \hat{X})$ is given by $|\bar{\bold{E}}, \bar{\tau}, X>$. The conjugate momentum is written as $(\hat{\bold{p}}_{\bar{\bold{E}}}, \hat{p}_{\bar{\tau}}, \hat{p}_{X})$. The Majorana fermions $\psi^{\mu}$ and $\lambda_{\hat{D}_T}^A$ are self-conjugate. Renormalized operators $\hat{\tilde{\psi}}^{\mu}:=\sqrt{\bar{E}_z^0}\hat{\psi}^{\mu}$ and $\hat{\tilde{\lambda}}_{\hat{D}_T}^{A}:=\sqrt{\bar{E}_{\bar{z}}^0}\hat{\lambda}_{\hat{D}_T}^{A}$ satisfy $\{ \hat{\tilde{\psi}}^{\mu}(\bar{\sigma}), \hat{\tilde{\psi}}^{\nu}(\bar{\sigma}') \}= \frac{1}{\bar{E}} \eta^{\mu \nu} \delta(\bar{\sigma}-\bar{\sigma}')$ and $\{ \hat{\tilde{\lambda}}_{\hat{D}_T}^{A}(\bar{\sigma}), \hat{\tilde{\lambda}}_{\hat{D}_T}^{B}(\bar{\sigma}') \}= \frac{1}{\bar{E}} \delta^{A B} \delta(\bar{\sigma}-\bar{\sigma}')$, respectively.  By defining creation and annihilation operators for $\tilde{\psi}^{\mu}$ as $\hat{\tilde{\psi}}^{\hat{\mu} \dagger}:= \frac{1}{\sqrt{2}}(\hat{\tilde{\psi}}^{\hat{\mu}}-i\hat{\tilde{\psi}}^{\hat{\mu}+\frac{d}{2}})$ and $\hat{\tilde{\psi}}^{\hat{\mu}}:= \frac{1}{\sqrt{2}}(\hat{\tilde{\psi}}^{\hat{\mu}}+i\hat{\tilde{\psi}}^{\hat{\mu}+\frac{d}{2}})$ where $\hat{\mu}=0, \cdots \frac{d}{2}-1$,
one obtains an algebra $\{ \hat{\tilde{\psi}}^{\hat{\mu}}(\bar{\sigma}), \hat{\tilde{\psi}}^{\hat{\nu} \dagger}(\bar{\sigma}') \}= \frac{1}{\bar{E}} \eta^{\hat{\mu} \hat{\nu}} \delta(\bar{\sigma}-\bar{\sigma}')$, $\{ \hat{\tilde{\psi}}^{\hat{\mu}}(\bar{\sigma}), \hat{\tilde{\psi}}^{\hat{\nu}}(\bar{\sigma}') \}= 0$, and $\{ \hat{\tilde{\psi}}^{\hat{\mu} \dagger}(\bar{\sigma}), \hat{\tilde{\psi}}^{\hat{\nu} \dagger}(\bar{\sigma}') \}=0$. The vacuum $|0>$ for this algebra is defined by $\hat{\tilde{\psi}}^{\hat{\mu}}(\bar{\sigma}) |0>=0$. The eigen state $|\tilde{\psi}>$, which satisfies $\hat{\tilde{\psi}}^{\hat{\mu}}(\bar{\sigma}) |\tilde{\psi}>= \tilde{\psi}^{\hat{\mu}}(\bar{\sigma}) |\tilde{\psi}>$, is given by 
$
e^{-\tilde{\psi} \cdot \hat{\tilde{\psi}}^{\dagger}} |0>
=
e^{- \int d\bar{\sigma} \bar{E} \tilde{\psi}_{\hat{\mu}}(\bar{\sigma}) \hat{\tilde{\psi}}^{\hat{\mu} \dagger}(\bar{\sigma})} |0>$. Then, the inner product is given by $<\tilde{\psi} | \tilde{\psi}'>=e^{\tilde{\psi}^{\dagger} \cdot \tilde{\psi}'}$, whereas the completeness relation is
$\int \mathcal{D}\tilde{\psi}^{\dagger} \mathcal{D}\tilde{\psi} |\tilde{\psi}>e^{-\tilde{\psi}^{\dagger} \cdot \tilde{\psi}}<\tilde{\psi}|=1$. The same is applied to $\tilde{\lambda}_{\hat{D}_T}^{A}(\bar{\tau})$.

Because (\ref{HPropagatorDefinition}) means that $\Delta_F$ is an inverse of $H$, $\Delta_F$ can be expressed by a matrix element of the operator $\hat{H}^{-1}$ as
\begin{eqnarray}
&&\Delta_F(\bar{\bold{E}}, \bar{\tau}, \bold{X}_{\hat{D}_T}(\bar{\tau}), \lambda_{\hat{D}_T}(\bar{\tau}); \; \bar{\bold{E}},' \bar{\tau},' \bold{X}'_{\hat{D}_T}(\bar{\tau}'), \lambda'_{\hat{D}_T}(\bar{\tau}'))
\nonumber \\
&=&
<\bar{\bold{E}}, \bar{\tau}, \bold{X}_{\hat{D}_T}(\bar{\tau}), \lambda_{\hat{D}_T}(\bar{\tau})| \hat{H}^{-1}(\hat{p}_{\bar{\tau}}, \hat{p}_{X}, \hat{\bold{X}}_{\hat{D}_T}(\bar{\tau}), \hat{\lambda}_{\hat{D}_T}(\bar{\tau}), \hat{\bar{\bold{E}}}) |\bar{\bold{E}},' \bar{\tau},' \bold{X}_{\hat{D}_T}'(\bar{\tau}'), \lambda'_{\hat{D}_T}(\bar{\tau}')>. \nonumber \\
&&
\label{HSuperInverseH2}
\end{eqnarray}
(\ref{IntegralFormula}) implies that
\begin{eqnarray}
&&\Delta_F(\bar{\bold{E}}, \bar{\tau}, \bold{X}_{\hat{D}_T}(\bar{\tau}), \lambda_{\hat{D}_T}(\bar{\tau}); \; \bar{\bold{E}},' \bar{\tau},' \bold{X}'_{\hat{D}_T}(\bar{\tau}'), \lambda'_{\hat{D}_T}(\bar{\tau}')) \nonumber \\
&=&
\int _0^{\infty} dT <\bar{\bold{E}}, \bar{\tau}, \bold{X}_{\hat{D}_T}(\bar{\tau}), \lambda_{\hat{D}_T}(\bar{\tau})|  e^{-T\hat{H}} |\bar{\bold{E}},' \bar{\tau},' \bold{X}_{\hat{D}_T}'(\bar{\tau}'), \lambda'_{\hat{D}_T}(\bar{\tau}')>.
\end{eqnarray}

In order to define two-point correlation functions that is invariant under the general coordinate transformations in the superstring geometry, we define in and out states as
\begin{eqnarray}
||\bold{X}_{\hat{D}_T i}, \lambda_{\hat{D}_T i} \,|\,\bold{E}_f, ; \bold{E}_i>_{in}
&:=& \int_{\bold{E}_i}^{\bold{E}_f} \mathcal{D}\bold{E}'|\bar{\bold{E}},'  \bar{\tau}=-\infty, \bold{X}_{\hat{D}_T i}, \lambda_{\hat{D}_T i} > \nonumber \\
<\bold{X}_{\hat{D}_T f}, \lambda_{\hat{D}_T f} \,|\,\bold{E}_f, ; \bold{E}_i||_{out}&:=& \int_{\bold{E}_i}^{\bold{E}_f} \mathcal{D} \bold{E} <\bar{\bold{E}}, \bar{\tau}=\infty, \bold{X}_{\hat{D}_T f}, \lambda_{\hat{D}_T f}|,
\end{eqnarray}
where $\bold{E}_i$ and $\bold{E}_f$ represent the super vierbeins of the supercylinders at $\bar{\tau}=\pm \infty$, respectively. When we insert asymptotic states, we integrate out $\bold{X}_{\hat{D}_T f}$, $\bold{X}_{\hat{D}_T i}$, $\lambda_{\hat{D}_T f}$, $\lambda_{\hat{D}_T i}$, $\bold{E}_f$ and $\bold{E}_i$ in the two-point correlation function for these states;
\begin{eqnarray}
&&\Delta_F(\bold{X}_{\hat{D}_T f}, \lambda_{\hat{D}_T f}; \bold{X}_{\hat{D}_T i}, \lambda_{\hat{D}_T i}|\bold{E}_f, ; \bold{E}_i) \nonumber \\
&=&
\int_{\bold{E}_i}^{\bold{E}_f} \mathcal{D}\bold{E}
\int_{\bold{E}_i}^{\bold{E}_f} \mathcal{D} \bold{E}' 
<\tilde{H}^{\bot}_{d(\mu \bar{\sigma} \bar{\theta})}
(\bar{\bold{E}}, \bar{\tau}=\infty, \bold{X}_{\hat{D}_T f}, \lambda_{\hat{D}_T f})
\nonumber \\
&&\tilde{H}^{\bot}_{d(\mu \bar{\sigma} \bar{\theta})}
(\bar{\bold{E}},'  \bar{\tau}=-\infty, \bold{X}_{\hat{D}_T i}, \lambda_{\hat{D}_T i} )>.
\end{eqnarray}
By inserting 
\begin{eqnarray}
1&=&
\int d\bar{\bold{E}}_{m} d\bar{\tau}_m d\bold{X}_{\hat{D}_T m}(\bar{\tau}_m)  d\lambda_{\hat{D}_T m}(\bar{\tau}_m) \nonumber \\
&&|\bar{\bold{E}}_{m}, \bar{\tau}_m, \bold{X}_{\hat{D}_T m}(\bar{\tau}_m), \lambda_{\hat{D}_T m}(\bar{\tau}_m) > 
e^{-\tilde{\psi}^{\dagger}_m \cdot \tilde{\psi}_m
-\tilde{\lambda}^{\dagger}_{\hat{D}_T m} \cdot \tilde{\lambda}_{\hat{D}_T m}}
<\bar{\bold{E}}_{m}, \bar{\tau}_m, \bold{X}_{\hat{D}_T m}(\bar{\tau}_m), \lambda_{\hat{D}_T m}(\bar{\tau}_m)  |
\nonumber \\ 
1&=&
\int dp_{\bar{\tau}}^i dp_{X}^i
| p_{\bar{\tau}}^i, p_{X}^i>
< p_{\bar{\tau}}^i, p_{X}^i|.
\end{eqnarray}
This can be written as\footnote{The correlation function is zero if $\bold{E}_i$ and $\bold{E}_f$ of the in state do not coincide with those of the out states, because of the delta functions in the seventh line.}
\begin{eqnarray}
&&\Delta_F(\bold{X}_{\hat{D}_T f}, \lambda_{\hat{D}_T f}; \bold{X}_{\hat{D}_T i}, \lambda_{\hat{D}_T i}|\bold{E}_f, ; \bold{E}_i) \nonumber \\
&:=&\int _0^{\infty} dT <\bold{X}_{\hat{D}_T f}, \lambda_{\hat{D}_T f} \,|\,\bold{E}_f, ; \bold{E}_i||_{out}  e^{-T\hat{H}} ||\bold{X}_{\hat{D}_T i}, \lambda_{\hat{D}_T i} \,|\,\bold{E}_f, ; \bold{E}_i>_{in}\nonumber \\
&=&
\int _0^{\infty} dT \lim_{N \to \infty}
\int_{\bold{E}_i}^{\bold{E}_f}\mathcal{D} \bold{E}
\int_{\bold{E}_i}^{\bold{E}_f}\mathcal{D} \bold{E}'
\prod_{m=1}^N \prod_{i=0}^N
\int d\bar{\bold{E}}_{m} d\bar{\tau}_m d\bold{X}_{\hat{D}_Tm}(\bar{\tau}_m) d\lambda_{\hat{D}_T m}(\bar{\tau}_m)   \nonumber \\
&&
e^{-\tilde{\psi}^{\dagger}_m \cdot \tilde{\psi}_m
-\tilde{\lambda}^{\dagger}_{\hat{D}_T m} \cdot \tilde{\lambda}_{\hat{D}_T m}}
<\bar{\bold{E}}_{i+1}, \bar{\tau}_{i+1}, \bold{X}_{\hat{D}_T i+1}(\bar{\tau}_{i+1}), \lambda_{\hat{D}_T i+1}(\bar{\tau}_{i+1})| e^{-\frac{1}{N}T \hat{H}} |\bar{\bold{E}}_{i}, \bar{\tau}_i, \bold{X}_{\hat{D}_T i}(\bar{\tau}_i), \lambda_{\hat{D}_T i}(\bar{\tau}_i)>
\nonumber \\
&=&
\int _0^{\infty} dT_0 \lim_{N \to \infty} \int d T_{N+1} 
\int_{\bold{E}_i}^{\bold{E}_f}\mathcal{D} \bold{E}
\int_{\bold{E}_i}^{\bold{E}_f}\mathcal{D} \bold{E}'
\prod_{m=1}^N \prod_{i=0}^N
\int d T_m  d\bar{\bold{E}}_{m} d\bar{\tau}_m d\bold{X}_{\hat{D}_Tm}(\bar{\tau}_m) d\lambda_{\hat{D}_T m} (\bar{\tau}_m)
 \nonumber \\
&& 
e^{-\tilde{\psi}^{\dagger}_m \cdot \tilde{\psi}_m
-\tilde{\lambda}^{\dagger}_{\hat{D}_T m} \cdot \tilde{\lambda}_{\hat{D}_T m}}<\bar{\tau}_{i+1}, \bold{X}_{\hat{D}_T i+1}(\bar{\tau}_{i+1}), \lambda_{\hat{D}_T i+1}(\bar{\tau}_{i+1})| e^{-\frac{1}{N}T_i \hat{H}}  |\bar{\tau}_i, \bold{X}_{\hat{D}_T i}(\bar{\tau}_i), \lambda_{\hat{D}_T i}(\bar{\tau}_i)> 
\nonumber \\
&&\delta(T_{i}-T_{i+1}) \delta(\bar{\bold{E}}_{i}-\bar{\bold{E}}_{i+1}))
\nonumber \\
&=&\int _0^{\infty} dT_0 \lim_{N \to \infty} \int d T_{N+1}
\int_{\bold{E}_i}^{\bold{E}_f}\mathcal{D} \bold{E}
\prod_{m=1}^N \prod_{i=0}^N
\int d T_m   d\bar{\tau}_m d\bold{X}_{\hat{D}_Tm}(\bar{\tau}_m) d\lambda_{\hat{D}_T m} (\bar{\tau}_m)  e^{-\tilde{\psi}^{\dagger}_m \cdot \tilde{\psi}_m
-\tilde{\lambda}^{\dagger}_{\hat{D}_T m} \cdot \tilde{\lambda}_{\hat{D}_T m}}
\nonumber \\
&&
\int  dp_{\bar{\tau}}^i dp_{X}^i
<\bar{\tau}_{i+1}, X_{i+1}| 
p_{\bar{\tau}}^i, p_{X}^i>
\nonumber \\
&&
<p_{\bar{\tau}}^i, p_{X}^i|
<\tilde{\psi}_{i+1}, \tilde{\lambda}_{\hat{D}_T i+1}|
e^{-\frac{1}{N}T_i \hat{H}}
|\tilde{\psi}_{i}, \tilde{\lambda}_{\hat{D}_T i}> |\bar{\tau}_i, X_i> \delta(T_{i}-T_{i+1}) 
\nonumber \\
&=&
\int _0^{\infty} dT_0 \lim_{N \to \infty} \int d T_{N+1} 
\int_{\bold{E}_i}^{\bold{E}_f}\mathcal{D} \bold{E}
\prod_{m=1}^N \prod_{i=0}^N
\int d T_m d\bar{\tau}_m d\bold{X}_{\hat{D}_Tm}(\bar{\tau}_m) d\lambda_{\hat{D}_T m}(\bar{\tau}_m)   e^{-\tilde{\psi}^{\dagger}_m \cdot \tilde{\psi}_m
-\tilde{\lambda}^{\dagger}_{\hat{D}_T m} \cdot \tilde{\lambda}_{\hat{D}_T m}}
\nonumber \\
&&
\int dp_{\bar{\tau}}^i dp_{X}^i
 e^{-\frac{1}{N}T_i H (p_{\bar{\tau}}^i, p_{X}^i, \bold{X}_{\hat{D}_T i}(\bar{\tau}_i), \lambda_{\hat{D}_T i}(\bar{\tau}_i), \bar{\bold{E}})}
e^{\tilde{\psi}^{\dagger}_{i+1} \cdot \tilde{\psi}_i
+\tilde{\lambda}^{\dagger}_{\hat{D}_T i+1} \cdot \tilde{\lambda}_{\hat{D}_T i}}
\delta(T_{i}-T_{i+1}) 
\nonumber \\
&&
e^{i(p_{\bar{\tau}}^i(\bar{\tau}_{i+1}-\bar{\tau}_{i})+p_{X}^i\cdot(X_{i+1}-X_{i}))}
\nonumber \\
&=&
\int _0^{\infty} dT_0 \lim_{N \to \infty} \int d T_{N+1}
\int_{\bold{E}_i}^{\bold{E}_f}\mathcal{D} \bold{E}
\prod_{m=1}^N \prod_{i=0}^N
\int d T_m  d\bar{\tau}_m d\bold{X}_{\hat{D}_Tm}(\bar{\tau}_m) d\lambda_{\hat{D}_T m} (\bar{\tau}_m)  
\int dp_{T_{i}}  dp_{\bar{\tau}}^i dp_{X}^i \nonumber \\
&&
\exp \Biggl(- \sum_{i=0}^{N} \triangle t \Bigl(-ip_{T_{i}} \frac{T_{i+1}-T_{i}}{\Delta t} 
+\tilde{\psi}^{\dagger}_{i+1} \cdot \frac{\tilde{\psi}_{i+1}-\tilde{\psi}_{i}}{\triangle t}
+\tilde{\lambda}^{\dagger}_{\hat{D}_T i+1} \cdot \frac{\tilde{\lambda}_{\hat{D}_T i+1}-\tilde{\lambda}_{\hat{D}_T i}}{\triangle t}\nonumber \\
&&
-ip_{\bar{\tau}}^{i} \frac{\bar{\tau}_{i+1}-\bar{\tau}_{i}}{\triangle t} -ip_{X}^{i}\cdot \frac{X_{i+1}-X_{i}}{\triangle t}
+T_i H(p_{\bar{\tau}}^{i}, p_{X}^{i}, \bold{X}_{\hat{D}_T i}(\bar{\tau}_{i}), \lambda_{\hat{D}_T i}(\bar{\tau}_{i}), \bar{\bold{E}}) \Bigr)\Biggr) \nonumber \\
&& e^{\tilde{\psi}^{\dagger}_{N+1} \cdot \tilde{\psi}_{N+1}
+\tilde{\lambda}^{\dagger}_{\hat{D}_T N+1} \cdot \tilde{\lambda}_{\hat{D}_T N+1}}
\nonumber \\
&=&
\int^{\bold{E}_f, \infty, \bold{X}_{\hat{D}_T f}, \lambda_{\hat{D}_T f} }_{\bold{E}_i, -\infty, \bold{X}_{\hat{D}_T i}, \lambda_{\hat{D}_T i} } 
\mathcal{D} T  \mathcal{D}\bold{E} \mathcal{D}\bar{\tau} \mathcal{D}\bold{X}_{\hat{D}_T}(\bar{\tau}) \mathcal{D}\lambda_{\hat{D}_T}(\bar{\tau}) 
\int 
\mathcal{D} p_T
\mathcal{D}p_{\bar{\tau}}  \mathcal{D}p_{X} 
\nonumber \\
&&e^{- \int_{-\infty}^{\infty} dt (-ip_{T} \cdot \frac{d}{dt} T  
+\tilde{\psi}^{\dagger} \cdot \frac{d}{dt}\tilde{\psi}
+\tilde{\lambda}_{\hat{D}_T}^{\dagger} \cdot \frac{d}{dt}\tilde{\lambda}_{\hat{D}_T}
-ip_{\bar{\tau}} \frac{d}{dt} \bar{\tau}-ip_{X} \cdot \frac{d}{dt} X+TH(p_{\bar{\tau}}, p_{X}, \bold{X}_{\hat{D}_T}(\bar{\tau}), \lambda_{\hat{D}_T}(\bar{\tau}), \bar{\bold{E}}))},
\end{eqnarray}
where $\bar{\bold{E}}_{\; 0}=\bar{\bold{E}}'$, $\bar{\tau}_0=-\infty$, $\bold{X}_{\hat{D}_T \; 0}=\bold{X}_{\hat{D}_T i}$, $\lambda_{\hat{D}_T \; 0}=\lambda_{\hat{D}_T i}$, $\bar{\bold{E}}_{N+1}=\bar{\bold{E}}$, $\bar{\tau}_{N+1}=\infty$, $\bold{X}_{\hat{D}_T N+1}=\bold{X}_{\hat{D}_T f}$, and $\lambda_{\hat{D}_T \; N+1}=\lambda_{\hat{D}_T f}$. $p_{X} \cdot \frac{d}{dt}X=\int d\bar{\sigma} \bar{e} p_{X}^{\mu} \frac{d}{dt} X_{\mu}$ and $\Delta t=\frac{1}{\sqrt{N}}$ as in the bosonic case. 
A trajectory of points $[\bar{{\boldsymbol \Sigma}}, \bold{X}_{\hat{D}_T}(\bar{\tau}), \lambda_{\hat{D}_T}(\bar{\tau}), \bar{\tau}]$ is necessarily continuous in $\mathfrak{M}_{D_T}$ so that the kernel $<\bar{\bold{E}}_{i+1}, \bar{\tau}_{i+1}, \bold{X}_{\hat{D}_T i+1}(\bar{\tau}_{i+1}), \lambda_{\hat{D}_T i+1}(\bar{\tau}_{i+1})| e^{-\frac{1}{N}T_i \hat{H}}  |\bar{\bold{E}}_{i}, \bar{\tau}_i, \bold{X}_{\hat{D}_T i}(\bar{\tau}_{i}), \lambda_{\hat{D}_T i}(\bar{\tau}_{i})> $ in the fourth line is non-zero when $N \to \infty$. If we integrate out $p_{\bar{\tau}}(t)$ and $p_{X}(t)$ by using the relation of the ADM formalism, the relation between  $\tilde{\psi}^{\mu}$ and $\psi^{\mu}$ and the relation between $\tilde{\lambda}_{\hat{D}_T}^A$ and $\lambda_{\hat{D}_T}^A$, we obtain
\begin{eqnarray}
&&\Delta_F(\bold{X}_{\hat{D}_T f}, \lambda_{\hat{D}_T f}; \bold{X}_{\hat{D}_T i}, \lambda_{\hat{D}_T i}|\bold{E}_f, ; \bold{E}_i) \nonumber \\
&=&
\int^{\bold{E}_f, \infty, \bold{X}_{\hat{D}_T f}, \lambda_{\hat{D}_T f} }_{\bold{E}_i, -\infty, \bold{X}_{\hat{D}_T i}, \lambda_{\hat{D}_T i} } 
\mathcal{D} T 
\mathcal{D}\bold{E} \mathcal{D}\bar{\tau} \mathcal{D}\bold{X}_{\hat{D}_T}(\bar{\tau}) \mathcal{D}\bold\lambda_{\hat{D}_T}(\bar{\tau}) \int 
\mathcal{D} p_T
\nonumber \\
&& \exp \Biggl(- \int_{-\infty}^{\infty} dt \Bigl(-i p_{T}(t) \frac{d}{dt} T(t)  +\zeta \bar{\rho}\frac{1}{T(t)}(\frac{d \bar{\tau}(t)}{dt})^2\nonumber \\
&&+\int d\bar{\sigma} \sqrt{\bar{h}} T(t) \biggl( \frac{1}{2 \bar{n}^2}(\frac{1}{T(t)}\frac{\partial}{\partial t} X^{\mu} -\bar{n}^{\bar{\sigma}} \partial_{\bar{\sigma}} X^{\mu}
+\frac{1}{2}\bar{n}^2  \bar{E}_{\bar{z}}^0 \psi_{\mu} \bar{\chi}_z )^2 
 +\frac{1}{2}\frac{1}{T(t)} \psi^{\mu} \bar{E}^{0}_{z} \frac{\partial}{\partial t} \psi_{\mu} \nonumber \\
&&+\frac{1}{2}\frac{1}{T(t)} \lambda_{\hat{D}_T}^{A} \bar{E}^{0}_{\bar{z}} \frac{\partial}{\partial t} \lambda_{\hat{D}_TA} \biggr) 
-\int d\bar{\sigma} d \bar{\theta}  \bar{\bold{E}} T(t) \frac{1}{2}\bold{X}_{\hat{D}_T}^{\mu}(\bar{\tau}) \tilde{\partial}_{z} \tilde{\bold{D}}_{\theta} \bold{X}_{\hat{D}_T \mu}(\bar{\tau})
\nonumber \\
&+&\int  d\bar{\sigma} \sqrt{\bar{h}} T(t) \frac{1}{2}
\lambda_{\hat{D}_T}^A(\bar{\tau}) \bar{E}_{\bar{z}}^1 \partial_{\bar{\sigma}} \lambda_{\hat{D}_TA}(\bar{\tau})
\Bigr) \Biggr)
\nonumber \\
&=&
\int^{\bold{E}_f, \infty, \bold{X}_{\hat{D}_T f}, \lambda_{\hat{D}_T f} }_{\bold{E}_i, -\infty, \bold{X}_{\hat{D}_T i}, \lambda_{\hat{D}_T i} } 
\mathcal{D} T 
\mathcal{D}\bold{E} \mathcal{D}\bar{\tau} \mathcal{D}\bold{X}_{\hat{D}_T}(\bar{\tau}) \mathcal{D}\lambda_{\hat{D}_T}(\bar{\tau}) \int 
\mathcal{D} p_T
\exp \Biggl(- \int_{-\infty}^{\infty} dt \Bigl(-i p_{T}(t) \frac{d}{dt} T(t)
\nonumber \\
&&  +\zeta \bar{\rho}\frac{1}{T(t)}(\frac{d \bar{\tau}(t)}{dt})^2
+\int d\bar{\sigma} d \bar{\theta}  \bar{\bold{E}} T(t) \frac{1}{2} \bar{\partial}'_{z} \bold{X}_{\hat{D}_T}^{\mu}(\bar{\tau})  \bar{\bold{D}}'_{\theta} \bold{X}_{\hat{D}_T \mu}(\bar{\tau})
+\int d\bar{\sigma} \sqrt{\bar{h}} T(t) \frac{1}{2}
\lambda_{\hat{D}_T}^A(\bar{\tau}) \bar{\partial}'_{\bar{z}} \lambda_{\hat{D}_TA}(\bar{\tau})
\Bigr) \Biggr). \nonumber \\
&&
\end{eqnarray}
When the last equality is obtained, we use (\ref{Hquad1}) and (\ref{Hquad2}).
In the last line, $\bar{\bold{D}} '_{\theta}$, $\bar{\partial}'_{z}$ and $\bar{\partial}'_{\bar{z}}$ are given by replacing $\frac{\partial}{\partial \bar{\tau}}$ with $\frac{1}{T(t)}\frac{\partial}{\partial t}$ in $\bar{\bold{D}}_{\theta}$, $\bar{\partial}_{z}$ and $\bar{\partial}_{\bar{z}}$, respectively. 
The path integral is defined over all possible trajectories with fixed boundary values, on the heterotic superstring manifold $\mathfrak{M}_{D_T}$.

By inserting
$\int \mathcal{D}c \mathcal{D}b
e^{\int_0^{1} dt \left(\frac{d b(t)}{dt} \frac{d c(t)}{dt}\right)
},$
where $b(t)$ and $c(t)$ are bc ghosts, we obtain 
\begin{eqnarray}
&&\Delta_F(\bold{X}_{\hat{D}_T f}, \lambda_{\hat{D}_T f}; \bold{X}_{\hat{D}_T i}, \lambda_{\hat{D}_T i}|\bold{E}_f, ; \bold{E}_i) 
\nonumber \\
&=&
\bold{Z}_0\int^{\bold{E}_f, \infty, \bold{X}_{\hat{D}_T f}, \lambda_{\hat{D}_T f} }_{\bold{E}_i, -\infty, \bold{X}_{\hat{D}_T i}, \lambda_{\hat{D}_T i} } 
\mathcal{D} T 
\mathcal{D}\bold{E} \mathcal{D}\bar{\tau} \mathcal{D}\bold{X}_{\hat{D}_T}(\bar{\tau}) \mathcal{D}\lambda_{\hat{D}_T}(\bar{\tau})
\mathcal{D}c \mathcal{D}b
\int 
\mathcal{D} p_T
\exp \Biggl(- \int_{-\infty}^{\infty} dt \Bigl(-i p_{T}(t) \frac{d}{dt} T(t) 
\nonumber \\
&&
+\frac{d b(t)}{dt} \frac{d (T(t) c(t))}{dt}
+\zeta \bar{\rho}\frac{1}{T(t)}(\frac{d \bar{\tau}(t)}{dt})^2
\nonumber \\
&&
+\int d\bar{\sigma} d \bar{\theta}  \bar{\bold{E}} T(t) \frac{1}{2} \bar{\partial}'_{z} \bold{X}_{\hat{D}_T}^{\mu}(\bar{\tau})  \bar{\bold{D}}'_{\theta} \bold{X}_{\hat{D}_T \mu}(\bar{\tau})
+\int d\bar{\sigma} \sqrt{\bar{h}} T(t) \frac{1}{2}
\lambda_{\hat{D}_T}^A(\bar{\tau}) \bar{\partial}'_{\bar{z}} \lambda_{\hat{D}_TA}(\bar{\tau})
\Bigr) \Biggr).
\end{eqnarray}
where we have redefined as $c(t) \to T(t) c(t)$. $\bold{Z}_0$ represents an overall constant factor, and we will rename it $\bold{Z}_1, \bold{Z}_2, \cdots$ when the factor changes in the following.
This path integral is obtained if 
\begin{equation}
F_1(t):=\frac{d}{dt}T(t)=0 \label{HsuperF1gauge}
\end{equation}
 gauge is chosen in 
\begin{eqnarray}
&&\Delta_F(\bold{X}_{\hat{D}_T f}, \lambda_{\hat{D}_T f}; \bold{X}_{\hat{D}_T i}, \lambda_{\hat{D}_T i}|\bold{E}_f, ; \bold{E}_i) 
\nonumber \\
&=&
\bold{Z}_1\int^{\bold{E}_f, \infty, \bold{X}_{\hat{D}_T f}, \lambda_{\hat{D}_T f} }_{\bold{E}_i, -\infty, \bold{X}_{\hat{D}_T i}, \lambda_{\hat{D}_T i} } 
\mathcal{D} T 
\mathcal{D}\bold{E} \mathcal{D}\bar{\tau} \mathcal{D}\bold{X}_{\hat{D}_T}(\bar{\tau}) \mathcal{D}\lambda_{\hat{D}_T}(\bar{\tau})
\int 
\exp \Biggl(- \int_{-\infty}^{\infty} dt \Bigl(
\nonumber \\
&&  
+\zeta \bar{\rho}\frac{1}{T(t)}(\frac{d \bar{\tau}(t)}{dt})^2
+\int d\bar{\sigma} d \bar{\theta}  \bar{\bold{E}} T(t) \frac{1}{2} \bar{\partial}'_{z} \bold{X}_{\hat{D}_T}^{\mu}(\bar{\tau})  \bar{\bold{D}}'_{\theta} \bold{X}_{\hat{D}_T \mu}(\bar{\tau})
+\int d\bar{\sigma} \sqrt{\bar{h}} T(t) \frac{1}{2}
\lambda_{\hat{D}_T}^A(\bar{\tau}) \bar{\partial}'_{\bar{z}} \lambda_{\hat{D}_TA}(\bar{\tau}) \Bigr) \Biggr),
\nonumber \\
&&
\label{Hsuperpathint}
\end{eqnarray}
which has a manifest one-dimensional diffeomorphism symmetry with respect to $t$, where $T(t)$ is transformed as an einbein \cite{Schwinger0}.

Under $\frac{d\bar{\tau}}{d\bar{\tau}'}=T(t)$, $T(t)$ disappears in (\ref{Hsuperpathint}) as in the bosonic case, and we obtain

\begin{eqnarray}
&&\Delta_F(\bold{X}_{\hat{D}_T f}, \lambda_{\hat{D}_T f}; \bold{X}_{\hat{D}_T i}, \lambda_{\hat{D}_T i}|\bold{E}_f, ; \bold{E}_i) 
\nonumber \\
&=&
\bold{Z}_2\int^{\bold{E}_f, \infty, \bold{X}_{\hat{D}_T f}, \lambda_{\hat{D}_T f} }_{\bold{E}_i, -\infty, \bold{X}_{\hat{D}_T i}, \lambda_{\hat{D}_T i} } 
\mathcal{D}\bold{E} \mathcal{D}\bar{\tau} \mathcal{D}\bold{X}_{\hat{D}_T}(\bar{\tau}) \mathcal{D}\lambda_{\hat{D}_T}(\bar{\tau})
\int 
\exp \Biggl(- \int_{-\infty}^{\infty} dt \Bigl(
\nonumber \\
&&  
+\zeta \bar{\rho}(\frac{d \bar{\tau}(t)}{dt})^2
+\int d\bar{\sigma} d \bar{\theta}  \bar{\bold{E}}\frac{1}{2} \bar{\partial}''_{z} \bold{X}_{\hat{D}_T}^{\mu}(\bar{\tau})  \bar{\bold{D}}''_{\theta} \bold{X}_{\hat{D}_T \mu}(\bar{\tau})
+\int d\bar{\sigma} \sqrt{\bar{h}}  \frac{1}{2}
\lambda_{\hat{D}_T}^A(\bar{\tau}) \bar{\partial}''_{\bar{z}} \lambda_{\hat{D}_TA}(\bar{\tau}) \Bigr) \Biggr),
\nonumber \\
&&
\label{Hsuperpathint2}
\end{eqnarray}
where $\bar{\bold{D}}''_{\theta}$, $\bar{\partial}''_{z}$ and $\bar{\partial}''_{\bar{z}}$ are given by replacing $\frac{\partial}{\partial \bar{\tau}}$ with $\frac{\partial}{\partial t}$ in $\bar{\bold{D}}_{\theta}$, $\bar{\partial}_{z}$ and $\bar{\partial}_{\bar{z}}$, respectively. This action is still invariant under the diffeomorphism with respect to t if $\bar{\tau}$ transforms in the same way as $t$.

If we choose a different gauge
\begin{equation}
F_2(t):=\bar{\tau}-t=0, \label{HsuperF2gauge}
\end{equation} 
in (\ref{Hsuperpathint2}), we obtain 
\begin{eqnarray}
&&\Delta_F(\bold{X}_{\hat{D}_T f}, \lambda_{\hat{D}_T f}; \bold{X}_{\hat{D}_T i}, \lambda_{\hat{D}_T i}|\bold{E}_f, ; \bold{E}_i) \nonumber \\
&=&
\bold{Z}_3\int^{\bold{E}_f, \infty, \bold{X}_{\hat{D}_T f}, \lambda_{\hat{D}_T f} }_{\bold{E}_i, -\infty, \bold{X}_{\hat{D}_T i}, \lambda_{\hat{D}_T i} } 
\mathcal{D}\bold{E} \mathcal{D}\bar{\tau} \mathcal{D}\bold{X}_{\hat{D}_T}(\bar{\tau}) \mathcal{D}\lambda_{\hat{D}_T}(\bar{\tau})
\int 
\mathcal{D} \alpha \mathcal{D}c \mathcal{D}b
\nonumber \\
&&
\exp \Biggl(- \int_{-\infty}^{\infty} dt \Bigl(\alpha(t) (\bar{\tau}-t) +b(t)c(t)(1-\frac{d \bar{\tau}(t)}{dt}) +\zeta \bar{\rho}(\frac{d \bar{\tau}(t)}{dt})^2 \nonumber \\
&&
+\int d\bar{\sigma} d \bar{\theta}  \bar{\bold{E}}  \frac{1}{2} \bar{\partial}''_{z} \bold{X}_{\hat{D}_T}^{\mu}(\bar{\tau})  \bar{\bold{D}}''_{\theta} \bold{X}_{\hat{D}_T \mu}(\bar{\tau})
+\int d\bar{\sigma} \sqrt{\bar{h}}  \frac{1}{2}
\lambda_{\hat{D}_T}^A(\bar{\tau}) \bar{\partial}''_{\bar{z}} \lambda_{\hat{D}_TA}(\bar{\tau})
\Bigr) \Biggr)\nonumber \\
&=&
\bold{Z}\int^{\bold{E}_f, \bold{X}_{\hat{D}_T f}, \lambda_{\hat{D}_T f} }_{\bold{E}_i, \bold{X}_{\hat{D}_T i}, \lambda_{\hat{D}_T i} } 
\mathcal{D}\bold{E}  \mathcal{D}\bold{X}_{\hat{D}_T}
 \mathcal{D}\lambda_{\hat{D}_T}
\int 
\exp \Biggl(- \int_{-\infty}^{\infty} d\bar{\tau} 
\Bigl(
\nonumber \\
&&
\frac{1}{4 \pi}\int d\bar{\sigma} \sqrt{\bar{h}} 
\zeta \bar{R}(\bar{\sigma}, \bar{\tau})
+\int d\bar{\sigma} d \bar{\theta}  \bar{\bold{E}}  \frac{1}{2} \bar{\partial}_{z} \bold{X}_{\hat{D}_T}^{\mu}  \bar{\bold{D}}_{\theta} \bold{X}_{\hat{D}_T \mu}
+\int d\bar{\sigma} \sqrt{\bar{h}} \frac{1}{2}
\lambda_{\hat{D}_T}^A \bar{\partial}_{\bar{z}} \lambda_{\hat{D}_TA}\Bigr) \Biggr). \nonumber \\
\end{eqnarray}
In the second equality, we have redefined as $c(t)(1-\frac{d \bar{\tau}(t)}{dt}) \to c(t)$ and integrated out the ghosts. The path integral is defined over all possible heterotic super Riemannian manifolds with fixed punctures in $\bold{R}^{d}$. By using the two-dimensional superdiffeomorphism and super Weyl invariance of the action, we obtain
\begin{eqnarray}
&&\Delta_F(\bold{X}_{\hat{D}_T f}, \lambda_{\hat{D}_T f}; \bold{X}_{\hat{D}_T i}, \lambda_{\hat{D}_T i}|\bold{E}_f, ; \bold{E}_i) \nonumber \\
&=&
\bold{Z}\int^{\bold{E}_f, \bold{X}_{\hat{D}_T f}, \lambda_{\hat{D}_T f} }_{\bold{E}_i, \bold{X}_{\hat{D}_T i}, \lambda_{\hat{D}_T i} } 
\mathcal{D}\bold{E}  \mathcal{D}\bold{X}_{\hat{D}_T} \mathcal{D}\lambda_{\hat{D}_T}
e^{- \zeta  \chi}
e^{-\int d^2\sigma d \theta  \bold{E} \frac{1}{2} \partial_{z} \bold{X}_{\hat{D}_T}^{\mu}  \bold{D}_{\theta} \bold{X}_{\hat{D}_T \mu}
-\int d^2\sigma \sqrt{h} \frac{1}{2}
\lambda_{\hat{D}_T}^A \partial_{\bar{z}} \lambda_{\hat{D}_TA}
}, \label{HcSuperLast} \nonumber \\
&&
\end{eqnarray}
where $\chi$ is the Euler number of the reduced space. By inserting asymptotic states to (\ref{HcSuperLast}) and renormalizing the metric, we obtain the perturbative all-order scattering amplitudes that possess the supermoduli in the $SO(32)$ and $E_8 \times E_8$ heterotic superstring theory for $T=$ $SO(32)$ and $E_8 \times E_8$, respectively\cite{textbook}. Especially, in superstring geometry, the consistency of the perturbation theory around the background (\ref{HSupersolution}) determines $d=10$ (the critical dimension).

\section{Conclusion}
\setcounter{equation}{0}
In this paper, we defined superstring geometry: spaces of superstrings including the interactions, their topologies, charts, and metrics. Especially, we can define spaces where the trajectories in asymptotic processes reproduce the moduli spaces of the super Riemann surfaces in target spaces. Based on the superstring geometry, we defined Einstein-Hilbert action coupled with gauge fields, and formulated superstring theory non-perturbatively by summing over metrics, and the gauge fields on superstring manifolds. This theory does not depend on backgrounds. The theory has a supersymmetry, as a part of the diffeomorphisms symmetry.

We have derived the all-order  perturbative scattering amplitudes that possess the super moduli in type IIA, type IIB and SO(32) type I superstring theory from the single theory, by expanding the action to the second order of the metric around  fixed backgrounds representing type IIA, type IIB and SO(32) type I perturbative vacua, respectively. Here, we explain some reasons for this in the point of view of symmetry. Because this expansion corresponds to see only one string state, we can move to a formalism of the first quantization, where the state is described by a trajectory in the superstring manifold $\mathfrak{M}_{D_T}$. By definition of the neighborhood, the effective action becomes local on a worldsheet. The $(\bar{\sigma}, \bar{\theta})$ supersymmetry of the action are dimensional reductions in $\bar{\tau}$ direction of the two-dimensional $\mathcal{N}=(1,1)$ local supersymmetry, where the number of supercharges of the transformations is the same as of the two-dimensional ones as in (\ref{OSUSYtrans}). Because we can choose a gauge where a trajectory $t$ coincides $\bar{\tau}$ by using an one-dimensional diffeomorphism transformation on the trajectory, 
the supersymmetry becomes the two-dimensional local  $\mathcal{N}=(1,1)$ supersymmetry of the perturbative superstring theory.


We have shown that a trajectory in an asymptotic process on $\mathfrak{M}_{D_T}$ is a worldsheet of a superstring with punctures in $M$. Macroscopically, such a worldsheet becomes a worldline of a superparticle in $M$, namely a trajectory in an asymptotic process on $M$. By the way, one way to identify a background as $M$ is to observe all the trajectories in asymptotic processes on the background. Because all the trajectories in asymptotic processes on $\mathfrak{M}_{D_T}$ become macroscopically those on $M$, we see that macroscopically, a superstring manifold $\mathfrak{M}_{D_T}$ becomes the space-time manifold $M$. Conversely, this means that if we look at space-time $M$ in a microscopic way, we see a superstring manifold $\mathfrak{M}_{D_T}$. On the other hand, we have shown that the effective theory of a part of fluctuations of the action on $\mathfrak{M}_{D_T}$ reduces to the perturbative superstring theory. Macroscopically, the perturbative superstring theory describes all the matter and gauge particles including graviton. That is, macroscopically, the fluctuations of $\mathfrak{M}_{D_T}$ become these particles. Conversely, if we observe particles in a microscopic way, we see superstrings, which are the fluctuations of $\mathfrak{M}_{D_T}$. Therefore, superstring manifolds unify matter and the space-time: macroscopically, the fluctuations of $\mathfrak{M}_{D_T}$ are particles and $\mathfrak{M}_{D_T}$ itself is the space-time.

\section{Discussion}
\setcounter{equation}{0}

The superstring geometry solution to the equations of motion of the theory in this paper, has the most simple superstring background, that is, the flat metric, the constant dilaton and the other zero backgrounds. We need to find superstring geometry solutions that have more general superstring backgrounds, namely, a metric, a NS-NS B-field, a dilaton, R-R fields, and  gauge fields on D-branes. We can identify superstring backgrounds of superstring geometry solutions by deriving superstring actions in the backgrounds from the fluctuations around the solutions, in the same way as in this paper.

We formulated the single theory that manifestly includes type IIA, type IIB, and $SO(32)$ type I superstrings. We also formulated another theory that manifestly includes the $SO(32)$ and $E_8 \times E_8$ heterotic superstrings based on superstring geometry. We expect that these two theories are equivalent because of the $SO(32)$ type I / hetero duality.

We derived the propagator of  fluctuations around the superstring geometry solution. Then, we moved to the first quantization formalism, and we derived the path-integral of the perturbative superstring action. This implies that we also derived the string states and the D-brane boundary states in the first quantization formalism. Next task is to derive a whole Hilbert space of the theory. We can identify string states and D-brane states in the Hilbert space corresponding to the string states and the D-brane boundary states in the first quantization formalism, by using the correspondence between the first and the second quantizations.

Because string backgrounds are included in configurations of the fields of the string geometry, we expect that instantons of the string geometry reduce to instantons of the string backgrounds and instanton effects of string geometry give non-perturbative effects in string theory where a string background changes to another.


\section*{Acknowledgements}
We would like to thank  
H. Aoki,
M. Fukuma,
Y. Hamada,
K. Hashimoto,
Y. Hosotani,
K. Hotta,
Y. Hyakutake,
N. Ishibashi,
K. Ishikawa,
G. Ishiki,
Y. Ito,
Y. Kaneko,
N. Kawamoto,
T. Kobayashi,
H. Kitamoto,
T. Kuroki,
H. Kyono,
K. Maruyoshi,
Y. Matsuo,
S. Matsuura,
S. Mizoguchi,
T. Morita,
S. Moriyama,
K. Ohta,
N. Ohta,
Y. Okawa,
T. Okuda,
T. Onogi,
M. Sakaguchi,
Y. Sakatani,
S. Seki,
S. Shiba,
H. Shimada,
S. Shimasaki,
K. Suehiro,
S. Sugimoto,
Y. Sugimoto,
S. Sugishita,
T. Suyama,
H. Suzuki,
T. Tada,
T. Takahashi,
T. Takayanagi,
K. Tsumura,
S. Yamaguchi,
K. Yamashiro,
Y. Yokokura, 
and especially 
S. Iso, H. Itoyama, H. Kawai, J. Nishimura, A. Tsuchiya, and T. Yoneya
for long and valuable discussions.

\appendix

\section{ADM formalism }
\setcounter{equation}{0}

The ADM decomposition of a two-dimensional metric is given by 
\begin{equation}
\bar{h}_{mn}=
\left(
\begin{array}{cc}
\bar{n}^2+ \bar{n}_{\bar{\sigma}} \bar{n}^{\bar{\sigma}} & \bar{n}_{\bar{\sigma}} \\
\bar{n}_{\bar{\sigma}} & \bar{e}^2
\end{array}
\right),
\end{equation}
where $\bar{n}$ is a lapse function and $\bar{n}_{\bar{\sigma}}$ is a shift vector. $\bar{e}^2$ is a metric on $\bar{\sigma}$ direction. $\bar{n}^{\bar{\sigma}}:=\bar{e}^{-2}\bar{n}_{\bar{\sigma}}$. As a result, we obtain $\sqrt{\bar{h}}=\bar{n}\bar{e}$ and 
\begin{equation}
\bar{h}^{mn}=
\left(
\begin{array}{cc}
\frac{1}{\bar{n}^2} & -\frac{\bar{n}^{\bar{\sigma}}}{\bar{n}^2} \\
-\frac{\bar{n}^{\bar{\sigma}}}{\bar{n}^2} & \bar{e}^{-2}+ \left(\frac{\bar{n}^{\bar{\sigma}}}{\bar{n}}\right)^2
\end{array}
\right).
\end{equation}
An action for scalar fields $X^{\mu}$ is decomposed as
\begin{eqnarray}
S&=&\int d\bar{\tau} d\bar{\sigma} \sqrt{\bar{h}}(\frac{1}{2} \bar{h}^{mn} \bar{\partial}_m X^{\mu} \bar{\partial}_n X_{\mu}) \nonumber \\
&=&
\int d\bar{\tau} d\bar{\sigma} (-i\bar{e} p_{X}^{\mu}\partial_{\bar{\tau}} X_{\mu}) + \int d\bar{\tau} H,
\end{eqnarray}
where
\begin{equation}
H=\int d\bar{\sigma} \left(
\bar{n}\bar{e} \left(
\frac{1}{2}(p_{X}^{\mu})^2 + \frac{1}{2}\bar{e}^{-2}(\partial_{\bar{\sigma}} X^{\mu})^2
\right)
+i\bar{e} \bar{n}^{\bar{\sigma}} p_{X}^{\mu}\partial_{\bar{\sigma}} X_{\mu}
\right).
\end{equation}
Actually, if $p_{X}^{\mu}$ is integrated out in the second line, we obtain the first line.

\section{Canonical commutation relations of Majorana fermions}
\setcounter{equation}{0}

An action for Majorana fermions $\psi_{\mu}$ in a two-dimensional curved space-time is given by 
\begin{equation}
S_F=\int d^2\bar{\sigma} \bar{E} (-\frac{1}{2} \bar{\psi}^{\mu}\bar{E}^{m}_{q}\gamma^{q} \bar{\partial}_m \psi_{\mu}),
\end{equation}
where $\bar{E}_{m}^{q}$ is a vierbein, whose determinant $\bar{E}$ satisfies $\bar{E}=\sqrt{\bar{h}}=\bar{n}\bar{e}$. We use $\gamma^0=\left(
\begin{array}{cc}
0 & -i \\
i & 0
\end{array}
\right)
$
and
$\gamma^1=\left(
\begin{array}{cc}
0 & 1 \\
1 & 0
\end{array}
\right)
$. 
$\bar{\psi}^{\mu}:=\psi^{\mu} \gamma^0$ and $p_{\psi}^{\mu}:=\delta S_F / \delta \bar{\partial}_0 \psi_{\mu}
=- \bar{E} \psi^{\mu}  \gamma^0 \gamma^{q} \bar{E}^{0}_{q}
=- \bar{E} \psi^{\mu}  \mbox{diag}(\bar{E}^0_0 -i\bar{E}^0_1, \bar{E}^0_0 +i\bar{E}^0_1)$. The canonical commutation relations are given by
\begin{equation}
\{\hat{p}_{\psi}^{\mu \alpha}(\bar{\sigma}, \bar{\tau}), \hat{\psi}^{\nu}_{\beta}(\bar{\sigma}', \bar{\tau}) \}
= \eta^{\mu\nu} \delta^{\alpha}_{\beta} \delta(\bar{\sigma}-\bar{\sigma}').
\end{equation}
Then we obtain
\begin{eqnarray}
&&\{(-\bar{E}^0_0+i\bar{E}^0_1) \hat{\psi}^{\mu}_{0}(\bar{\sigma}, \bar{\tau}) , \hat{\psi}^{\nu}_{0}(\bar{\sigma}', \bar{\tau}) \}
=\frac{1}{\bar{E}} \eta^{\mu\nu}  \delta(\bar{\sigma}-\bar{\sigma}') \nonumber \\
&&\{(-\bar{E}^0_0-i\bar{E}^0_1) \hat{\psi}^{\mu}_{1}(\bar{\sigma}, \bar{\tau}) , \hat{\psi}^{\nu}_{1}(\bar{\sigma}', \bar{\tau}) \}
=\frac{1}{\bar{E}}\eta^{\mu\nu}  \delta(\bar{\sigma}-\bar{\sigma}') \nonumber \\
&&\{\hat{\psi}^{\mu}_{0}(\bar{\sigma}, \bar{\tau}) , \hat{\psi}^{\nu}_{1}(\bar{\sigma}', \bar{\tau}) \}
=0.
\end{eqnarray}
If we normalize as
$\tilde{\psi}^{\mu}_0(\bar{\sigma}, \bar{\tau}):=\sqrt{-\bar{E}^0_0+i\bar{E}^0_1}\psi^{\mu}_0(\bar{\sigma}, \bar{\tau})$,
and 
$\tilde{\psi}^{\mu}_1(\bar{\sigma}, \bar{\tau}):=\sqrt{-\bar{E}^0_0-i\bar{E}^0_1}\psi^{\mu}_1(\bar{\sigma}, \bar{\tau})$,
we obtain
\begin{equation}
\{\hat{\tilde{\psi}}^{\mu}_{\alpha}(\bar{\sigma}, \bar{\tau}), \hat{\tilde{\psi}}^{\nu}_{\beta}(\bar{\sigma}', \bar{\tau}) \}
= \frac{1}{\bar{E}} \eta^{\mu\nu}  \delta_{\alpha \beta} \delta(\bar{\sigma}-\bar{\sigma}').
\end{equation}


\vspace*{0cm}





\begin{thebibliography}{99}
\bibitem{HMS}M. Kontsevich, ``Homological algebra of mirror symmetry," In Proceedings of the International Congress of Mathematicians, Vol. 1, 2 (1995) Birkhauser, alg-geom/9411018





\bibitem{MSato}M. Sato, 
``Moduli Space in Homological Mirror Symmetry,''
Adv. Math. Phys. \textbf{2019}, 1693102 (2019)




\bibitem{Gromov}M. Gromov, ``Pseudo holomorphic curves in symplectic manifolds," Invent. math. {\bf 82} (1985) 307




\bibitem{Polyfold1}H. Hofer, K. Wysocki, E. Zehnder, ``Applications of Polyfold Theory I: The Polyfolds of Gromov-Witten Theory," arXiv:1107.2097 [math.SG]




\bibitem{WittenCubic}E. Witten, ``Noncommutative Geometry and String Field Theory," Nucl. Phys. {\bf B268} (1986) 253





























\bibitem{RiemannianGeometry}W. Klingenberg, ``Riemannian Geometry," De Gruyter Studies in Mathematics 1, de Gruyter; 2nd Rev ed. (April 6, 1995)
















\bibitem{KricheverNovikov1}I. M. Krichever, S. P. Novikov, ``Algebras of Virasoro type, Riemann surfaces and the structure of soliton theory," Funct. Anal. Appl. {\bf 21} (1987) 126

\bibitem{KricheverNovikov2}I. M. Krichever, S. P. Novikov, ``Virasoro-type algebras, Riemann surfaces and strings in Minkowsky space," Funct. Anal. Appl. {\bf 21} (1987) 294


\bibitem{Polyfold2}H. Hofer, K. Wysocki, E. Zehnder, ``Polyfold and Fredholm Theory,"  arXiv:1707.08941 [math.FA]

\bibitem{Polyfold3}H. Hofer, ``A General Fredholm Theory and Applications," arXiv:math/0509366 [math.SG]





\bibitem{Majumdar}S. D. Majumdar, ``A class of exact solutions of Einstein's field equations,"  Phys. Rev. {\bf 72} (1947) 390



\bibitem{Papapetrou}A. Papapetrou, ``A static solution of the equations of the gravitational field for an arbitrary charge distribution," Proc. Roy. Irish Acad. {\bf A 51}  (1948) 191




\bibitem{Schwinger0}E.S. Fradkin, D.M. Gitman, ``Path integral representation for the relativistic particle propagators and BFV quantization," Phys. Rev. {\bf D44} (1991) 3230



\bibitem{textbook}J. Polchinski, ``String Theory Vol. 1, 2" Cambridge University Press, Cambridge, UK, 1998

















\bibitem{NotesOnSupermanifolds}E. Witten, ``Notes On Supermanifolds and Integration," arXiv:1209.2199 [hep-th] 

\bibitem{WittenSupermoduli}E. Witten, ``Notes On Super Riemann Surfaces And Their Moduli," arXiv:1209.2459 [hep-th] 

\bibitem{SuperPeriod}E. Witten, ``The Super Period Matrix With Ramond Punctures," J. Geom. Phys. {\bf 92}  (2015) 210





\bibitem{Howe}P. S. Howe, ``Super Weyl Transformations in Two-Dimensions," J. Phys. {\bf A12} (1979) 393

\bibitem{DHokerPhong}Eric D'Hoker, D. H. Phong, ``Conformal scalar fields and chiral splitting on super Riemann surfaces," Comm. Math. Phys. {\bf 125} 3 (1989) 469



\bibitem{SuperStringAction}L. Brink, P. Di Vecchia, P. S. Howe, `` A Locally Supersymmetric and Reparametrization Invariant Action for the Spinning String," Phys. Lett. {\bf 65B} (1976) 471











\bibitem{Smolin}L. Smolin, ``M theory as a matrix extension of Chern-Simons theory," Nucl. Phys. {\bf B591} (2000) 227

\bibitem{AzumaIsoKawaiOhwashi}T. Azuma, S. Iso, H. Kawai, Y. Ohwashi, ``Supermatrix models," Nucl. Phys. {\bf B610} (2001) 251

\bibitem{OkudaTakayanagi}T. Okuda, T. Takayanagi, ``Ghost D-branes," JHEP {\bf 0603} (2006) 062

\bibitem{DijkgraafHeidenreichJeffersonVafa}R. Dijkgraaf, B. Heidenreich, P. Jefferson, C. Vafa, ``Negative Branes, Supergroups and the Signature of Spacetime," arXiv:1603.05665 [hep-th] 




\bibitem{IKKT} N. Ishibashi, H. Kawai, Y. Kitazawa, A. Tsuchiya, ``A Large-N Reduced Model as Superstring," Nucl. Phys. {\bf B498} (1997) 467




\bibitem{BFSS}T. Banks, W. Fischler, S.H. Shenker, L. Susskind, ``M Theory As A Matrix Model: A Conjecture," Phys. Rev. {\bf D55} (1997) 5112




\bibitem{Motl} L. Motl, `` Proposals on nonperturbative superstring interactions," hep-th/9701025

\bibitem{BS}T. Banks and N. Seiberg, ``Strings from Matrices," Nucl. Phys. {\bf B497} (1997) 41

\bibitem{DVV}R. Dijkgraaf, E. Verlinde and H. Verlinde, ``Matrix String Theory," Nucl. Phys. {\bf B500} (1997) 43

\bibitem{DM}R. Dijkgraaf and L. Motl, ``Matrix string theory, contact terms, and superstring field theory," hep-th/0309238






\bibitem{HanadaKawaiKimura}M.Hanada, H. Kawai, Y. Kimura, ``Describing curved spaces by matrices," Prog. Theor. Phys. {\bf 114} (2006) 1295





\bibitem{Hooft}G. 't Hooft, ``A PLANAR DIAGRAM THEORY FOR STRONG INTERACTIONS,"  Nucl. Phys. {\bf B72} (1974) 461

\bibitem{BrezinItzyksonParisiZuber}E. Brezin, C. Itzykson, G. Parisi and J.B. Zuber, ``Planar Diagrams," Commun. Math. Phys. {\bf 59} (1978) 35

\bibitem{KazakovMigdal}V.A. Kazakov and A.A. Migdal, ``RECENT PROGRESS IN THE THEORY OF NONCRITICAL STRINGS," Nucl. Phys. {\bf B311} (1988) 171

\bibitem{DistlerKawai}J. Distler and H. Kawai, ``Conformal Field Theory and 2-D Quantum Gravity or Who's Afraid of Joseph Liouville?," Nucl. Phys. {\bf B321} (1989) 509

\bibitem{DasJevicki}S. R. Das and A. Jevicki, ``STRING FIELD THEORY AND PHYSICAL INTERPRETATION OF D = 1 STRINGS," Mod. Phys. Lett {\bf A5} (1990) 1639

\bibitem{GrossMiljkovic}D. J. Gross and N. Miljkovic, ``A NONPERTURBATIVE SOLUTION OF D = 1 STRING THEORY," Phys. Lett. {\bf B238} (1990) 217

\bibitem{GinspargZinn-Justin}P. H. Ginsparg and J. Zinn-Justin, ``2-D GRAVITY + 1-D MATTER," Phys. Lett. {\bf B240} (1990) 333

\bibitem{Itoyama}L. Alvarez-Gaume, H. Itoyama, J.L. Manes and A. Zadra, ``Superloop Equations and Two Dimensional Supergravity," Int. J. Mod. Phys. {\bf A7} (1992) 5337

\bibitem{GinspargMoore}For a review, \\
P. Ginsparg and G. Moore, ``Lectures on 2D gravity and 2D string theory," hep-th/9304011 \\
I. Klebanov, ``String Theory in Two Dimensions," hep-th/9108019


\bibitem{RM}T. Eguchi and H. Kawai, Phys. Rev. Lett. {\bf 48} (1982) 1063.\\
      G. Parisi, Phys. Lett. {\bf 112B} (1982) 463.\\
      D. Gross and Y. Kitazawa, Nucl. Phys. {\bf B206} (1982) 440.\\
      G. Bhanot, U. Heller and H. Neuberger, Phys. Lett. {\bf 113B} (1982) 47.\\
      S. Das and S. Wadia, Phys. Lett. {\bf 117B} (1982) 228.

\bibitem{BrooksMuhammadGates}R. Brooks, F. Muhammad, S.J. Gates, ``Unidexterous D=2 Supersymmetry in Superspace," Nucl. Phys. {\bf B268} (1986) 599

\bibitem{HullWitten}C.M. Hull, E. Witten, ``Supersymmetric Sigma Models and the Heterotic String," Phys. Lett. {\bf 160B} (1985) 398

\bibitem{BergshoeffSezginNishino}E. Bergshoeff, E. Sezgin, H. Nishino, ``Heterotic Models and Conformal Supergravity in Two-dimensions," Phys. Lett. {\bf 166B} (1986) 141











\end{thebibliography}
\end{document}